\documentclass[iop]{emulateapj}

\usepackage[utf8]{inputenc}
\usepackage{graphicx}
\usepackage{amsmath}
\usepackage{mathrsfs}
\usepackage{natbib}
\usepackage{array}

\graphicspath{{figures/}}

\slugcomment{Preprint version submitted to ApJ: ArXiv low resolution}

\shorttitle{Dust capture and long-lived density enhancements triggered by 2D vortices}
\shortauthors{C. Surville et al.}

\begin{document}

\title{Dust capture and long-lived density enhancements triggered by vortices in 2D protoplanetary disks}

\author{Cl\'{e}ment Surville, Lucio Mayer}
\affil{Institute for Computational Science, University of Zurich, Winterthurerstrasse 190, 8057 Zurich, Switzerland}
\email{clement.surville@physik.uzh.ch}
\and 
\author{Douglas N. C. Lin}
\affil{Department of Astronomy and Astrophysics, University of California, Santa Cruz, CA 95064, USA}

\begin{abstract}
We study dust capture by vortices and its long-term consequences in global two-fluid inviscid disk simulations using a new polar grid code RoSSBi that allows to model the background gas flow with very high stability and accuracy thanks to a well-balanced scheme. We perform the longest integrations so far, several hundred disk orbits, at the highest resolution attainable in global simulations of disks with dust, namely 2048x4096 grid points. This allows to study the dust evolution well beyond vortex dissipation. We vary a wide range of parameters, most notably the dust-to-gas ratio in the initial setup varies in the range $10^{-3}$ to $0.1$. Irrespective of the initial dust-to-gas ratio we find rapid concentration of the dust inside vortices, reaching dust-to-gas ratios of order unity inside the vortex. We present an analytical model that describes very well the dust capture process inside vortices, finding consistent results for all dust-to-gas ratios. A vortex streaming instability develops which causes invariably vortex destruction. After vortex dissipation large-scale dust-rings encompassing a disk annulus form in most cases, which sustain very high dust concentration, approaching ratios of order unity. The rings are long livedlasting as long as the duration of the simulations. They also develop a streaming instability, which manifests itself in eddies at various scales within which the dust forms compact high density clumps. Such clumps would be unstable to gravitational collapse in absence of strong dissipation by viscous forces. When vortices are particularly long lived, rings do not form but dust clumps inside vortices become then long lived features and would likely undergo collapse by gravitational instability. Rings encompass almost an Earth mass of solid material, while even larger masses of dust do accumulate inside vortices in the earlier stage. We argue that rapid planetesimal formation would occur in the dust clumps inside the vortices as well as in the post-vortex ring. Planetesimals would have a wide range of sizes given the turbulent nature of the density fluctuations, with potentially important implications for the formation of terrestrial planets as well as for the cores of gas giants. Even higher resolution will be needed to study robustly the streaming instability in the rings as well as in the low Stokes number regime, while self-gravity will be added in the future to explore the eventual collapse of the dust clumps.
\end{abstract}


\section{ Introduction }
\label{Sect_Intro}

	In the context of planet formation, the grain growth from micrometre or millimetre size to metre or kilometre size is of crucial interest. The outcome is the existence of planetesimals in the disk, which are the building blocks of rocky planets and of the core of giant planets in the core-accretion scenario. Whether planetesimals form from coagulation of dust grains through collisions or via gravitational instability of the dust layer, reaching a critically high concentration of solids within the protoplanetary disk, is a necessary preliminary step. Furthermore, even in models which postulate specific triggers of local dust concentration, such as transport along pressure gradients across spiral arms in gravitationally unstable disks \citep{Rice2006, Gibbons2014} or the two-fluid streaming instabilities \citep{Johansen2007, Kowalik2013}, the initial conditions often considered by numerical models assume that some initial dust density enhancement has already occurred beyond the canonical $10^{-2}$ dust-to-gas ratio, in order to enhnance the effect of the aforementioned processes. Finally, grain evolution as a result of coagulation, fragmentation, sticking, or bouncing, is also very sensitive, if not boosted, by the local dust density enhancements in the disk. For all these reasons, it is mandatory to explore different ways to enhance the local contribution of solids.

	The capture of solid material by vortices has been proposed since decades as an efficient mechanism to this end \citep{Barge1995, Tanga1996, Chavanis2000, Tanga2000, Heng2010}. Recent observational papers also argue for the evidence of dusty asymmetric structures in disks, attributed to vortex capture \citep{VanderMarel2013, Perez2014, Hashimoto2015}. The existence of vortices is justified by their frequent formation in the disk, even in the 'dead zones', as products of the Rossby wave instability (hereafter RWI) \citep{Lovelace1999}, or the baroclinic instability \citep{Klahr2003}. Nevertheless their survival could be limited by the effective disk viscosity \citep{Godon1999}, or by 3D instabilities like the elliptical instability \citep{Lesur2009} or the magnetorotational instability (MRI) \citep{Lyra2011, Lyra2012}. However, whatever the destruction process, the numerical simulations of these studies show that vortices can survive at least hundreds of orbits inside the disk. Many important questions remain unanswered so far. For example, is hundreds of disk orbits a sufficiently long timescale to change the dust distribution significantly even starting from very small initial dust-to-gas ratios? How exactly a vortex dissipates in global disk models? What is the legacy of vortex dissipation, namely is there some imprint on the dust distribution that persists beyond vortex dissipation?

	To answer these questions, detailed models of dust capture in vortices and simulations of the evolution of the dusty structures are necessary. There were some attempts to deal with the stability of dusty vortex models, in particular by \cite{Chang2010}, but this analytical study is limited by a overly simplified treatment of the drag force, in particular by neglecting the back reaction onto the gas, and also by using a prescribed vortex model. We present here a set of 2D two-fluid global disk simulations with an imposed vortex reaching very high resolution and exploring a wide range of parameters in the dust properties, including some models starting from dust-to-gas ratios much smaller than usually assumed in the literature. The global nature of our simulations allows to follow the evolution of the dust component for very long timescales, up to more than a thousand disk orbits, which puts us in the position to draw quite robust conclusions on the long-term effect on the dust distribution eventually triggered by vortex formation. In addition, we use the results of our numerical simulations to develop an accurate dynamical model of dust capture in a vortex, which enables us to understand the main results of the simulations in simple physical terms and hence corroborate them further. With our calculations we aim at providing a more realistic and general scenario for the role of vortices in driving dust evolution in protoplanetary disks.

	The paper is organized as follows. The first section is devoted to setting the disk model and the physical context, describing the numerical method, and comparing a test case to previous studies. In the second section, we derive an analytical model of dust capture in a vorticity profile, and present new estimates of dust density evolution and capture timescale. An exhaustive numerical study is conducted to explore the parameter space of particle size, initial dust-to-gas ratio, and vortex size, and the results are described Section \ref{Sect_Results}. We discuss and explain the different processes highlighted by the simulations in Section \ref{Sect_Discussion}. Finally, the last section deals with the outcomes of this scenario, and the possible implications on our understanding of planetesimal and planet formation.

\section{ Disk model and methods }

	We consider the classical 2D protoplanetary disk model, based on the minimum mass solar nebula (MMSN). The disk profile is a power law for gas density and temperature, $\sigma_0(r) \propto \left(r/r_0\right)^{\beta_\sigma}$ and $T_0(r) \propto \left(r/r_0\right)^{\beta_T}$. In this study, the power law exponents are fixed to standard values, $\beta_\sigma = -1.5$ and $\beta_T = -0.5$.
	
	We neglect any disk self gravity, thus the gravitational force exerted on the disk reduces to the Keplerian force of the central solar mass star. This imposes an angular velocity profile with the following radial dependence: $\Omega_k(r) \propto \left(r/r_0\right)^{\beta_\Omega}$, with $\beta_\Omega = -3/2$. 
	
	The gas thermodynamics follow the perfect gas model, with an adiabatic index $\gamma = 1.4$, thus the pressure has the dependence: $P_0(r) \propto \left(r/r_0\right)^{\beta_\sigma + \beta_T}$. The normalisation of all these quantities is then obtained given a reference radius, fixed at $r_0=7.5 \; AU$. The gas density at the reference radius is $\sigma_0(r_0) = 83 \; g.cm^{-2}$, and the pressure is set up by the prescription of a disk scale height. We define the isothermal disk scale height with $H_0^2(r) = \left[P_0(r)/\sigma_0(r)\right]/\Omega_k^2(r)$. Then, setting the disk aspect ratio with $H_0(r_0) = 0.05 \; r_0$ defines the pressure normalisation.

	Finally, we consider the presence of solid particles, mixed in the gas. To simplify, the typical dust density distribution in the disk will follow the MMSN model, with $\sigma_p(r) \sim \epsilon \sigma_0(r)$. The global dust-to-gas ratio, $\epsilon$, is variable in this study. The commonly used value $\epsilon = 10^{-2}$ is obtained from models considering all the solid material. However, when considering only one particle size, one can reasonably expect the dust-to-gas ratio of this particular dust population to be much smaller. For this reason, we will change $\epsilon$ from $10^{-4}$ to $10^{-2}$.

\subsection{ Equations of motion }

	The time evolution of the gas conserved variables is described by the inviscid Euler equations. The full compressible evolution of mass, momentum, and energy is studied in a cylindrical coordinate system. To relate density and velocity to the evolution of pressure, we use the total energy of an adiabatic gas given by
\begin{equation}
	E = \frac{P}{\gamma -1} + \frac{1}{2} \sigma \vec{V}^2 \: ,
\end{equation}
	and solve the energy equation under the adiabatic assumption.
	
	Concerning the evolution of the particle fluid, we use the pressure less fluid approximation. The compressible Euler equations are solved, but we remove pressure terms and energy equation as the particle fluid has no internal energy. The two fluids are coupled by a drag force, $\vec{f}_{aero}$, which influences the particle phase evolution as well as the gas evolution. The expression of the drag force is given following the work of \cite{Nakagawa1986}. We will limit our study to the Epstein regime.

	The main set of equations that we will use in the analytical part of this study is the density and velocity evolution. From the compressible Euler equations, one obtains for the gas,
\begin{align}
\label{Equ_Vel_Gas}
	& \partial_t \sigma_g + U_g \partial_r \sigma_g + \frac{V_g}{r}\partial_\theta \sigma_g + \sigma_g \vec{\nabla} \cdot \vec{V}_g = 0 \: ,\\
	& \partial_t U_g + U_g \partial_r U_g + \frac{V_g}{r}\partial_\theta U_g = \frac{V_g^2}{r} - r\Omega_k^2(r) - \sigma_g^{-1} \partial_r P - \frac{\vec{f}_{aero} \cdot \vec{e}_r}{\sigma_g} \: , \\
	& \partial_t V_g + U_g \partial_r V_g + \frac{V_g}{r}\partial_\theta V_g = - \frac{U_g V_g}{r} - \sigma_g^{-1} \frac{1}{r}\partial_\theta P - \frac{\vec{f}_{aero} \cdot \vec{e}_\theta}{\sigma_g} \: ,
\end{align}
with $U_g$ and $V_g$ the radial and azimuthal components of the gas velocity field, respectively.

We obtain for the solid particle phase,
\begin{align}
\label{Equ_Vel_Part}
	& \partial_t \sigma_p + U_p \partial_r \sigma_p + \frac{V_p}{r}\partial_\theta \sigma_p + \sigma_p \vec{\nabla} \cdot \vec{V}_p = 0 \: ,\\
	& \partial_t U_p + U_p \partial_r U_p + \frac{V_p}{r}\partial_\theta U_p = \frac{V_p^2}{r} - r\Omega_k^2(r)+ \frac{\vec{f}_{aero} \cdot \vec{e}_r}{\sigma_p} \: , \\
	& \partial_t V_p + U_p \partial_r V_p + \frac{V_p}{r}\partial_\theta V_p = - \frac{U_p V_p}{r} + \frac{\vec{f}_{aero} \cdot \vec{e}_\theta}{\sigma_p} \: , 
\end{align}
with $U_p$ and $V_p$ the radial and azimuthal components of the dust fluid velocity field, respectively.

The drag force, $\vec{f}_{aero}$ can be parametrized with \citep{Weidenschilling1977}
\begin{equation}
	\vec{f}_{aero} = - \sigma_p \Omega_k(r) {S_t}^{-1} \left( \vec{V}_p - \vec{V}_g \right) \: ,
\end{equation}
using the Stokes number for the Epstein regime:
\begin{equation}
\label{Equ_Stokes_number}
	S_t = \frac{\pi}{2} \rho_s r_s \sigma_g^{-1} \: .
\end{equation}

	In the Epstein regime, the particle radius $r_s$ is smaller than the mean free path of the gas molecules. It is a relevant approach since we aim at studying the evolution of small sized grains, which are the most common ones in early stages of the disk evolution, during which planet formation occurs. Considering our disk setup, the mean free path of the gas molecules is about two metres; it is larger than ten centimetres down to a few $AU$.

	We use the internal grain density $\rho_s=3$ $g.cm^{-3}$, to mimic an ice poor dust grain. These parameters define the largest particle size for which $S_t<1$, equalling $r_s^{max} = 18 \;cm$. For smaller particle sizes, the pressure less fluid approximation is valid. In fact, inter particle collisions are negligible since the collisions are dominated by gas molecules and solid grains interaction. We would like to warn against to confuse the Epstein regime with the $S_t<1$ condition. The Epstein regime is still valid for some $S_t$ values larger than unity. In fact, we obtain $S_t=5.6$ for a metre radius particle, the boundary between Epstein and Stokes regimes. But we will not investigate this domain, and limit our study to $S_t<1$ grains, for which it is accepted that the fluid approach is relevant.

\subsection{ Numerical method }

	The compressible Euler equations are solved using a 2D finite volume approach. We use an updated version of the code RoSSBi developed by C. Surville and introduced in \cite{Surville2015} for a pure gas 2D vortex study. The numerical scheme is based on an exact Riemann solver for the gas phase \citep{Toro1999}, and a pressure less Riemann solver based on \cite{Paardekooper2006} for the dust phase. We use a parabolic limited interpolation to find the right and left states at the cell interfaces. Then the time integration is done with a direct second order Runge Kutta method (RK2). 

	To obtain high accuracy, radial and azimuthal flux (multidimensional unsplit approach) as well as source terms are summed at each step of the Runge Kutta integration. We compared the scheme with the classical method which uses independent advection (MUSCL) and source term integration (RK2), and we obtain a more stable and precise scheme. It is also due to the well-balanced scheme on which the code RoSSBi is based, consisting in equalling the flux and source term errors of the background flow. The details of the code RoSSBi will be available in an upcoming publication.

	We use a polar grid with $N_r$ linearly spaced radial grid cells and $N_\theta$ cells in the azimuthal direction. We performed global simulations of the $2 \pi$ domain with a disk extent $[2/3, \: 4/3]\: r_0$ and resolutions varying from $(N_r, \: N_\theta) = (512, \: 1024)$ to $(N_r, \: N_\theta) = (2048, \: 4096)$. Boundary conditions are periodic in the azimuthal direction. For the radial direction, we use custom zero flux conditions based on the radial profile of the background disk. A damping of the residuals between the background disk and the solution allows to cancel almost perfectly any wave reflection. These radial conditions will be detailed in the future paper on the code RoSSBi.

	To perform long term integration, on typically 1000 orbits, and to use high resolution in some simulations, the code is designed with a hybrid MPI/OpenMP method. The disk is split into $2^{2n}$ subdomains (with $n=2,3,4$) which are distributed to multicore nodes. Each node solves the Euler equations with OpenMP parallelism, and exchange only overlapping ghost cells with neighbouring nodes (on a 2D node grid). We achieve very good scaling up to $256$ nodes with $16$ threads, so a total of $4096$ threads for the highest resolution, on the Piz Daint Cray XE30, the flagship supercomputer at the Swiss National Supercomputing Center, amont the 10 fastest in the world (see Top500 website).

\subsection{ Fiducial case and comparison with previous work }
\label{Sect_Test_case}

	We begin our study by performing and analyzing a fiducial simulation whose setup and resolution is similar to others used in the literature. It should be considered as a useful representative case to illustrate the phenomenology of a dusty gas disk in presence of a vortex. After describing the setup we will illustrate the main results and compare with existing work. Later in section 4 we will present the actual suite of simulations, for which this fiducial run serves as a starting point. Already with the latter, however, we will highlight the importance of continuing the simulations well beyond the dissipation of the vortex in order to understand the response of the dust.

The initial dust density profile is set to
\begin{equation}
	\sigma_p(r)|_{_{t=0}} = \epsilon \sigma_0(r) \: ,
\end{equation}
with the global dust-to-gas ratio $\epsilon=10^{-2}$. The radius of the dust particles is $r_s=0.7\: cm$, giving at $r_0$ a Stokes number of $S_t = 0.04$.

\begin{figure*}
	\begin{center}
	\begin{tabular}{ccc}
	\scriptsize{Gas density} & \scriptsize{Rossby number} & \scriptsize{Dust density}\\
	\includegraphics[height=4.8cm, trim=4.5mm 0mm 4.5mm 0cm, clip=true]{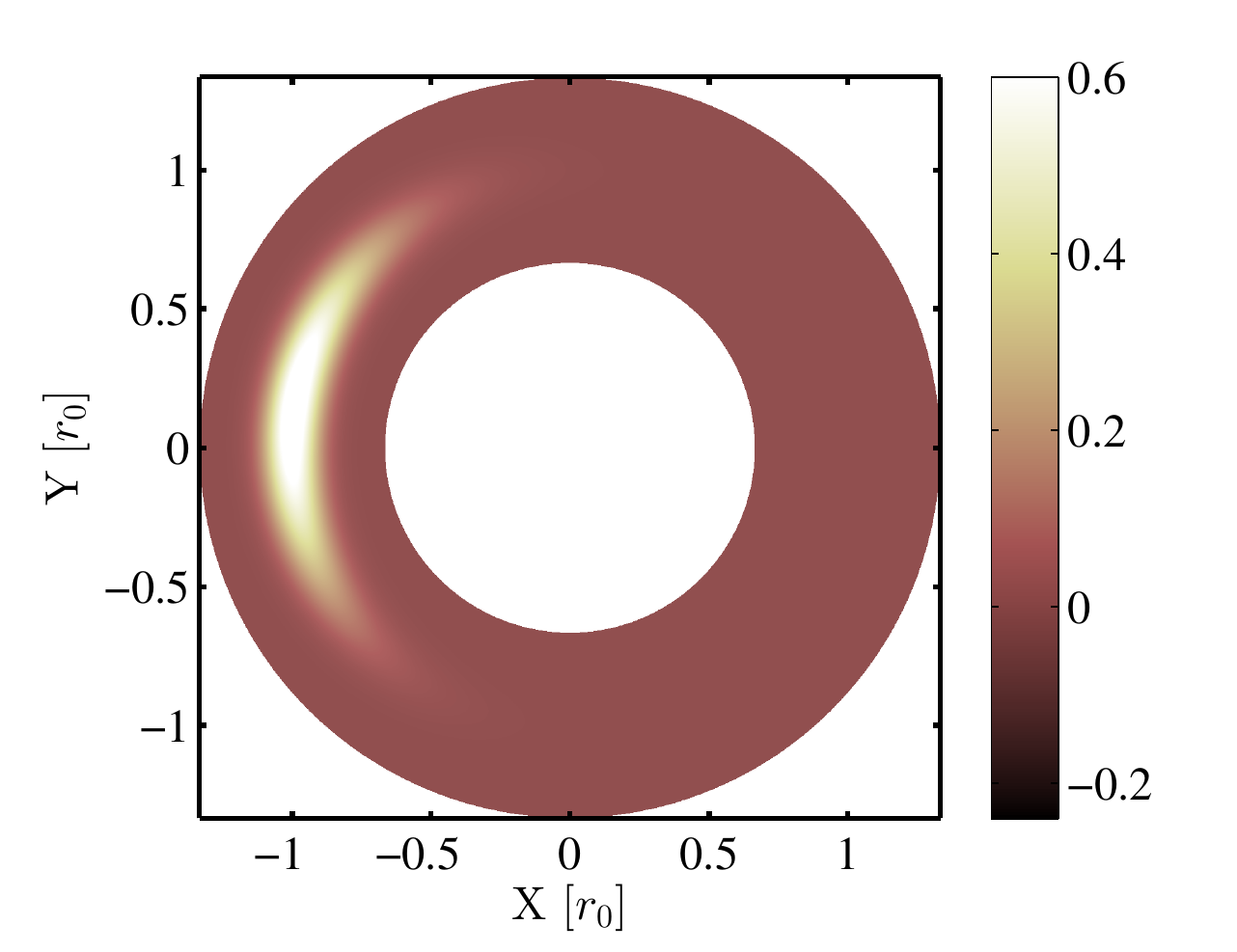} &
	\includegraphics[height=4.8cm, trim=4.5mm 0mm 4.5mm 0cm, clip=true]{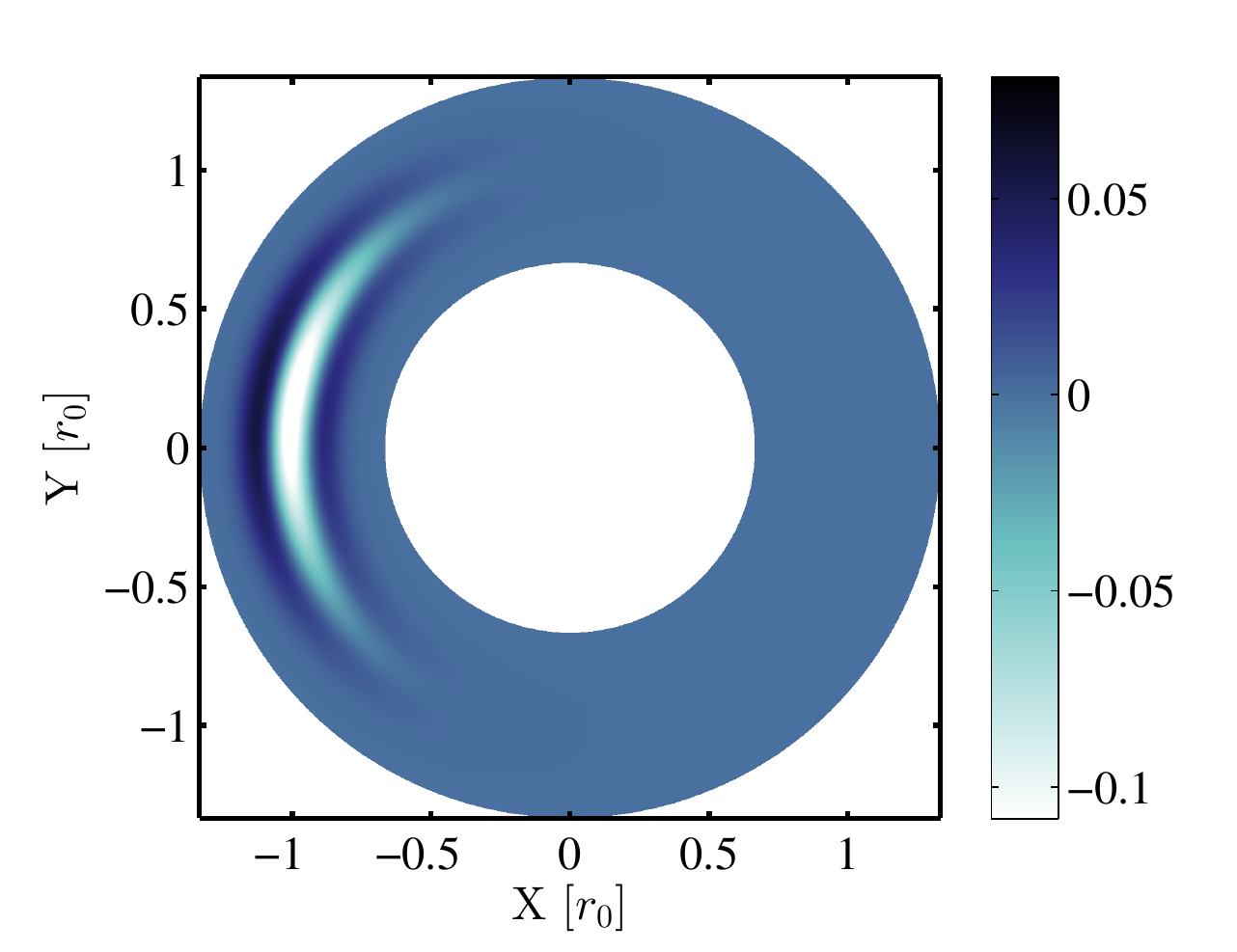} &
	\includegraphics[height=4.8cm, trim=4.5mm 0mm 4.5mm 0cm, clip=true]{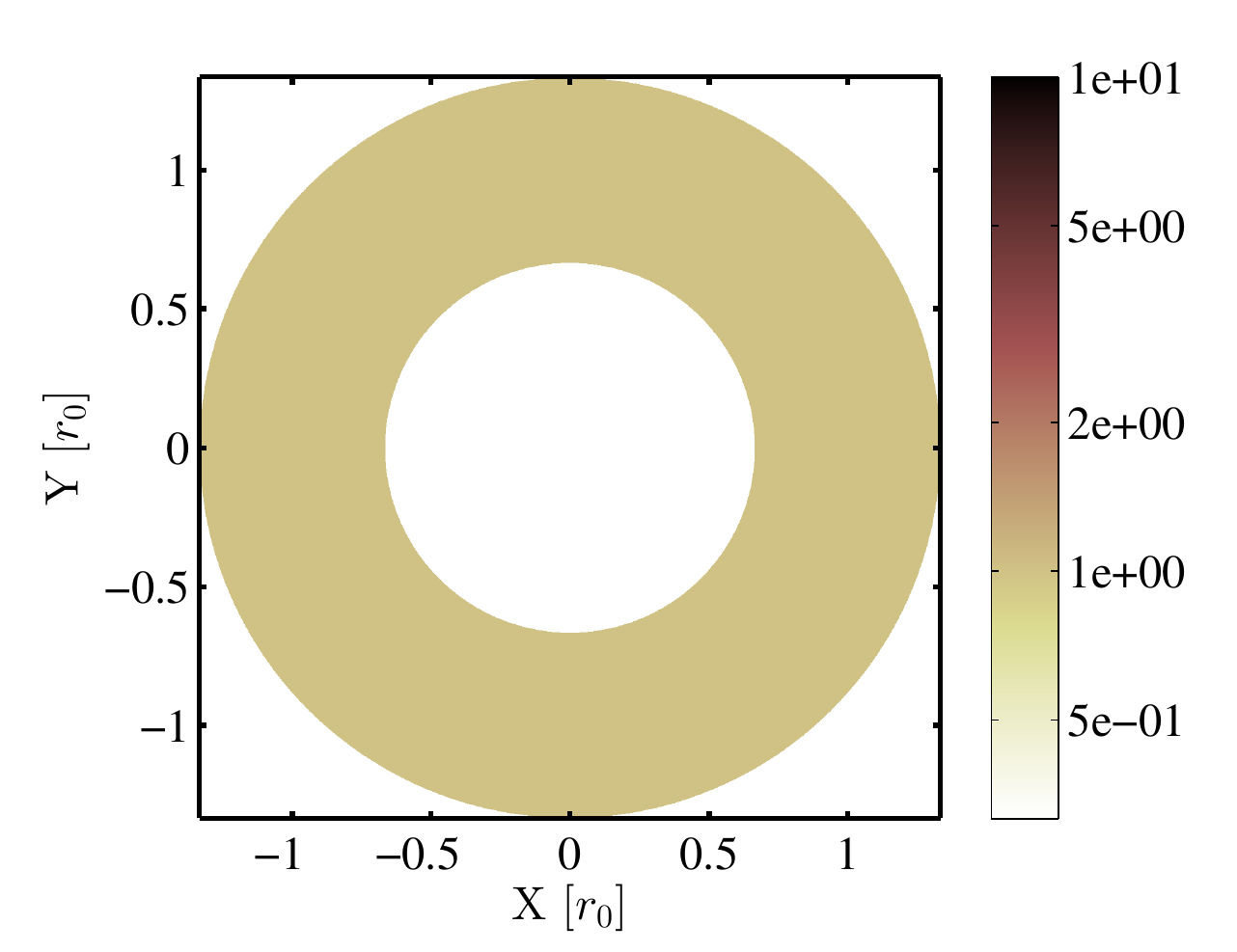}
	\end{tabular}
	\caption{\label{Initial_Cartesian} Initial conditions of the global disk simulations. The gas density (relative to the disk background), the gas Rossby number, and the initial unperturbed dust density (relative to the disk background) are shown {\it{from left to right}} respectively. A gas vortex model is imposed at $r=r_0$ and $\theta=3$.}
      \end{center}
\end{figure*}

	The initial gas vortex solution is issued from the Gaussian model described in \cite{Surville2015}. We chose vortex parameters that are found to be common in this study:
\begin{equation}
	(R_0, \: \chi_r, \: \chi_\theta)=(-0.13, \: 0.1, \: 6.5) \: .
\end{equation}
The vortex is initially located at $r_0$. The resulting vortex belongs to the incompressible family described in this paper, characterized by small wave excitation and a quasi-steady evolution with reduced radial migration. This ensures not to affect the effective Stokes number of the particles. We show Figure \ref{Initial_Cartesian} global color maps of the initial condition. One clearly observes the vortex solution of the gas density and vorticity (Rossby number) on the left and on the middle panels, respectively. This initial gas condition relaxes smoothly in the disk, during the first rotations, allowing us to cleanly follow the dust capture inside the vortex. The dust density (right panel) is unperturbed and thus equals the background profile.

\begin{figure*}
	\begin{center}
	\begin{tabular}{cccc}
	\scriptsize{Gas density} & \scriptsize{Dust density} & \scriptsize{Rossby number} & \\
	\includegraphics[height=4.5cm, trim=6mm 0cm 0cm 0cm, clip=true]{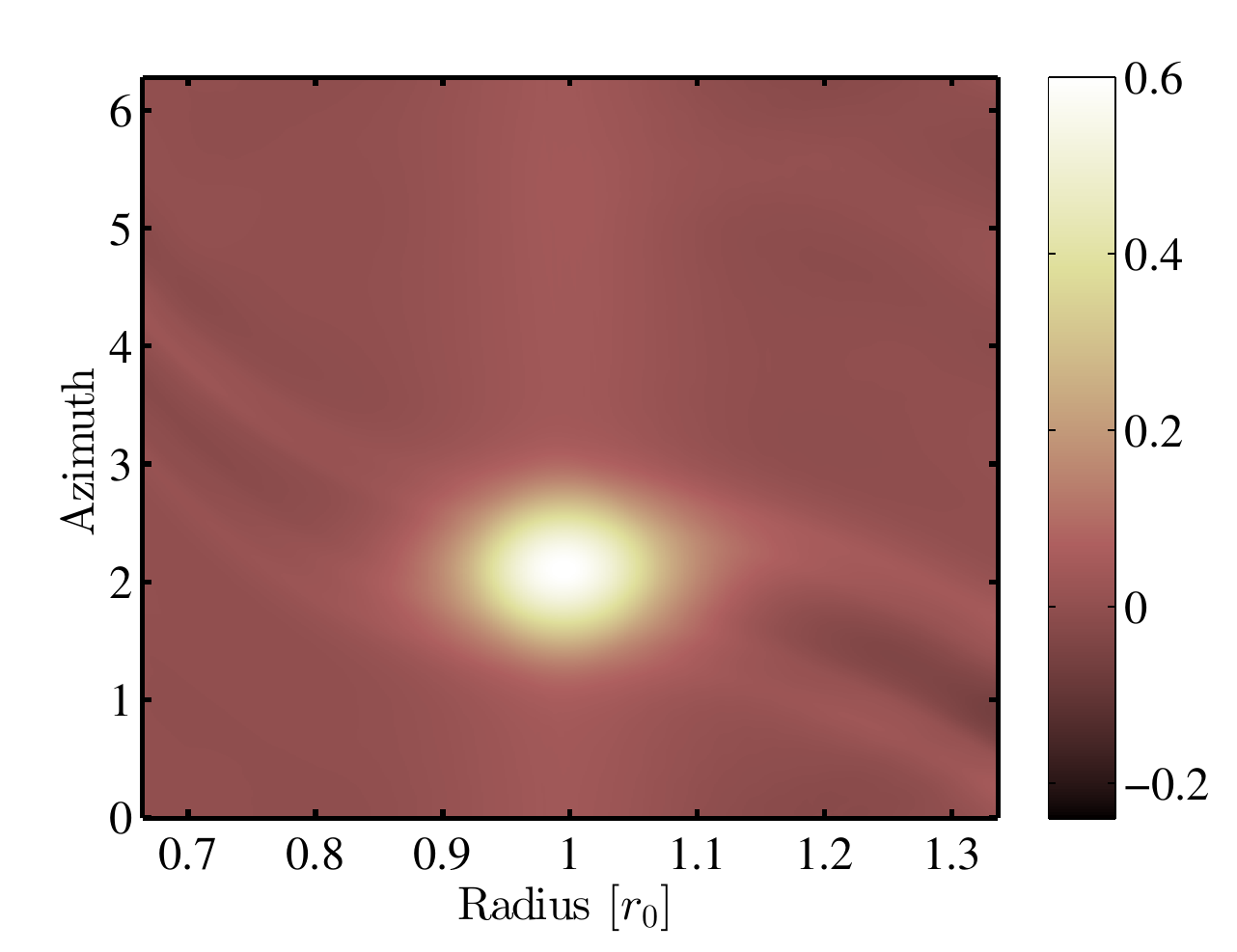} &
	\includegraphics[height=4.5cm, trim=6mm 0cm 0cm 0cm, clip=true]{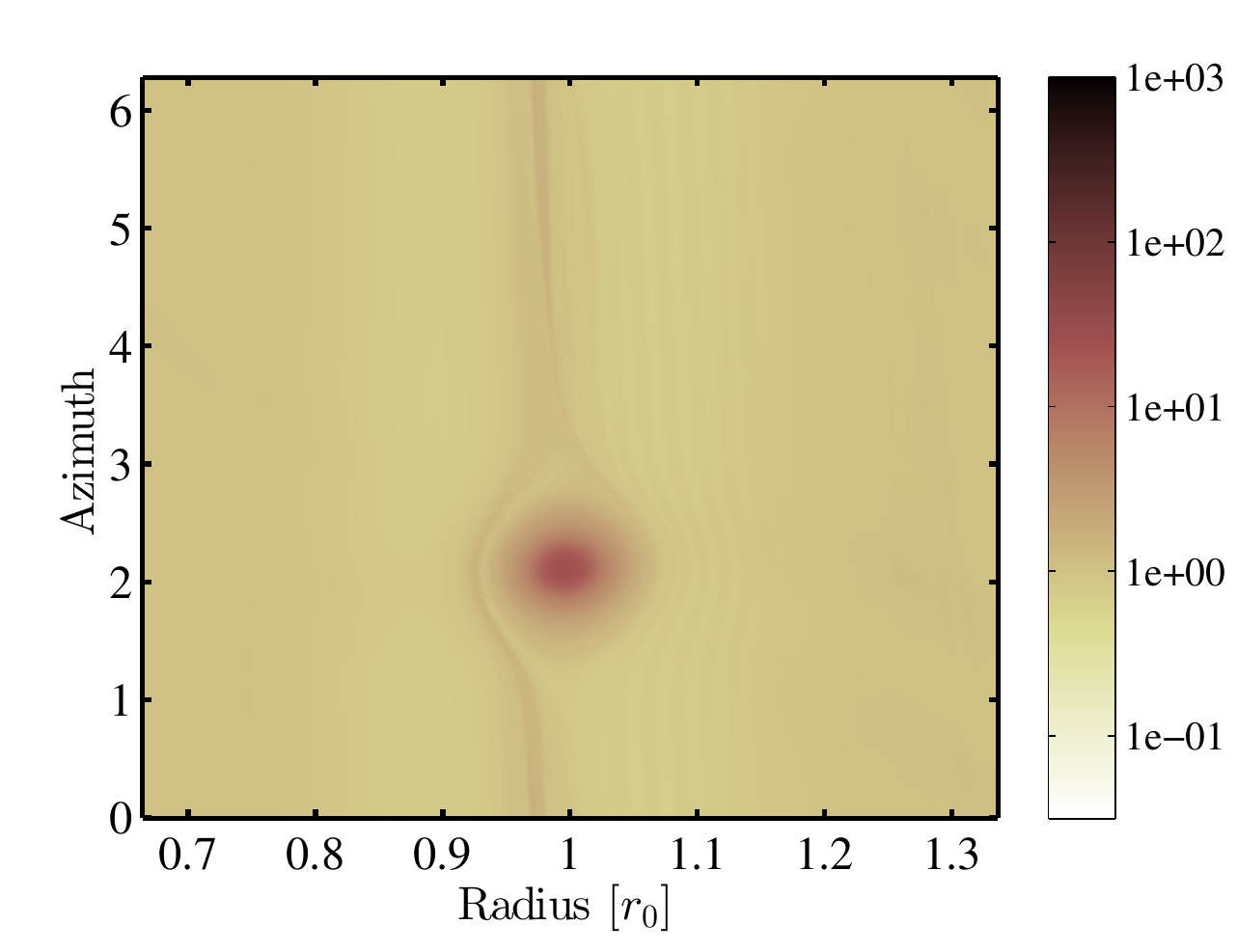} &
	\includegraphics[height=4.5cm, trim=6mm 0cm 0cm 0cm, clip=true]{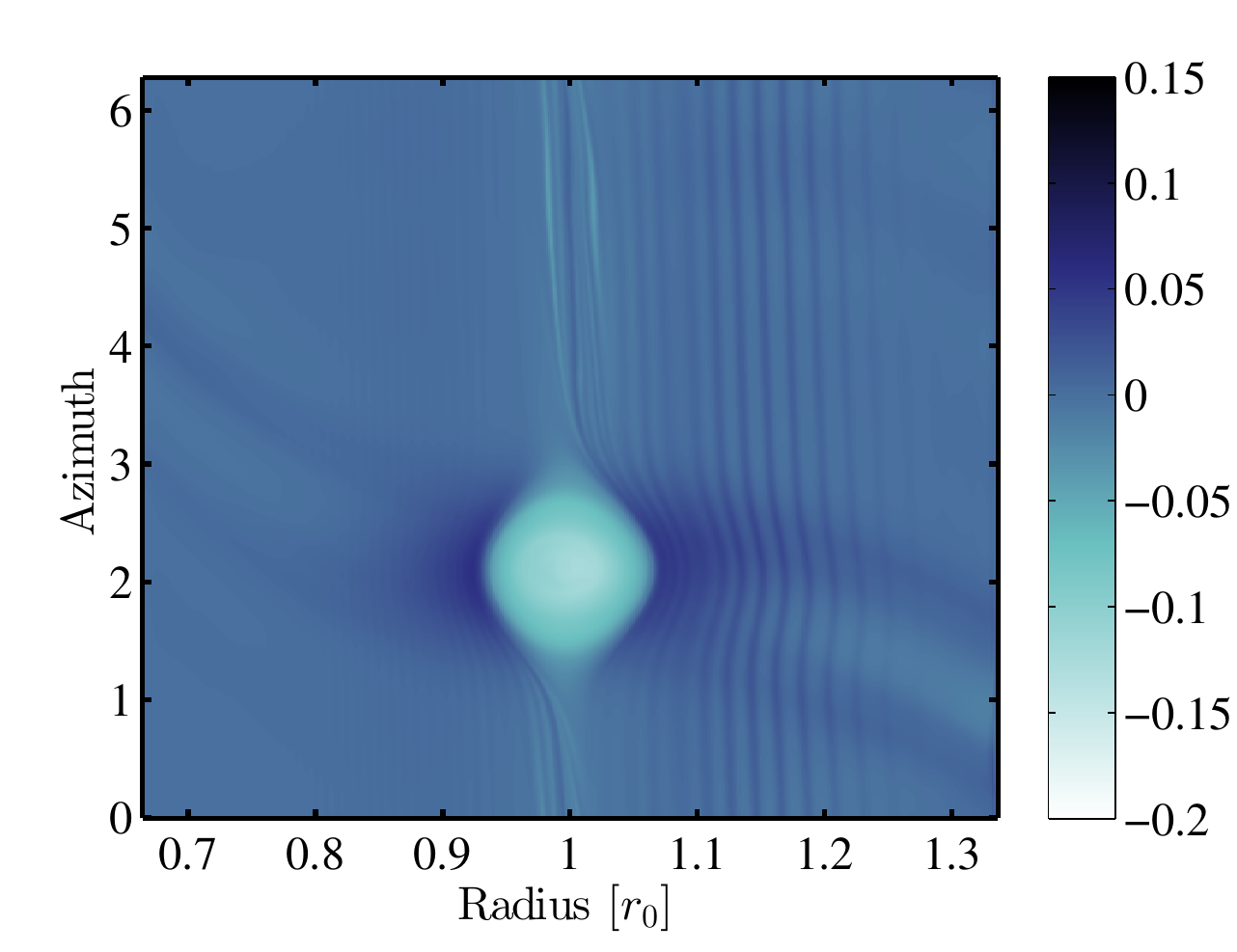} & \scriptsize{(a)}\\

	\includegraphics[height=4.5cm, trim=6mm 0cm 0cm 0cm, clip=true]{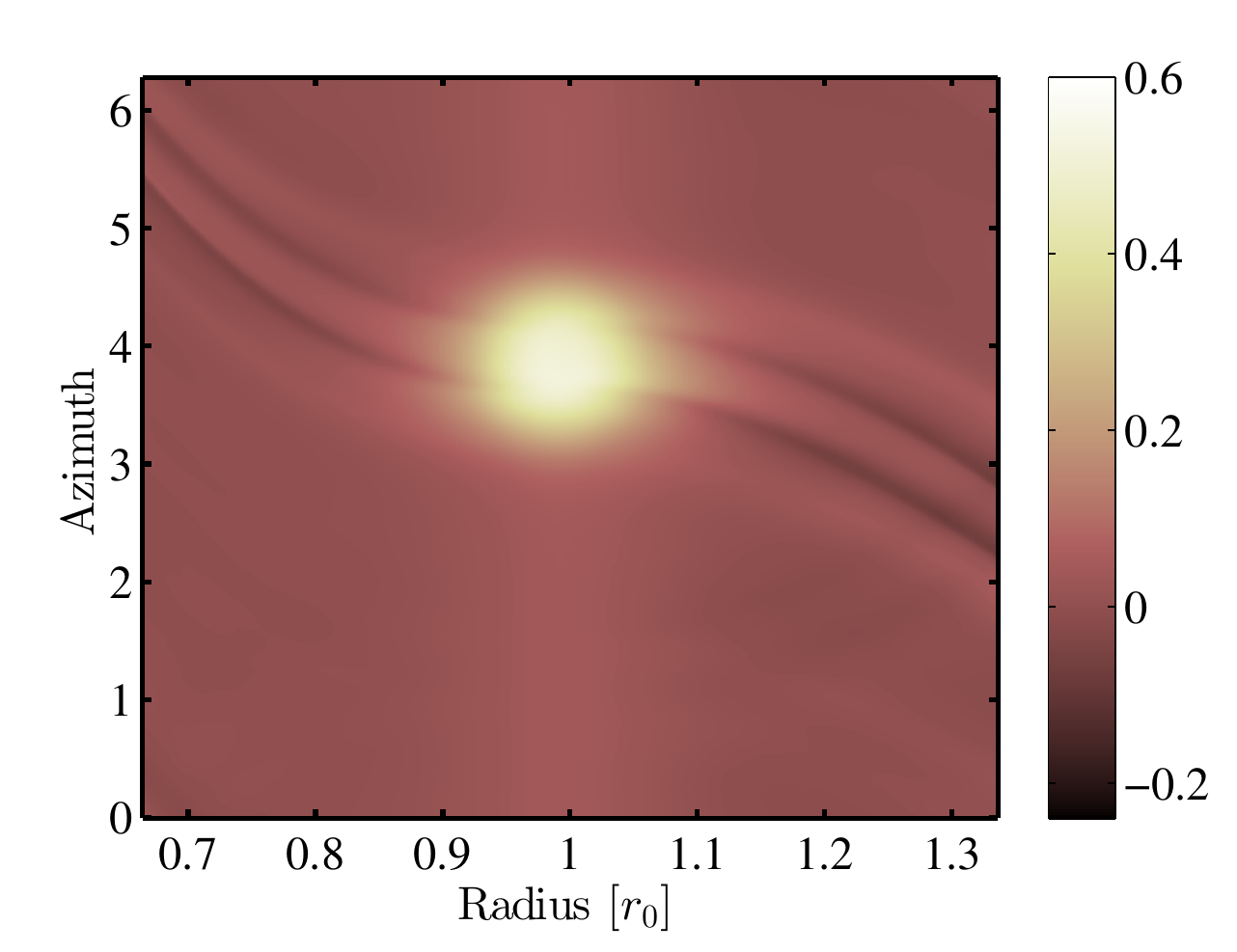} &
	\includegraphics[height=4.5cm, trim=6mm 0cm 0cm 0cm, clip=true]{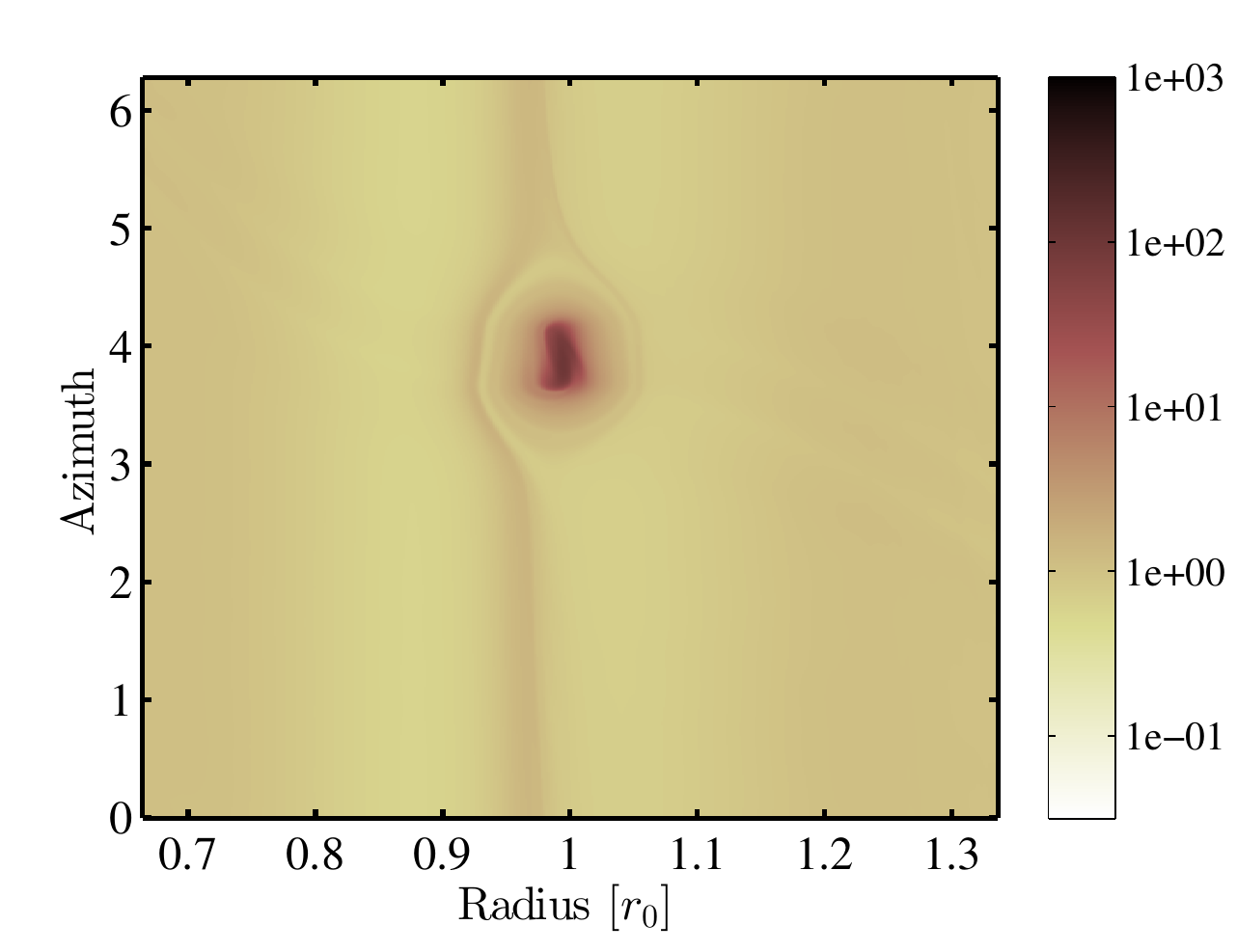} &
	\includegraphics[height=4.5cm, trim=6mm 0cm 0cm 0cm, clip=true]{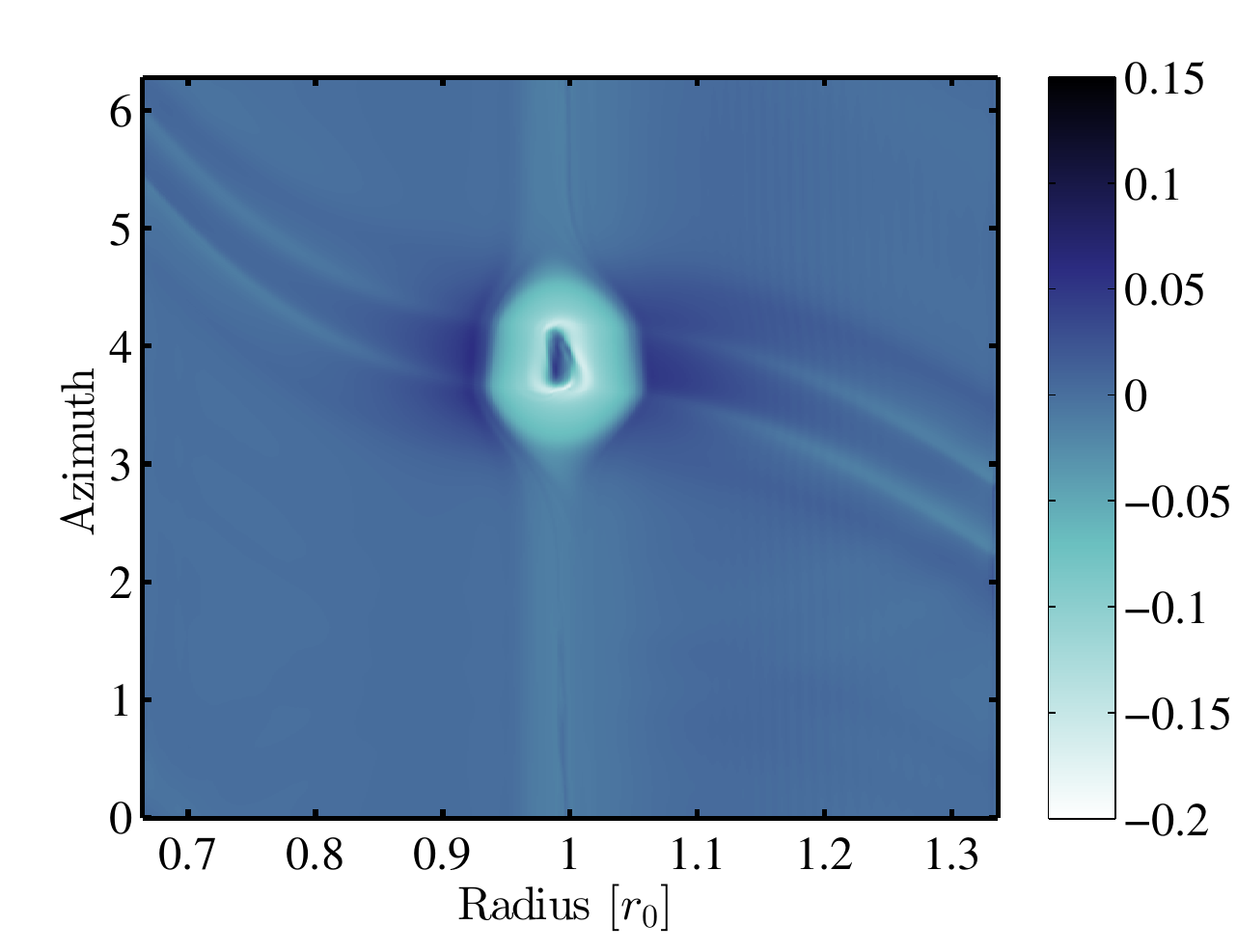} & \scriptsize{(b)}\\

	\includegraphics[height=4.5cm, trim=6mm 0cm 0cm 0cm, clip=true]{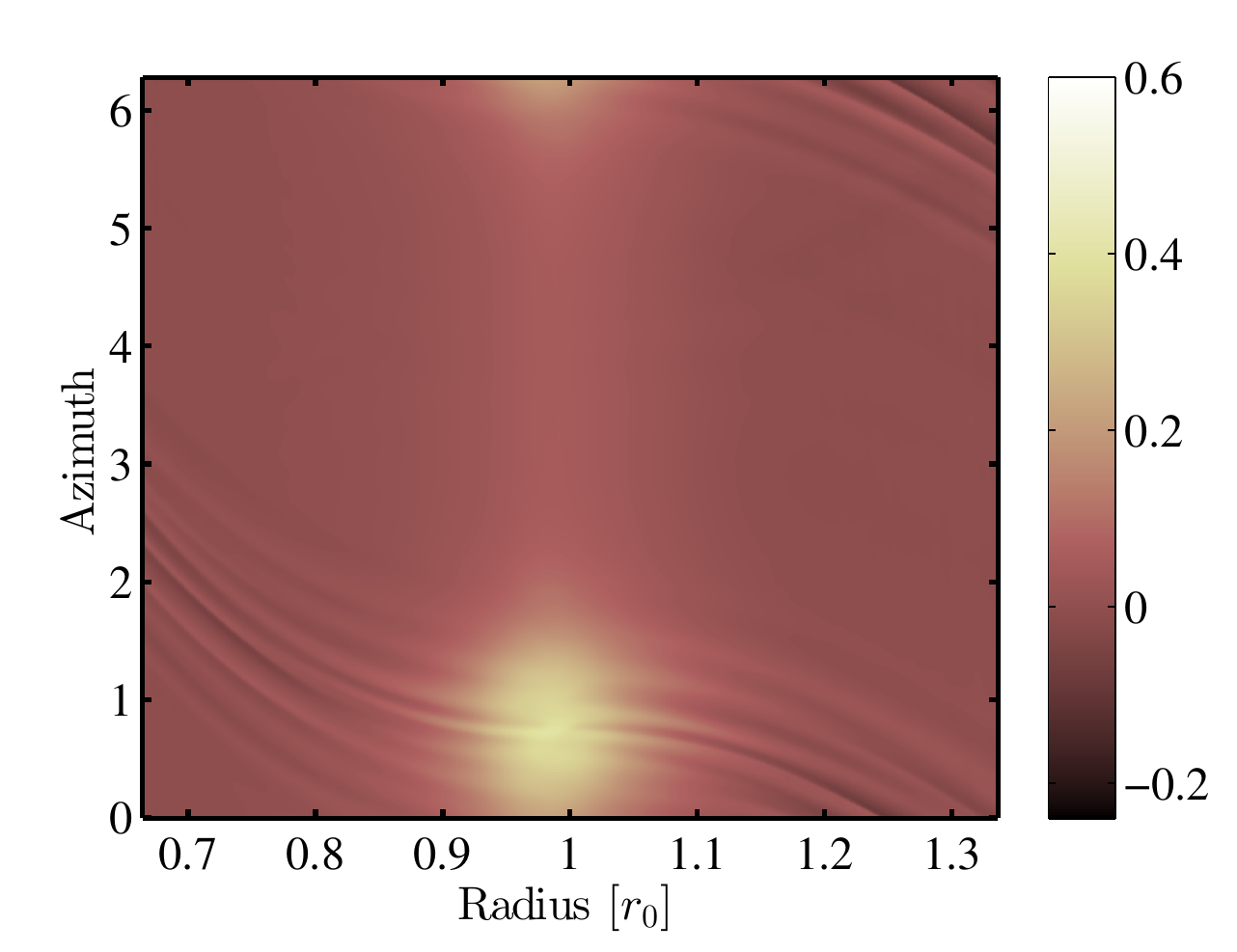} &
	\includegraphics[height=4.5cm, trim=6mm 0cm 0cm 0cm, clip=true]{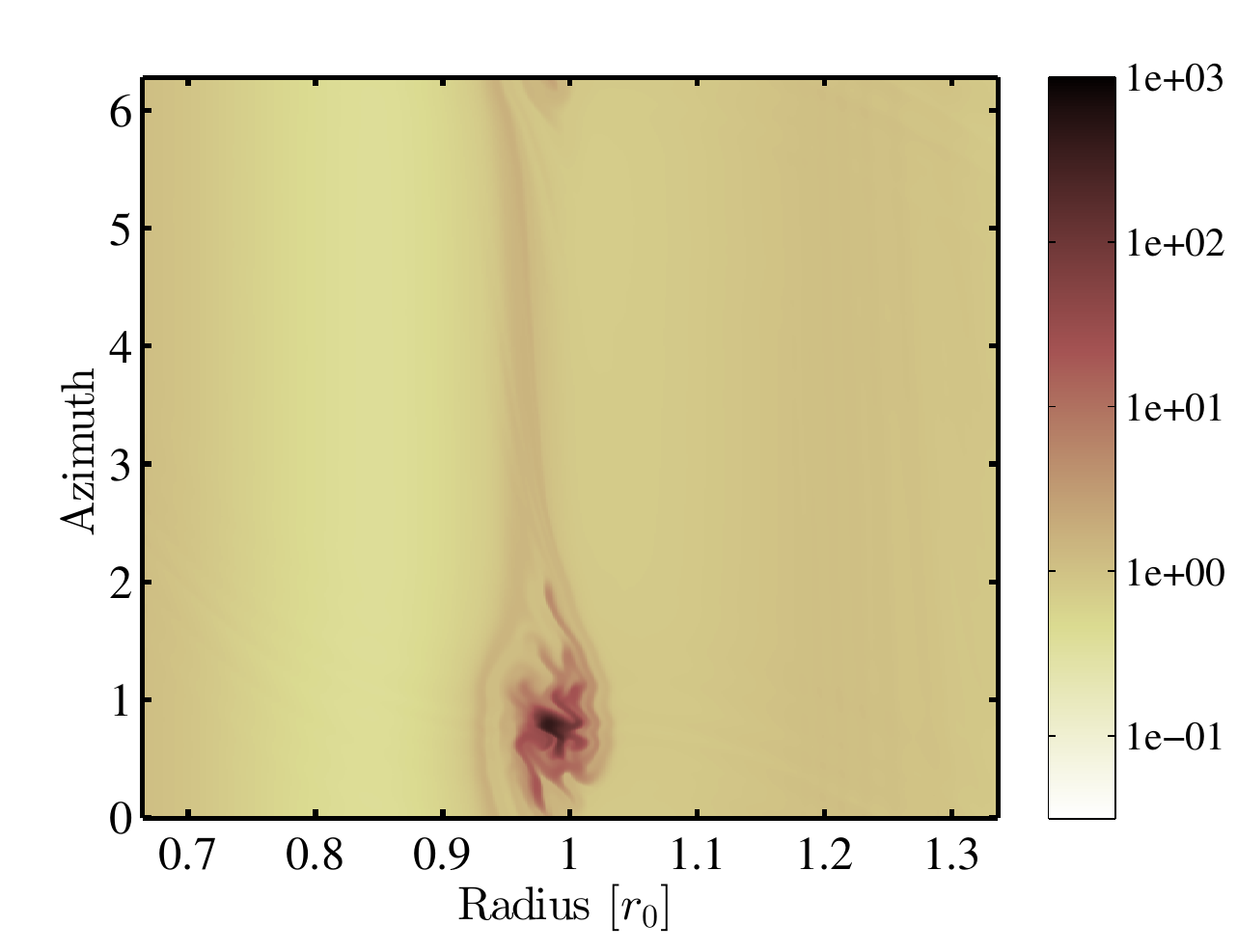} &
	\includegraphics[height=4.5cm, trim=6mm 0cm 0cm 0cm, clip=true]{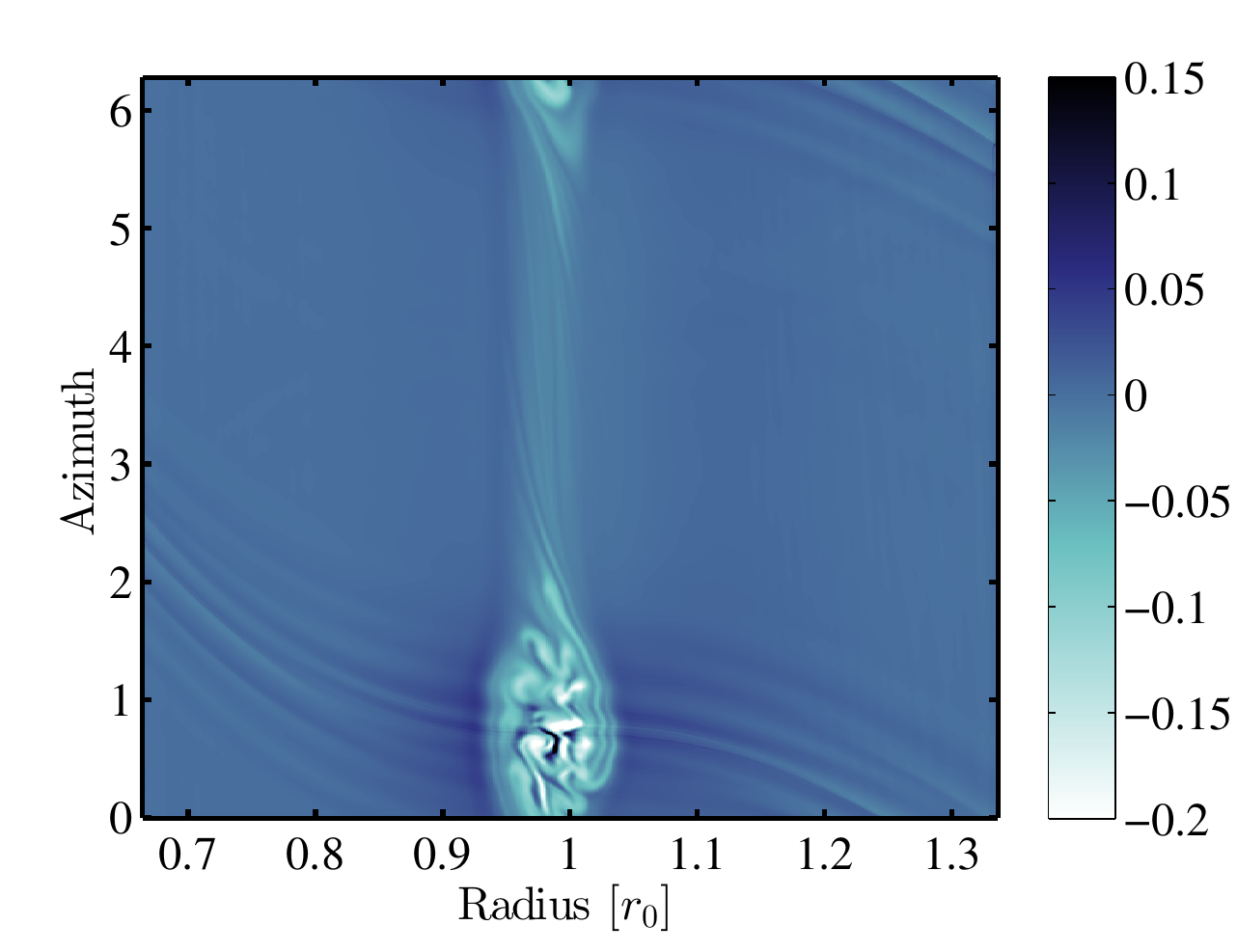} & \scriptsize{(c)}\\

	\includegraphics[height=4.5cm, trim=6mm 0cm 0cm 0cm, clip=true]{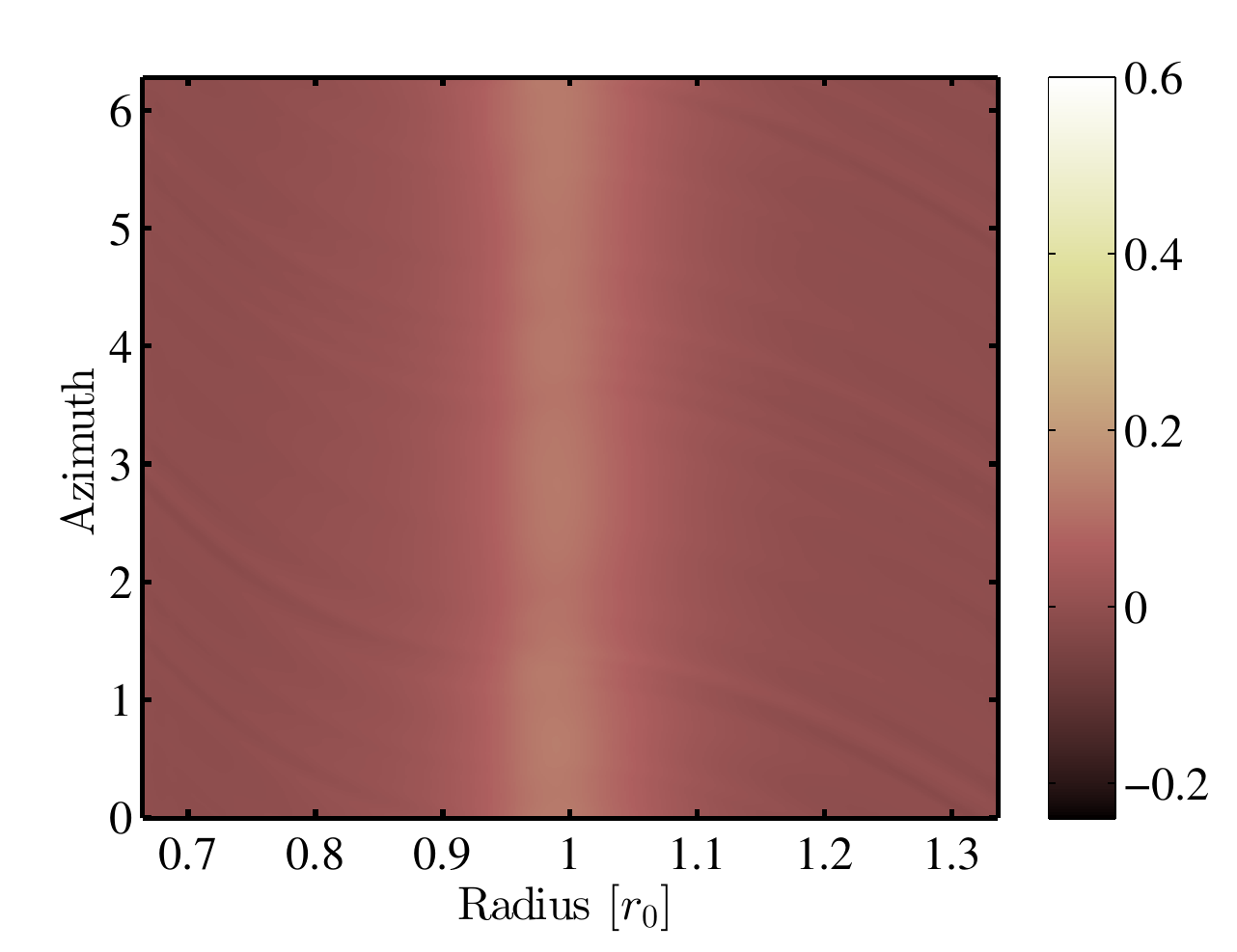} &
	\includegraphics[height=4.5cm, trim=6mm 0cm 0cm 0cm, clip=true]{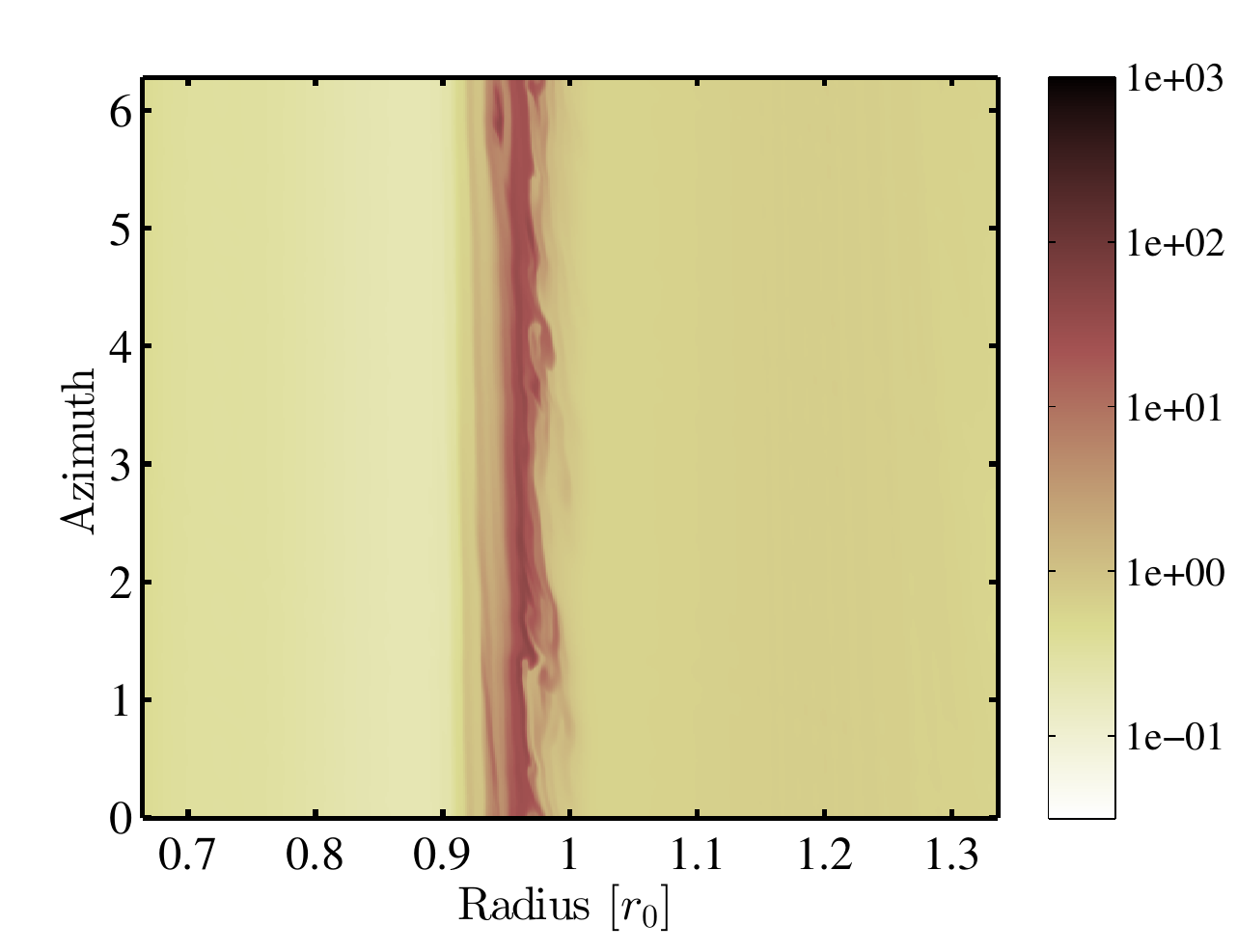} &
	\includegraphics[height=4.5cm, trim=6mm 0cm 0cm 0cm, clip=true]{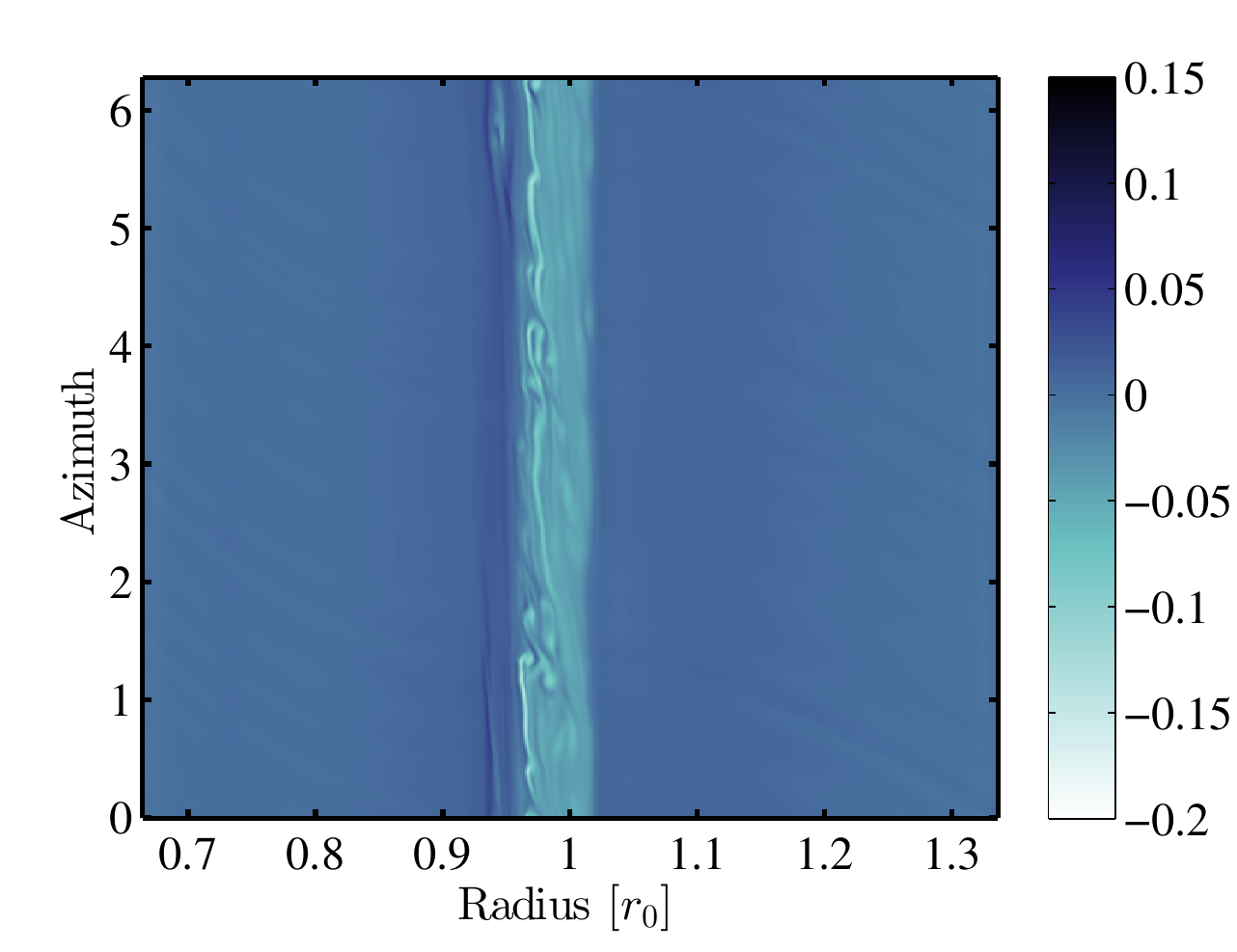} & \scriptsize{(d)}

	\end{tabular}
	\caption{\label{Evo_Typical} Time evolution of a vortex in the fiducial simulation described in Section \ref{Sect_Test_case}. {\it{From left to right}}: color maps in the $(r,\theta)$ plane of the relative gas density $\sigma_g/\sigma_0(r) -1$, of the relative dust density $\sigma_p/\left[\epsilon \sigma_0(r)\right]$ in $\log_{10}$, and of the Rossby number $\omega/\left[2\Omega_0(r)\right]$ (see Sec. \ref{Sect_Analytical_model}). The global dust-to-gas ratio is $\epsilon=10^{-2}$. Each row, (a) to (d), represents different states of evolution: after $t=50$, $t=100$, $t=200$, and $t=400$ rotations respectively.}
      \end{center}
\end{figure*}

	The simulation was performed with the small resolution $(N_r, \: N_\theta) = (512, \: 1024)$, during $500$ disk rotations (one rotation period equals $2\pi \Omega_0^{-1}(r_0)$). This choice is justified to obtain approximately the same accuracy as in other studies, like \cite{Bae2015} or \cite{Johansen2004} in terms of number of cells per disk scale height. Figure \ref{Evo_Typical} presents gas density, dust density, and gas vorticity at different times.

	First row (a) is shown the disk state after 50 rotations. The gas structure is very similar to the initial condition. The vorticity as well as the density profiles are not yet affected by the presence of the dust. However the dust density increases inside the vortex, with a local dust-to-gas ratio at the center already as large as $0.16$. This amount of dust particle captured inside the vortex comes from the co-orbital region, producing a slight depletion of dust in a ring between $0.9$ to $1.1 \: r_0$.

	After $100$ rotations, row (b), the impact of the dust phase onto the gas is evident. The vorticity near the vortex center is reduced, leading to a local cancellation of the Rossby number. This 'hollow' vortex profile excites sound waves that are also visible in the density map. The dust density has a compact profile inside the vortex, and increases up to $100$ times the initial value. The resulting local dust-to-gas ratio is $0.68$, because the gas density at the vortex center is 1.48 times larger than the disk background. The outer parts of the vortex are still similar to the earlier states.

	A longer evolution highlights that the vortex undergoes an instability. It is noteworthy, after 200 rotations, row (c), that small size vortical structures appear all inside the vortex. Some regions of large Rossby number are generated, surprisingly superimposed on the dust over densities. In particular a dust bubble of density $450$ times larger than the background persists. The local dust-to-gas ratio reached is larger than $3$. One can notice a slight over density in the gas phase, associated to this bubble. But the overall gas density shows that the vortex is vanishing.

	The final state of this evolution is the formation of a ring inside the disk, as here after 400 rotations, row (d). This quasi axisymmetric ring has some sub-structures, but they are dissipated during the time evolution. The dust ring still has an averaged density increase of a factor $40$. As the gas density is at a very low level over the background, the dust-to-gas ratio is still of order $0.4$ inside the ring. 

	As a conclusion of this test case, we identify three successive processes during the evolution of a vortex in a dusty disk:
\begin{itemize}
	\item the capture of dust inside the vortex, with a fast increase in the dust-to-gas ratio,
	\item the formation of a 'hollow' vortex, and a possible subsequent vortex instability,
	\item the release of dust in a ring shaped region, where a dust-to-gas ratio much larger than the background disk value, eventually larger than unity, can still persist.
\end{itemize}

	The first process is the most evident, and was the most studied since decade. It is observed and described with numerical simulation in global setups of the Rossby Wave Instability, in cylindrical coordinates \citep{Inaba2006, Meheut2012}, or even in Cartesian \citep{Lyra2009}. Vortices are formed as a consequence of the saturation of the RWI. Increase of the dust-to-gas ratio is measured inside the vortex, up to a value close to unity. However, the disk evolution is only followed typically on a few hundred disk orbits, and longer term evolution is unknown.

	Vortices excited by the RWI of a planetary gap opened by a massive planet are also observed to capture dust and to increase the dust-to-gas ratio locally to unity. In \cite{Bae2015}, a very similar method (FARGO) is used to perform two-fluid global simulations on $200$ orbits, using a resolution of $(N_r, \: N_\theta)=(512, \: 1024)$. Even numerical methods based on SPH approach can resolve this process of capture at the gap edge \citep{Ayliffe2012, Fouchet2010, Gonzalez2012}, or at the vortex center \citep{Ataiee2013, Lyra2008a}. 

	Furthermore, small scale vorticity as it exists in a turbulent state of disks is also capable of capturing and accumulating dust particles, increasing the dust-to-gas ratio. For example in \cite{Fromang2005}, the turbulence is driven by a 3D MRI, whereas in \cite{Gibbons2015}, the turbulent flow is triggered by the gravitational instability.

	However, among the first studies to use an analytical model to impose a vortex in the disk is \cite{Johansen2004}. Using an approach similar to ours, dust capture in a GNG vortex model \citep{Goodman1987} is studied using shearing box simulations with the PENCIL code in the two-fluid approximation. Unfortunately, the authors only followed the survival of the vortex on $\sim 10$ orbits, but already observed dust accumulation inside the vortex.

	The second process that we highlight is only possible under the treatment of the back reaction drag onto the gas. It is not commonly considered in the literature, but the major paper observing vortex instability is \cite{Fu2014}. In this study, vortices are a by-product of the instability of the planetary gap, resolved at a high resolution $(N_r, \: N_\theta) = (6144,\: 6144)$ in a global disk. Our simulations at $(N_r, \: N_\theta) = (2048,\: 4096)$ presented Section \ref{Sect_Results} have the equivalent local resolution, due to a smaller disk extent. When drag back reaction is on, the authors observe the vortex instability that looks similar to Figure \ref{Evo_Typical} (c). They also explore small initial dust-to-gas ratios $\epsilon = 10^{-2}$ to $10^{-3}$. Similarly, \cite{Raettig2015} present capture of dust treated as super particles, in a vortex generated by the baroclinic instability in a local shearing box. Particle feedback here also produces the disruption of the vortex. Finally, in \cite{Crnkovic-Rubsamen2015}, the vortex is set to a GNG model in the shearing box, and when the Stokes number $S_t$ of the particles is close to unity, a vortex instability is visible. However, these authors only explore the $\epsilon>10^{-2}$ regime, favourable for drag feedback. 

	Surprisingly, the dust ring formation that we observe in the last phase of our fiducial run is never mentioned. Firstly, we observe its formation after $300$ disk rotations, that is in many studies after the end of the simulations. Secondly, it is a global feature that may not be well described by local shearing box simulations. In this paper, we aim at providing a complete explanation of this process, and a description of which context is favourable to its formation.
	
	As most of the literature concerning dusty vortices focuses on the first phase of dust capture, we will discuss in the next section a new analytical approach to explain this aspect of evolution in detail in order to produce a benchmark model for the simulation results.

\section{ Analytical model for dust capture }
\label{Sect_Analytical_model}

	The purpose of this section is to provide a detailed analytical description of the first phase of the evolution: the dust capture inside the vortex. Some analyses have been done in the past \citep{Chang2010, Lyra2013, Mittal2015} but usually using a particle approach to find the capture time of a given dust particle. Although these studies are of interest, they are less relevant in the context of the bi-fluid approximation. This is why we propose a new analytical model for the dust capture that involves the time evolution of the particle fluid density.

	To this end, we start from the equations of time evolution of the velocity fields for the gas and the particle fluids (see Equations \ref{Equ_Vel_Gas} and \ref{Equ_Vel_Part}).We assume that the gas can deviate from the equilibrium solution by some radial and azimuthal velocity components, $u_g$ and $v_g$, which are not necessary small. Thus the gas velocity reads:
\begin{equation}
\begin{aligned}
	& U_g = u_g \: , \\
	& V_g = r \Omega_k(r) + \Delta V(r) + v_g \: .
\end{aligned}
\end{equation}

	The sub-Keplerian motion of the steady state is expressed in the term due to the global pressure gradient:
\begin{equation}
	\Delta V(r) = r \Omega_k(r) \left[ \left( 1 + \frac{\beta_P}{\gamma} \frac{{c_s}_0^2(r)}{r^2 \Omega_k^2(r)} \right)^{1/2} - 1 \right] \: .
\end{equation}

	By doing the same approach for the particle fluid, we suppose that the velocity field deviates from the pure Keplerian motion by some components $u_p$ and $v_p$:
\begin{equation}
\begin{aligned}
	& U_p = u_p \: , \\
	& V_p = r \Omega_k(r) + v_p \: .
\end{aligned}
\end{equation}

	By doing so, we are in the context used in \cite{Surville2015} to find a vortex solution for the gas. By assuming that $u_g$ and $v_g$ are vortex components, we search for the corresponding $u_p$ and $v_p$ of the dust fluid that respect a kind of equilibrium with the gas inside the vortex.

\subsection{ Evolution of the particle density }

	To this end, we need a linear solution of the coupled systems of velocity fields, for the gas and for the particles. Assuming a quasi-steady solution at the orbit $r=r_0$, which means that the operator $\partial_t + \Omega_k(r_0) \partial_\theta =0$, and conserving the first order terms, we obtain from the velocity equations of the particle phase:
\begin{align}
	& 0 = 2 \Omega_k(r) v_p - \Omega_k(r) S_t^{-1} \left(u_p - u_g\right) \: , \\
	& u_p \partial_r r \Omega_k(r) = - u_p \Omega_k(r) - \Omega_k(r) S_t^{-1} \left(v_p - \Delta V(r) - v_g \right) \: , 
\end{align}
	from which we derive:
\begin{eqnarray}
	\left(u_p - u_g \right)& = &  2 S_t v_p \: , \\
	\left(v_p - \Delta V(r) - v_g \right) & = & -2 S_t (1 + \beta_\Omega/2) u_p \: .
\end{eqnarray}

	This method is justified by the assuming a quasi-steady evolution. In fact, gas and dust fluids relax dynamically on timescales faster than the local orbital period. It is also justified from the linearized approach, which is valid for quasi elliptic vortex solutions \citep{Kida1981, Goodman1987}.

	Finally, we obtain the components of the velocity field of the particle fluid:
\begin{eqnarray}
\label{Eq_dust_velocity}
	u_p & = & A \left[ \Delta V(r) + v_g + \frac{u_g}{2 S_t} \right] \: , \\
	v_p & = & A \left[ \frac{\Delta V(r) + v_g}{2 S_t} -  (1 + \beta_\Omega/2) u_g \right] \: .
\end{eqnarray}

	Here, we have introduced a useful parameter that describes the degree of coupling between the two fluids:
\begin{equation}
	A = \frac{2 S_t}{1 + 4 ( 1 + \beta_\Omega/2) S_t^2} \: ,
\end{equation}
	which is, for Keplerian disks with $\beta_\Omega=-3/2$, the well-known parameter $2S_t /(1 + S_t^2)$.

	As the components $u_p$ and $v_p$ of the velocity field of the particles do not cancel in the general case, we can expect some mass flux through the disk, that will change the density profile as time goes. If we search for the evolution of a local maximum of the particle density, $\sigma_p^{max}$, we can set that $\partial_r \sigma_p^{max} = \partial_\theta \sigma_p^{max} = 0$. Then the continuity equation of the particle phase becomes:
\begin{equation}
	\partial_t \sigma_p^{max} = -\frac{\sigma_p^{max}}{r} \left( \partial_r r u_p + \partial_\theta v_p \right) \: .
\end{equation}

	Injecting the particle velocity field, we finally obtain the equation of evolution of the dust density maximum:
\begin{equation}
	\partial_t \sigma_p^{max} = - A \left[ \frac{1}{r}\partial_r r \Delta V(r) + \omega + \frac{\vec{\nabla} \cdot \vec{V}_g}{2 S_t} -\frac{\beta_\Omega}{2 r } \partial_\theta u_g \right] \sigma_p^{max} \: .
\end{equation}

	We have introduced the vorticity of the gas deviation, $\omega = r^{-1} \partial_r (r v_g) - r^{-1} \partial_\theta u_g$, which we expect to be the vorticity of an anticyclonic vortex. It is shown in \cite{Surville2015} that vortices usually have near zero compressibility at the center, and that the term $r^{-1} \partial_\theta u_g$ is small compared to vorticity, in particular for vortices of the incompressible family.

	Thus the evolution of a maximum of particle density simply follows the equation
\begin{equation}
	\partial_t \sigma_p^{max} = - A \left[ \frac{1}{r}\partial_r r \Delta V(r) + \omega \right] \sigma_p^{max} \: .
\end{equation}

	The first term is a growth of dust density due to the global pressure gradient that imposes a constant head wind for the particles, and then generates a radial flux of mass responsible for this growth. One can obtain this solution from the linear classical axisymmetric study of \cite{Nakagawa1986}.

	On top of this global and systematic growth, we reveal how vortices enhance dust density. The maximum at the vortex center changes with an exponential growth rate proportional to the vorticity of the vortex. As an anticyclonic vortex has negative vorticity $\omega$, the dust density increases. A cyclonic vortex would deplete the local dust density. However, cyclonic vortices remain stable only on very short timescales, and they are not expected to participate to the planetesimal formation.

\subsection{ Coupled evolution of vorticity and dust density }

	When we take into account the back reaction of the drag force onto the gas, the evolution of the gas velocity field is affected by the presence of the dust. In particular, this has a major impact on the evolution of the gas vorticity. The drag force creates a vorticity source term which reads:

\begin{equation}
	\vec{S}_{drag} = - \vec{\nabla} \times \frac{\vec{f}_{aero}}{\sigma_g} \: .
\end{equation}

	If we consider the evolution of the vorticity at the vortex center, we can assume that the local dust-to-gas ratio is a maximum. Thus we can neglect the spatial derivatives of $\sigma_p/\sigma_g$. This assumption simplifies the above equation and gives:

\begin{equation}
	\vec{S}_{drag} = \frac{\sigma_p}{\sigma_g} \Omega_k(r) {S_t}^{-1} \left[\frac{\beta_\Omega}{r} \left(V_p - V_g \right) \vec{e}_z + \vec{\nabla} \times \left(\vec{V}_p - \vec{V}_g\right) \right] \: .
\end{equation}

	From the previous section, we already set the expression of $\left(\vec{V}_p - \vec{V}_g\right)$ as a function of the gas velocity field. After inserting it into the above equation and some algebra, one obtains the vertical component of the drag force source term of vorticity:
\begin{equation}
\begin{aligned}
	S_{drag} = & - 2 \left(1+\frac{\beta_\Omega}{2}\right) \frac{\sigma_p}{\sigma_g} A \Omega_k(r)  \left[ \frac{1}{r}\partial_r r \Delta V(r) + \omega  \right. \\
		& + \frac{\beta_\Omega}{r}\left(\Delta V(r) + v_g \right)  
		+ \frac{1}{2 S_t}\left(\frac{1}{r}\partial_r(r u_g) +  \frac{\beta_\Omega}{r} u_g \right) \\
		& + \left. \frac{1}{2 S_t (1 +\beta_\Omega/2)}\frac{1}{r}\partial_\theta v_g	\right] \: .
\end{aligned}
\end{equation}

	This complex equation can be simplified at lowest order by assuming that, at the vortex center, $u_g$ and $v_g$ cancel as well as the derivatives appearing in the equation. This comes out from the Gaussian vortex model described in \cite{Surville2015}, and was confirmed with numerical results by these authors. This vorticity source term thus reduces to:
\begin{equation}
	S_{drag} = - 2 \left(1+\frac{\beta_\Omega}{2}\right)\frac{\sigma_p}{\sigma_g} A \Omega_k(r) 
		\left[ \omega  + \frac{1}{r}\partial_r r \Delta V(r) + \frac{\beta_\Omega}{r}\Delta V(r) \right] \: .
\end{equation}

	Like in the evolution of the maximum dust density, the vorticity source term is proportional to $A$ which parametrizes the degree of coupling of the particles to the gas. It is also proportional to the local dust-to-gas ratio. As the maximum dust density will increase in time, this source term of vorticity will also increase due to this factor. Finally, the term in brackets shows that the vorticity change is partly due to the vortex component $\omega$ and due to the disk gradients via $\Delta V(r)$.
	
	However, these last terms affect all the disk regions in a similar way, and not only the vortex region. The same remark can be done on the evolution of the dust density. We will thus only consider changes due to the presence of a vortex in the disk. This leads to the main system of equations of our study:
\begin{equation}
\begin{aligned}
	& \partial_t \sigma_p^{max}  = - A  \omega \sigma_p^{max} \: , \\
	& \partial_t \omega  = - 2 \left(1+\frac{\beta_\Omega}{2}\right) A \Omega_k(r) \frac{\sigma_p^{max}}{\sigma_g} \omega \: .
\end{aligned}
\end{equation}

\subsection{ Time invariant and achievable dust-to-gas ratio }
\label{Sect_time_invariant}

	We search for a combined solution of the time evolution of the dust density maximum and the vorticity at the vortex center, orbiting at $r=r_0$. To do so, we will introduce useful parameters without unity:
\begin{itemize}
	\item the particle density maximum $\sigma_p^*$, scaled to the background dust density via $\sigma_p^{max}=\epsilon \sigma_0(r) \sigma_p^*$,
	\item the Rossby number at the vortex center, $R_0 = \omega/[2 \Omega_k(r_0)]$,
	\item the time in orbital periods at $r_0$, $T= t \Omega_k(r_0)/(2 \pi)$,
	\item the gas density at the vortex center scaled to the disk background, $\sigma_g^* = \sigma_g/ \sigma_0(r_0)$.
\end{itemize}

	In a first approximation, we will consider this latest quantity almost constant in time, and depending on the vortex shape. These variables allow to write the system in a clearer way, assuming an anticyclonic vortex ($R_0<0$):
\begin{equation}
\begin{aligned}
	& \partial_T \sigma_p^*  = 4 \pi A |R_0| \sigma_p^* \: , \\
	& \partial_T |R_0|   = - 4 \pi \left(1+\frac{\beta_\Omega}{2}\right) A \frac{\epsilon}{\sigma_g^*} |R_0| \sigma_p^* \: .
\end{aligned}
\end{equation}
	This system describes the coupled evolution of the maximum dust density enhancement compared to the disk background with the Rossby number at the vortex center, as a function of vortex rotations. If we note by $\tilde{\epsilon}= \left(1+\beta_\Omega/2\right) \epsilon / \sigma_g^*$, a modified dust-to-gas ratio, we obtain:
\begin{equation}
\label{Sys_evo_1}
\begin{aligned}
	& \partial_T \sigma_p^*  = 4 \pi A |R_0| \sigma_p^* \: , \\
	& \partial_T |R_0|   = - 4 \pi A \tilde{\epsilon} |R_0| \sigma_p^* \: .
\end{aligned}
\end{equation}

	One can highlight the existence of a time invariant, characterizing the evolution of the system:
\begin{equation}
\label{Eq_Time_invariant}
	I_0 = \tilde{\epsilon} \sigma_p^* + |R_0| \: .
\end{equation}

	As at the initial time $T=0$, the dust phase has a density equal to the disk background, i.e. $\sigma_p^*=1$, we obtain $I_0=\tilde{\epsilon} + |R_0|_{T=0}$. This gives an estimate of the maximum density enhancement possible in a given vortex. In fact, the evolution equation of $|R_0|$ shows that it tends to zero as time goes, because the right hand side term is negative. When the Rossby number at the vortex center is zero, the dust density reaches a maximal value given by:
\begin{equation}
\label{Eq_density_saturation}
	\sigma_p^*(T\to \infty) =  \frac{I_0}{\tilde{\epsilon}} = 1 + \frac{|R_0|_{T=0}}{\tilde{\epsilon}} \: .
\end{equation}

	For example, for a typical vortex of initial $R_0=-0.15$, that is a common value in disks \citep{Surville2015}, and an initial dust-to-gas ratio at the vortex center $\epsilon / \sigma_g^*=10^{-2}$, one can estimate an asymptotic dust enhancement of $\sigma_p^* = 61$, equivalent to an asymptotic dust-to-gas ratio in the vortex of $0.61$. 
	
	From this analysis, the final dust-to-gas ratio is roughly given by $|R_0|_{T=0}/(1+\beta_\Omega/2)=4|R_0|_{T=0}$ with $\beta_\Omega=-3/2$, when the global dust-to-gas ratio $\epsilon$ is under $10^{-2}$. Thus it strongly depends on the vortex parameters, and no longer on the dust properties. As the streaming instability or the gravitational instability may require a dust-to-gas ratio of the order of unity, one can observe that the capturing vortex needs a high vorticity, with $R_0 \sim -0.25$. But our linear analysis does not consider the vortex instability triggered when the vorticity at the center goes to zero. These 'hollow' vortices are subjects to drastically increase the local dust-to-gas ratio, because high vorticity regions are created in the vortex during the instability. We can expect lower Rossby number vortices ($R_0\sim -0.15$) to create high dust-to-gas ratio by this process. This will be checked with nonlinear simulations in the next sections.

\subsection{ Solution of the coupled equations and predictions }

\begin{figure}
	\begin{center}
	\begin{tabular}{c}
	\includegraphics[height=5.5cm]{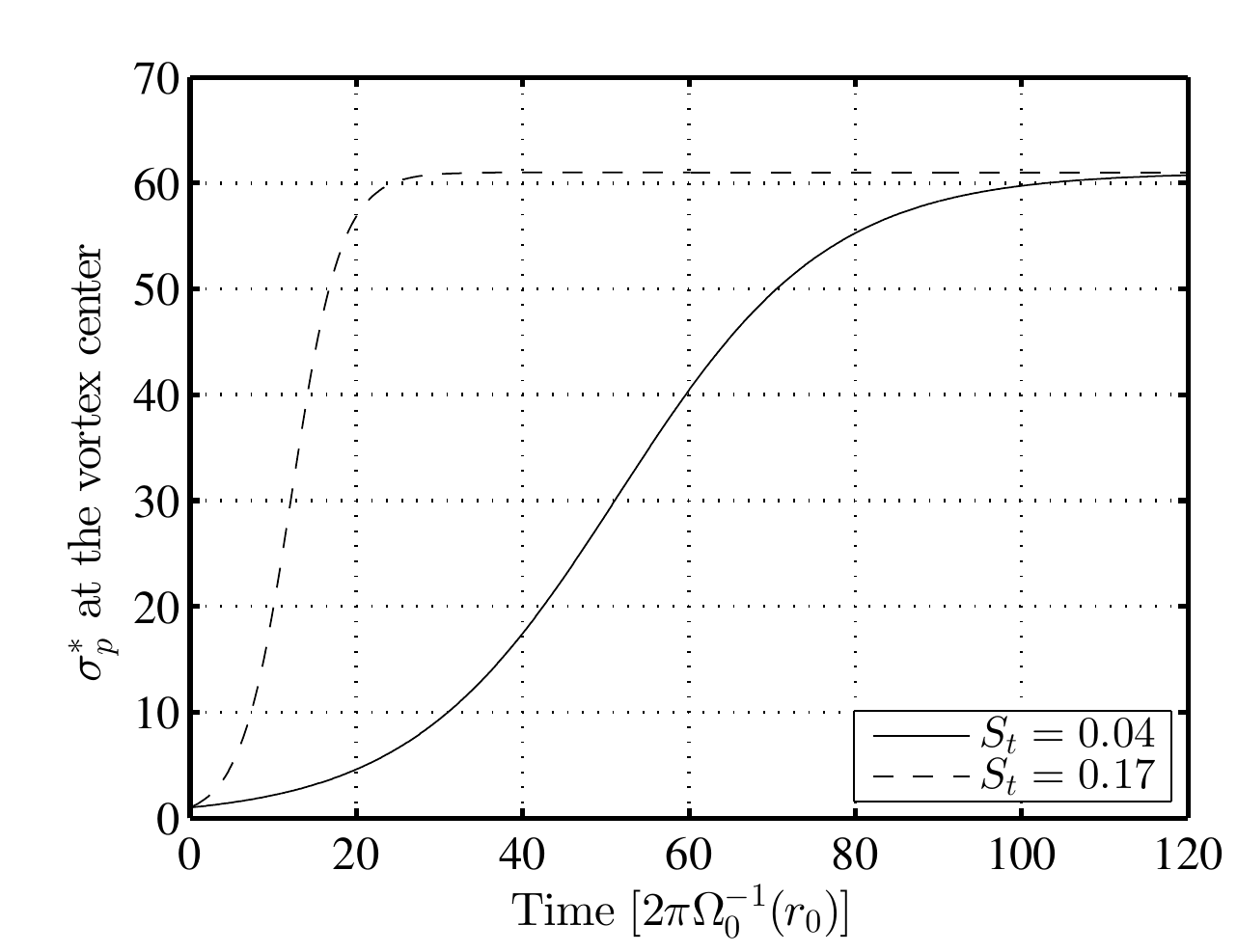} \\
	\includegraphics[height=5.5cm]{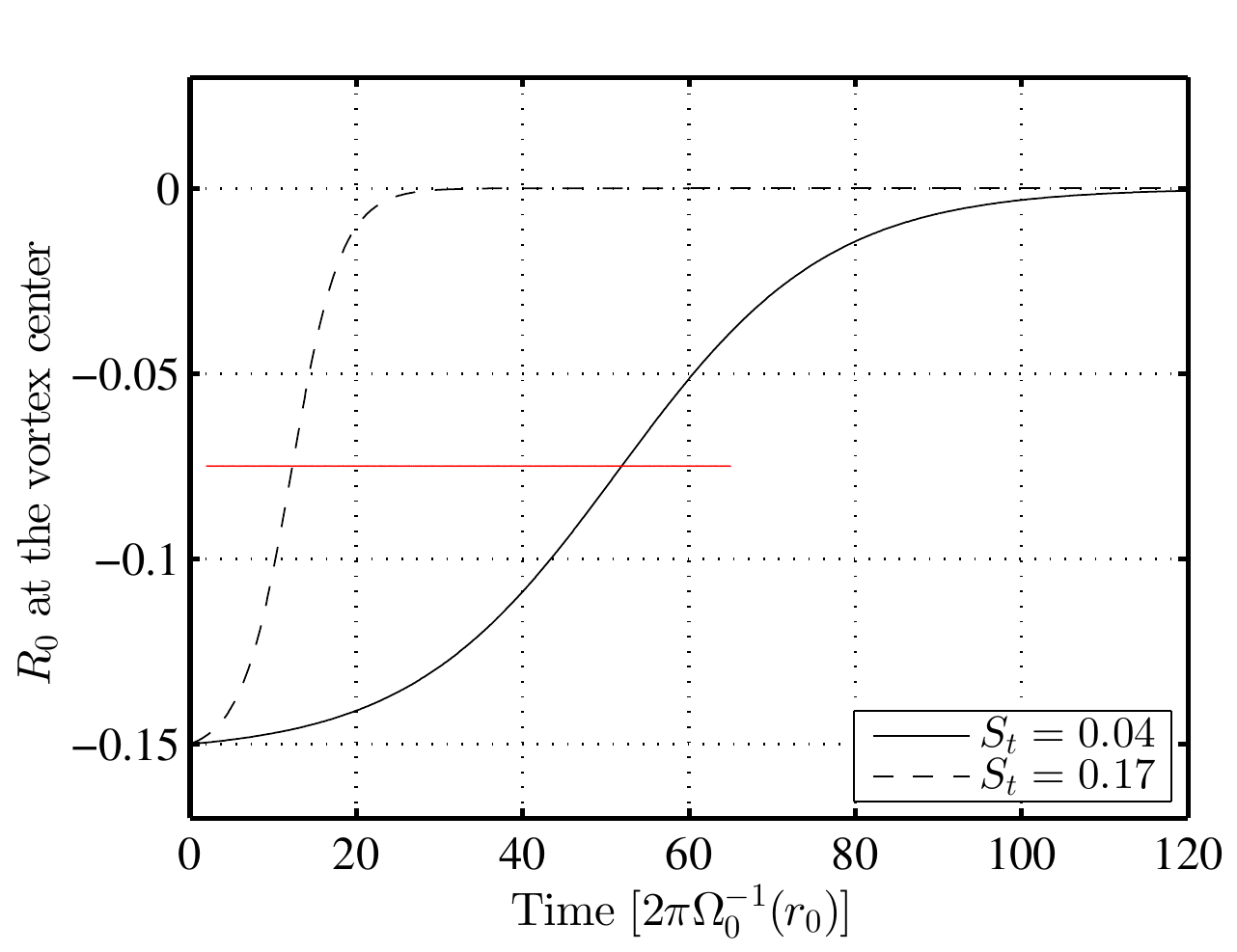}
	\end{tabular}
	\caption{\label{Evo_example} {\it{Top}}: Model of the time evolution of the dust density at the vortex center, $\sigma_p^*$, in a vortex of $R_0=-0.15$, for dust particles of Stokes number $S_t=0.04$ (solid line) and $S_t=0.17$ (dashed line). Here the  global dust-to-gas ratio is $\epsilon=10^{-2}$. {\it{Bottom}}: Evolution of the Rossby number at the vortex center obtained from the model, corresponding to the dust capture shown on the top panel. The horizontal red line defines the time when half the initial value of the Rossby number is reached. It defines the $\tau_{1/2}$ timescale of dust capture. }
      \end{center}
\end{figure}
%
\begin{figure}
	\begin{center}
	\begin{tabular}{c}
	\includegraphics[height=5.5cm]{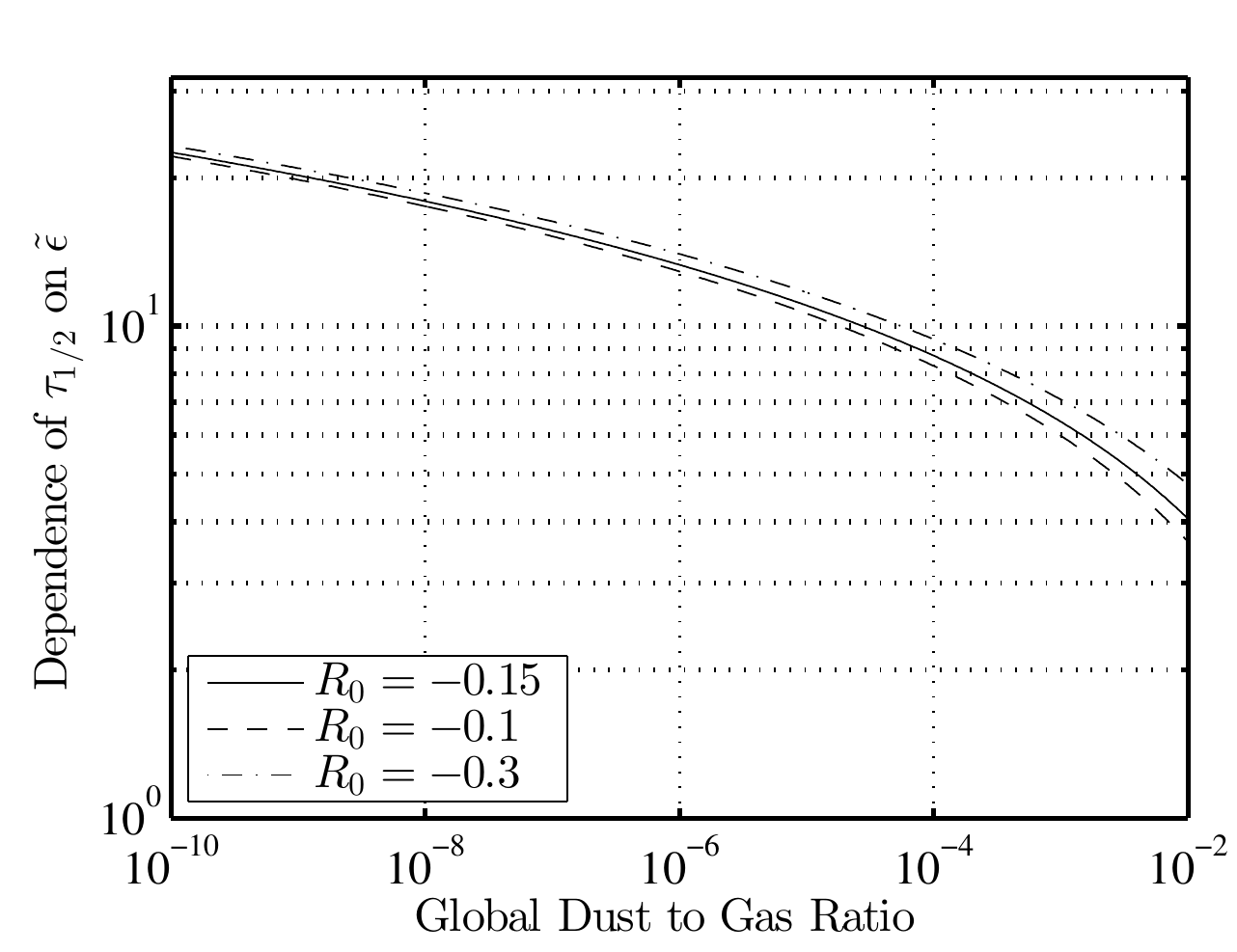} \\
	\includegraphics[height=5.5cm]{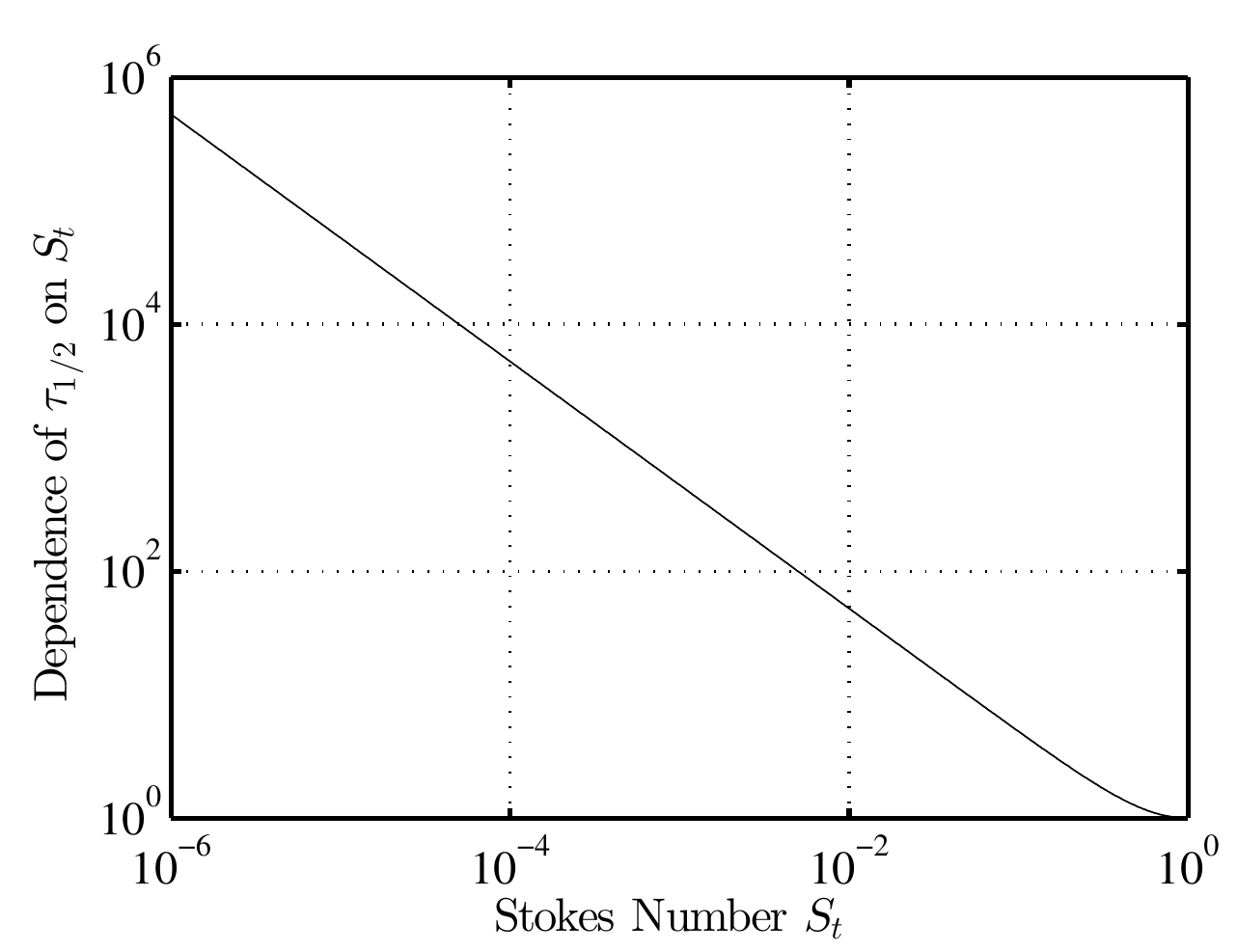}
	\end{tabular}
	\caption{\label{Tau_param} Dependence of the dust capture timescale $\tau_{1/2}$ on the global dust-to-gas ration $\epsilon$ ({\it{top}}) and on the Stokes number $S_t$ ({\it{bottom}}). On the top panel we show the effect of different Rossby numbers, $R_0$, concluding that it does not affect significantly the timescale of capture. }
      \end{center}
\end{figure}

	Even if, for a given vortex, the maximal dust density enhancement reached at very long timescales is independent on the particle Stokes number, the typical time on which the vorticity reduces is a function of $S_t$. To determine this dependence, we need a solution of the evolution system \ref{Sys_evo_1}, as a function of the orbital time $T$. If we write $\tilde{\sigma}_p = \tilde{\epsilon} \sigma_p^*$, one obtains a relatively simple system of equations:
\begin{equation}
\label{Sys_evo_2}
\begin{aligned}
	& \partial_T \tilde{\sigma}_p  = 4 \pi A |R_0| \tilde{\sigma}_p \: , \\
	& \partial_T |R_0|   = - 4 \pi A |R_0| \tilde{\sigma}_p \: .
\end{aligned}
\end{equation}

	This shows that the two quantities $\tilde{\sigma}_p$ and $|R_0|$ have opposite time variations. Because of this property, they depend on the same unknown function of time $f(T)$. Secondly, the time invariant of the system $I_0$ (Eq. \ref{Eq_Time_invariant}) can be written as $I_0 = \tilde{\sigma}_p + |R_0|$. From these two properties, one can search for solutions in the form
\begin{equation}
\begin{aligned}
	& \tilde{\sigma}_p  = \tilde{\epsilon} + |R_0|_{T=0} f(T) \: , \\
	& |R_0|   = |R_0|_{T=0} - |R_0|_{T=0} f(T) \: ,
\end{aligned}
\end{equation}
	using the relation $\tilde{\sigma}_p(T=0) = \tilde{\epsilon}$.

	By inserting these two solutions in the system Eq. \ref{Sys_evo_2}, one obtains a nonlinear differential equation of which the time evolution function $f(T)$ is a solution:
\begin{equation}
	\frac{1}{4\pi A} \partial_T f(T) = \tilde{\epsilon} + \left[|R_0|_{T=0} - \tilde{\epsilon}\right] f(T) - |R_0|_{T=0} f^2(T) \: .
\end{equation}

	Using the Maple software, we find a solution of real values in the form:
\begin{equation}
\label{Time_function_f}
	f(T) = \frac{1- exp(-4\pi A I_0 T)}{1 + |R_0|_{T=0} \tilde{\epsilon}^{-1} exp(-4\pi A I_0 T)} \: .
\end{equation}

	This function has a zero value at $T=0$ and saturates asymptotically to $f(T)=1$ when time goes to infinity. Using different Stokes numbers in a vortex of $|R_0|_{T=0}=0.15$, we show Figure \ref{Evo_example} the typical evolution of the particle density and of the Rossby number at the vortex center obtained from our model, top and bottom respectively. It comes out that the saturation is reached much slowly when the Stokes number is small. But the behaviour of the evolution may also depend on the global dust-to-gas ratio as well as the initial Rossby number. To find out these analytical dependences, we need to define a timescale of the saturation. To this end we set to $\tau_{1/2}$ the time when $f(T)=1/2$, which corresponds to the time when half the initial vorticity is released from the vortex center (red line).

	Inverting $f(\tau_{1/2})=1/2$ in the equation Eq. \ref{Time_function_f}, we obtain the capture timescale:
\begin{equation}
\label{Equ_tau_half}
	\tau_{1/2} = \frac{1}{4 \pi} A^{-1} |R_0|_{T=0}^{-1} \left(1 + \frac{\tilde{\epsilon}}{|R_0|_{T=0}}\right)^{-1} \ln \left(1 + \frac{I_0}{\tilde{\epsilon}}\right) \: ,
\end{equation}
in unit of orbital period $2\pi \; \Omega_k^{-1}(r_0)$.

	We plot on Figure \ref{Tau_param} the dependence of $\tau_{1/2}$ on the different parameters of the capture. On top, the variation of the global dust-to-gas ratio, $\epsilon$, shows that a change of eight orders of magnitude slows down the capture process by only a factor close to $5$. The different lines highlight the negligible impact of the initial Rossby number on the timescale of capture.
	
	On bottom, we show the variation of $\tau_{1/2}$ when changing the Stokes number of the particles. As it comes out from the equation Eq. \ref{Equ_tau_half}, it varies as $A^{-1}(S_t)$, giving this $loglog$ relation with $S_t$ . In the limit of small particles, or small Stokes numbers, we obtain $\tau_{1/2} \propto S_t^{-1}$, which means that the capture process is mostly dominated by the particle size.

	After a period of $2 \times \tau_{1/2}$, the Rossby number at the vortex center is close to zero, which is the time when the vortex instability seems to start in the fiducial run (Section \ref{Sect_Test_case}). In this qualitative way, we can predict the life time of a vortex in a dusty disk. Interestingly, in the last figure of their paper, \cite{Fu2014} show the gas vortex lifetime as function of the particle size and the global dust-to-gas ratio, measured on their simulations. These results have a very similar trend as our analytical results show Figure \ref{Tau_param}. The timescale $\tau_{1/2}$ being dominated by its dependence on the Stokes number, we can argue that the vortex life time is limited by the effect of the capture of the largest dust particles. Even if the proportion of this dust population is very small ($\epsilon=10^{-8}$ for example), it will dominate the evolution of the gas vorticity. In fact an eventual capture of particles five times smaller, but in such a proportion that $\epsilon=10^{-2}$, will happen on the same timescale. This effect of different dust size populations will be further investigated using numerical simulations in a future work.

	The predictions of our model as well as the time evolution of the particle density and the Rossby number during the capture process need to be compared with the results of the numerical simulations, which is the topic of the next section. The comparison should be limited to the regime of validity of the linear analysis, namely before the time when the vorticity at the vortex center cancels. Later on, a two-fluid instability develops inside the vortex, as we discuss later (see Section \ref{Sect_Test_case} and Figure \ref{Evo_Typical}). This induces nonlinear evolution of the coupled flow, and is therefore beyond the reach of our analytical model.

\section{ The Simulation Suite : Results }
\label{Sect_Results}

\begin{figure}
	\begin{center}
	\begin{tabular}{c}
	\includegraphics[height=5.5cm]{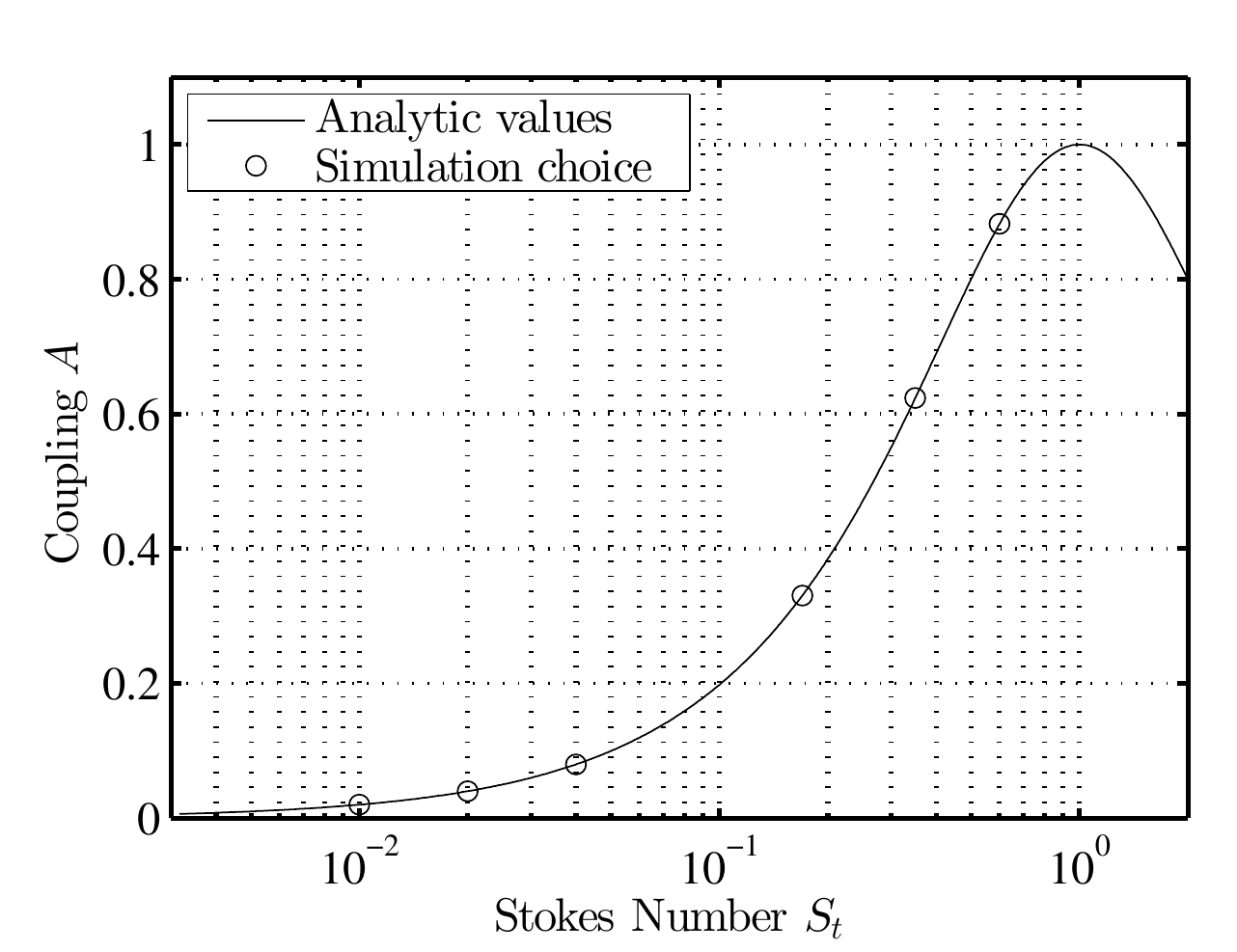}
	\end{tabular}
	\caption{\label{Choice_St} Dependence of the drag force coupling term $A$ on the Stokes number. The circles identify six values used in the simulation setups, namely $S_t=0.01, \:0.02, \:0.04, \:0.17, \:0.35,$ and $0.60$. In particular, the four last ones give almost evenly spaced values of the coupling parameter $A$.}
      \end{center}
\end{figure}

	We performed a set of simulations at a numerical resolution of $(N_r, \: N_\theta) = (2048, \: 4096)$, exploring the parameter spaces of Stokes numbers $S_t$ and global dust-to-gas ratios $\epsilon$. The initial vortex is the same Gaussian model as in Section \ref{Sect_Test_case}, i.e. $(R_0, \: \chi_r, \: \chi_\theta)=(-0.13, \: 0.1, \: 6.5)$. We chose the Stokes numbers to obtain evenly spaced values of the coupling parameter $A$. We present Figure \ref{Choice_St} the different employed values of $S_t$. We will focus on three principal choices: $S_t=0.04$, $S_t=0.17$, and $S_t=0.35$; the other ones have been used mainly for testing and produce coherent results. The global dust-to-gas ratio varies in $\epsilon=10^{-4}$, $\epsilon=10^{-3}$, and $\epsilon=10^{-2}$. We followed the coupled dusty flow during $500$ disk rotations. Some additional runs were carried out for significantly longer timescales ($>1200$ rotations).

\subsection{ Effect of the Stokes number }

\begin{figure*}
	\begin{center}
	\begin{tabular}{ccc}
	\scriptsize{$S_t=0.04$} & \scriptsize{$S_t=0.17$} & \scriptsize{$S_t=0.35$} \\
	\includegraphics[height=4.5cm, trim=4mm 0cm 0cm 0cm, clip=true]{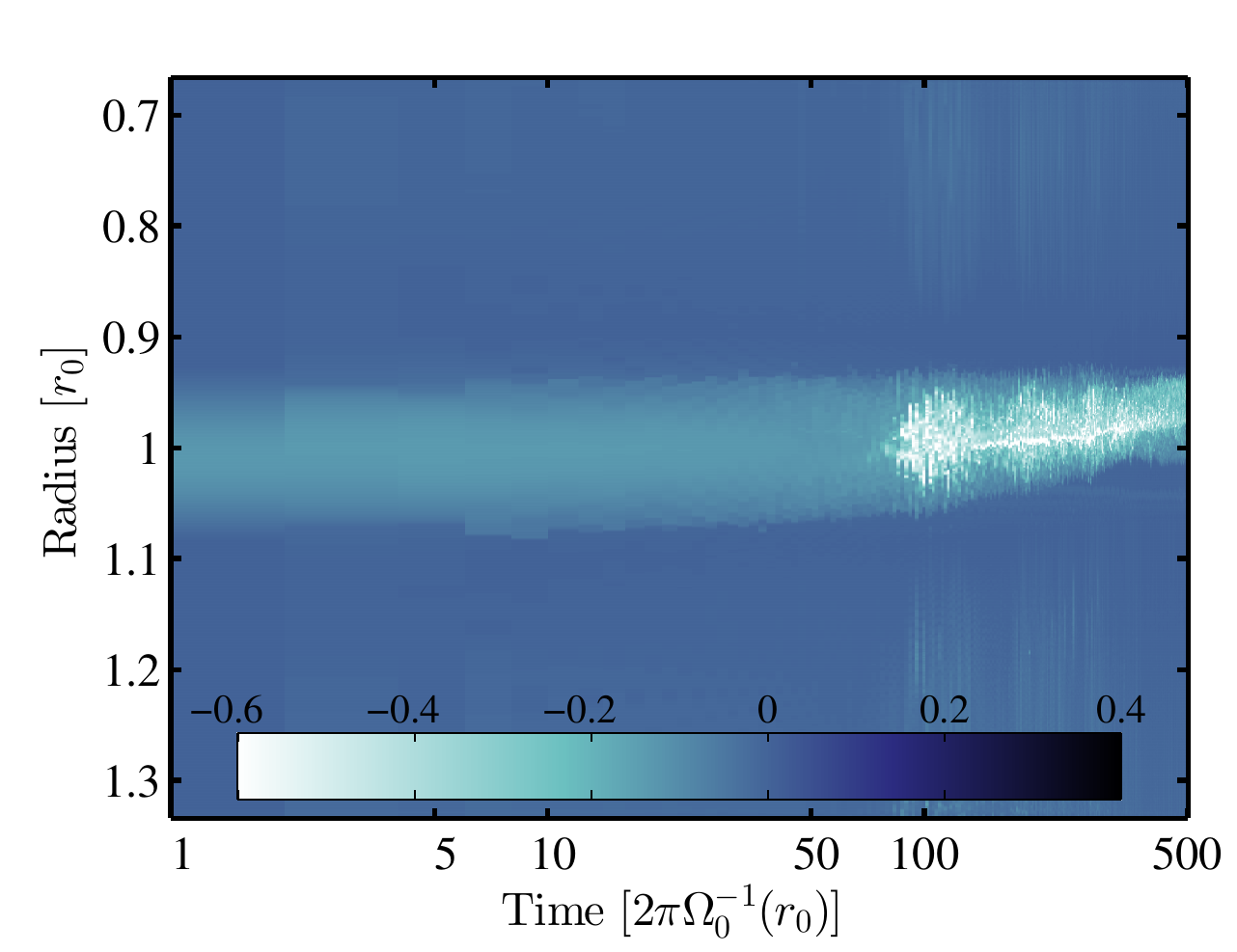} &
	\includegraphics[height=4.5cm, trim=4mm 0cm 0cm 0cm, clip=true]{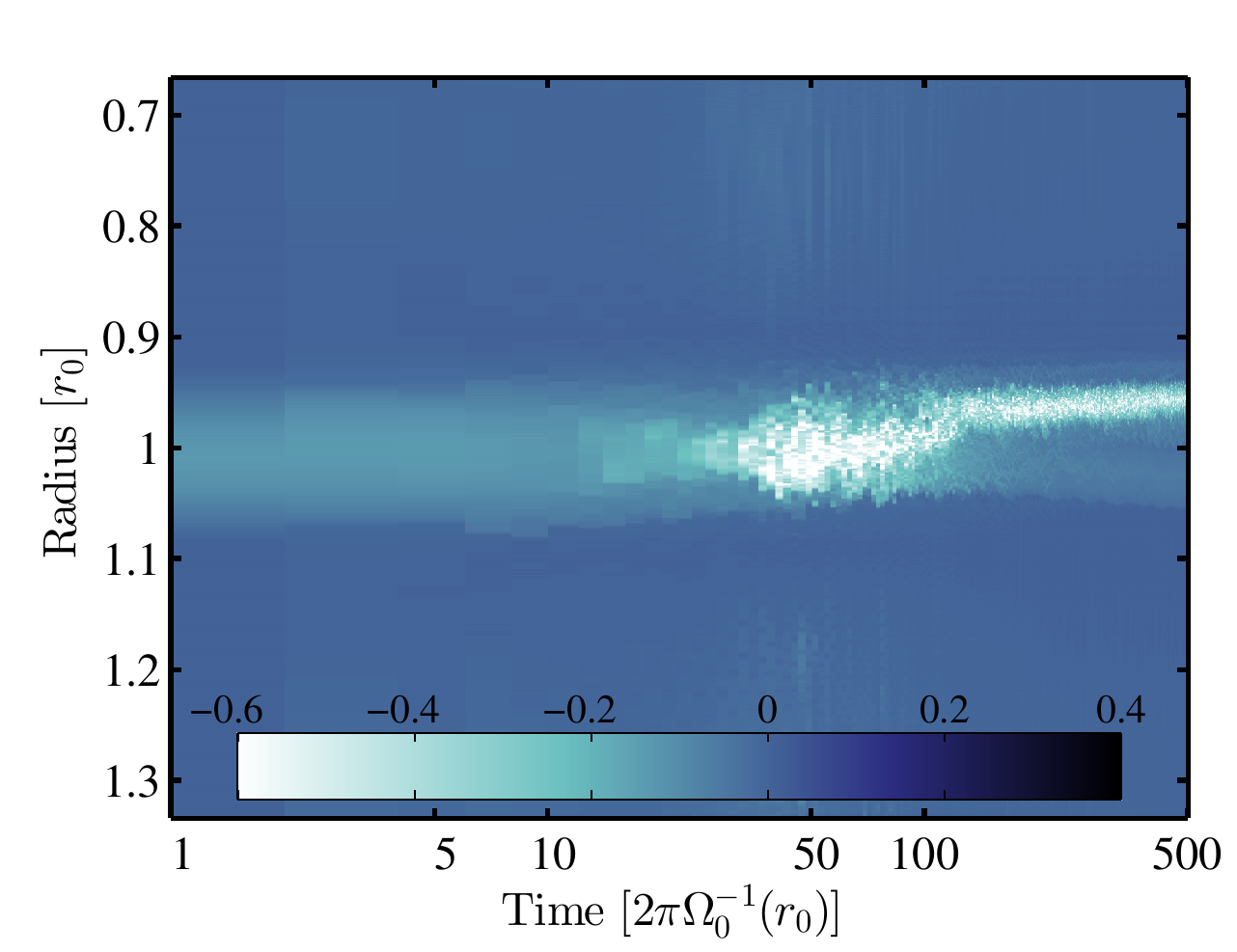} &
	\includegraphics[height=4.5cm, trim=4mm 0cm 0cm 0cm, clip=true]{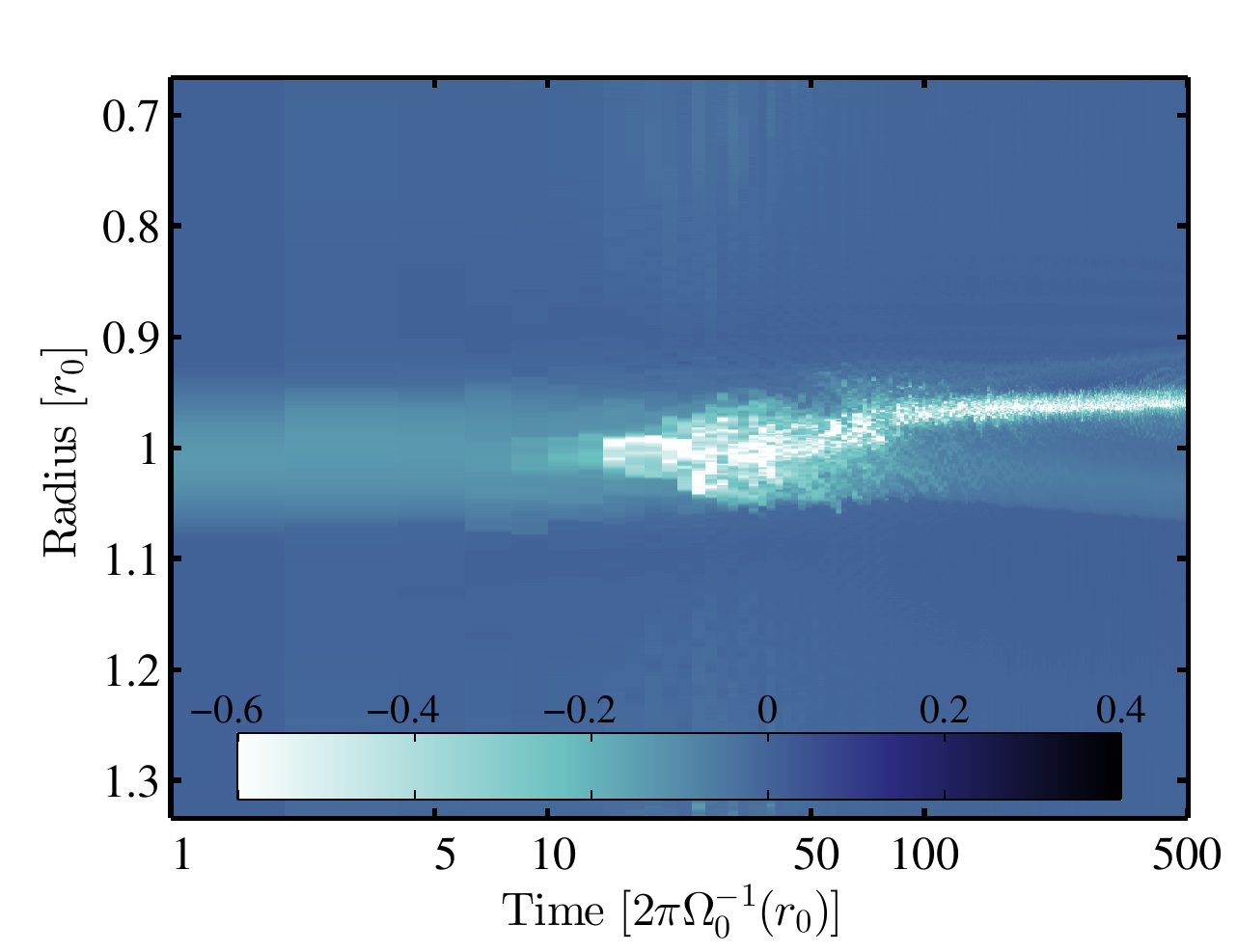} \\

	\includegraphics[height=4.5cm, trim=4mm 0cm 0cm 0cm, clip=true]{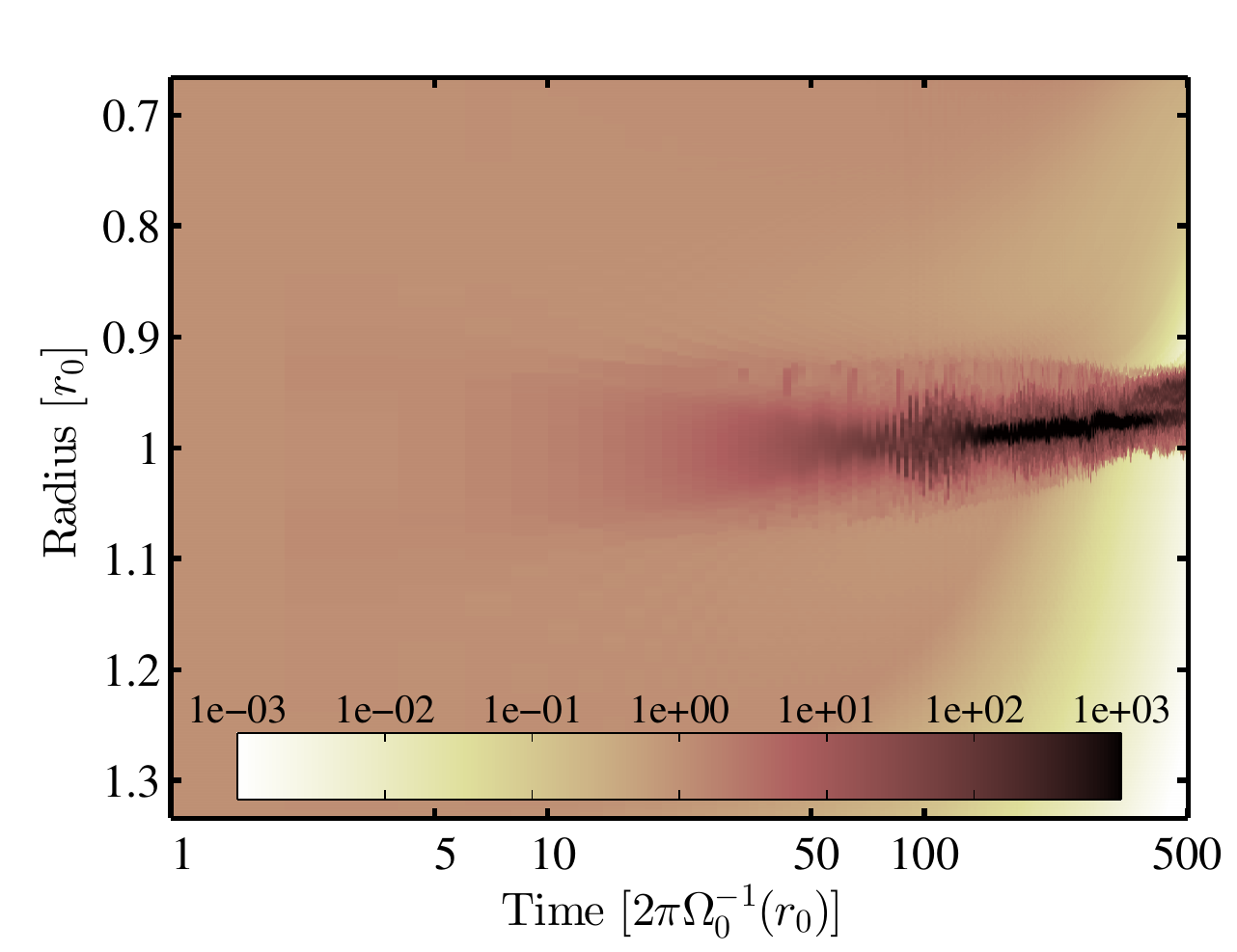} &
	\includegraphics[height=4.5cm, trim=4mm 0cm 0cm 0cm, clip=true]{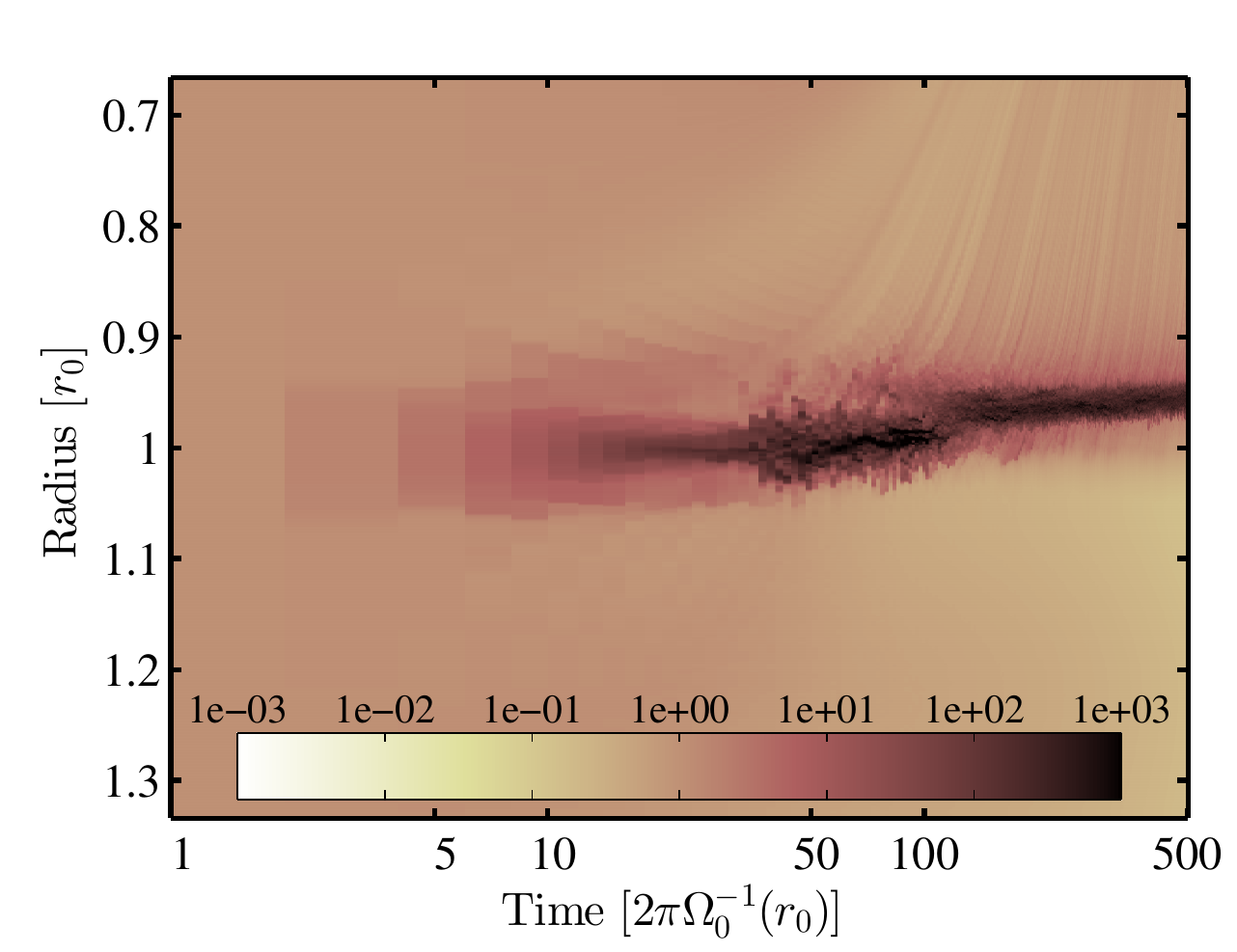} &
	\includegraphics[height=4.5cm, trim=4mm 0cm 0cm 0cm, clip=true]{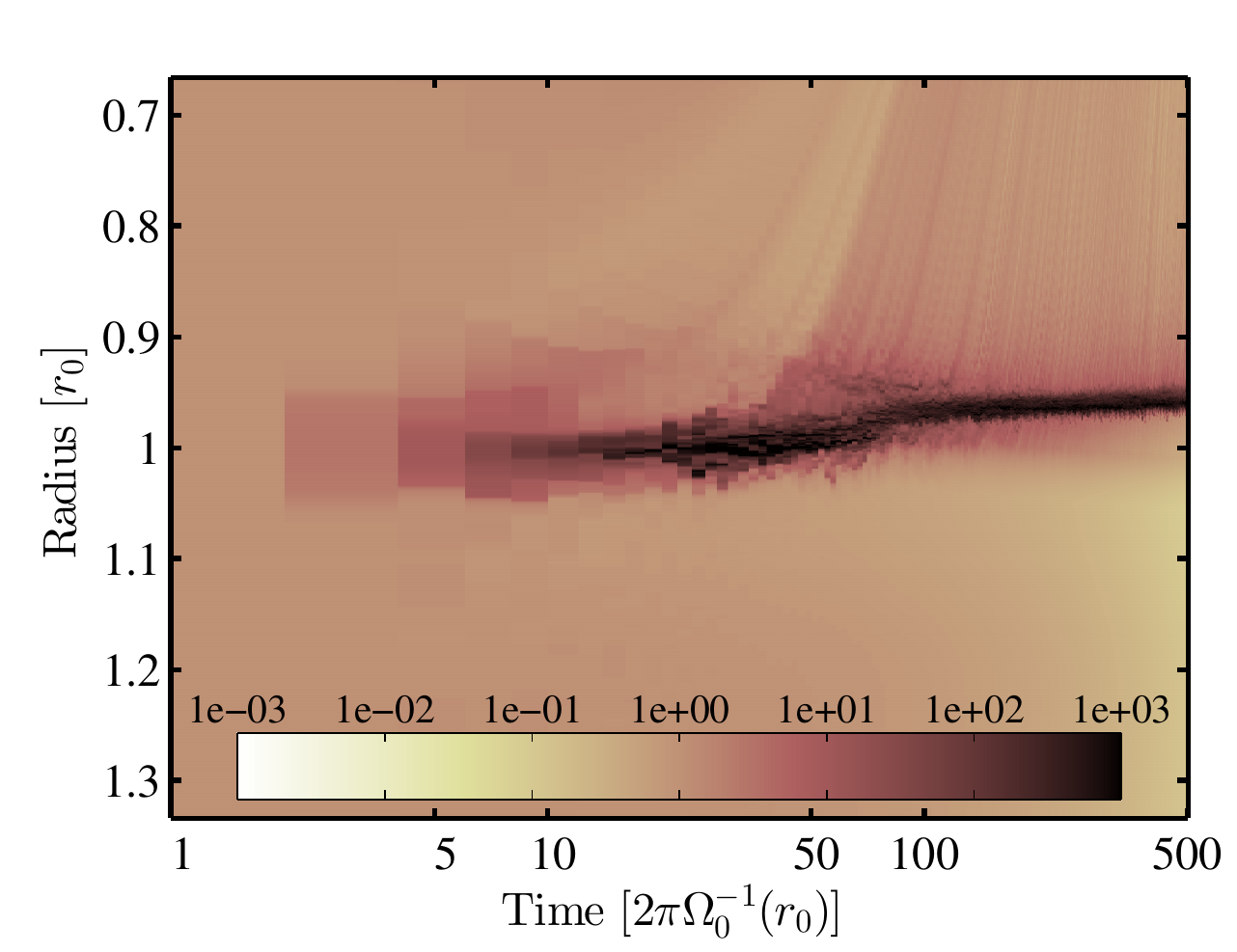} \\

	\end{tabular}
	\caption{\label{Time_evo_epsi2} Time evolution of the radial minimum of the Rossby number ({\it{top}}) and the maximum of the dust density ({\it{bottom}}) for different values of the Stokes number. The global dust-to-gas ratio is $\epsilon=10^{-2}$. {\it{From left to right}}: $S_t=0.04$,  $S_t=0.17$, and $S_t=0.35$ respectively. }
      \end{center}
\end{figure*}

	Here we investigate the impact of the particle size on the long term evolution of the disk, fixing the global dust-to-gas ratio to the standard $\epsilon=10^{-2}$ value. In order to display the evolution of the system in a concise manner, we present the time evolution of the maximum value of the dust density $\sigma_p^*=\sigma_p/\left[\epsilon \sigma_0(r)\right]$ and of the minimum of the Rossby number. These azimuthal extrema are evaluated for each radial position in the disk, and then plotted in a color map as a function of time. The sampling in time is every two disk rotations.

	We summarize Figure \ref{Time_evo_epsi2} the results for $S_t=0.04$, $S_t=0.17$, and $S_t=0.35$, from left to right respectively. Top row is reported the Rossby number and bottom row the particle density, in a logarithmic color scale.

	For these three simulations, we observe a first phase of capture, when the particle density increases in the vortex region and at the vortex center. The minimum Rossby number is not significantly modified. This is a drawback of plotting the minimum of $R_0$. In fact the vorticity is reduced at the vortex center, in accordance with the capture model. It is verified on the $(r, \; \theta)$ maps, as in Figure \ref{Evo_Typical}, row (b). This phase holds for a longer time when reducing the Stokes number. This is in agreement with our model of the capture, and the timescales compare well with the predicted $\tau_{1/2}$. Indeed we measure a capture time approximately equal to $10$, $25$, and $90$ rotations for Stokes numbers decreasing.

	During a second phase, we observe a generation of vorticity inside the vortex. It starts near the vortex center, and as time goes, the whole vortex is perturbed with large vorticity regions. It is clearly visible for the two cases with the largest Stokes numbers. This phase of instability holds during $\sim 250$ rotations for $S_t=0.04$, about $80$ rotations for $S_t=0.17$, and only $\sim 50$ rotations for $S_t=0.35$. During this phase, the particle density increases up to much larger values than the ones reached after the capture phase. Indeed the large vorticity generated inside the vortex speeds up the local capture and thus increases the dust density, as predicted by the model, see Eq. \ref{Sys_evo_1} first line. This unstable phase has a big impact on the local dust-to-gas ratio achieved in the vortex and persists for longer time when the particle size reduces.

	After this unstable phase, we observe the destruction of the vortex and the formation of a dust ring, for the three cases. It happens quickly for the large particles ($S_t=0.35$), but only at $t=400$ rotations for the $S_t=0.04$ case. A sharp region of the disk remains with large vorticity and large dust density. The radial size of this ring increases when reducing the Stokes number. The configuration of this ring is stable in time, although it is not a steady structure, and it lasts for several hundred rotations, until the end of the simulation. We only observe a slow radial drift attributed to the global disk accretion driven by the drag force, which is maintained by the radial gradient of the background pressure. 

	During the evolution of these cases we have identified the three phases proposed in the fiducial case, when $\epsilon = 10^{-2}$. The capture, the vortex instability, and the ring formation happen on timescales that become shorter when the Stokes number of the particles goes close to unity. Even the spatial dimensions on which these processes happen reduce when $S_t \to 1$. In the next section we check if the global dust-to-gas ratio affects these observations.

\subsection{ Effect of the global dust-to-gas ratio }

\begin{figure*}
	\begin{center}
	\begin{tabular}{ccc}
	\scriptsize{$S_t=0.04$} & \scriptsize{$S_t=0.17$} & \scriptsize{$S_t=0.35$} \\
	\includegraphics[height=4.5cm, trim=4mm 0cm 0cm 0cm, clip=true]{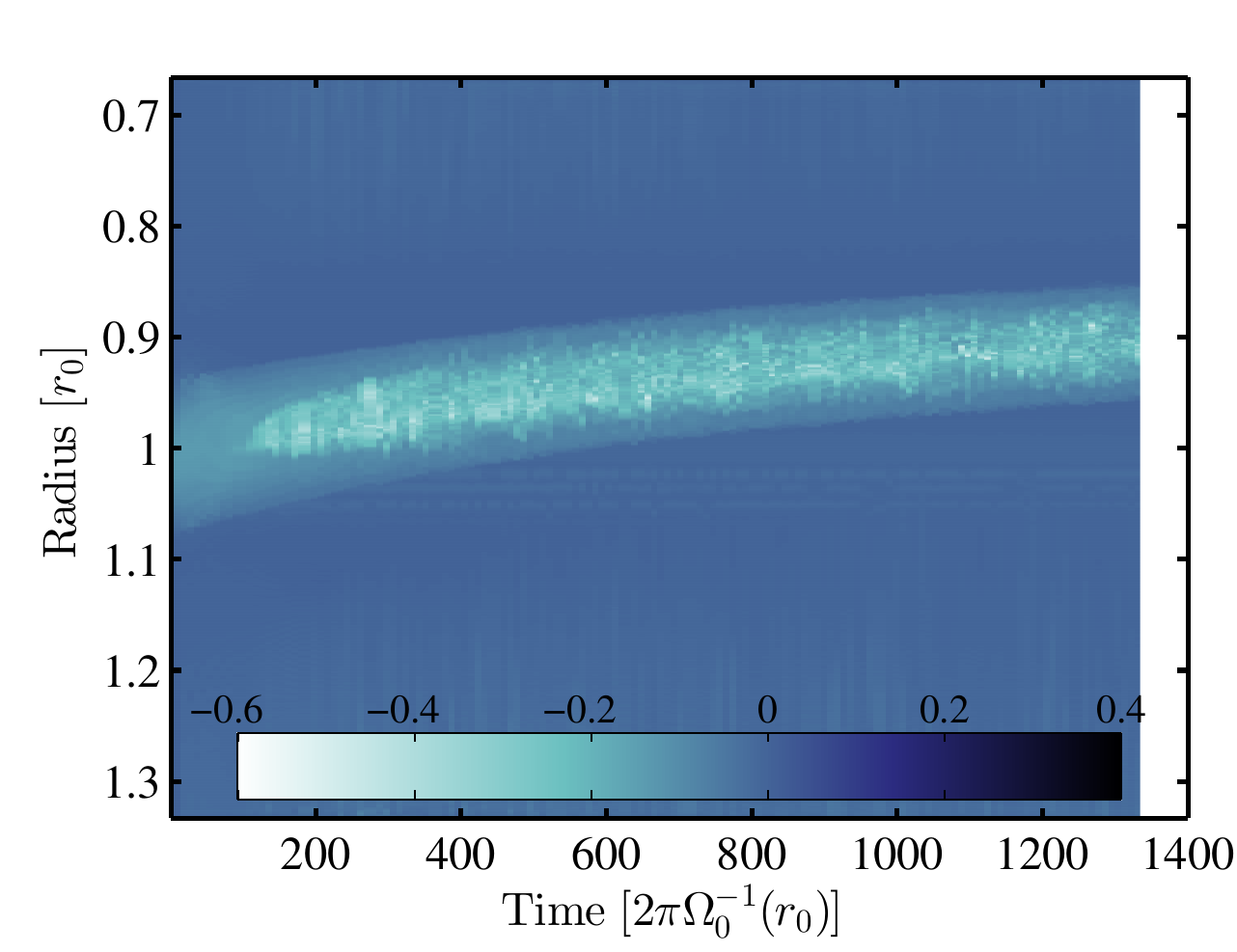} &
	\includegraphics[height=4.5cm, trim=4mm 0cm 0cm 0cm, clip=true]{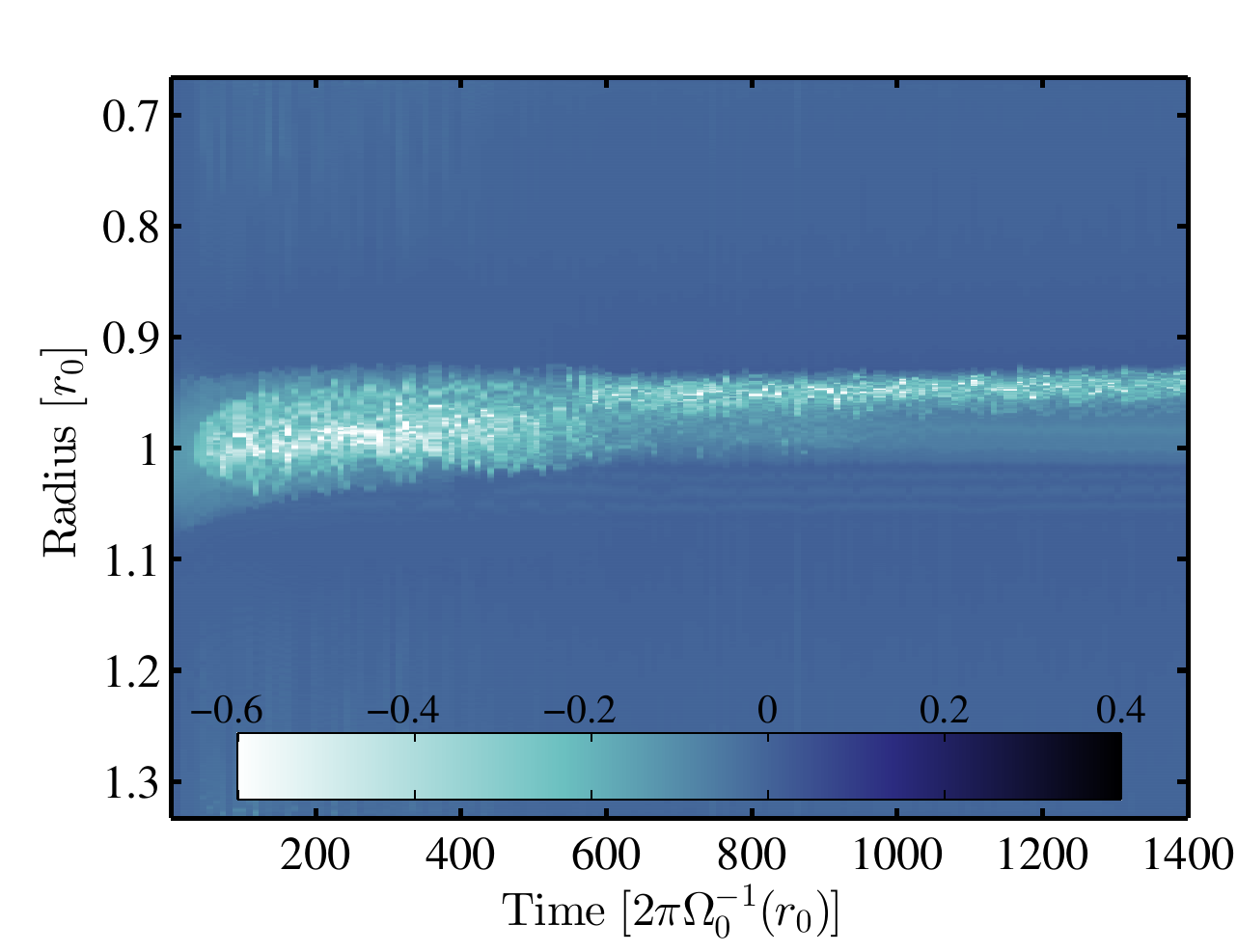} &
	\includegraphics[height=4.5cm, trim=4mm 0cm 0cm 0cm, clip=true]{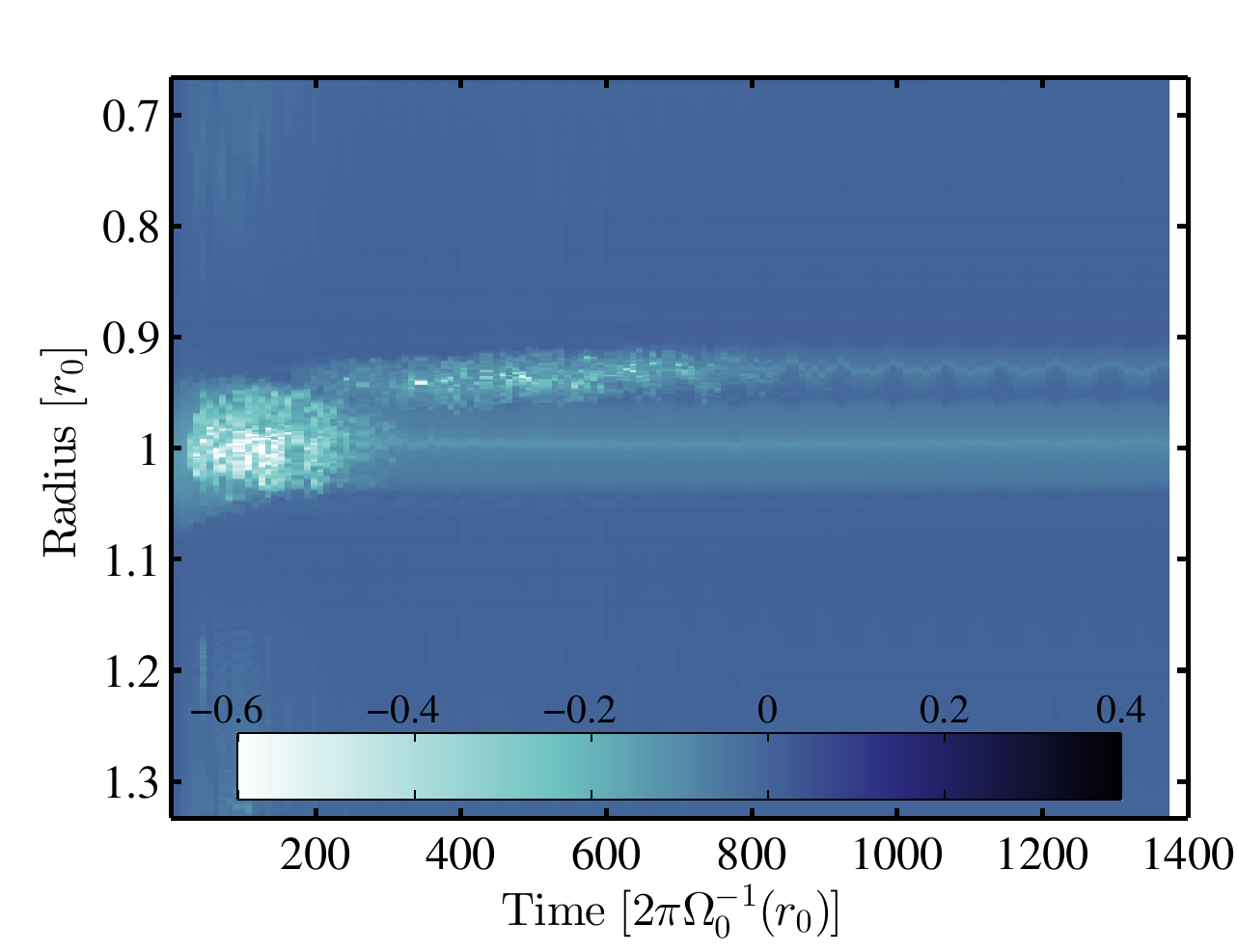} \\

	\includegraphics[height=4.5cm, trim=4mm 0cm 0cm 0cm, clip=true]{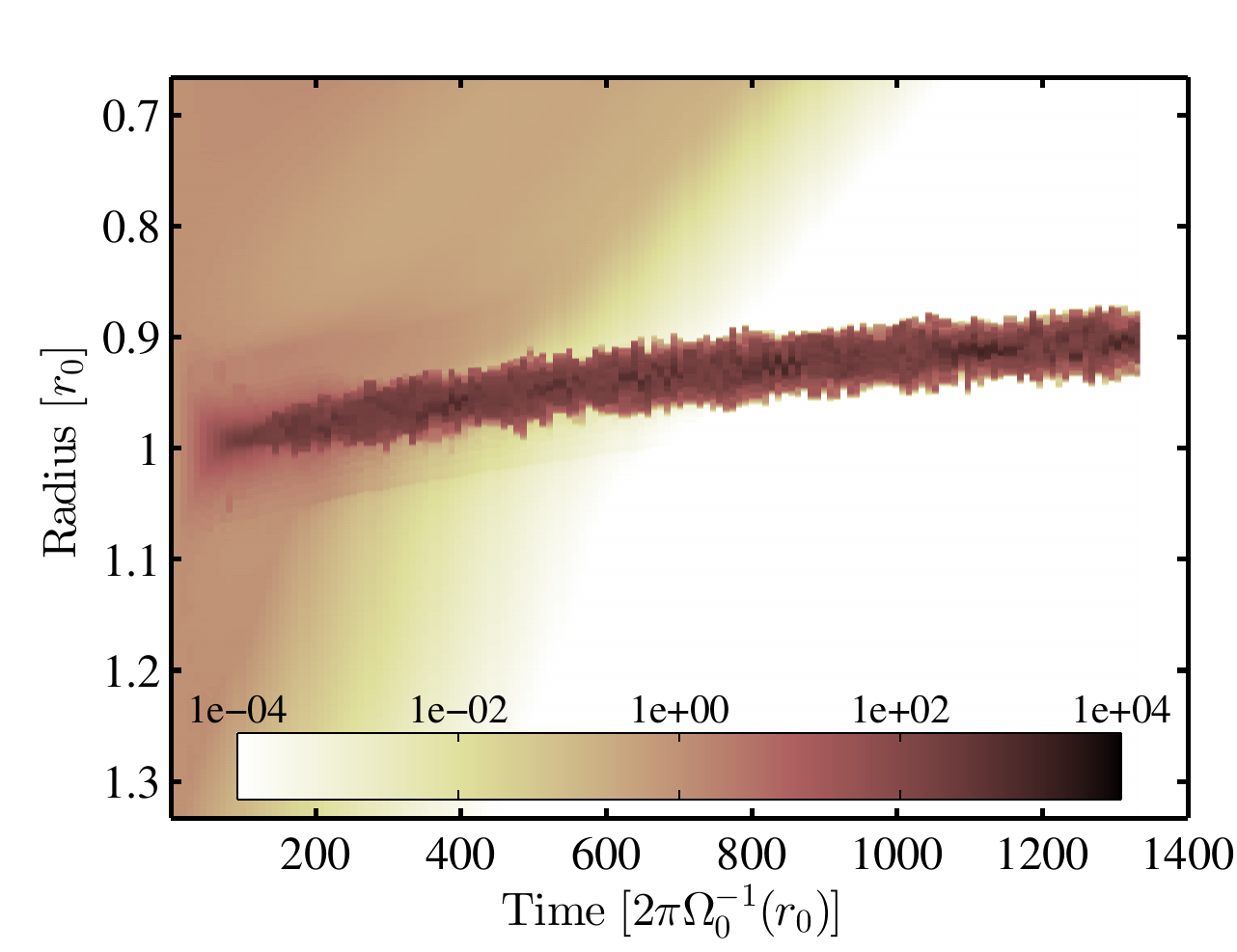} &
	\includegraphics[height=4.5cm, trim=4mm 0cm 0cm 0cm, clip=true]{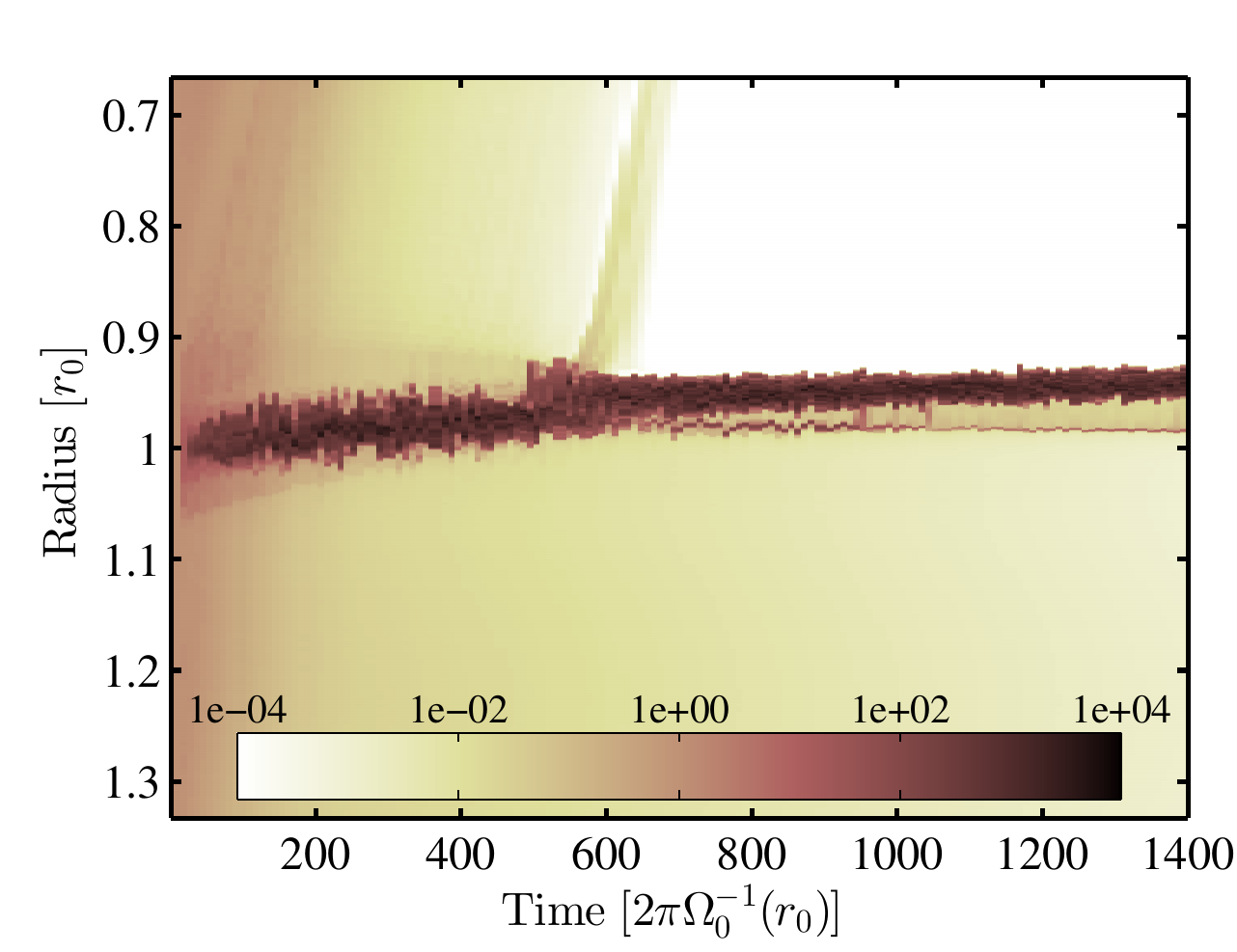} &
	\includegraphics[height=4.5cm, trim=4mm 0cm 0cm 0cm, clip=true]{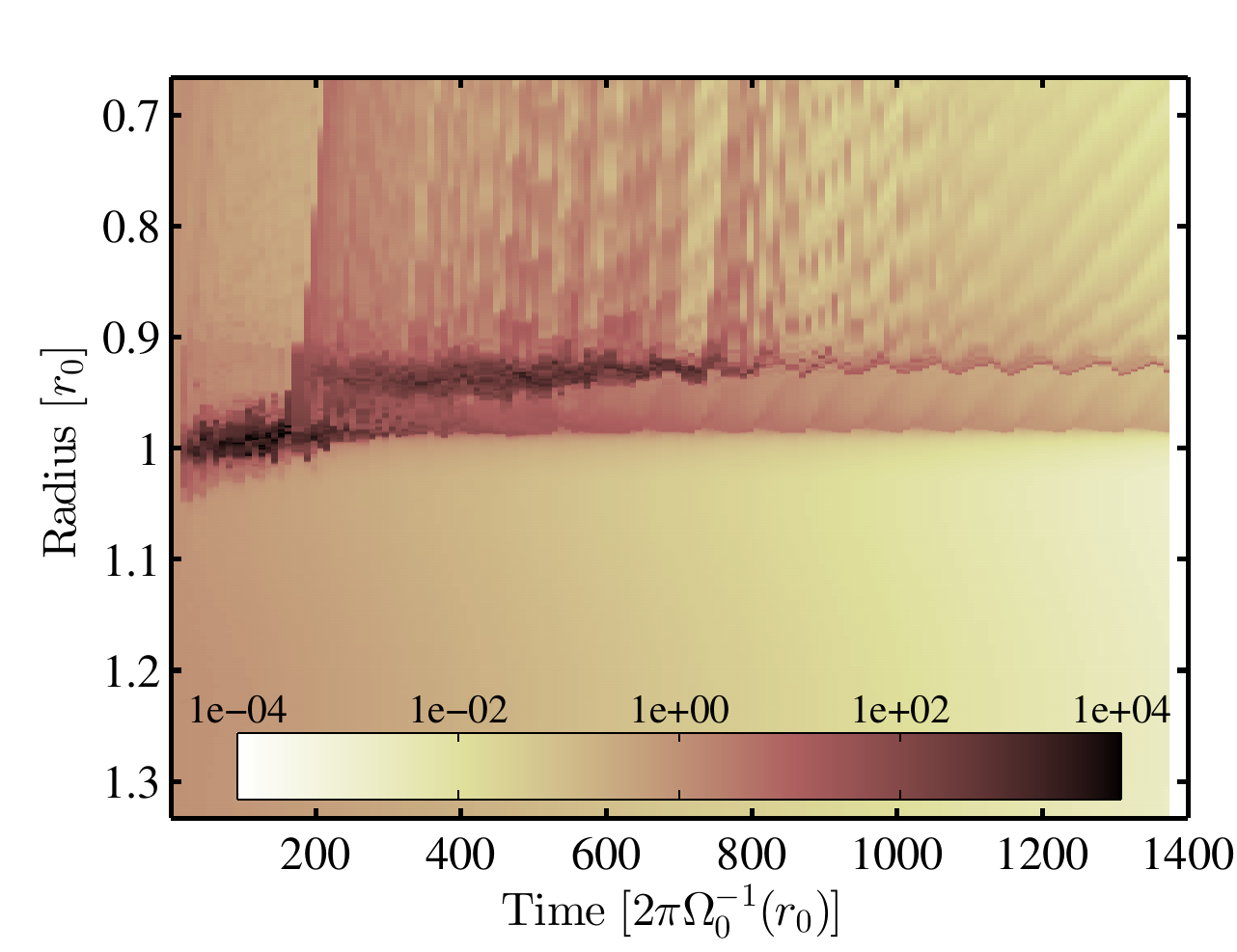} \\

	\end{tabular}
	\caption{\label{Time_evo_epsi3_long} Long term evolution of the radial minimum of the Rossby number ({\it{top}}) and the maximum of the dust density ({\it{bottom}}) for different values of the Stokes number. The global dust-to-gas ratio is $\epsilon=10^{-3}$. {\it{From left to right}}: $S_t=0.04$,  $S_t=0.17$, and $S_t=0.35$ respectively. }
      \end{center}
\end{figure*}

\begin{figure*}
	\begin{center}
	\begin{tabular}{ccc}
	\scriptsize{$S_t=0.04$} & \scriptsize{$S_t=0.17$} & \scriptsize{$S_t=0.35$} \\
	\includegraphics[height=4.5cm, trim=4mm 0cm 0cm 0cm, clip=true]{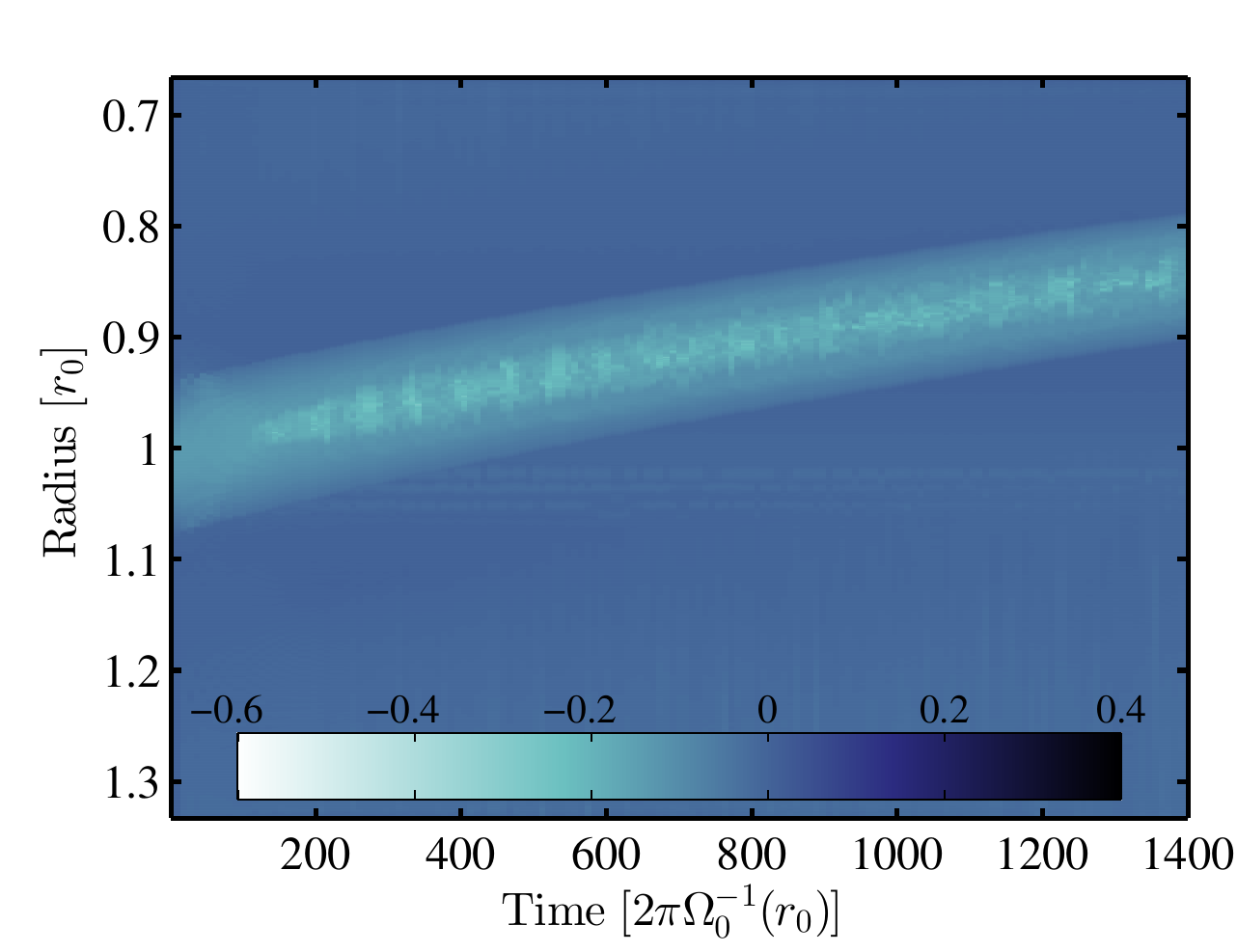} &
	\includegraphics[height=4.5cm, trim=4mm 0cm 0cm 0cm, clip=true]{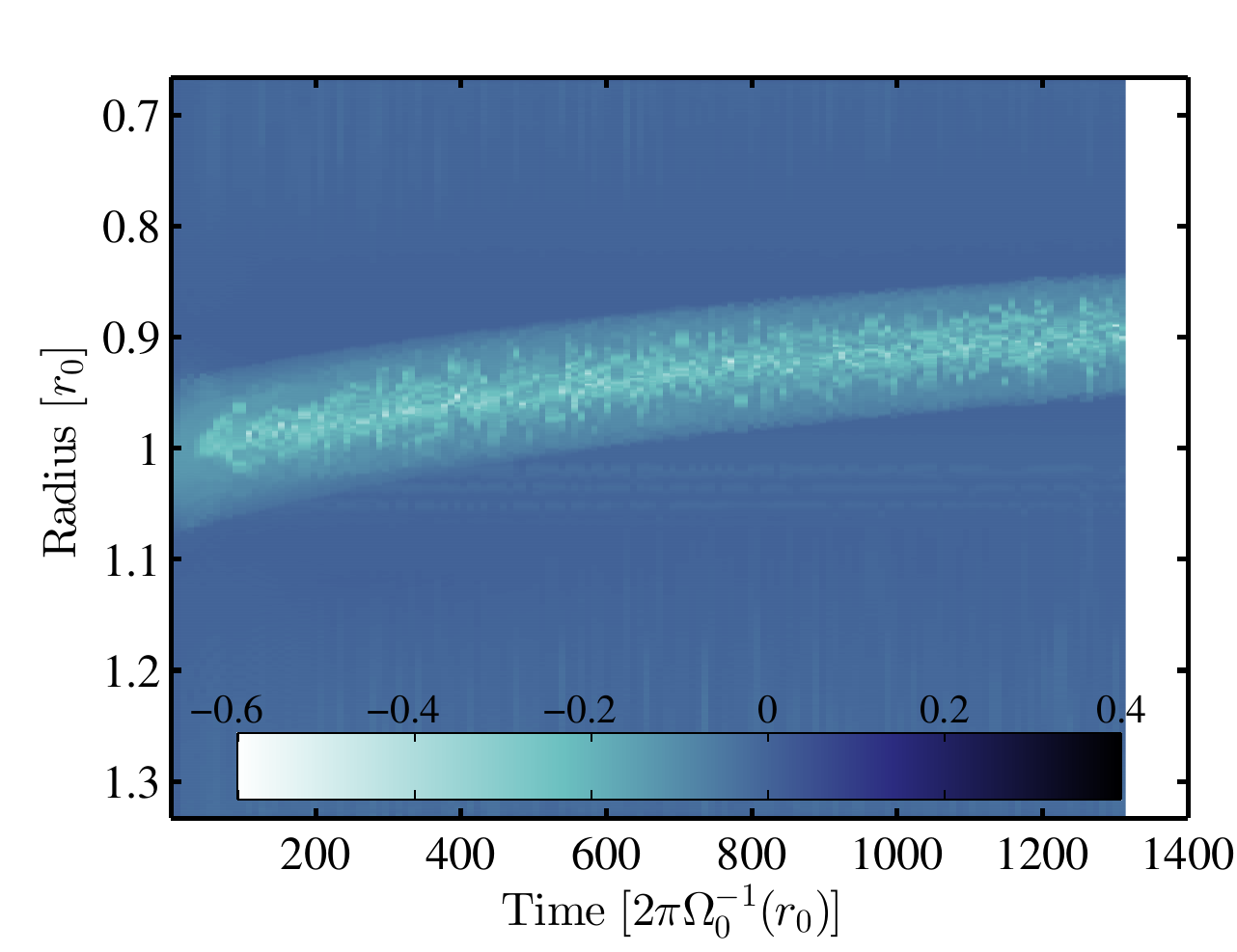} &
	\includegraphics[height=4.5cm, trim=4mm 0cm 0cm 0cm, clip=true]{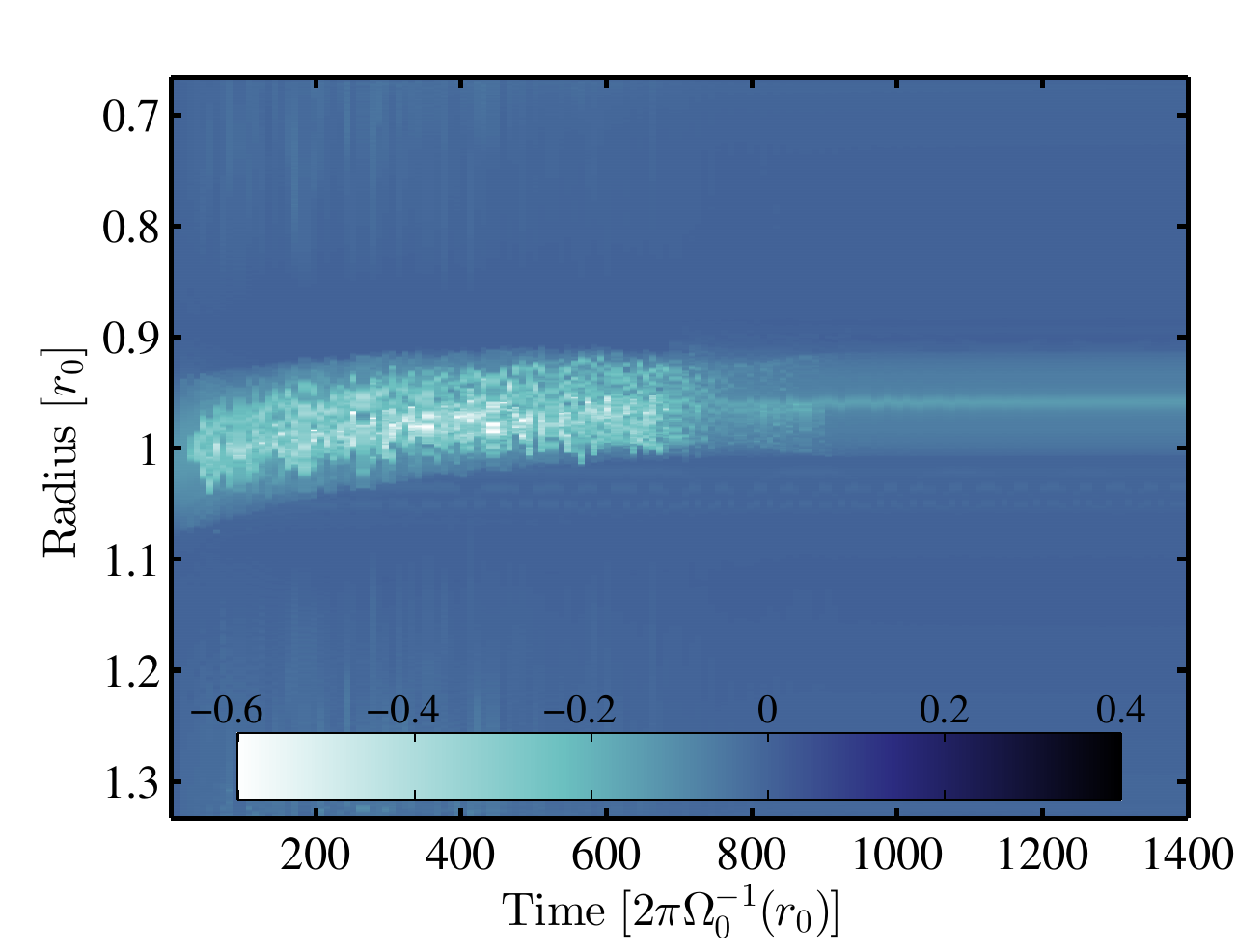} \\

	\includegraphics[height=4.5cm, trim=4mm 0cm 0cm 0cm, clip=true]{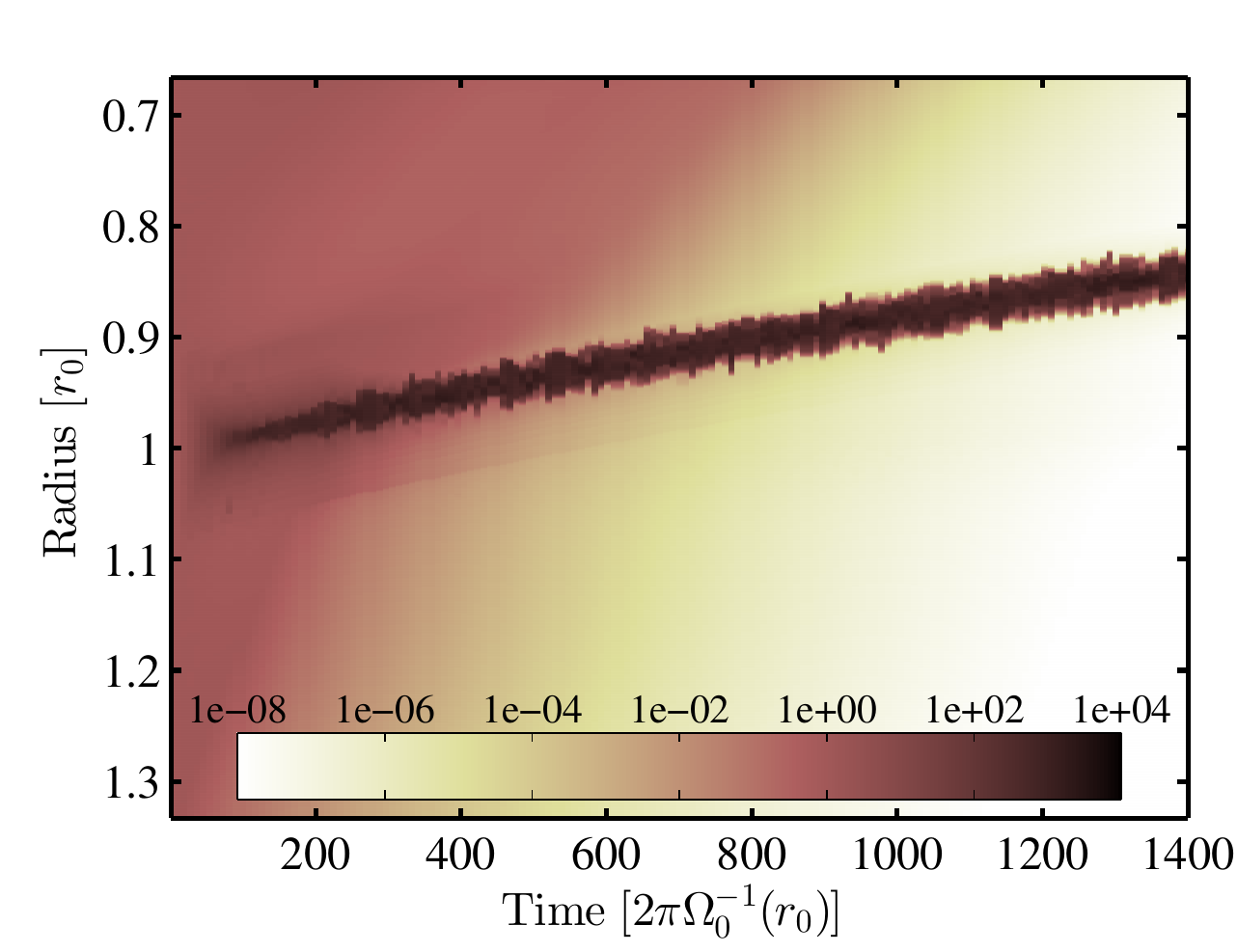} &
	\includegraphics[height=4.5cm, trim=4mm 0cm 0cm 0cm, clip=true]{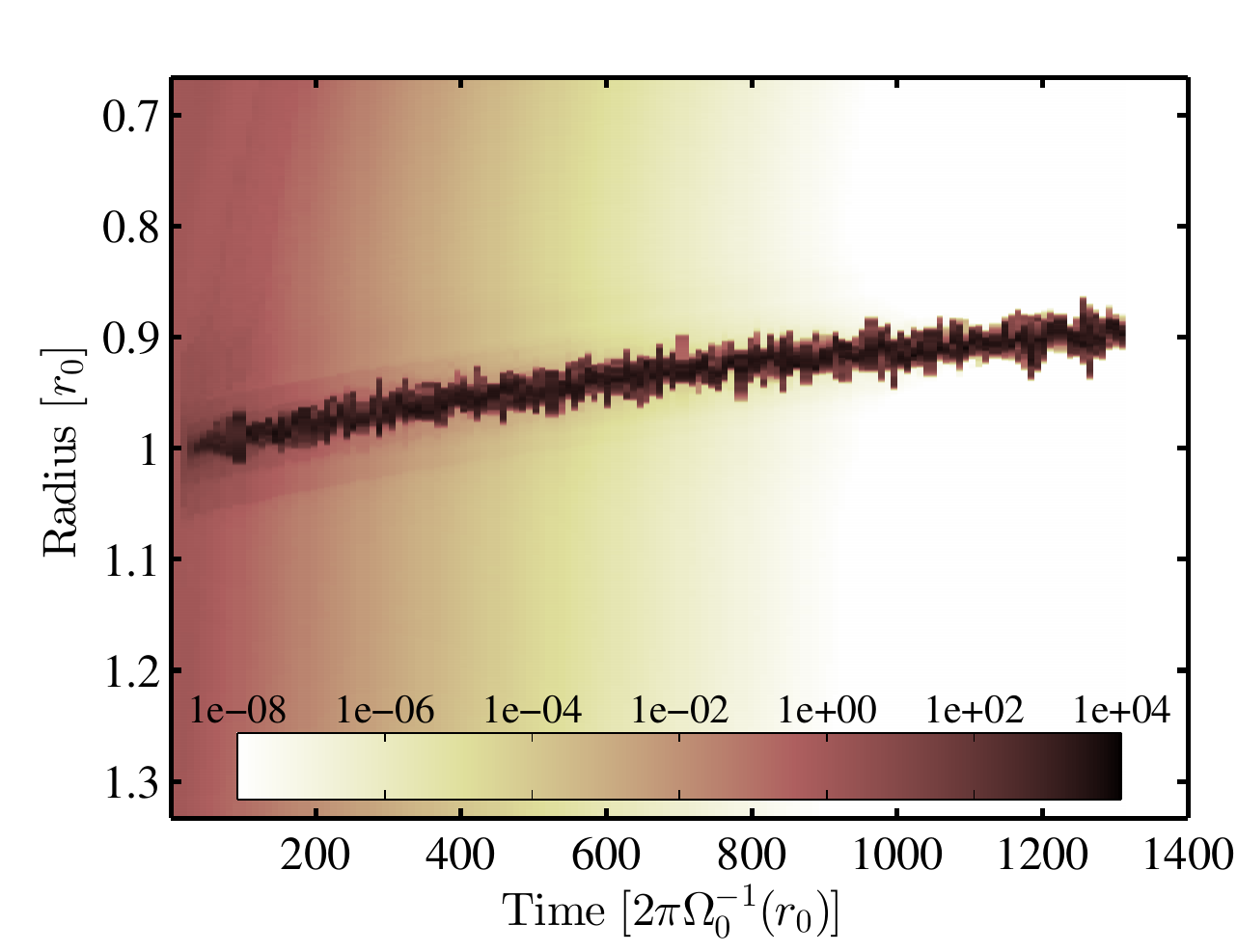} &
	\includegraphics[height=4.5cm, trim=4mm 0cm 0cm 0cm, clip=true]{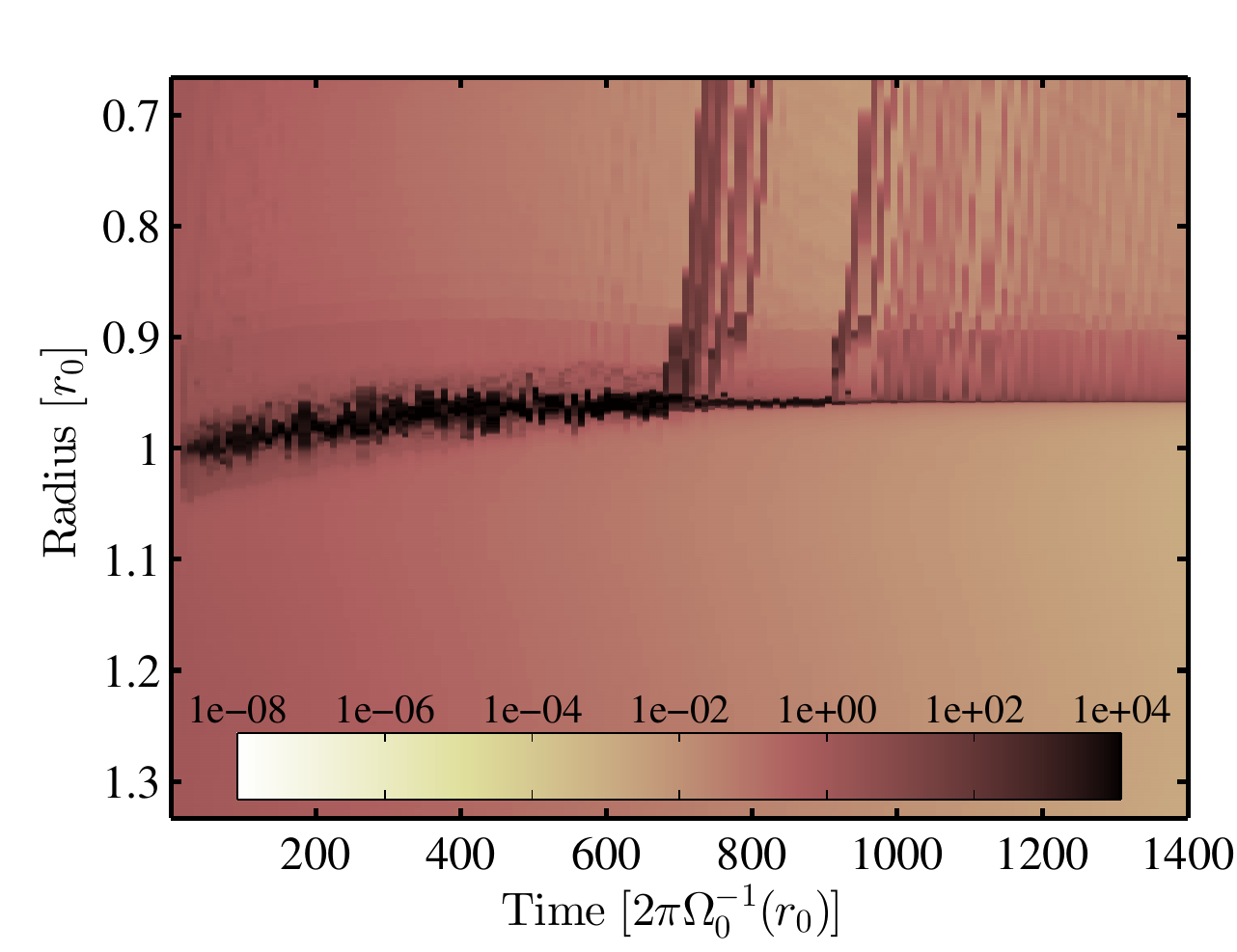} \\

	\end{tabular}
	\caption{\label{Time_evo_epsi4_long} Long term evolution of the radial minimum of the Rossby number ({\it{top}}) and the maximum of the dust density ({\it{bottom}}) for different values of the Stokes number. The global dust-to-gas ratio is $\epsilon=10^{-4}$. {\it{From left to right}}: $S_t=0.04$,  $S_t=0.17$, and $S_t=0.35$ respectively. }
      \end{center}
\end{figure*}

	Using the same initial vortex as well as the three Stokes numbers, we performed long term evolution ($\sim 1400$ rotations) with initial dust-to-gas ratios $\epsilon = 10^{-3}$ and $\epsilon = 10^{-4}$. The results are shown Figure \ref{Time_evo_epsi3_long} and \ref{Time_evo_epsi4_long} respectively.

	Whatever the initial dust-to-gas ratio, we observe the dust capture for all the particle sizes. For $\epsilon=10^{-3}$, it happens approximately on $25$, $40$, and $120$ rotations when decreasing the Stokes numbers. These timescales are comparable to the analytical estimates from the model. As expected, the reduced initial dust-to-gas ratio delays the duration of the capture. For $\epsilon=10^{-4}$, we measure approximately a capture timescale of $30$, $50$, and $150$ rotations for $S_t=0.35$, $S_t=0.17$, and $S_t=0.04$ respectively. It is almost twice the duration of this phase measured in the case with $\epsilon=10^{-2}$, which is in agreement with the $\tau_{1/2}$ dependence on $\epsilon$ (see Figure \ref{Tau_param}, top).

	When $\epsilon=10^{-3}$, the vortex instability phase lasts for a longer time than with $\epsilon=10^{-2}$. The unstable dusty vortex resists during about $250$ rotations for $S_t=0.35$ and $450$ rotations for $S_t=0.17$, and after, a ring is formed in the $0.9<r<0.95$ $r_0$ region. A small dust overdensity is also left at $r \simeq 0.98$, which is the position of the vortex center. However, in the $S_t=0.04$ case, no ring formation happens before the end of the simulation. When $\epsilon = 10^{-4}$, the only particle size for which we obtain a ring formation is $S_t=0.35$, after about $700$ rotations in the course of which the unstable dusty vortex survives. In this case the dust ring is very narrow, and we identify less structures than in the largest $\epsilon$ values.

	When we reduce the global dust-to-gas ratio, the captured dust inside the vortex is more and more confined, with large density gradients. The dust density contrast between the center and the outer parts of the vortex is much larger in the small $\epsilon$ cases than in the classical $\epsilon=10^{-2}$ case. At the same time, the vorticity perturbations generated inside the vortex are also smaller and restricted to a region closer to the vortex center as the global dust-to-gas ratio decreases. It is clearly visible when we compare the $S_t=0.04$ cases.

	However, we can identify a very slow spreading or expansion of the dust overdensity inside the vortex, that could lead to a very late ring formation. Comparing the duration of the vortex unstable phase between $S_t=0.17$ and $S_t=0.35$, in the $\epsilon=10^{-3}$ case, we obtain a factor close to $2$. As a consequence, if this ratio of timescales is similar when $\epsilon=10^{-4}$, a ring formation could be possible for $S_t=0.17$ after $2\times 700=1400$ rotations, which is after the end of the simulation. It could be much later in the case with $S_t=0.04$. This speculation has to be confirmed in a future study; but it shows that long term disk evolution is crucial in this context.
	 
	Finally, we stress out that during the vortex unstable phase, we observe an inward radial migration of the dusty structure. This vortex migration was already discovered and described in \cite{Paardekooper2010} and also in \cite{Surville2012, Surville2013}. When $\epsilon=10^{-4}$ it is almost at a constant rate, whereas the migration slows down when $\epsilon=10^{-3}$. However, this migration is still slow, with the maximum rate in the case with $\epsilon=10^{-4}$ and $S_t=0.04$, equal to $-1.1 \times 10^{-4}$ $r_0$ per period. The impact on the local value of the Stokes number is very limited in this case, estimated at $\sim 20\%$ after $1400$ orbits. This vortex migration induces a negligible effect on the coupling between gas and dust in our simulations. 

	To summarize, we observe the particle capture inside the vortex for all cases, on timescales compatible with the analytical model. The results also show that when the dust-to-gas ratio reduces, small particles stay within the unstable vortex for a long time, and the ring formation seems more and more unlikely, but very long term simulations are needed to confirm this statement. However the local dust density is enhanced by many orders of magnitude, even when starting with $\epsilon=10^{-4}$, whatever the Stokes number of the particles. The confinement of the dust in compact regions of small extension is obtained for all cases, which is very promising as concerns the possible gravitational evolution of dusty disks.
	
	In the next section, we will explore the impact of the vortex size, in particular its radial extent, on the vortex instability phase and on the ring formation.

\subsection{ Effect of the vortex size }

	We studied the disk evolution for the same set of $S_t$ and $\epsilon$ parameters as above, but using a different initial vortex:
\begin{equation}
	(R_0, \: \chi_r, \: \chi_\theta)=(-0.13, \: 0.06, \: 6.5) \: .
\end{equation}
	This vortex has the same Rossby number and aspect ratio as the first one, but has a smaller radial extent. This reduction by a factor $\sim 1.7$ modifies the maximal gas density at the vortex center. For the first vortex, the maximal gas density was $\sim 1.6$ times larger than the background, reducing the effective Stokes numbers at the vortex center by a factor $0.63$. For this second vortex, the maximal gas density is $\sim 1.2$ times larger than the background disk, implying that the effective Stokes numbers are $\sim 1.33$ times larger than with the first vortex. This can affect a little the evolution of the system, but we argue that it will not be a major effect, and most of the differences will be due to the radical change in the vortex size.

\begin{figure*}
	\begin{center}
	\begin{tabular}{ccc}
	\scriptsize{$S_t=0.04$} & \scriptsize{$S_t=0.17$} & \scriptsize{$S_t=0.35$} \\
	\includegraphics[height=4.5cm, trim=4mm 0cm 0cm 0cm, clip=true]{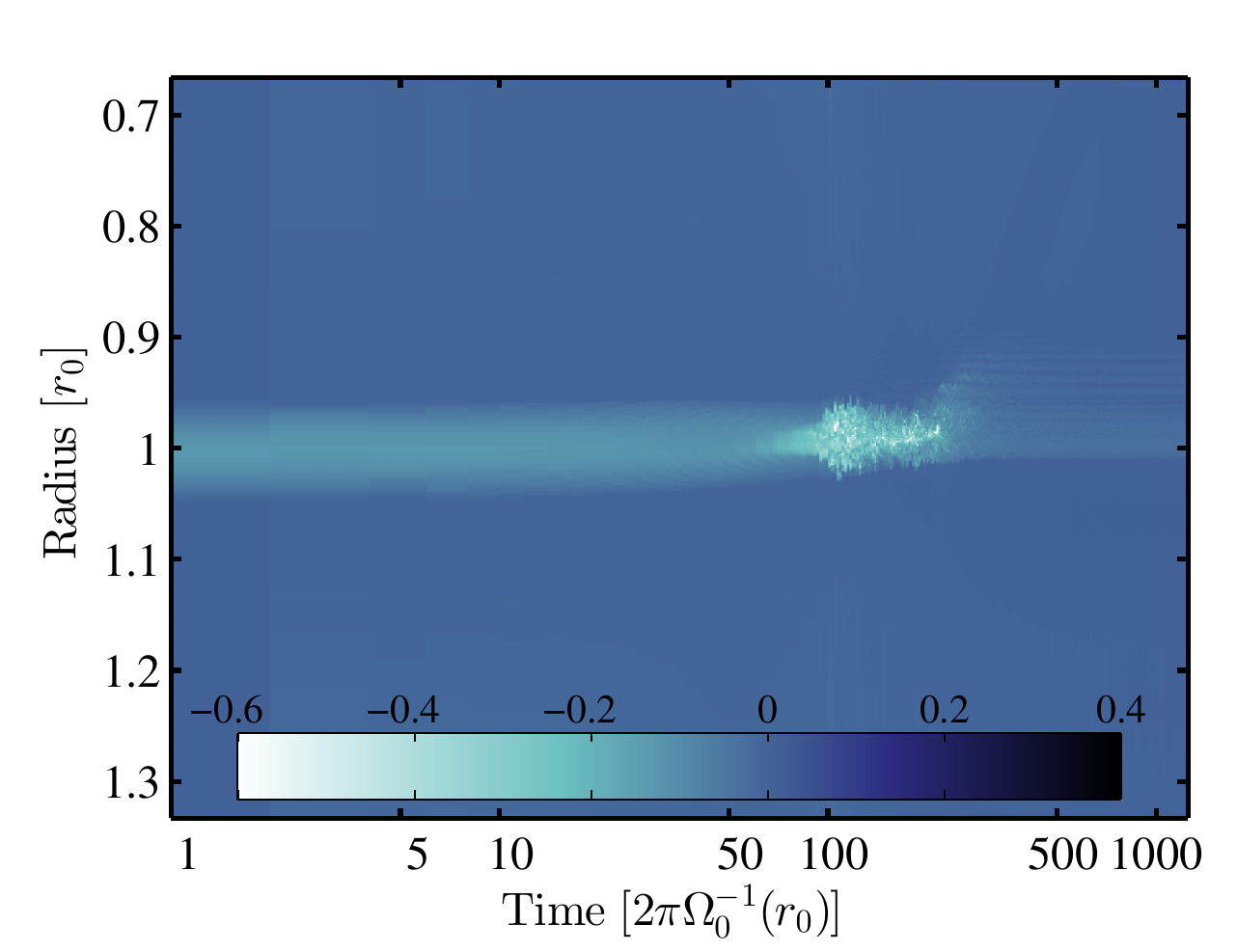} &
	\includegraphics[height=4.5cm, trim=4mm 0cm 0cm 0cm, clip=true]{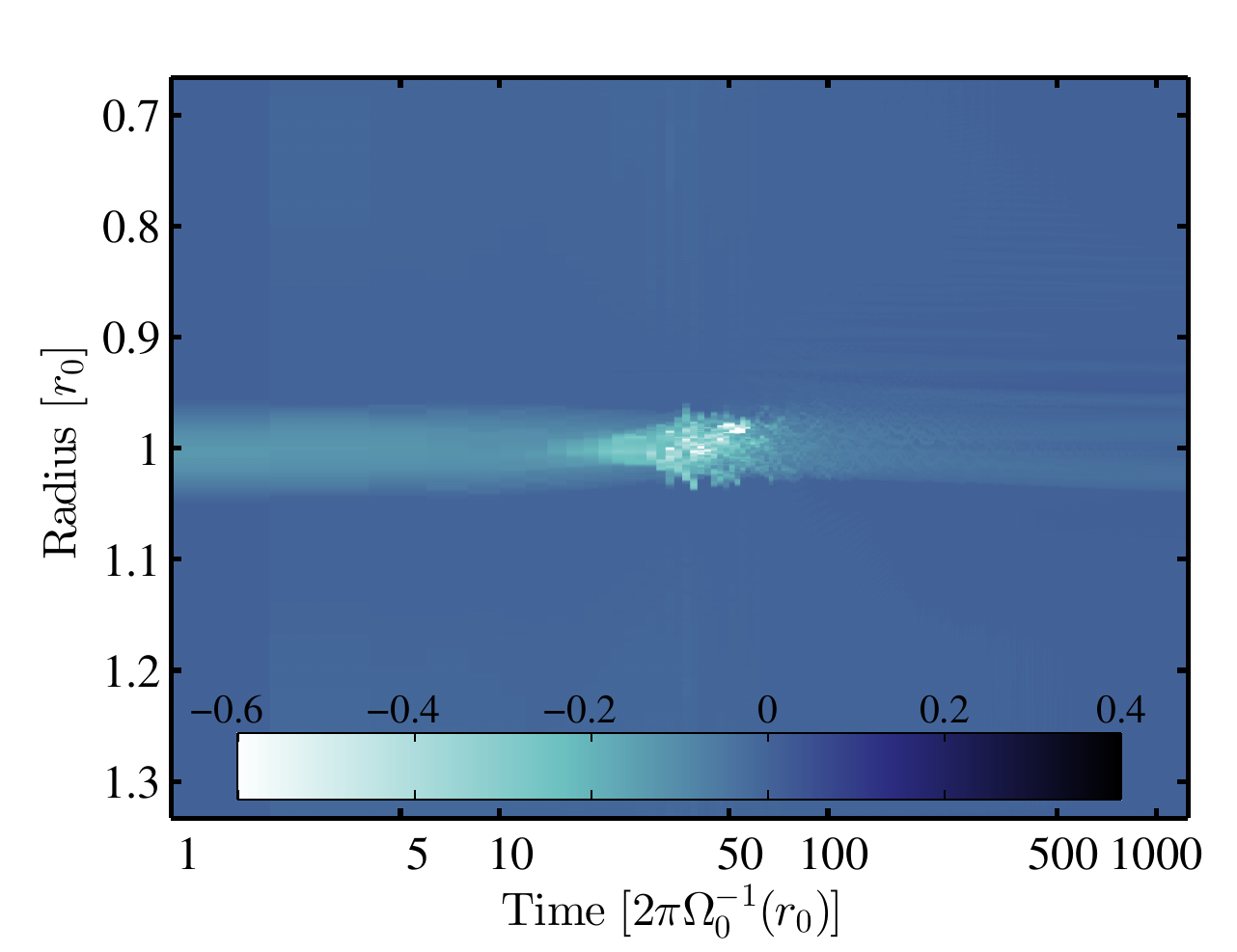} &
	\includegraphics[height=4.5cm, trim=4mm 0cm 0cm 0cm, clip=true]{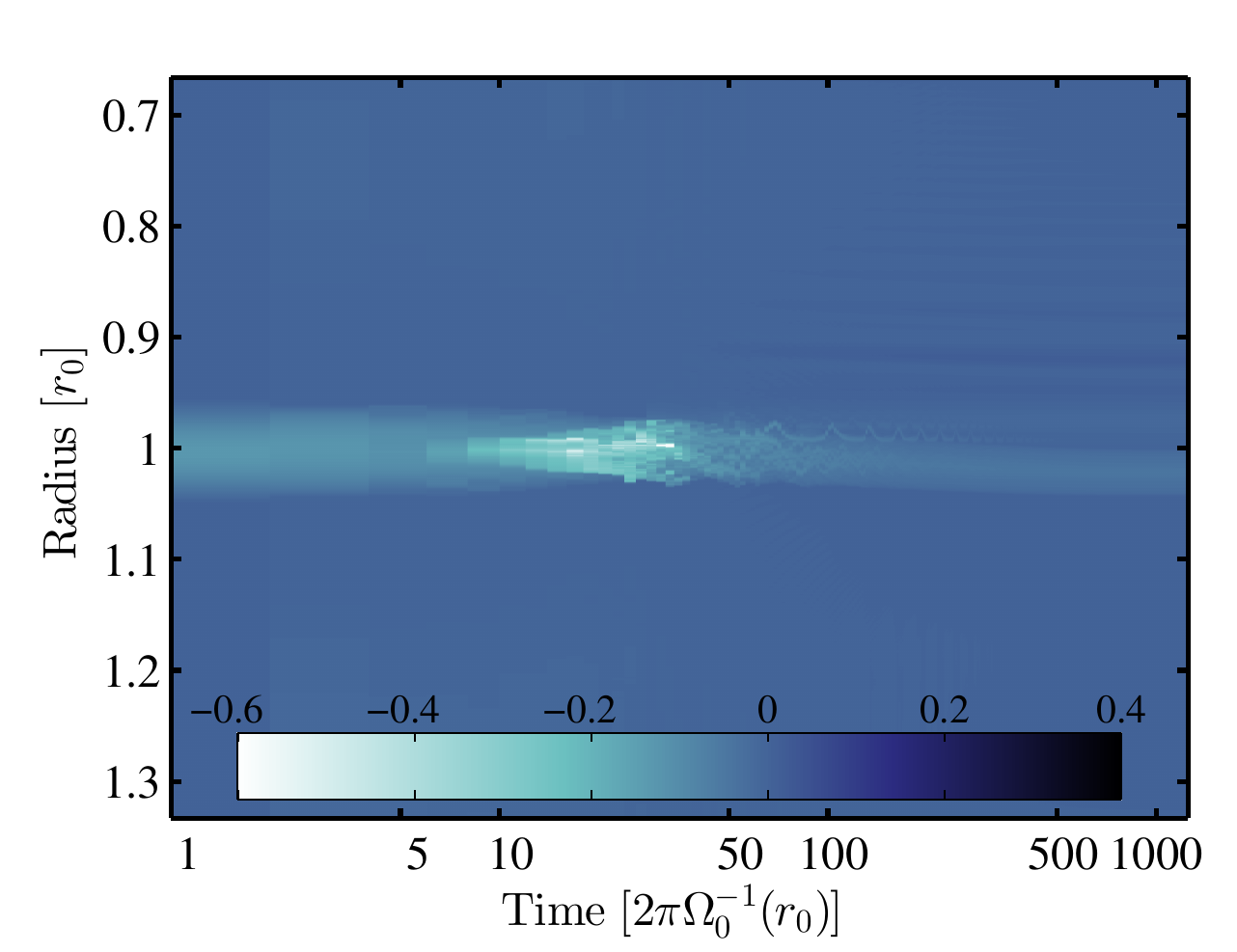} \\

	\includegraphics[height=4.5cm, trim=4mm 0cm 0cm 0cm, clip=true]{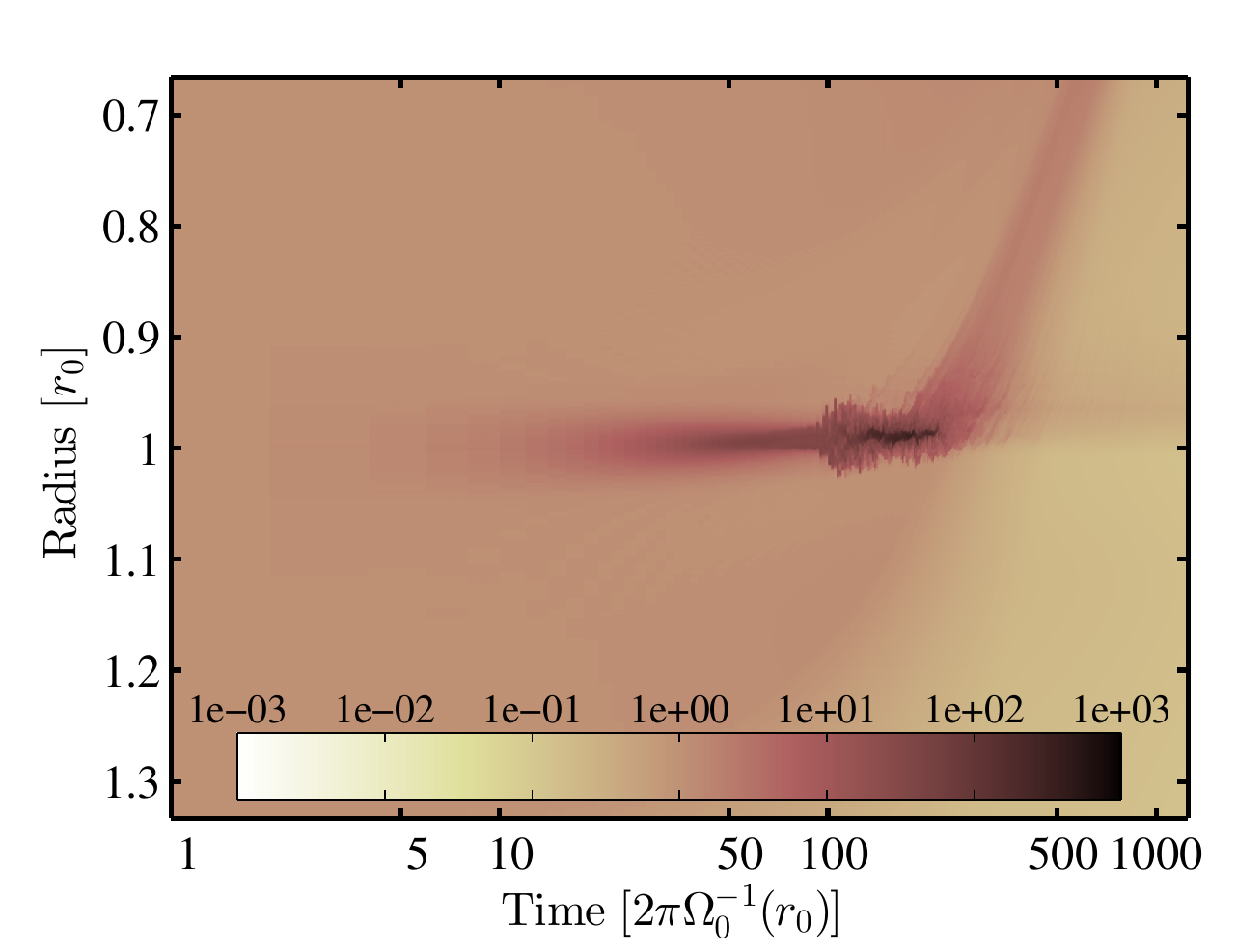} &
	\includegraphics[height=4.5cm, trim=4mm 0cm 0cm 0cm, clip=true]{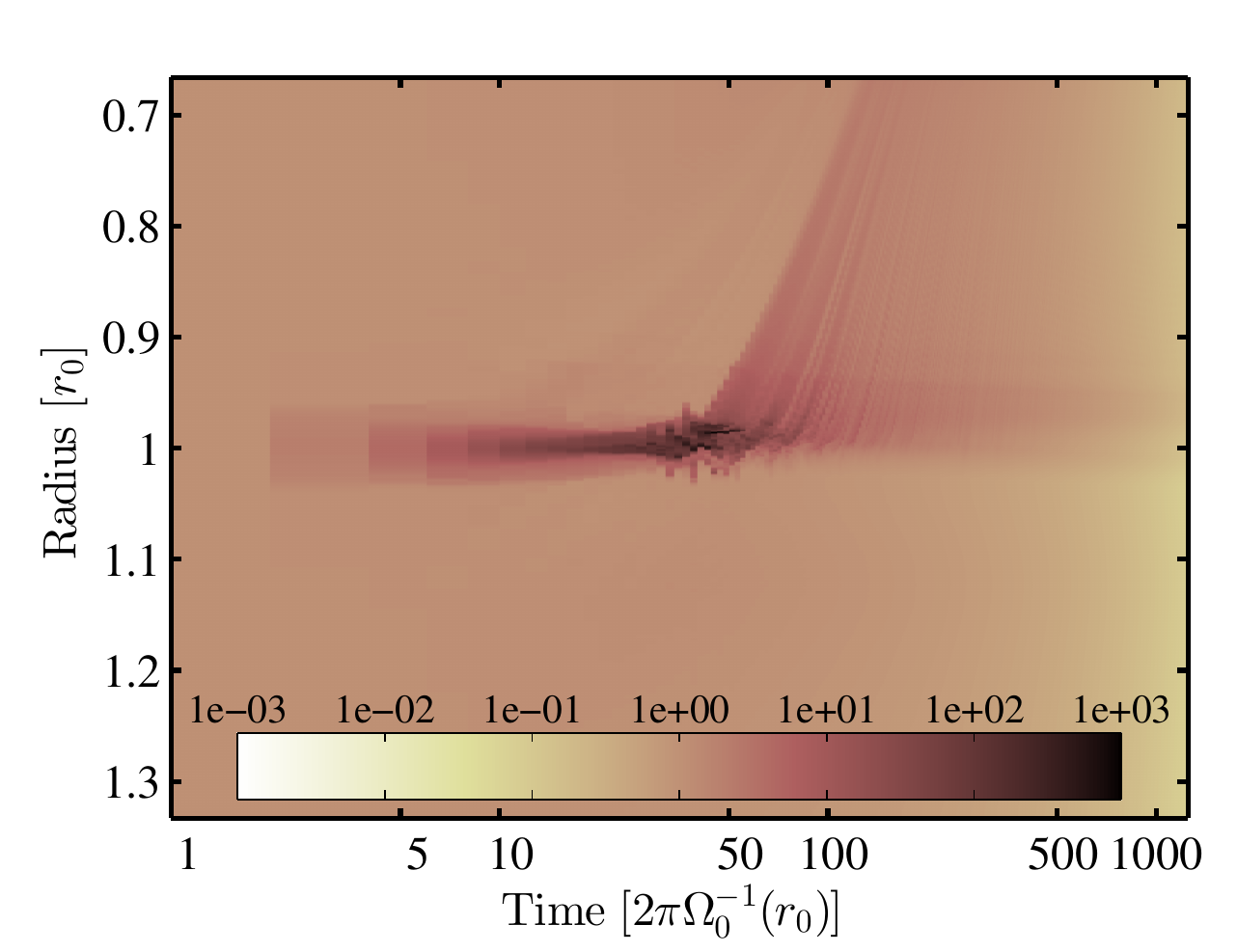} &
	\includegraphics[height=4.5cm, trim=4mm 0cm 0cm 0cm, clip=true]{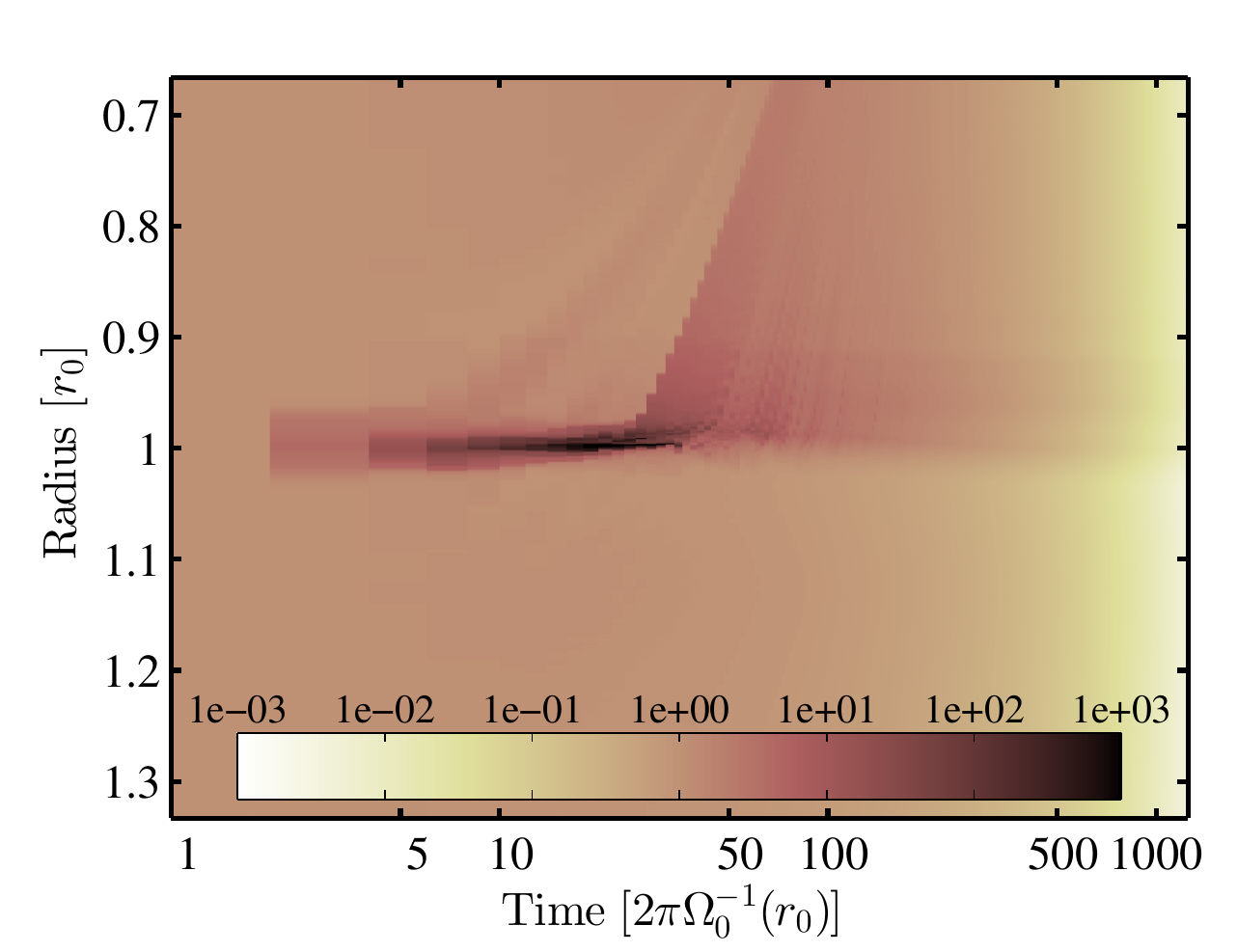} \\

	\end{tabular}
	\caption{\label{Time_evo_epsi2_vortex_2} Time evolution of the radial minimum of the Rossby number ({\it{top}}) and the maximum of the dust density ({\it{bottom}}) for different values of the Stokes number, for the small vortex with $(R_0, \: \chi_r, \: \chi_\theta)=(-0.13, \: 0.06, \: 6.5)$. The global dust-to-gas ratio is $\epsilon=10^{-2}$. {\it{From left to right}}: $S_t=0.04$,  $S_t=0.17$, and $S_t=0.35$ respectively. }
      \end{center}
\end{figure*}

	The time evolution of the minimal Rossby number and of the maximal dust density are shown Figure \ref{Time_evo_epsi2_vortex_2} in the case with $\epsilon = 10^{-2}$. When comparing the results with the case of the large vortex (Figure \ref{Time_evo_epsi2}), we observe a first phase of capture at the vortex center for all $S_t$, on timescales slightly shorter (about $1.3$ times), and compatible with the $1.33$ increase of the effective Stokes numbers. After the capture, we observe a vortex instability but with a vorticity generation much smaller than in the case of the large vortex. In particular, the survival of the unstable vortex is reduced to $\sim 50$ orbits for $S_t=0.04$, $\sim 35$ orbits for $S_t=0.17$, and only $\sim 20$ orbits for $S_t=0.35$. Small vortices are more easily destroyed by the instability than large ones. 
	
	As a result, we observe a release of the dust in the disk, after the end of the vortex instability phase. For any particle size no ring is formed, contrary to the outcome in the large vortex cases. The dust density enhancement is here a transient process, occurring on timescales limited approximately to four times the capture timescales predicted by the model. After that, the dust density reduces to values close to the background level. 

\begin{figure*}
	\begin{center}
	\begin{tabular}{ccc}
	\scriptsize{$S_t=0.04$} & \scriptsize{$S_t=0.17$} & \scriptsize{$S_t=0.35$} \\
	\includegraphics[height=4.5cm, trim=4mm 0cm 0cm 0cm, clip=true]{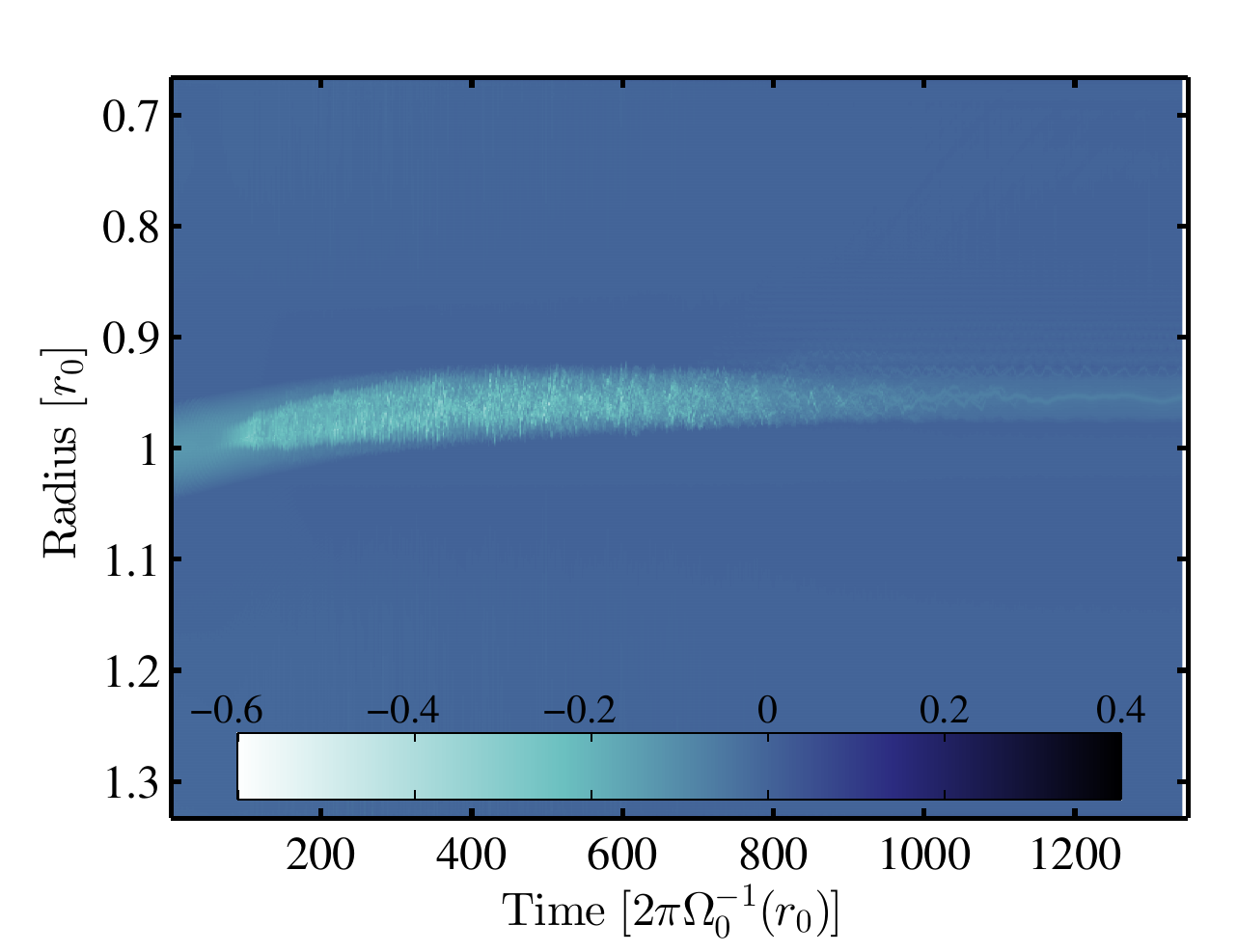} &
	\includegraphics[height=4.5cm, trim=4mm 0cm 0cm 0cm, clip=true]{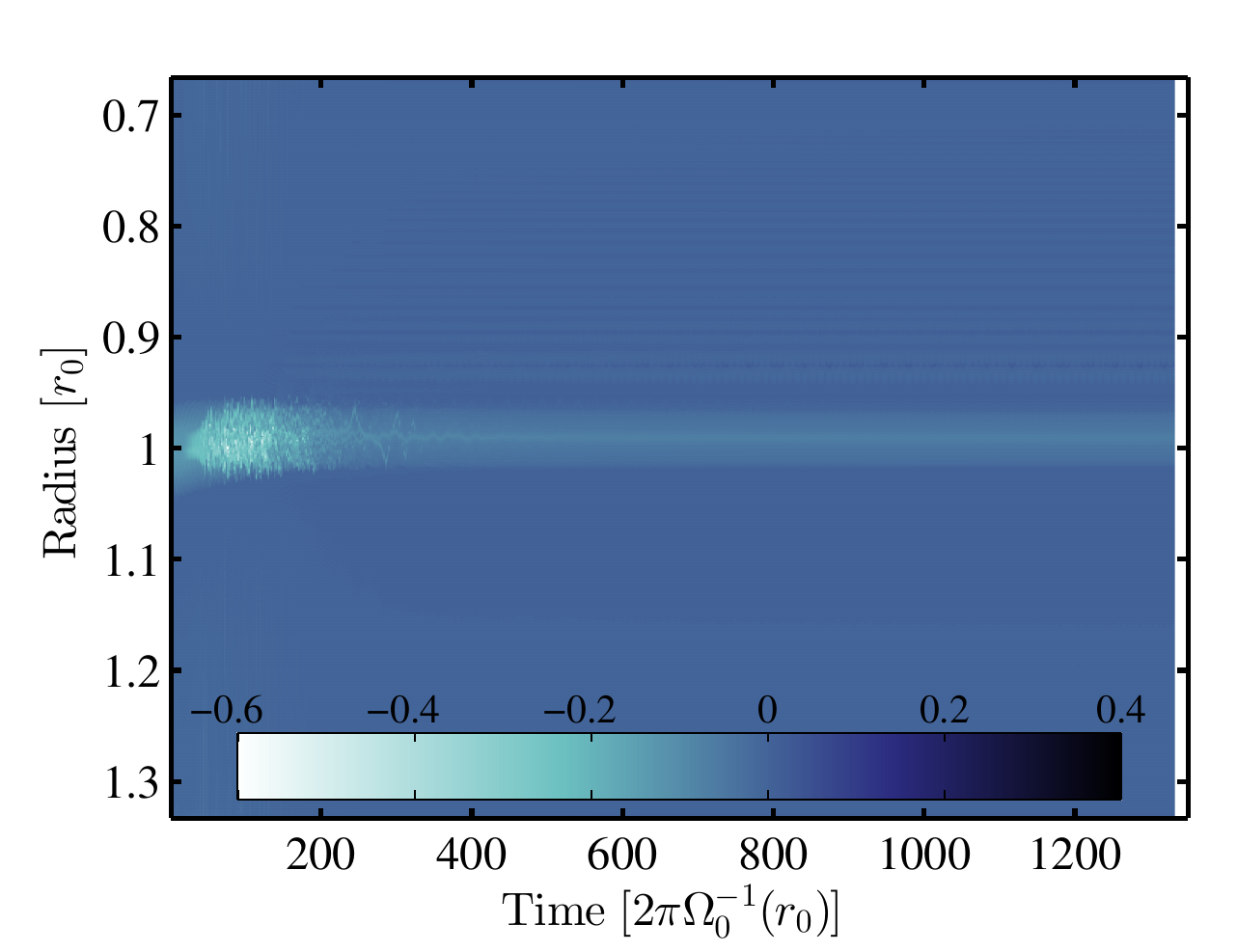} &
	\includegraphics[height=4.5cm, trim=4mm 0cm 0cm 0cm, clip=true]{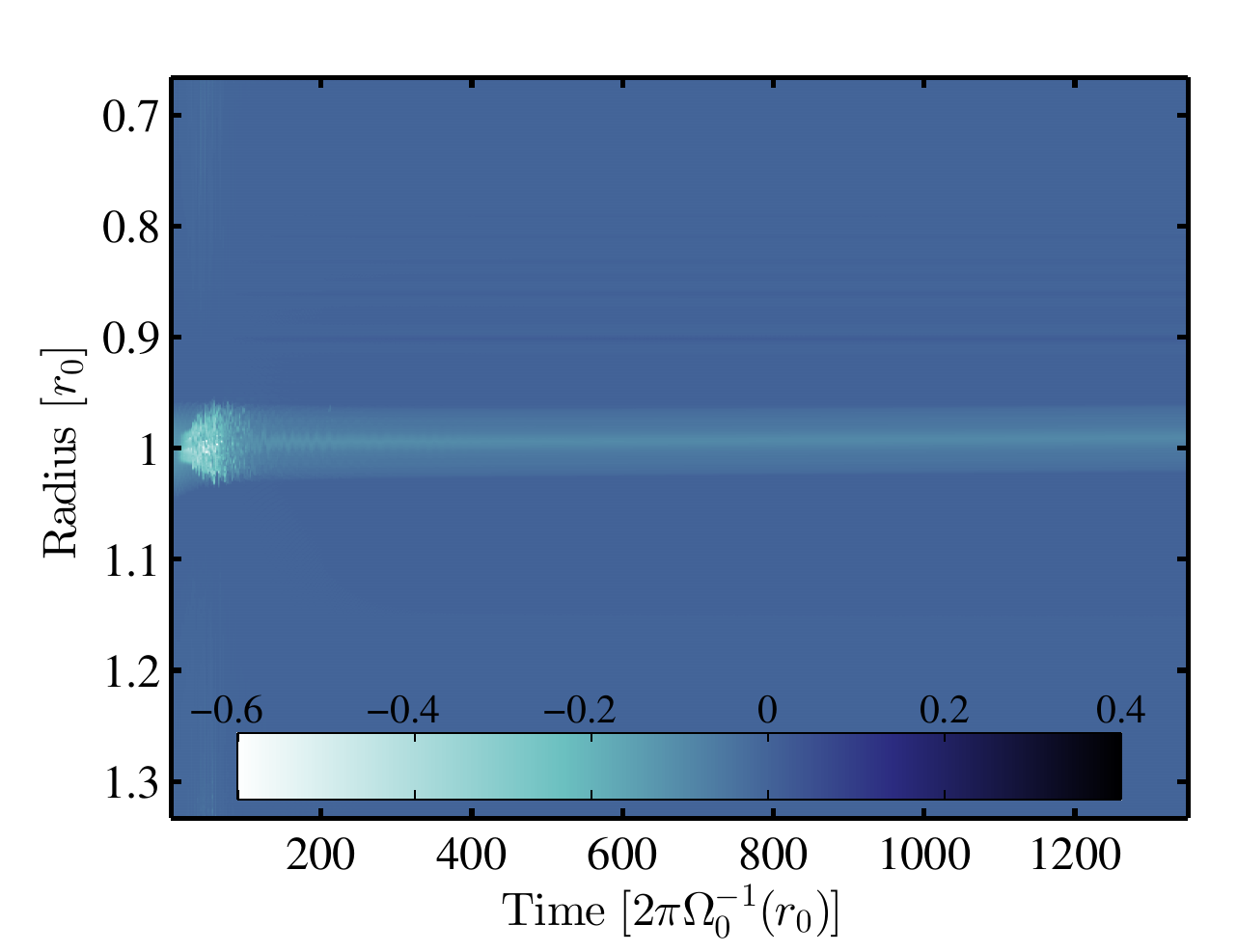} \\

	\includegraphics[height=4.5cm, trim=4mm 0cm 0cm 0cm, clip=true]{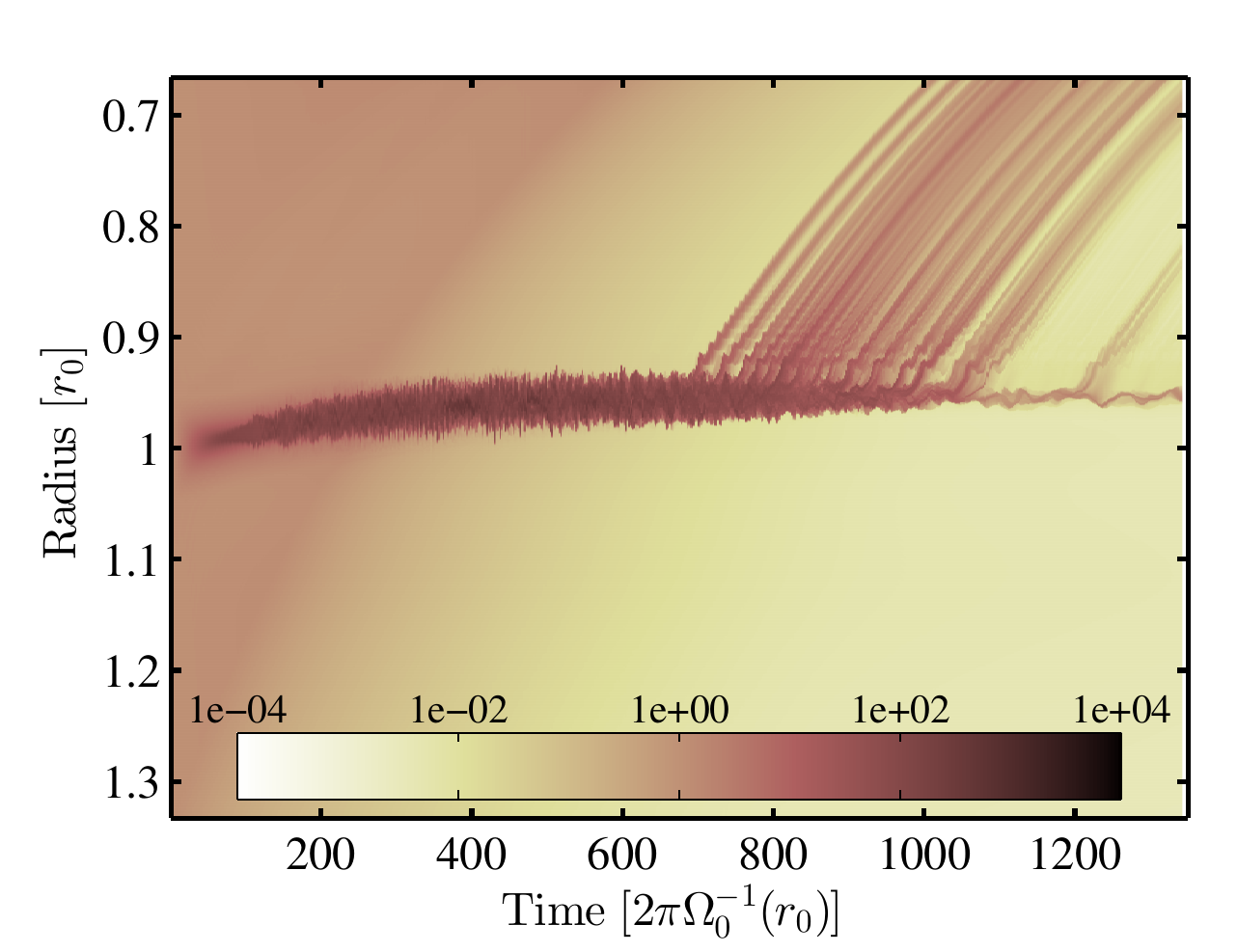} &
	\includegraphics[height=4.5cm, trim=4mm 0cm 0cm 0cm, clip=true]{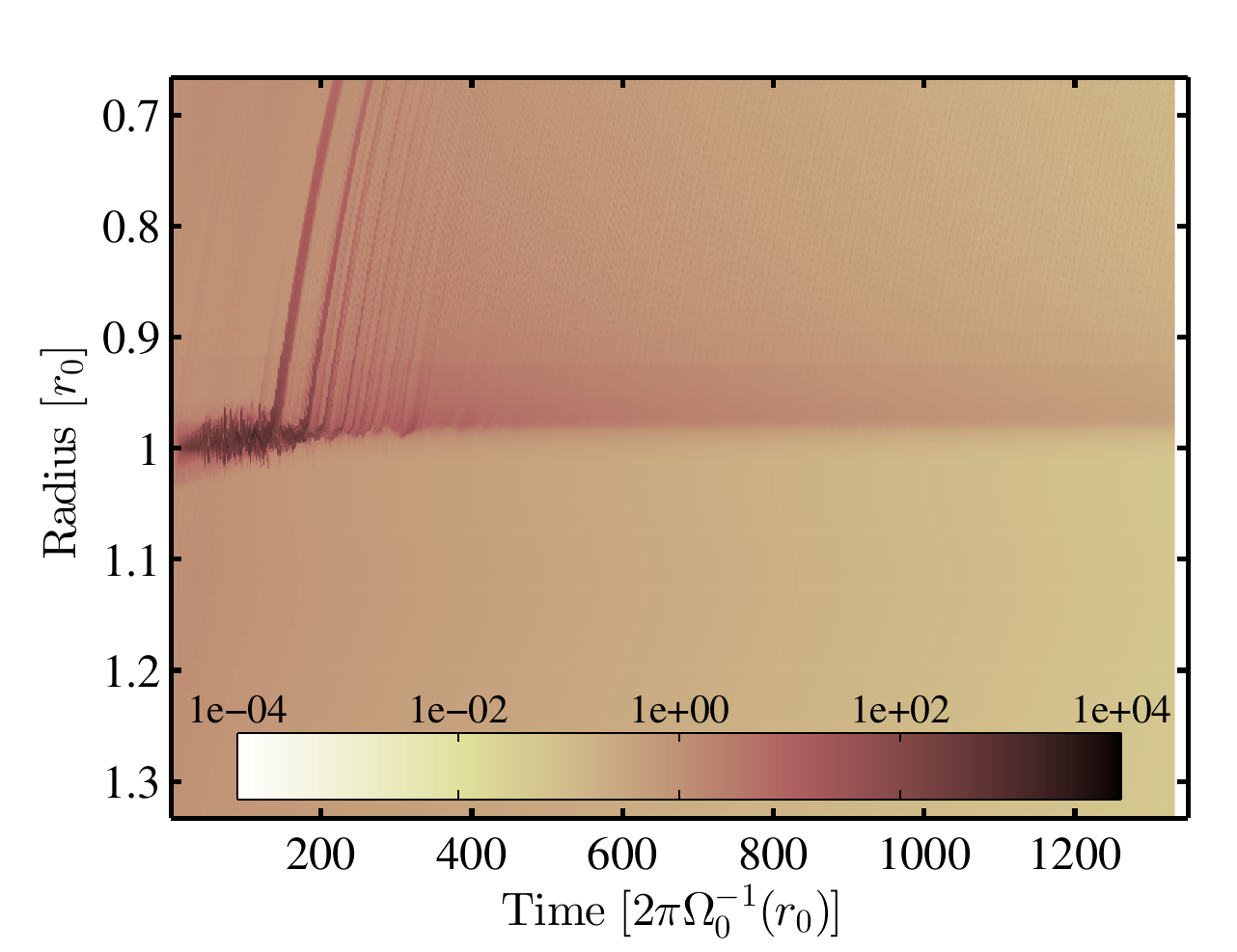} &
	\includegraphics[height=4.5cm, trim=4mm 0cm 0cm 0cm, clip=true]{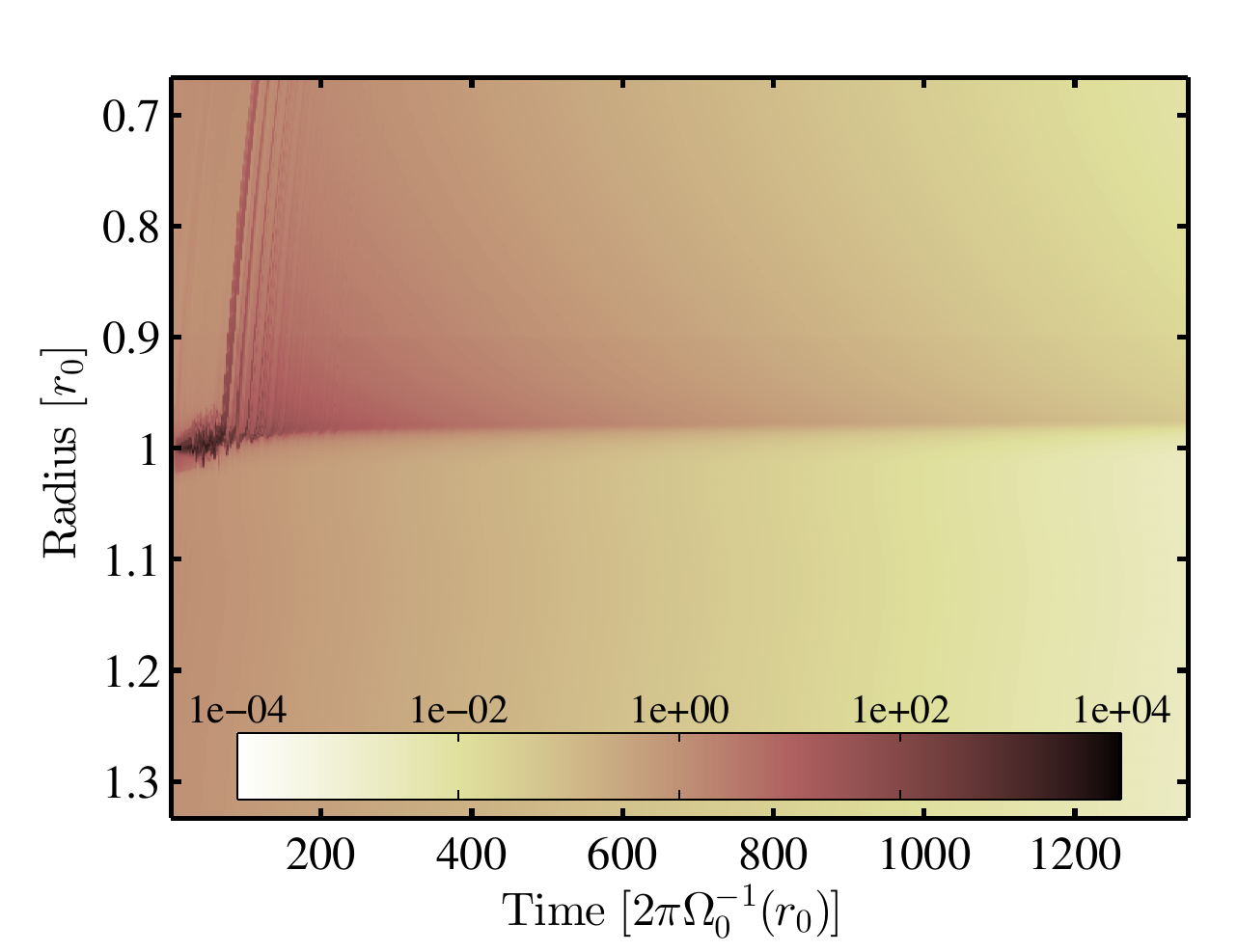} \\

	\end{tabular}
	\caption{\label{Time_evo_epsi3_vortex_2} Time evolution of the radial minimum of the Rossby number ({\it{top}}) and the maximum of the dust density ({\it{bottom}}) for different values of the Stokes number, for the small vortex with $(R_0, \: \chi_r, \: \chi_\theta)=(-0.13, \: 0.06, \: 6.5)$. The global dust-to-gas ratio is $\epsilon=10^{-3}$. {\it{From left to right}}: $S_t=0.04$,  $S_t=0.17$, and $S_t=0.35$ respectively. }
      \end{center}
\end{figure*}

\begin{figure*}
	\begin{center}
	\begin{tabular}{ccc}
	\scriptsize{$S_t=0.04$} & \scriptsize{$S_t=0.17$} & \scriptsize{$S_t=0.35$} \\
	\includegraphics[height=4.5cm, trim=4mm 0cm 0cm 0cm, clip=true]{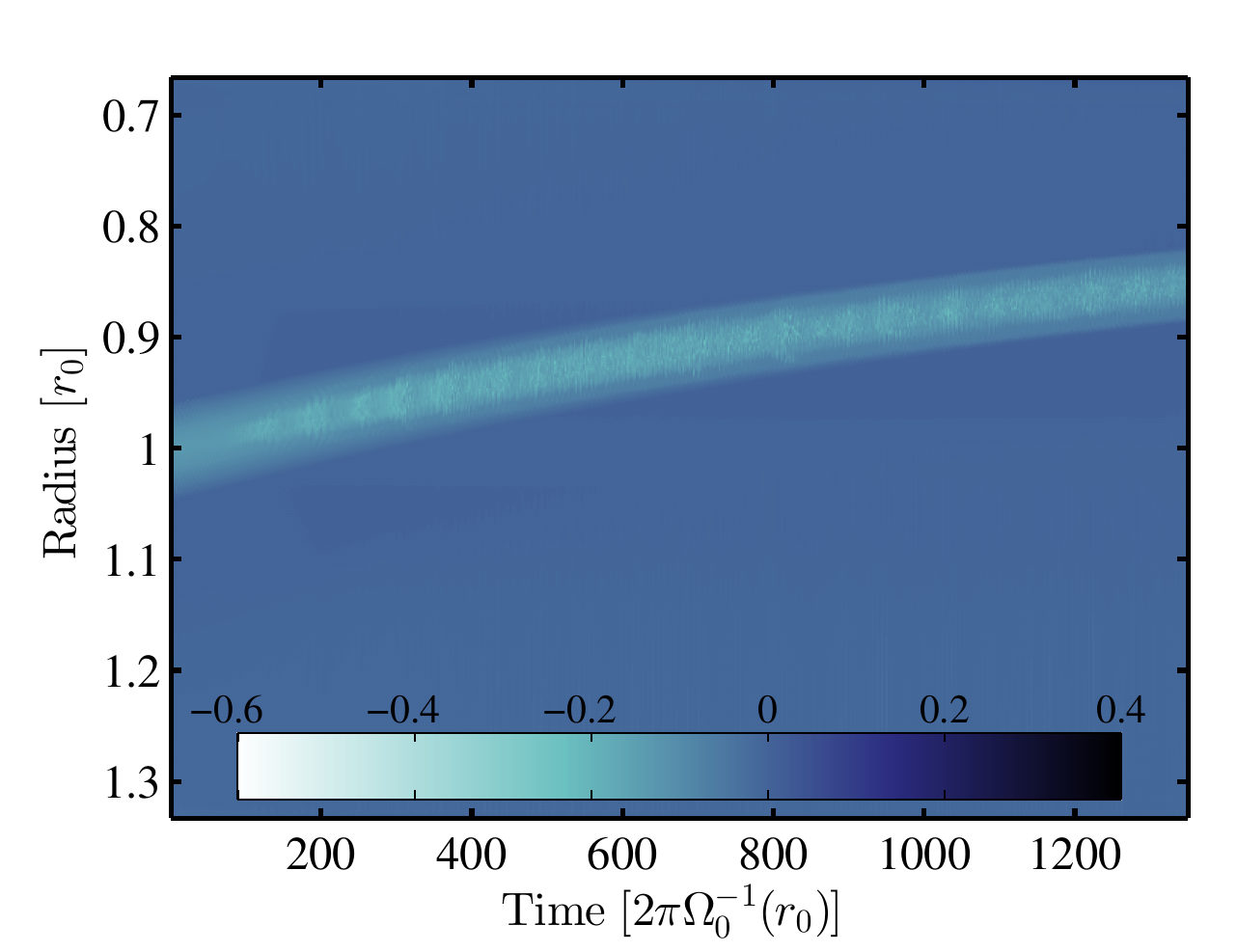} &
	\includegraphics[height=4.5cm, trim=4mm 0cm 0cm 0cm, clip=true]{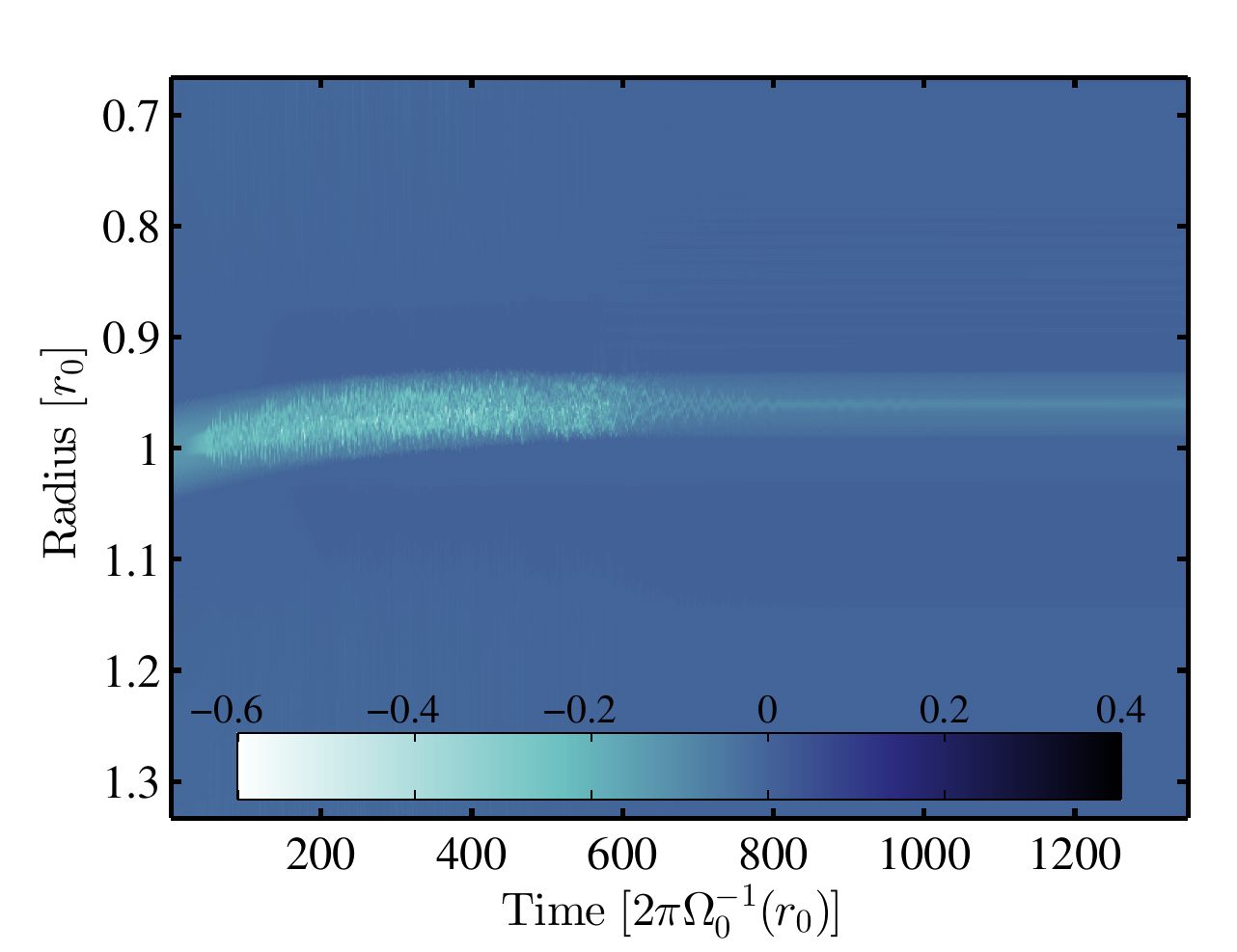} &
	\includegraphics[height=4.5cm, trim=4mm 0cm 0cm 0cm, clip=true]{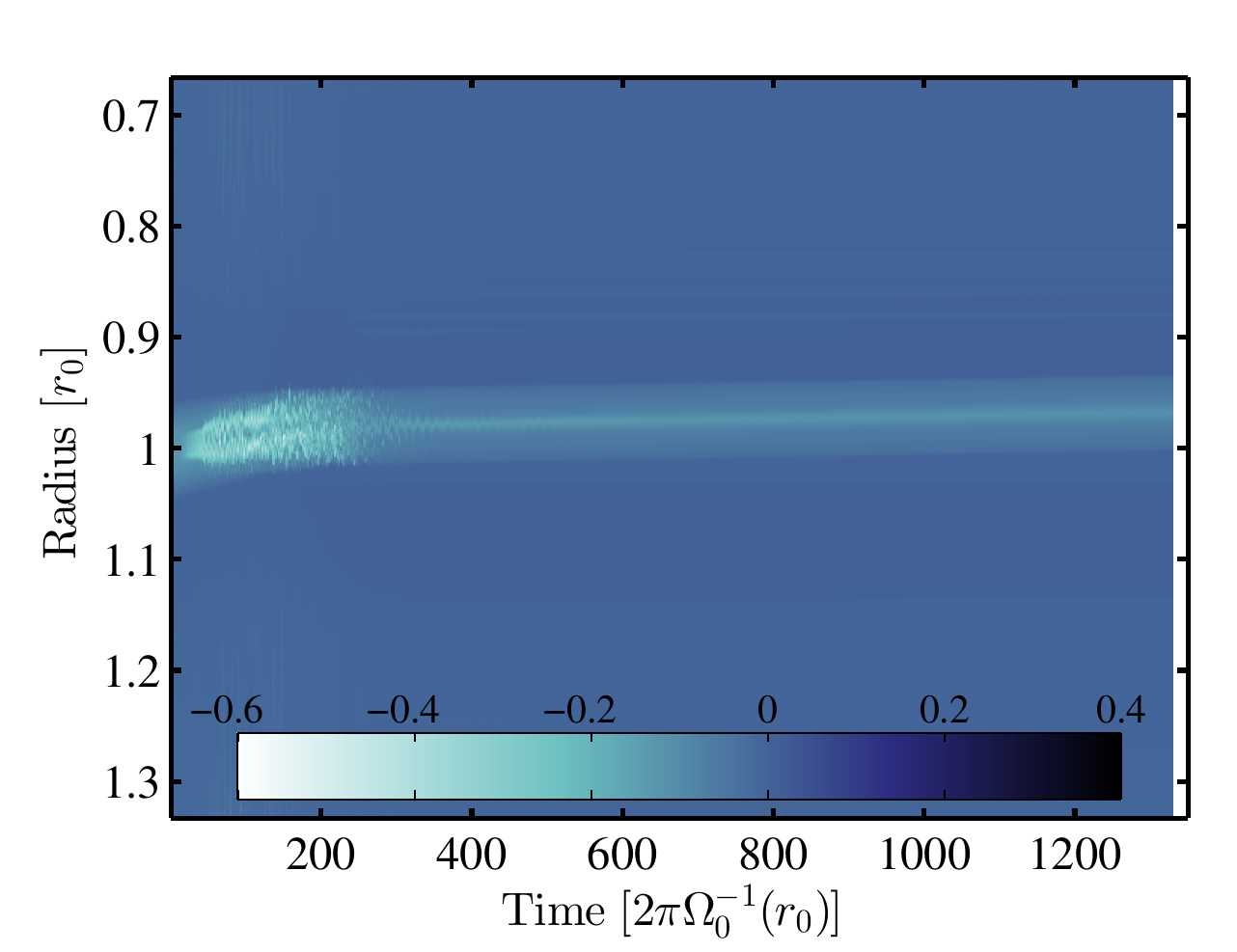} \\

	\includegraphics[height=4.5cm, trim=4mm 0cm 0cm 0cm, clip=true]{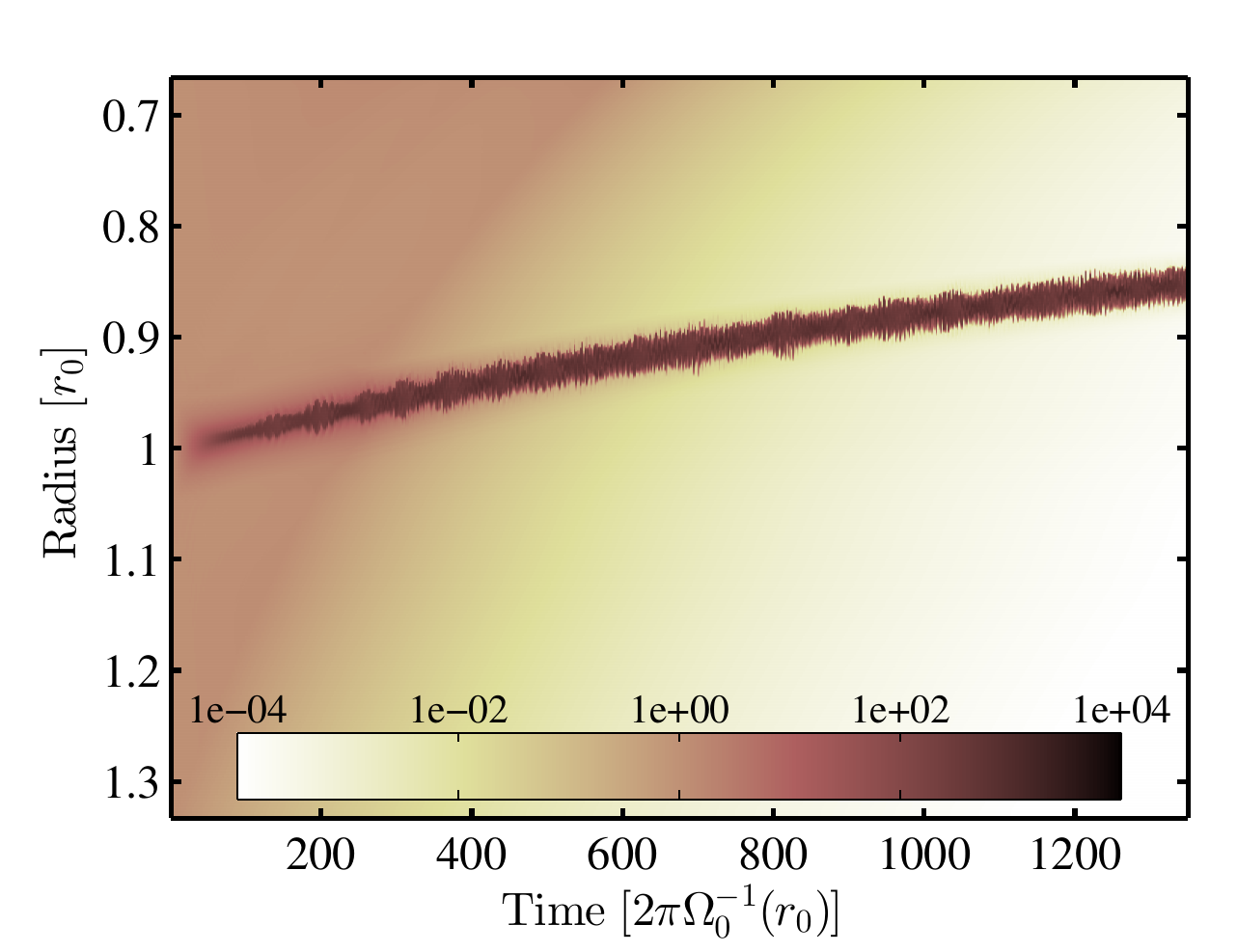} &
	\includegraphics[height=4.5cm, trim=4mm 0cm 0cm 0cm, clip=true]{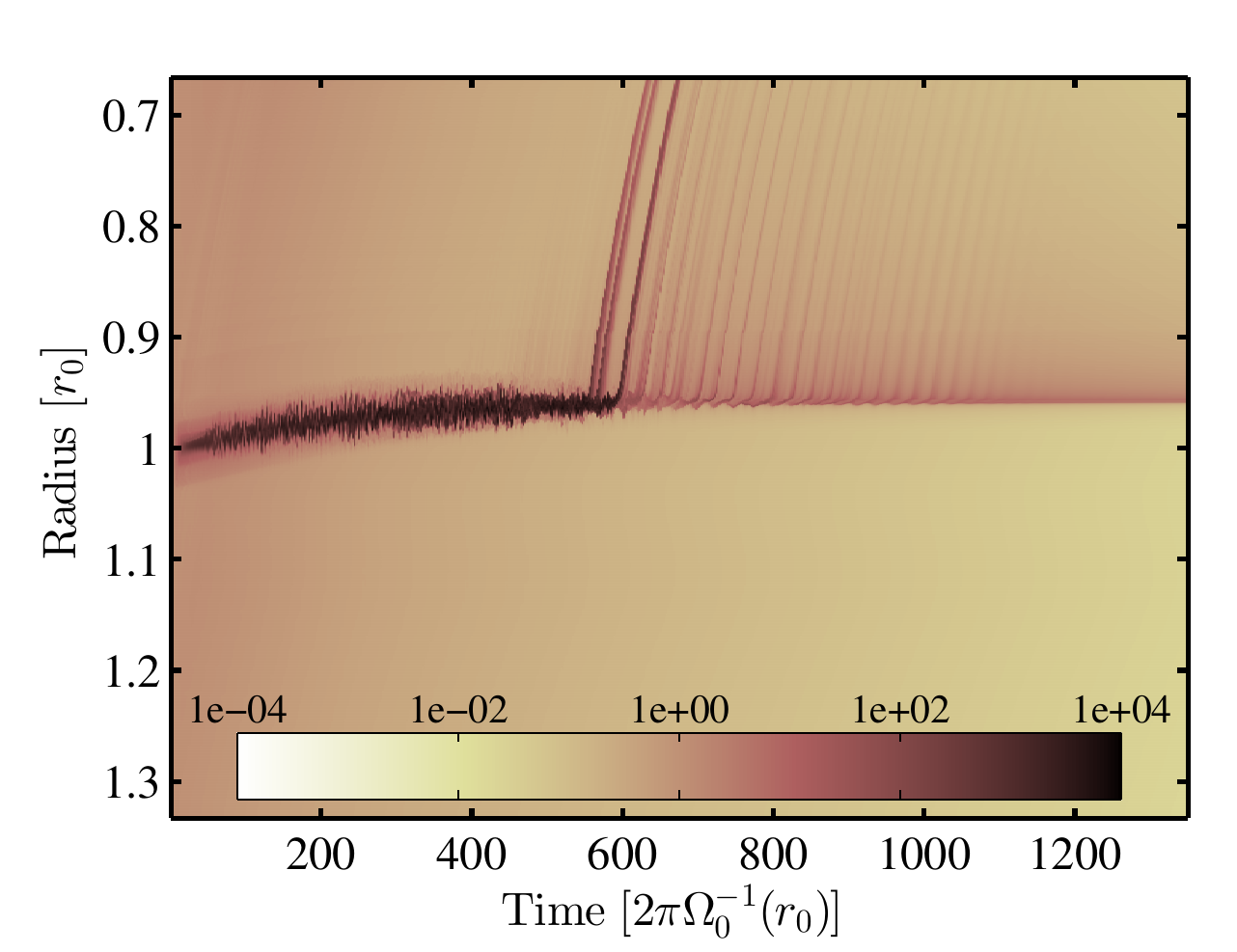} &
	\includegraphics[height=4.5cm, trim=4mm 0cm 0cm 0cm, clip=true]{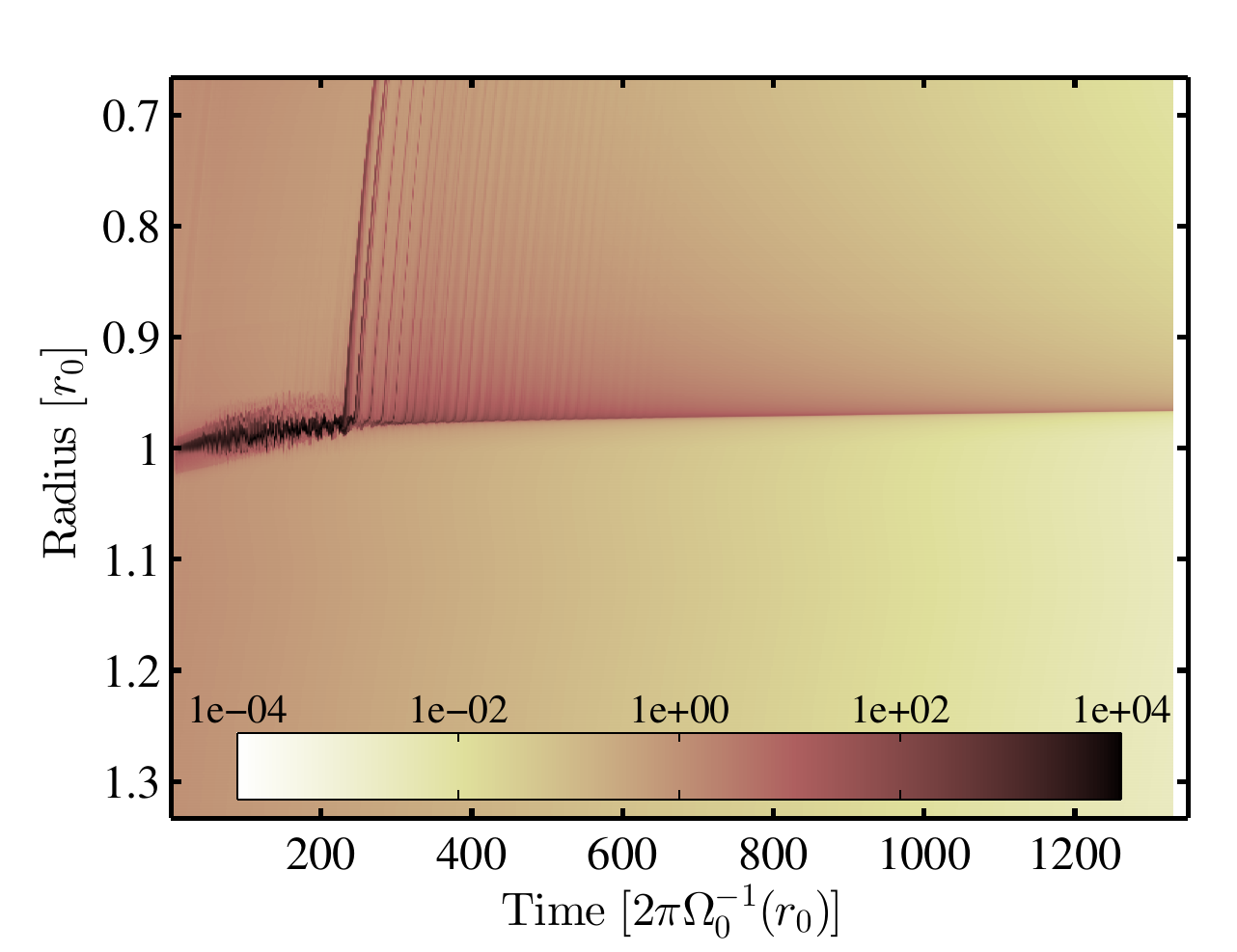} \\

	\end{tabular}
	\caption{\label{Time_evo_epsi4_vortex_2} Time evolution of the radial minimum of the Rossby number ({\it{top}}) and the maximum of the dust density ({\it{bottom}}) for different values of the Stokes number, for the small vortex with $(R_0, \: \chi_r, \: \chi_\theta)=(-0.13, \: 0.06, \: 6.5)$. The global dust-to-gas ratio is $\epsilon=10^{-4}$. {\it{From left to right}}: $S_t=0.04$, $S_t=0.17$, and $S_t=0.35$ respectively. }
      \end{center}
\end{figure*}

	On figures \ref{Time_evo_epsi3_vortex_2} and \ref{Time_evo_epsi4_vortex_2} we present the results obtained with $\epsilon = 10^{-3}$ and $\epsilon = 10^{-4}$ respectively. The evolution in the case when $\epsilon=10^{-4}$ and $S_t=0.04$ is comparable to the one obtained with the large vortex (Figure \ref{Time_evo_epsi4_long}, left). The dust is confined inside the unstable vortex, after capture, and this dusty structure lasts during the whole simulation. The radial migration has also the same rate. However, the maximal dust density reached is only $\sim 400$ larger than the disk background. This dust density enhancement is $\sim 3$ times smaller than the one obtained with the large vortex.

	Concerning the other cases, the maximal dust density created inside the vortex during the instability is comparable to that obtained with the large vortex. However the duration of the unstable phase is much shorter, and we obtain the vortex destruction in all these cases. It shows that even with lower initial dust-to-gas ratios, the dusty vortex is unlikely to survive for a long time when its size is reduced. The confined dust is released inside the rest of the disk.

	The most striking difference of the results obtained by reducing the vortex size is the absence of a final dust ring. The only cases where a tiny ring is observed is when $\epsilon=10^{-4}$ (Figure \ref{Time_evo_epsi4_vortex_2}), but the dust density enhancement is only by a factor of ten. It is much smaller than in the case of the large vortex, where the dust density in the ring is by at least two orders of magnitude larger than the disk background. In fact, no ring of gas is produced after the vortex instability, and even when it is the case, the gas density contrast is very small. As a result, the particles are not confined in a near zero pressure gradient region. We will investigate this process of ring formation in the next section.

	As a conclusion, we emphasize that the radial vortex size has a significant impact on the disk evolution. Even though the particle capture happens on a timescale only defined by the Rossby number, the Stokes number and the initial dust-to-gas ratio, smaller vortices are more efficiently destroyed by the vortex instability. This is an interesting outcome because, then, the large amount of particles captured by the vortex is released inside the disk. However, if the vortex size is too small, these released particles are not confined in a dust ring and are dispersed in the inner parts of the disk.

	The next section is devoted to propose a detailed explanation of the different regimes highlighted by these results. We will also discuss the possible implications on disk evolution and planetesimal formation scenarios.

\section{ Discussion }
\label{Sect_Discussion}

	This section is devoted to providing physical insight into the various phases of dust evolution already sketched in Section \ref{Sect_Test_case} with the fiducial run and analyzed further in the previous section. In particular, we will discuss, in order, the linear capture phase, the vortex instability phase concurrent with vortex dissipation, for which we will show that it is a form of streaming instability, and the dust ring formation phase after vortex dissipation. We will also discuss in detail the conditions that affect the survival of vortices in the disk, and why in certain cases vortices do persist till the end of the simulations. Finally, we will discuss the effect of resolution on our numerical results and conclude with the caveats associated with the lack of viscous forces associated with MHD phenomena in accretion disks.

\subsection{ Linear capture phase }
	
	We observe the accumulation of dust inside the vortex for all the different cases, whatever the Stokes number, the initial dust-to-gas ratio, or the vortex size. Our first qualitative comparison of the timescales with the analytical capture model showed a good agreement. Here we will analyse in more details the confrontation of the linear analysis with the numerical results.

\begin{figure}
	\begin{center}
	\begin{tabular}{c}
	\scriptsize{$\epsilon=10^{-2}$} \\
	\includegraphics[height=5.5cm, trim=0mm 0cm 0cm 0cm, clip=true]{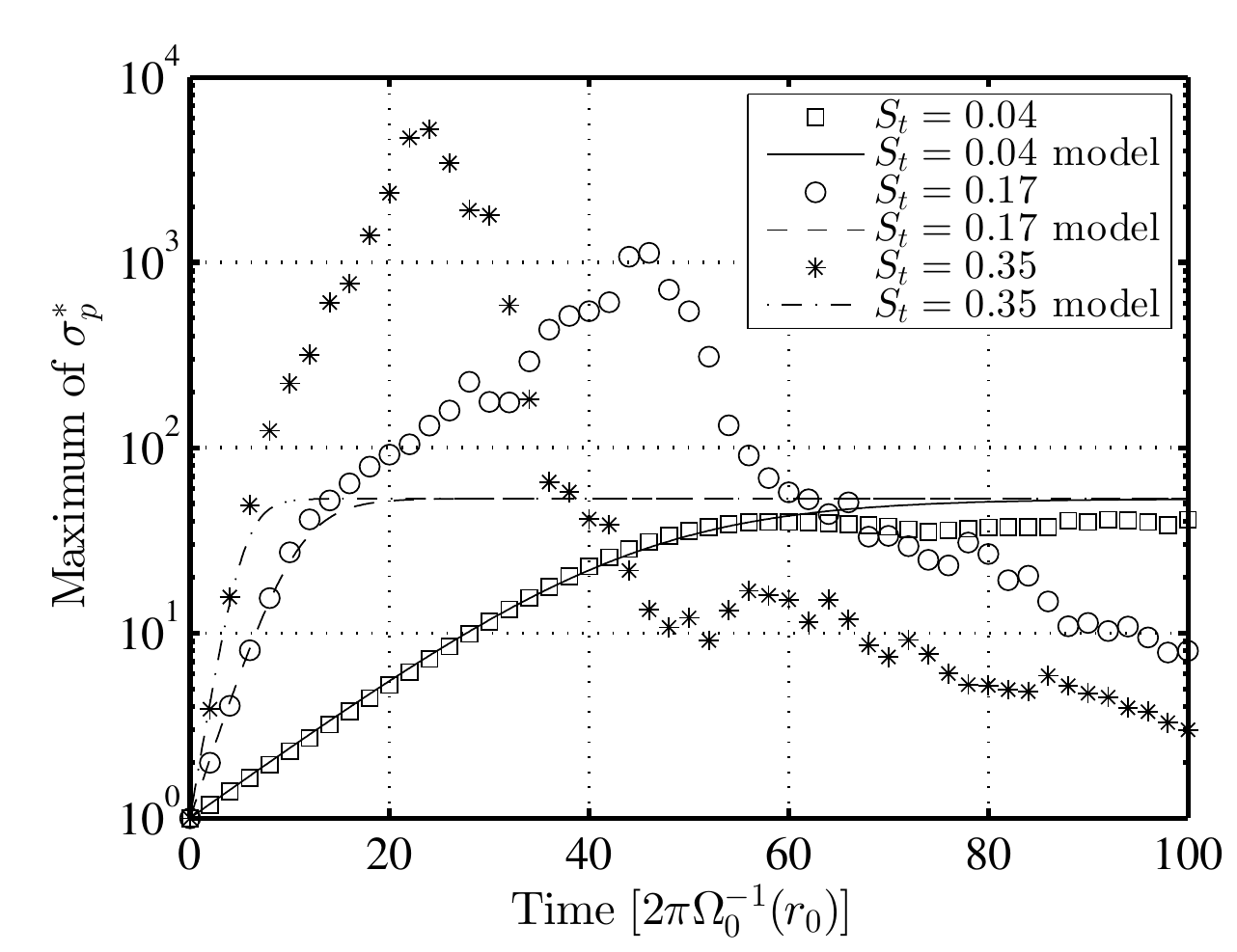} \\ \\
	\scriptsize{$\epsilon=10^{-3}$} \\
	\includegraphics[height=5.5cm, trim=0mm 0cm 0cm 0cm, clip=true]{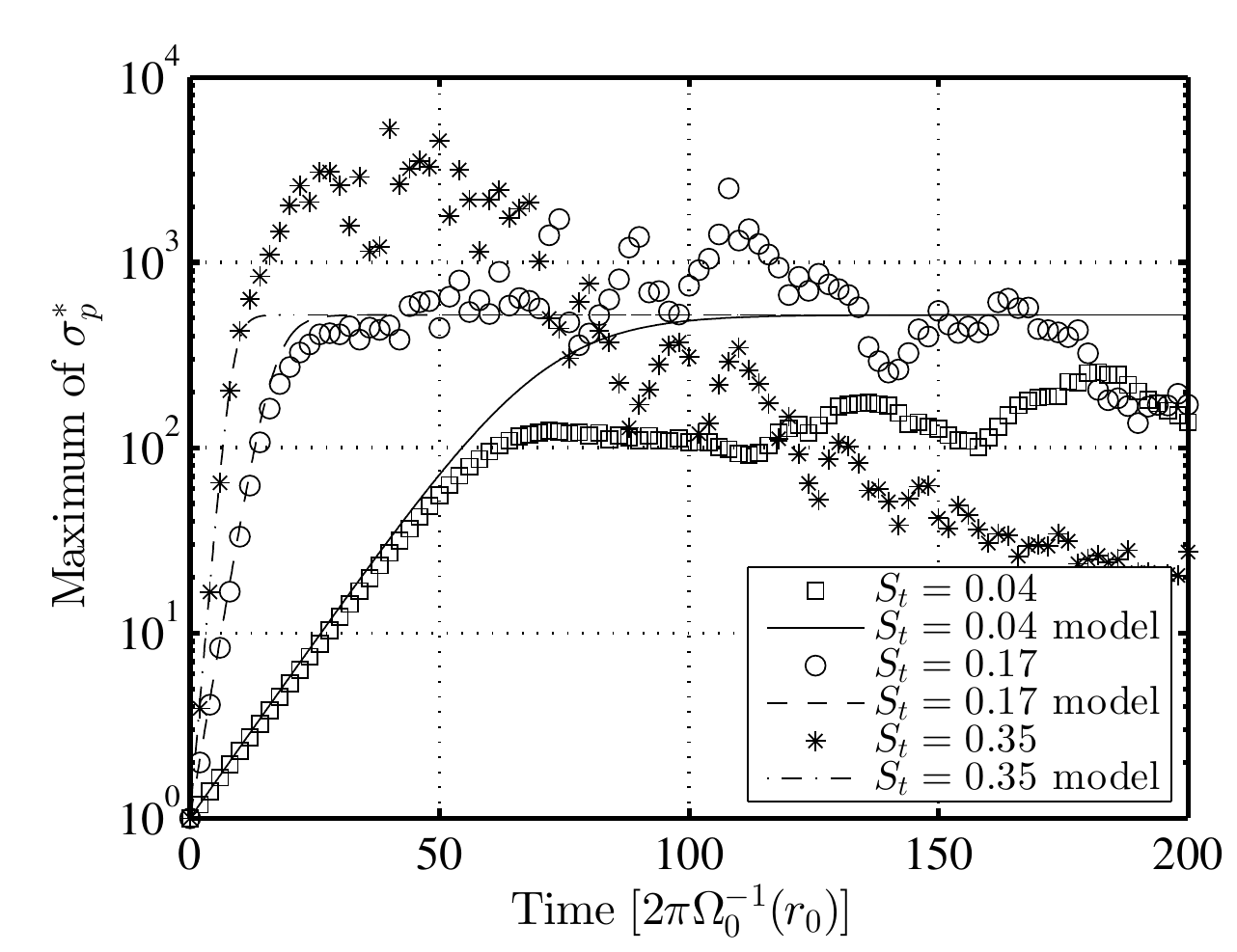} \\ \\
	\scriptsize{$\epsilon=10^{-4}$} \\
	\includegraphics[height=5.5cm, trim=0mm 0cm 0cm 0cm, clip=true]{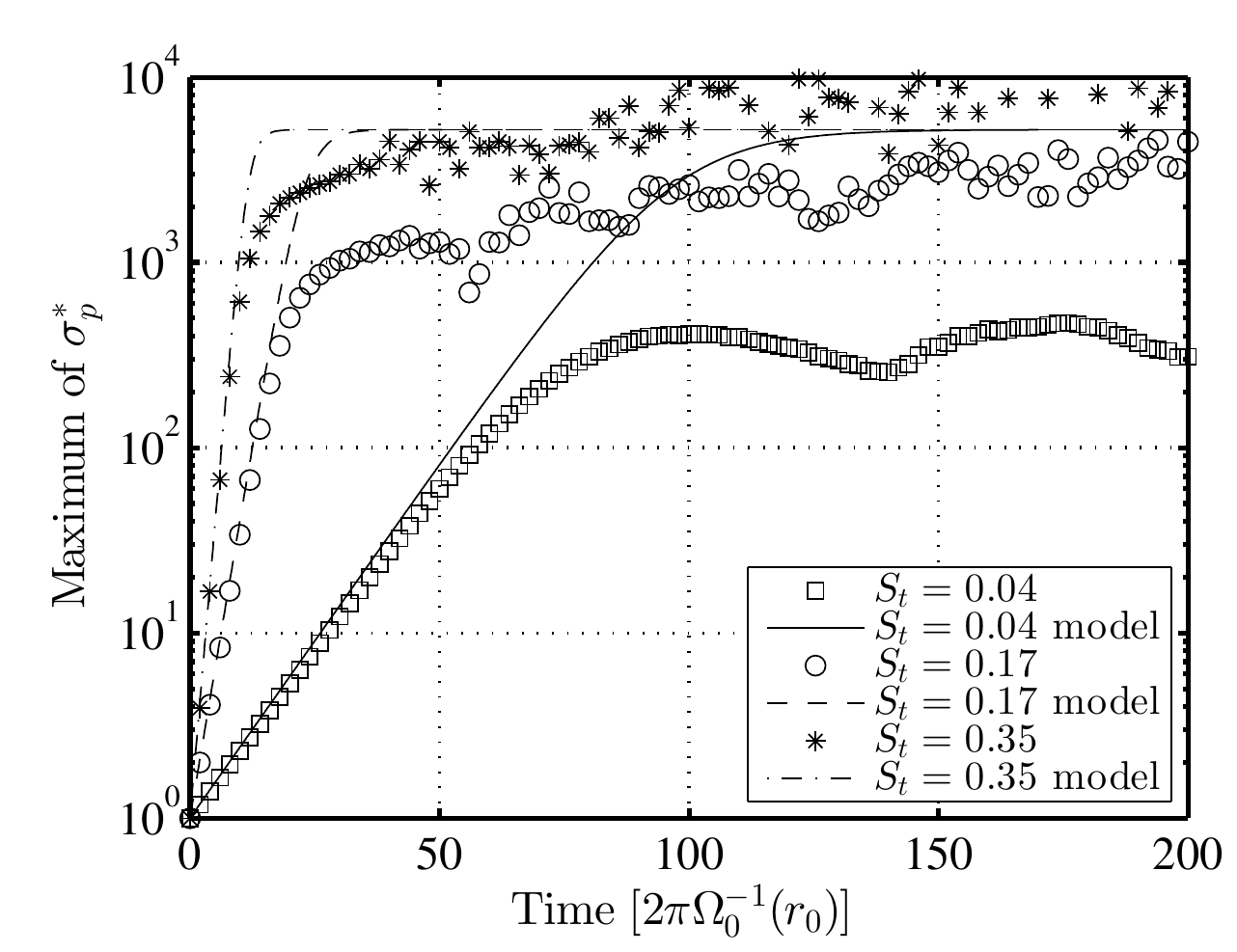} 
	\end{tabular}
	\caption{\label{Linear_Capture_compar_Vortex_2} Evolution of the dust density maximum inside the small vortex with $(R_0, \: \chi_r, \: \chi_\theta)=(-0.13, \: 0.06, \: 6.5)$, for different dust-to-gas ratios, $\epsilon = 10^{-2}, \:10^{-3},$ and $10^{-4}$ {\it{from top to bottom}} respectively. In each panel we compare the simulation results for different Stokes numbers with the dust capture model (lines). }
      \end{center}
\end{figure}

	One main new result obtained with the capture model is the variation of the maximum of the dust density at the vortex center as a function of the orbital time (see Eq. \ref{Sys_evo_2}). As a consequence, we measure for each simulation the maximum of dust density and compute the variable $\sigma_p^* = \sigma_p/[\epsilon \sigma_0(r)]$, and also the corresponding analytical model. The results are presented Figure \ref{Linear_Capture_compar_Vortex_2} for the big vortex, with on the left the main set of simulations discussed during the last section (Section \ref{Sect_Results}), and on the right the same setups but at lower resolution: $(N_r, \: N_\theta) = (512, \: 1024)$.

\begin{figure*}
	\begin{center}
	\begin{tabular}{ccc}
	\scriptsize{$(N_r, \: N_\theta) = (2048, \: 4096)$} & \scriptsize{$(N_r, \: N_\theta) = (512, \: 1024)$} & \\
	\includegraphics[height=5.5cm]{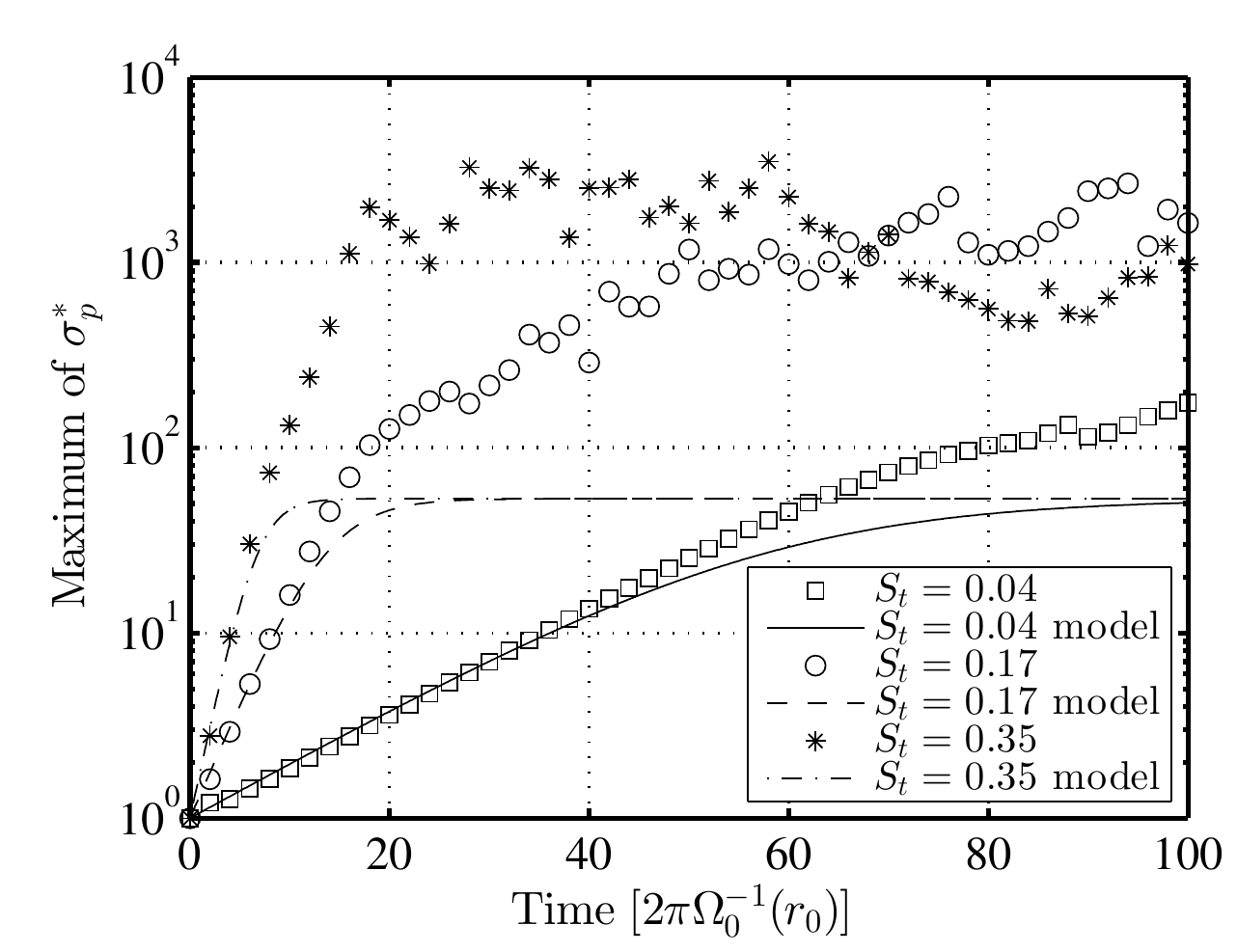} &
	\includegraphics[height=5.5cm]{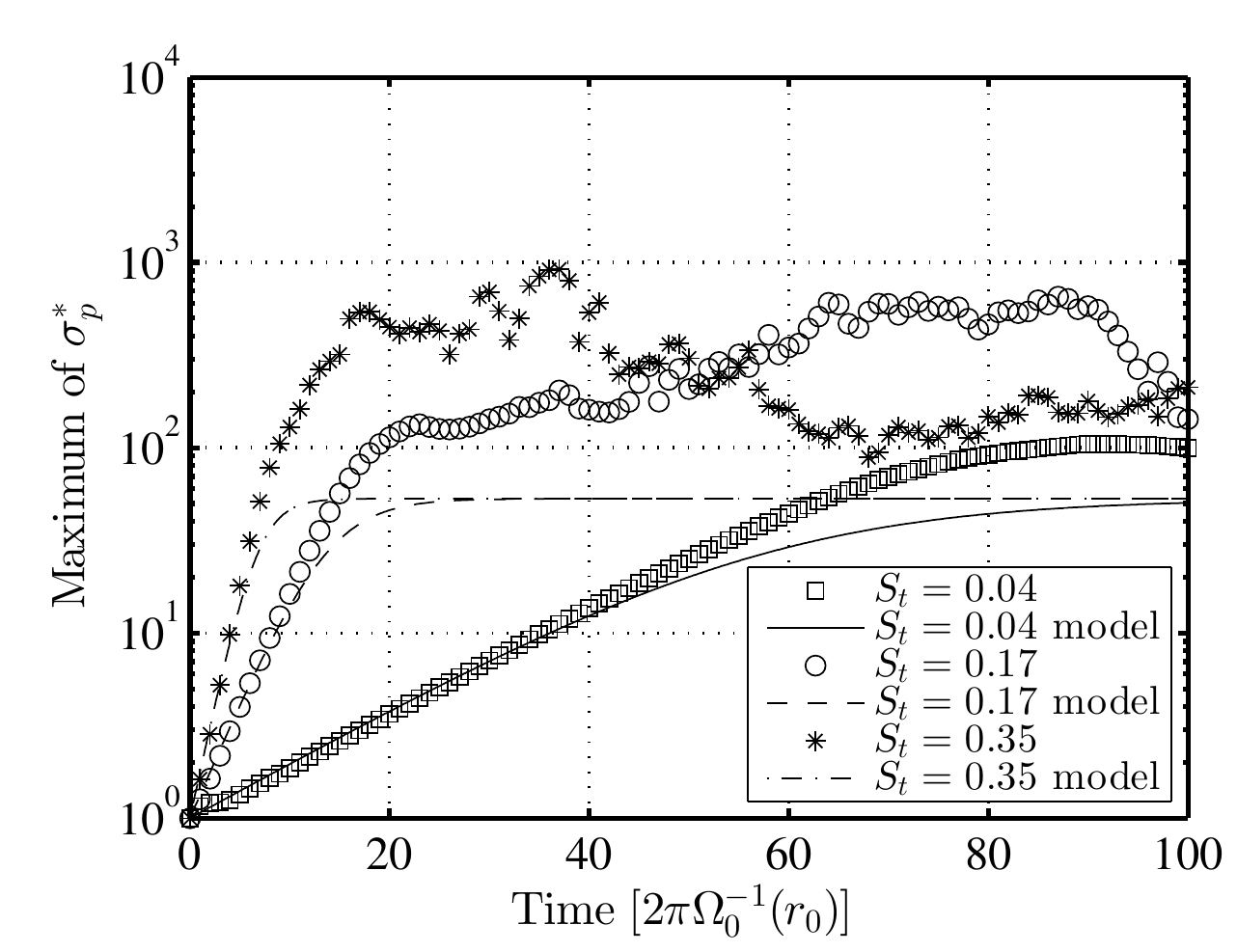} &
	\scriptsize{(a)} \\
	\includegraphics[height=5.5cm]{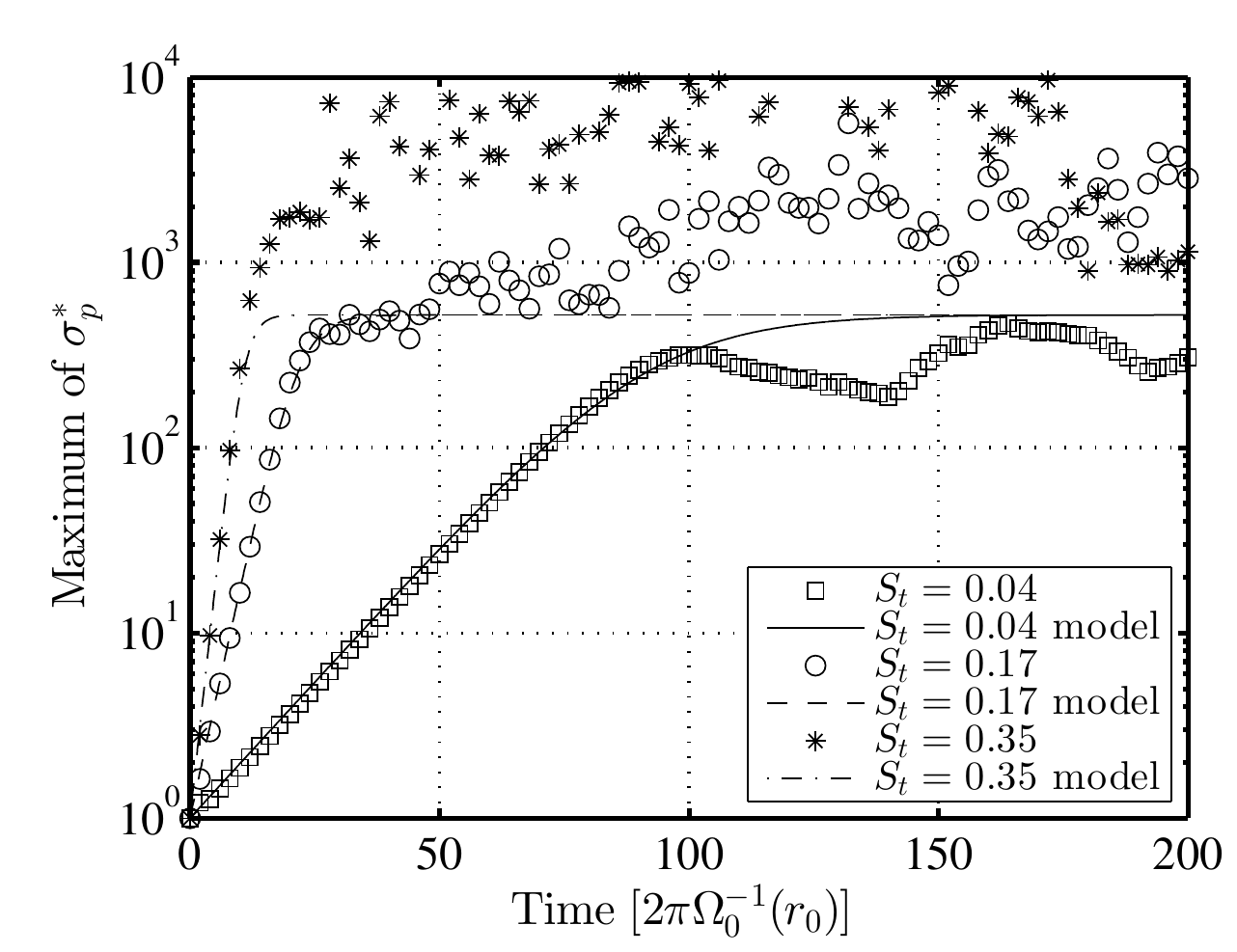} &
	\includegraphics[height=5.5cm]{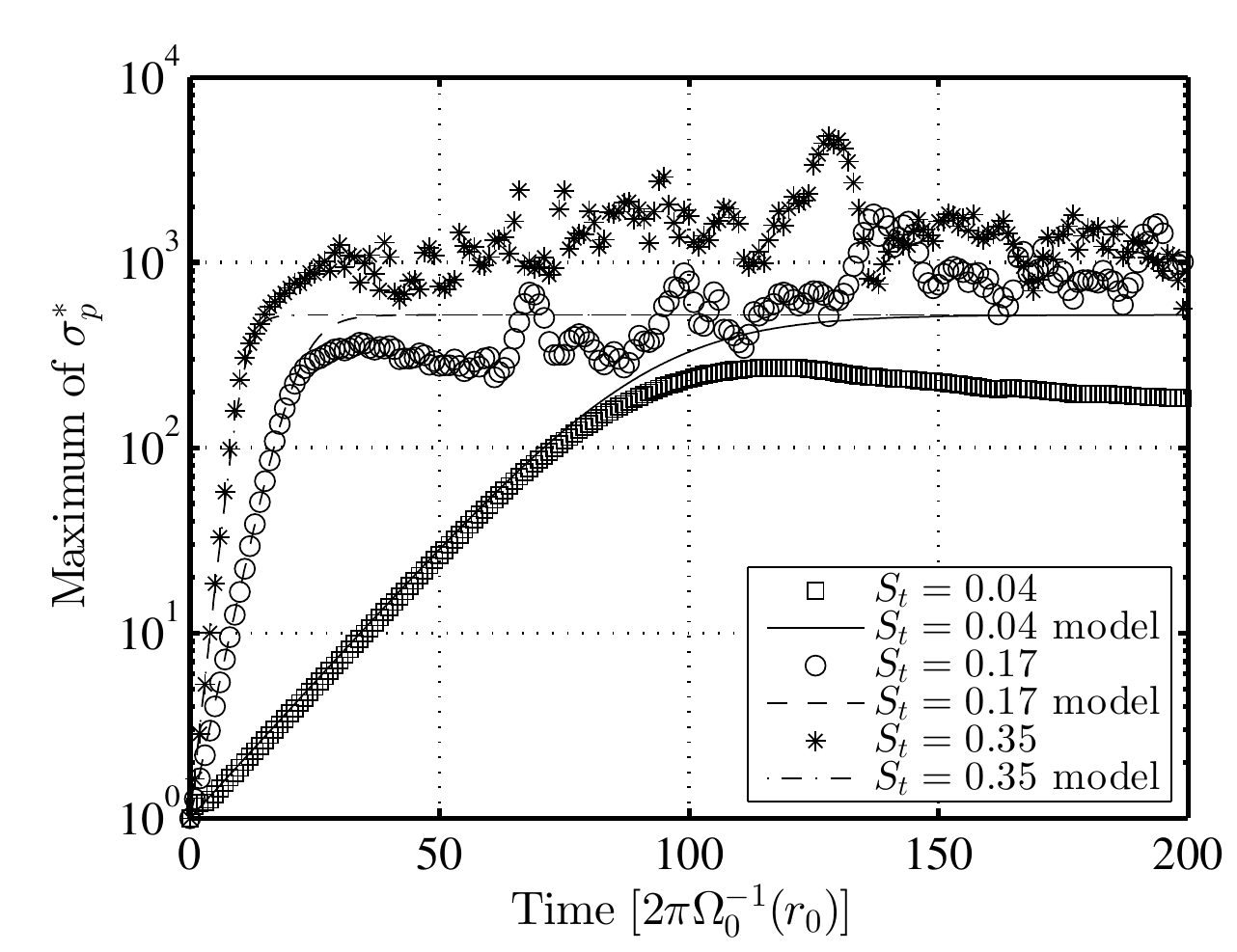} &
	\scriptsize{(b)} \\
	\includegraphics[height=5.5cm]{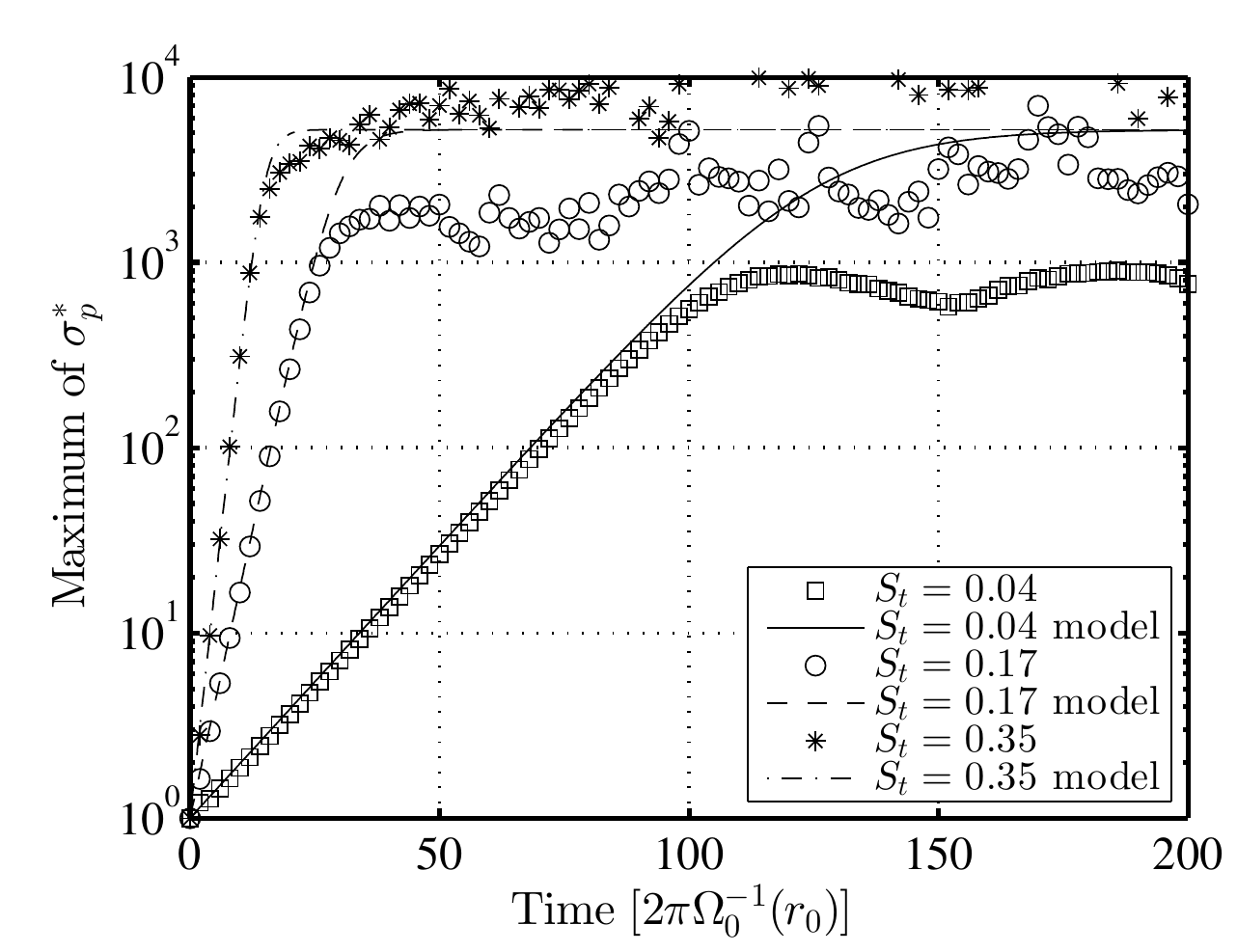} &
	\includegraphics[height=5.5cm]{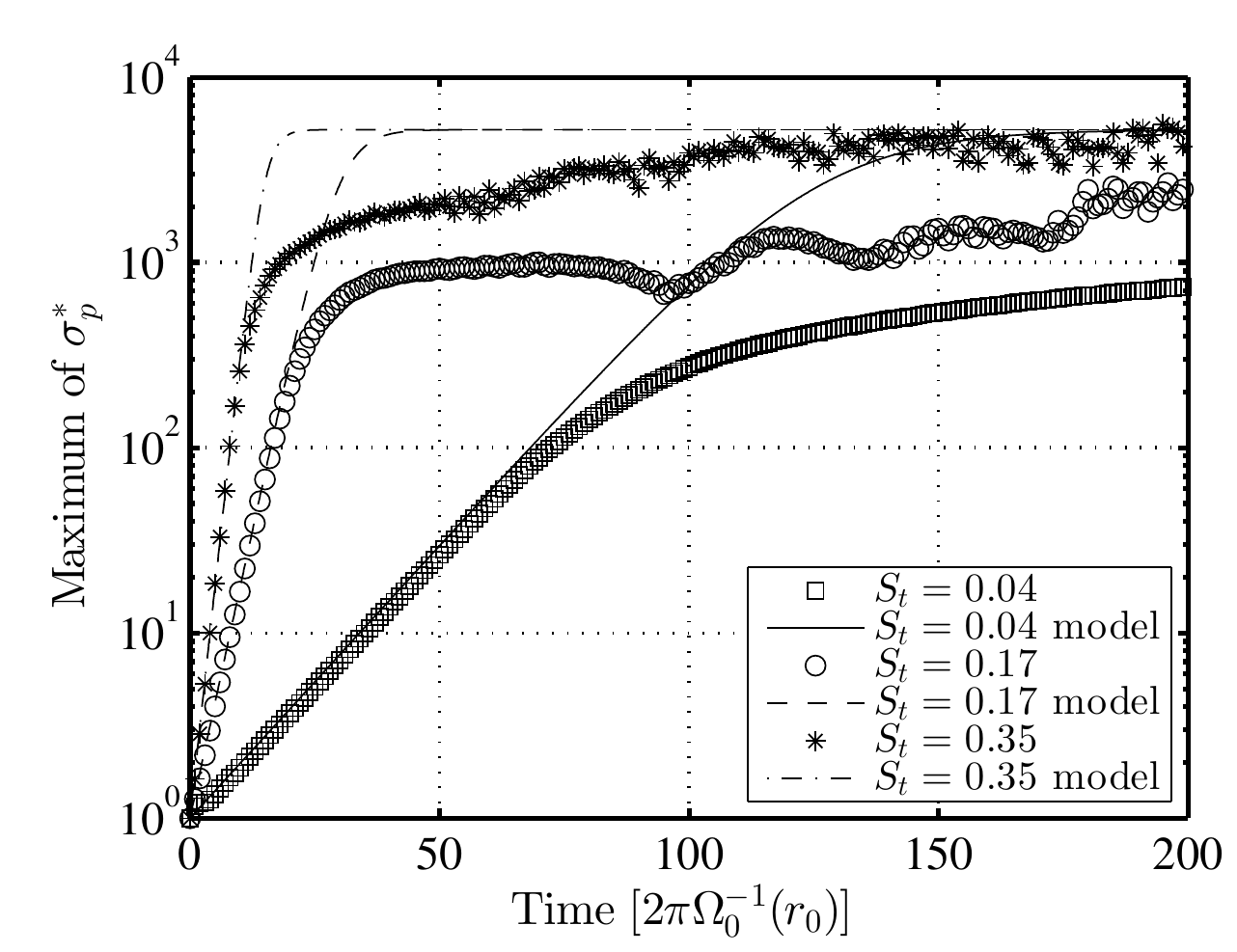} &
	\scriptsize{(c)} \\
	\end{tabular}
	\caption{\label{Linear_Capture_compar} Evolution of the dust density maximum inside the big vortex with $(R_0, \: \chi_r, \: \chi_\theta)=(-0.13, \: 0.1, \: 6.5)$, for different dust-to-gas ratios, $\epsilon = 10^{-2}, \:10^{-3},$ and $10^{-4}$ {\it{from top to bottom}} respectively. In each panel we compare the simulation results for different Stokes numbers (symbols) with the dust capture model (lines). To identify the effect of the numerical resolution, we show high resolution setups ({\it{left}}) and the same ones at low resolution ({\it{right}}). }
      \end{center}
\end{figure*}

	The Stokes number of particles at the vortex center is modified from the background value by the presence of the vortex, thus following Eq. \ref{Equ_Stokes_number} we obtain an effective Stokes number $S_t^{eff}=0.6\; S_t$. By using this effective Stokes number, we construct the model of evolution of the maximal dust density for each case (lines on the plots). The agreement with the numerical data is very good during the exponential phase, and the model describes with accuracy the growth rate of the dust density at the vortex center. This linear capture is resolved even with the lower resolution, here four times smaller in each direction. For the cases with $\epsilon = 10^{-2}$ and $\epsilon = 10^{-3}$, rows (a) and (b) respectively, the evolution before the saturation is similar for the two resolutions. However, in the last case ($\epsilon = 10^{-4}$), the data from the low resolution simulations differ from the model more quickly and the maximal values at saturation ($t \sim 50$ rot.) are $2$ to $4$ times smaller than in the high resolution cases. As the global dust-to-gas ratio reduces, the  gradients of dust density inside the vortex are steeper, and the maximal value is more than $3$ orders of magnitude larger than in the rest of the disk. High resolution simulations are thus necessary to follow the capture with accuracy.

	The saturation process, due to the vorticity pumping generated by the drag force back reaction onto the gas, is well estimated and confirmed by the simulations with $\epsilon=10^{-4}$. It is also resolved in the case $\epsilon=10^{-3}$ with $S_t=0.04$ and $S_t=0.17$, the latter case fitting particularly well the analytical model until $50$ rotations. After this phase, the vortex becomes unstable as we have seen in the previous section. The subsequent generation of large vorticity inside the vortex thus speeds up again the particle capture, responsible for the late frequent increases of dust density. However, during the first hundred rotations, the maximal dust density reached is in agreement with the asymptotic estimate of our model (see Section \ref{Sect_time_invariant}) and equals $\sigma_p^*=1 + |R_0|_{T=0}/\tilde{\epsilon}$, as shown Eq. \ref{Eq_density_saturation}. For this vortex, we can express this value as a local dust-to-gas ratio
%
\begin{equation}
\label{Eq_Local_D_G_ratio_Insta}
\frac{\sigma_p}{\sigma_g} \sim \epsilon \left( 1+ \frac{|R_0|_{T=0}}{ \tilde{\epsilon}} \right) \max \left(\frac{\sigma_0(r)}{\sigma_g} \right) = 0.32 \: ,
\end{equation}
the $\max$ function being the gas density at the vortex center relative to the background. We find out this value in other studies, in particular in \cite{Crnkovic-Rubsamen2015} where the authors conclude that vortex instability happens when the local dust-to-gas ratio is in the range $30$ to $50 \%$ and independently on the dust size. For the first time, we propose a physical interpretation of this range, as the values depend on the Rossby number of the vortex $R_0$ and on the initial dust-to-gas ratio $\epsilon$. For $\epsilon<10^{-2}$, the range $30$ to $50 \%$ corresponds to $-0.2<R_0<-0.12$, which is the typical domain of Rossby numbers that vortices have in 2D disks \citep{Surville2015}. Our model also confirms that it is not affected by the Stokes number of the particles.

	When the global dust-to-gas ratio is $\epsilon=10^{-2}$, we obtain larger values of the dust density maximum than the saturation of the model predicts. In this case, the vortex instability generates very quickly large vorticity regions within the vortex. As a consequence, the dust is still captured even if the Rossby number at the center is zero. Actually, this capture happens close to the center, in a kind of ring where large Rossby numbers are located. This process confirms that large vorticity regions are the location of dust accumulation, as predicted by our model, and are responsible for the variations of the maximal dust densities reported on the plots.

	To investigate the effect of the vortex size, we did the same analysis for the small vortex with $(R_0, \: \chi_r, \: \chi_\theta)=(-0.13, \: 0.06, \: 6.5)$. In this case the effective Stokes number of particles is $S_t^{eff}=0.75\; S_t$, because of the different value of the gas density at the vortex center. Again, using this effective Stokes number, we compare analytical model with the numerical results on Figure \ref{Linear_Capture_compar_Vortex_2}.

	One obtains a very similar behaviour as with the large vortex. The initial growth rate of the maximal dust density is in good accordance with the model. The saturation is also resolved for $\epsilon = 10^{-3}$, in particular when $S_t=0.17$, and for $\epsilon = 10^{-4}$, for the largest values of $S_t$. However, compared with the results obtained with the large vortex, the saturated value in the case $S_t=0.04$ is twice smaller, and about one order of magnitude under the predicted value. This discrepancy is probably due to a coarser effective numerical resolution on the vortex, as there are less grid cells on the small vortex than on the large one, the total number of grid cells being the same in the disk, $(N_r, \: N_\theta) = (2048, \: 4096)$. This confirm the necessity of using high resolution simulations to investigate the evolution of dusty disks. Using the analytical estimate of the critical local dust-to-gas ratio Eq. \ref{Eq_Local_D_G_ratio_Insta}, we obtain for this vortex $0.42$, due to a smaller gas density at the vortex center. It is again in agreement with previous studies.

	We show that our dust capture model is very accurate in the first stages of evolution of the dusty vortex, during the accumulation of particles within the vortex. The resulting vorticity damping at the vortex center is also in agreement with the simulations, leading in most cases, in particular when $\epsilon \le 10^{-3}$, to the good estimate of the saturated value of the dust density. By this process, we explain the observed local dust-to-gas ratio after which vortex instability happens, identified in recent studies. We are able to propose a theoretical formulation by Eq. \ref{Eq_Local_D_G_ratio_Insta}. The dust size has no impact on this criterion, but the vortex size modifies it through the gas density maximum at the center, that varies approximately as the square of the radial size \citep{Surville2015} for a given $R_0$. Finally, the capture happens also in regions of generation of vorticity, and is fasten as well as amplified by the vortex instability. This regime is the topic of the next section.

\subsection{ Vortex instability }
\label{Sect_Vort_Insta}

	After the first phase of linear capture, every simulation shows the development of a vortex instability. This phenomenon was already emphasized in \cite{Fu2014} but the authors were unsure about the nature of this instability. Here we aim at giving a more detailed picture of this process, even if an analytical description is out of the scope of our paper. We agree with the conclusions of \cite{Fu2014} on the fact that it is not a Streaming Instability, in the classical meaning \citep{Youdin2005, Kowalik2013} as the vertical coupling is impossible in our setup, nor a heavy-core instability \citep{Chang2010} as only the gas phase is affected. The dust only follows the unstable evolution of the gas flow through the effect of drag interaction. Moreover, there is no density gradient at the vortex center compatible with the development of this heavy-core instability \citep{Chang2010}. 
	
\begin{figure*}
	\begin{center}
	\begin{tabular}{cc}
	\scriptsize{$t=32$ rotations} & \scriptsize{$t=46$ rotations} \\
	\includegraphics[height=5.5cm, trim=0mm 0cm 0cm 0cm, clip=true]{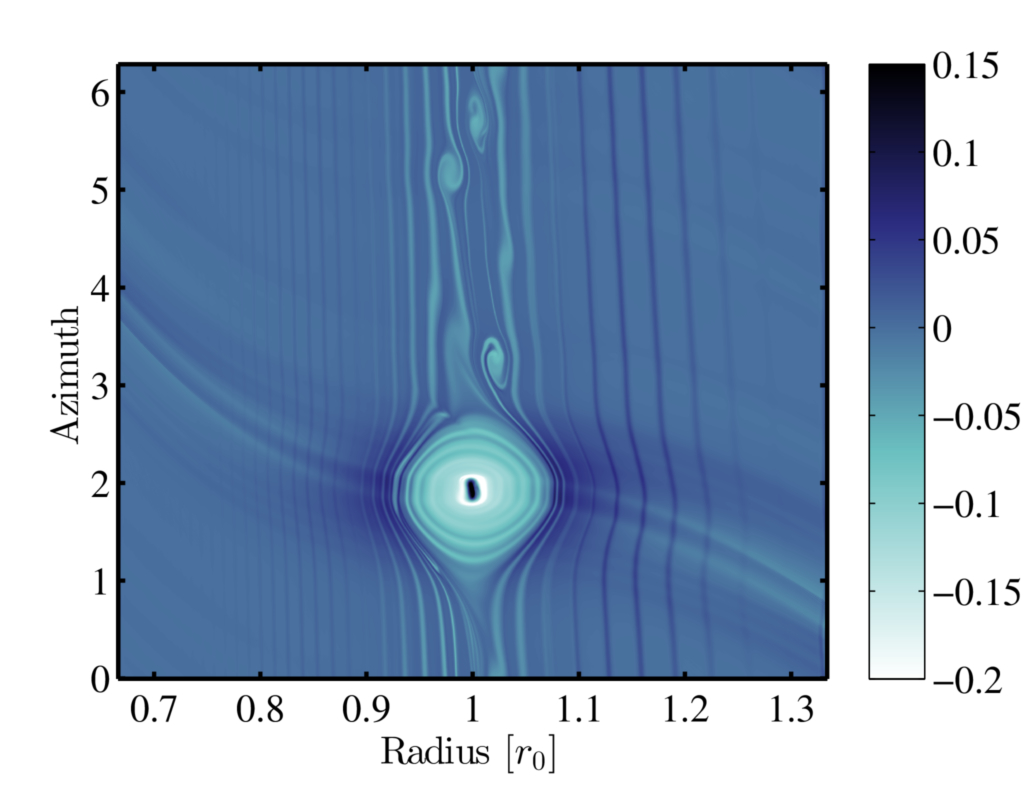} &
	\includegraphics[height=5.5cm, trim=0mm 0cm 0cm 0cm, clip=true]{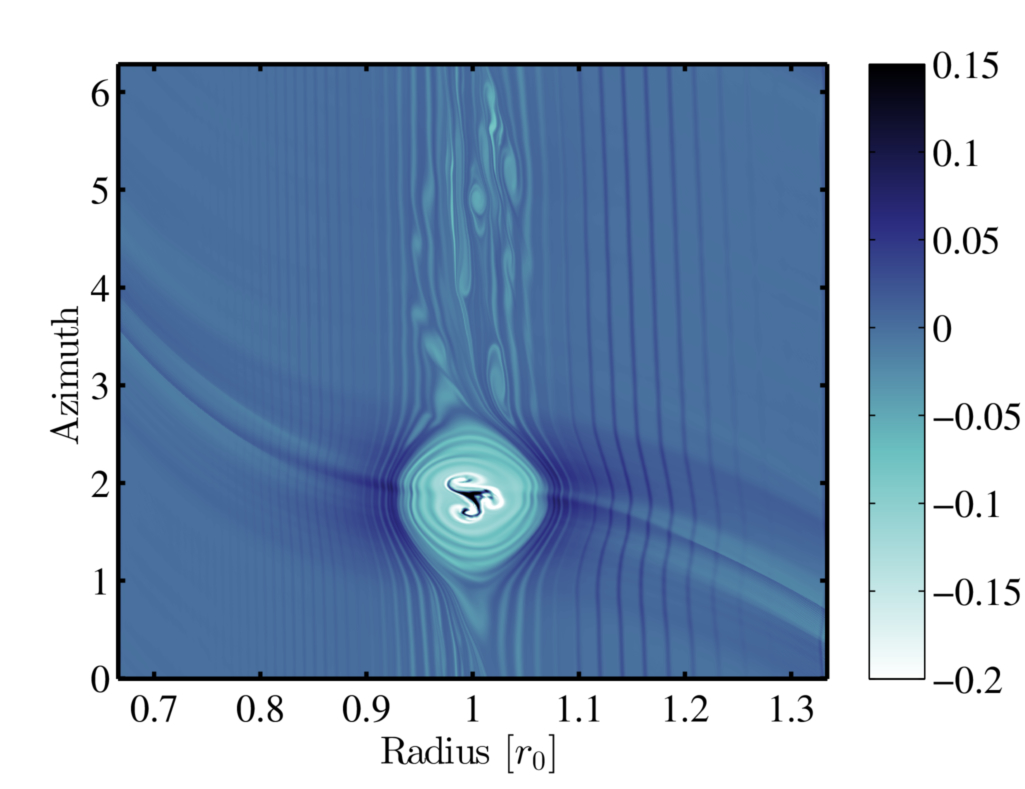} \\ \\
	\scriptsize{$t=68$ rotations} & \scriptsize{$t=138$ rotations} \\
	\includegraphics[height=5.5cm, trim=0mm 0cm 0cm 0cm, clip=true]{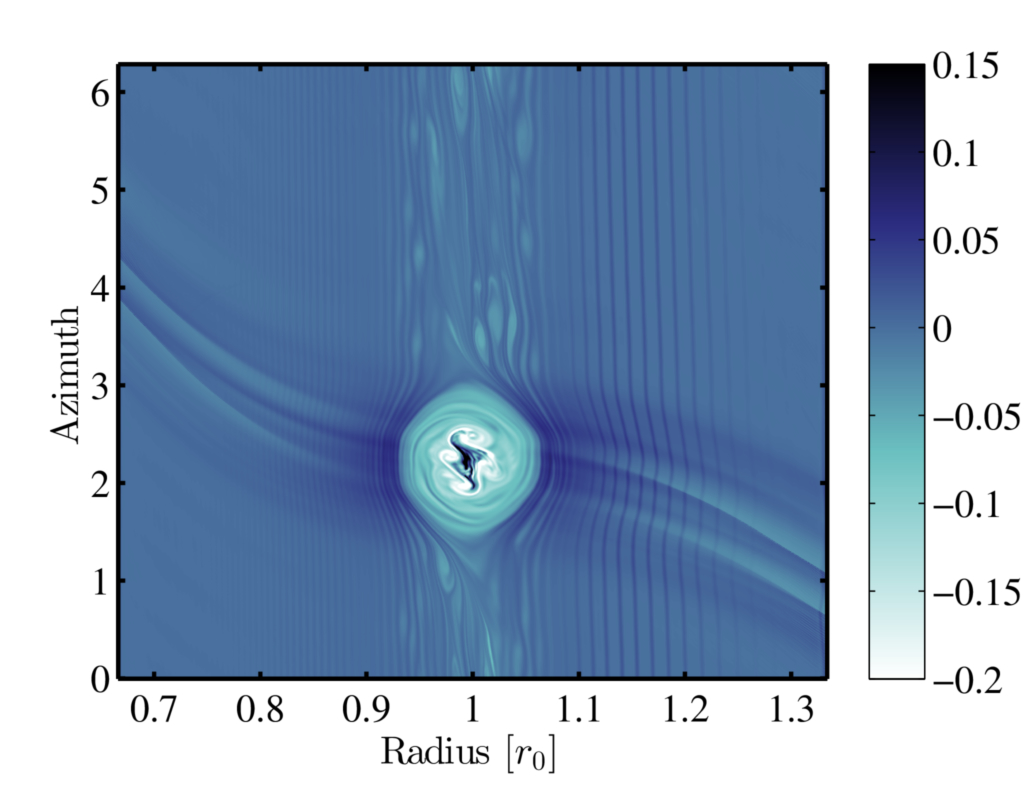} &
	\includegraphics[height=5.5cm, trim=0mm 0cm 0cm 0cm, clip=true]{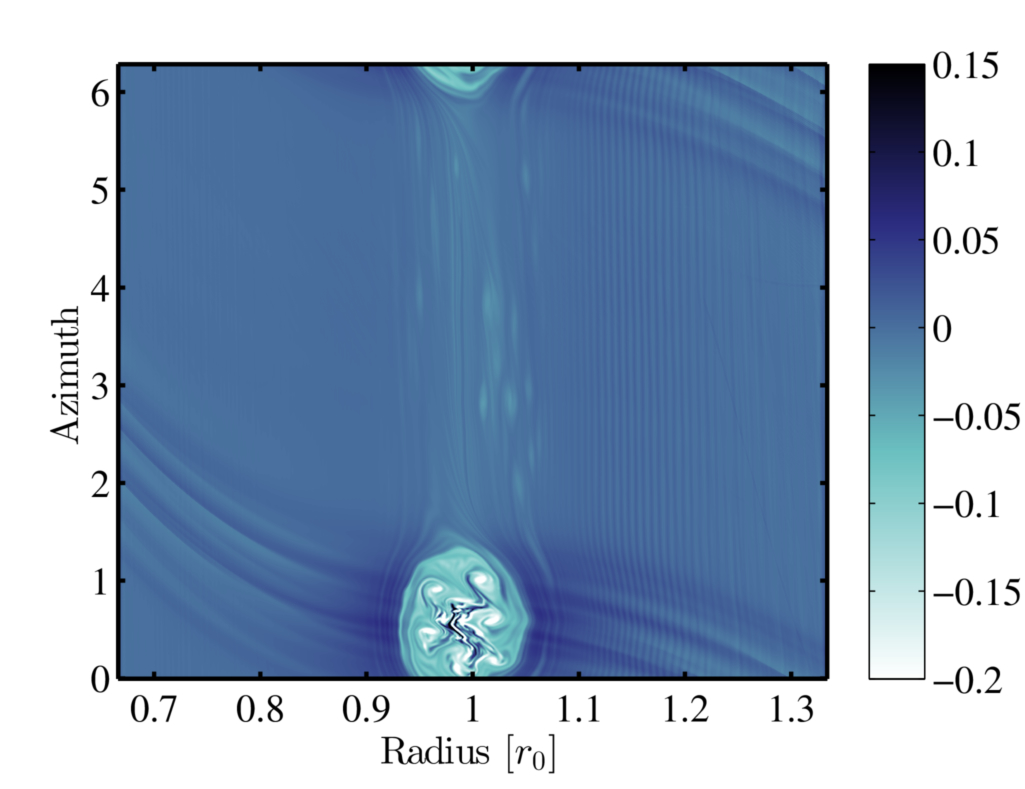} 
	\end{tabular}
	\caption{\label{Insta_vortex} Vortex instability for particles of $S_t=0.17$ in a disk of global dust-to-gas ratio $\epsilon=10^{-3}$. {\it{From left to right, top to bottom}}: We show the Rossby number of the gas after $t=32$, $t=46$, $t=68$, and $t=138$ rotations, respectively.  After the Rossby number at the center becomes zero, the gas shows a successive growth of modes $m=3$, $m=4$, and $m=6$ inside the vortex. During this process, the dust-to-gas ratio in the vortex is close to unity.}
      \end{center}
\end{figure*}

	However, the possibility of a 2D streaming instability in the $(r, \, \theta)$ plane has been argued in \cite{Raettig2015}, in their local study (shearing sheet) of capture of dust in baroclinic vortices. Even with their particle approach for describing the dust dynamics, they could capture the instability with a numerical resolution of $288 \times 288$ in a box of $4 H_0 \times 16 H_0$. It would correspond to a resolution of $936 \times 2262$ in a global disk as we use. Thus our simulations are performed with about twice more resolution than in this paper.

	The instability begins when the vorticity at the vortex center is close to the background value, i.e. when the Rossby number at the center cancels. At that time, the local dust-to-gas ratio is relatively large, typically greater than $0.1$, in agreement with Eq. \ref{Eq_Local_D_G_ratio_Insta}. The spatial distribution of dust is also still very symmetric to the vortex center. However, the drag force exerted onto the gas has a significant amplitude. Any small perturbation in the dust density, position, or in dust velocities will affect directly the gas motion. For example, a small increase of density will increase the local vorticity, as predicted by the capture model. Though, due to the drag force, a small phase shift exists between the dust and gas velocities, in the local vortex frame. Gas regions affected by dust variations will then modify retroactively in a slightly delayed manner the dust dynamics. Step by step, this will amplify any perturbation in the dust phase, making the instability grow. This process is only possible when the drag force amplitude is large enough compared to other forces acting onto the gas (Coriolis and pressure forces). It is a streaming effect, that we could call a "vortex streaming instability". 
	
	We can give arguments to this qualitative description by exploring in detail the results of one simulation setup: $\epsilon=10^{-3}$, $S_t=0.17$ in the large vortex. Figure \ref{Insta_vortex} illustrates the development of the instability, by showing the gas vorticity at different times of the evolution, $t=32$, $t=46$, $t=68$, and $t=138$ rotations. At $t=32$ orbits, one can already observe a significant elongation of the region at the vortex center where the Rossby number cancels. Later, at $t=46$ orbits, a clear cross pattern is visible, revealing a mode $m=3$ in a local elliptic frame centerd in the vortex. Successively, higher modes settle, with $m=4$ at $t=68$ and $m=6$ at $t=138$ rotations. We thus have the growth and saturation of successively higher modes in the vortex flow. During this simulation, as the local dust-to-gas ratio is larger than $0.1$ inside most of the vortex region, the vortex streaming instability can develop in the whole vortex. Later on, it leads to the spreading of the gas and the dust in a dusty ring. 

\subsubsection{ Comparison with a KH like instability: the RWI }

	The growth of different modes in the vortex could also be interpreted as the development of another type of instability. In particular, we can consider the possibility that some single fluid instability such as the Kelvin-Helmoltz instability (KHI) may be at play, as a vortex is partakes to a shearing flow. A simple way to test this possibility is to compare the dusty evolution of a setup similar to KHI for disks, namely the Rossby Waves Instability (RWI).

	During the process of the linear RWI, a radially stable profile of vorticity becomes unstable to azimuthal modes. These modes grow with a rate depending mostly on the value of the Rossby number in the initial perturbation, and on the mode number. Thus a change in the gas vorticity, density, and pressure profiles will modify both the growth rate of a given mode, but also the mode selection. The coupling of a dust phase with the gas induces a time dependent change in the vorticity profile of the gas, as shown previously, but neither significantly in the gas pressure nor density profiles. As the RWI linear evolution is well known, we can precisely find the impact of the drag coupling in the evolution of the instability.

	 The initial axisymmetric profile of the setup is a Gaussian bump of density, with a polytropic relation between density and pressure: 

\begin{align}
	\sigma_g(r) & = \sigma_0(r)\left[ 1 + f_{gauss}(r) \right] \: , \\
	P(r) & = P_0(r) \left[ 1 + f_{gauss}(r) \right]^\gamma \: , \\
	f_{gauss}(r) & = f_a ~ exp \left( - \frac{(r-r_0)^2}{\Delta_r^2}\right) \: ,
\end{align}
where we have set up the bump amplitude to $f_a=15\%$ and its radial width to $\Delta_r=0.06$ $r_0$. The initial radial velocity is null, and the initial azimuthal velocity is given by the centrifugal balance:
	
\begin{equation}
	V_g^2(r) = V_k^2(r) + \frac{r}{\sigma_g(r)} \partial_r P(r) \: .
\end{equation}

	This creates an initial axisymmetric profile of vorticity with a minimum Rossby number $R_0(r_0)=-0.07$, slightly smaller than the typical Rossby numbers in vortices, which range mostly in $-0.2<R_0<-0.1$. But our choice insures a slow growth of the perturbing modes, allowing the dust phase to modify the vorticity profile during the linear growth phase of the RWI.

	The initial state in then perturbed by a random noise of small Gaussian perturbations the amplitude of which is $1\%$ the value of $f_a$. This random noise is superimposed on the density and pressure profile, leading to an isothermal triggering noise. A wide number of azimuthal modes is generated by this noise. It allows to let the instability select the most unstable modes.

	The dust phase initial state is simply an unperturbed power law profile, like in the setups with a vortex. To avoid any difference between dust and gas background velocities, due to the global radial pressure gradient, we impose a constant density background $\sigma_0(r) = \sigma_0(r_0)$, as well as a constant temperature $T_0(r) = T_0(r_0)$. This unrealistic disk profile is only a case study, and allows for a simpler dust/gas evolution, and enables a cleaner analysis of the dusty RWI evolution.

	We performed two simulations, with a numerical resolution $(N_r, N_\theta)=(512, 1024)$. In the first case, we evolve only the gas phase, to obtain the RWI evolution of the initial state. In the second case, we evolve the coupled dust/gas flow, with the same initial gas state (including the same noise). The initial dust-to-gas ratio is in this case $\epsilon = 10^{-2}$. The Stokes number at $r_0$ of the background disk is $S_t = 0.17$, and considering the variation of the gas density profile due to the Gaussian initial state, the effective Stokes number in the bump is $S_t{^{eff}} = 0.148 ~$.

	The dusty evolution of the RWI was already explored in previous papers \citep{Inaba2006, Lyra2009, Meheut2012}. However, the typical  used initial gas setup generates a very fast instability, with a typical growth rate of one orbital period, $2\pi \Omega_0^{-1}(r_0)$. Thus after less than $10$ orbits, the bump is clumped in a chain of $5$ to $7$ vortices, depending on the most unstable mode. On the other hand, the time needed for the drag force to really affect the vorticity profile, and thus the linear evolution of the RWI, is longer, usually $3$ to $4$ times depending on the Stokes number of the particles. These authors ended up with a system where dust capture happens inside the vortices, and not enough inside the bump, during the growth phase of the RWI. They did not measure any significant change when considering the dusty case.

	This is the reason why we have chosen a tiny initial gas state to trigger the RWI, so that the saturation happens in more than $50$ orbits. In the same time, the dust-to-gas ratio and the Stokes number used in the dusty case generate a capture time of the particles of about $25$ orbits. We insure that the drag coupling will act before the saturation of the RWI, and we will be able to measure its impact on the growth of the modes.

\begin{figure}
	\begin{center}
	\begin{tabular}{c}
	\includegraphics[height=5.5cm]{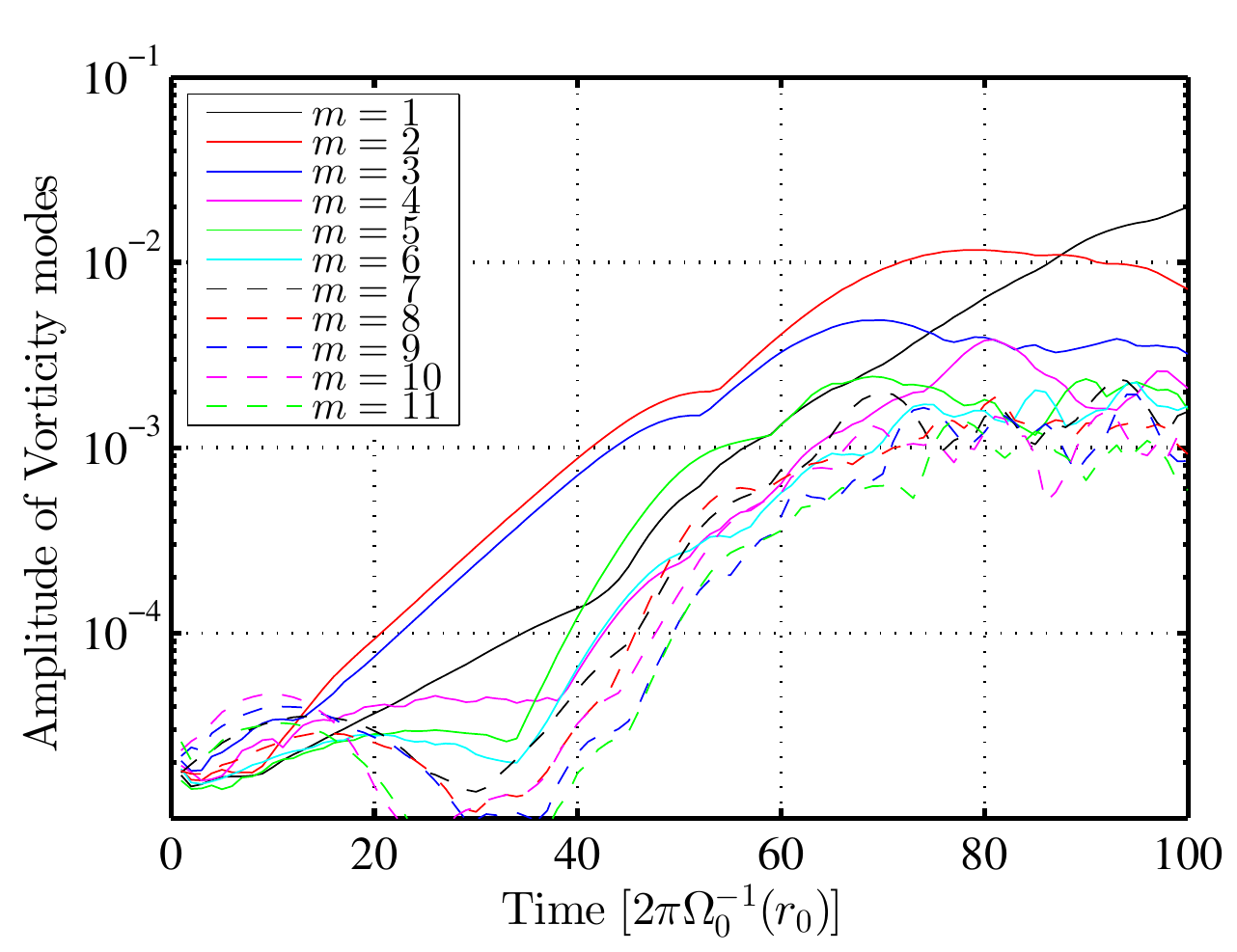} \\
	\includegraphics[height=5.5cm]{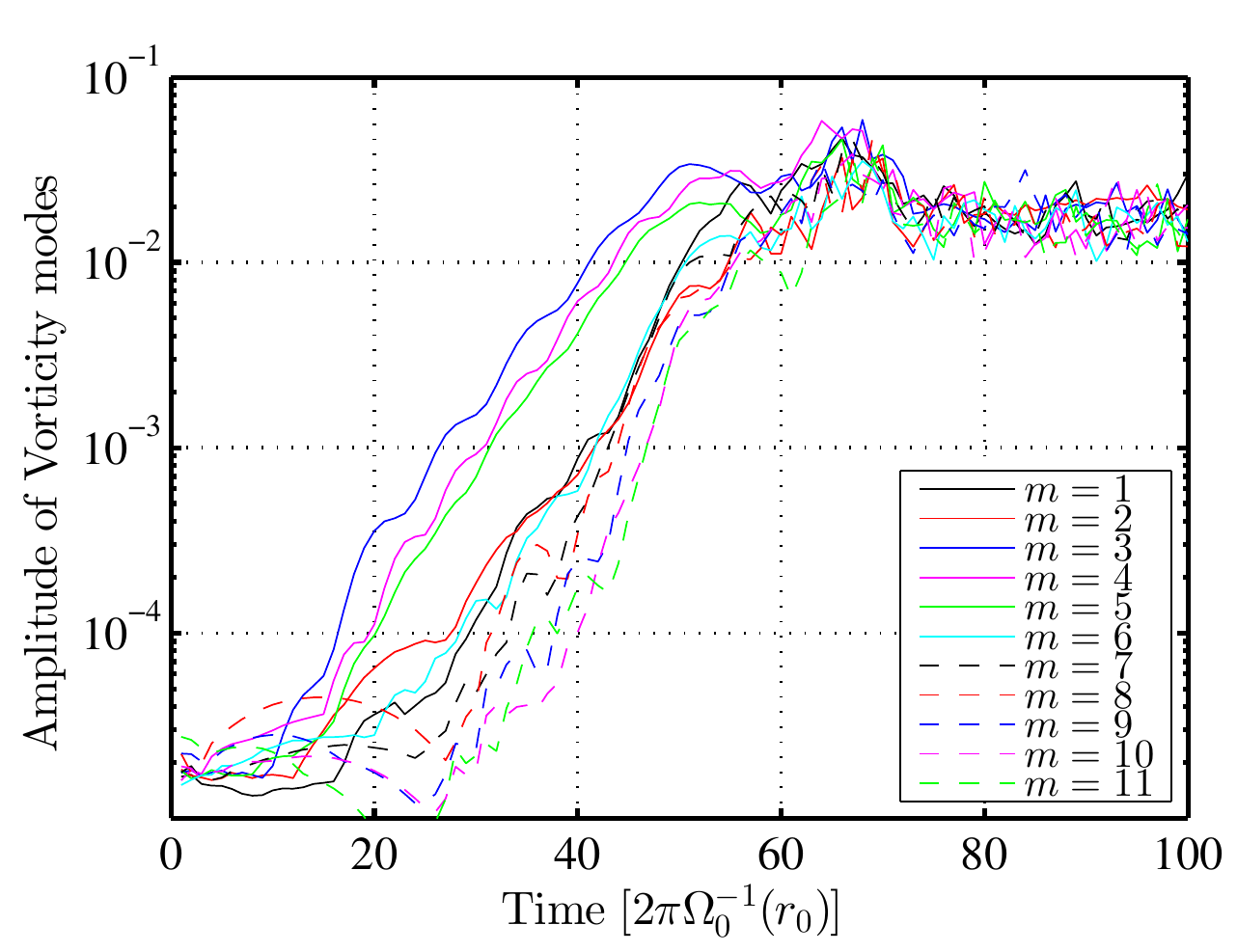}
	\end{tabular}
	\caption{\label{Modes_vorticity} Evolution of the amplitude of the first eleven azimuthal modes of the gas vorticity during the RWI. {\it{Top}}: In the case of a pure gas setup, the modes $m=2$ and $m=3$ are the first to grow.  {\it{Bottom}}: In the case of a dusty flow setup, the modes $m=3$, $m=4$, and $m=4$ are the first to grow with a larger rate.  }
      \end{center}
\end{figure}

\begin{figure}
	\begin{center}
	\includegraphics[height=5.5cm]{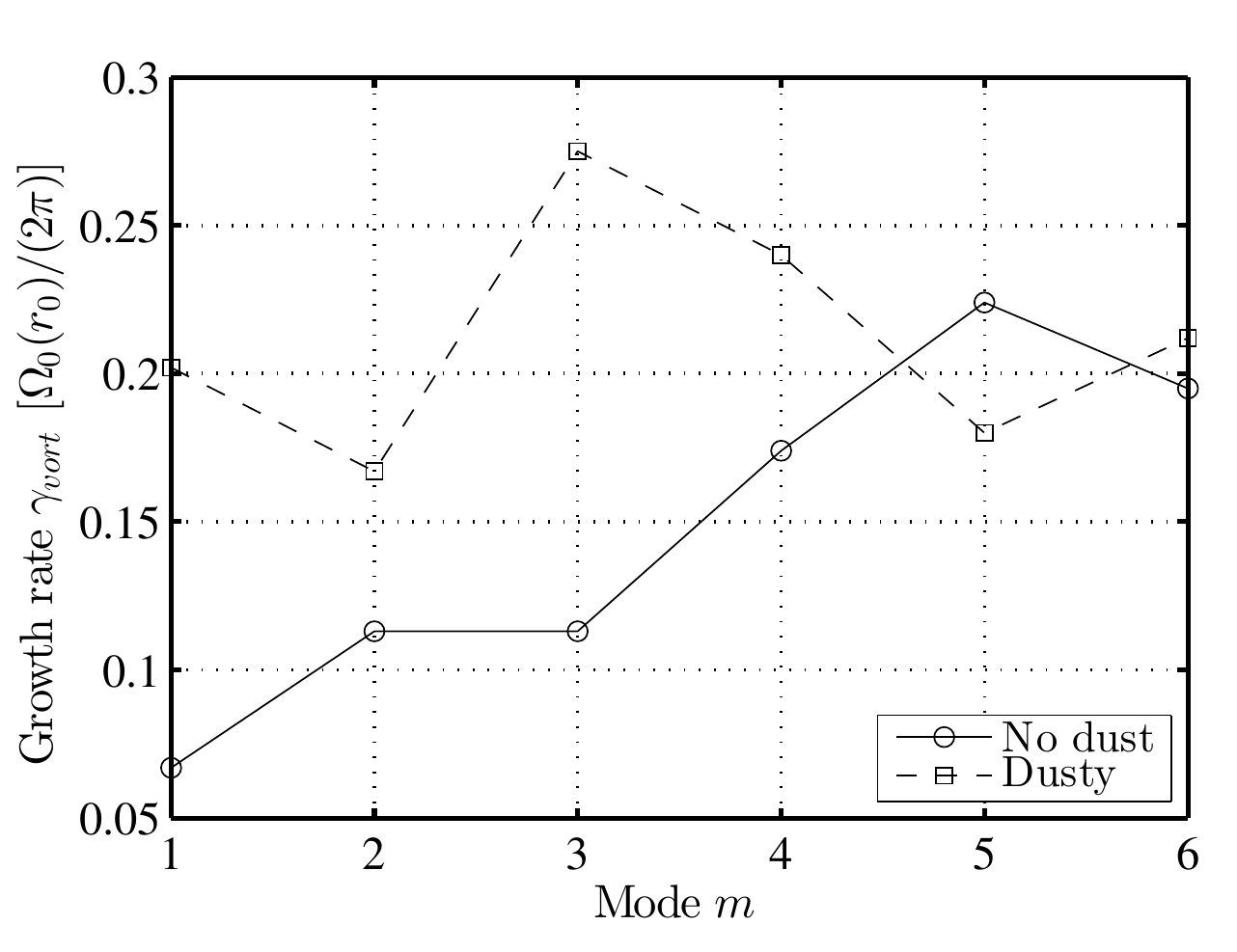}
	\caption{\label{Modes_vorticity_compar} Comparison of the growth rate of the first six azimuthal modes of the gas vorticity during the RWI, for the pure gas case (solid line and circles) and for the dusty case (dashed line and squares).}
      \end{center}
\end{figure}

	We tracked the time evolution of the $11$ first azimuthal modes of the gas vorticity in the non dusty and the dusty cases. The amplitude of the modes is presented in a semilog plot Figure \ref{Modes_vorticity}. In the pure gas setup, on the top, we see the growth of two main modes, $m=2$ and $m=3$, occurring between $t=10$ to $t=70$ orbits. The lowest mode, $m=1$, also grows until $t=40$ orbits at a slow rate, and then at a faster rate. This date corresponds to the excitation of modes with larger $m$. However, the saturated amplitude of these modes ($m > 4$) is by one order of magnitude smaller than the first unstable modes.

	In the dusty case (bottom plot), the mode structure is different. Three main unstable modes grow rapidly from $t=10$ to $t=50$ orbits: $m=3$, $m=4$, and $m=5$. Soon after, at $t=30$ rotations, the growth of all the other tracked modes is visible; all the modes seem to grow at a similar growth rate. This difference in mode growth rates between the pure gas and the dusty cases has been observed in \cite{Lyra2009}, but not tracked quantitatively as here. On the other side, \cite{Hendrix2014} show that the KHI mode structure is not significantly affected by the coupling with the dust. On Figure \ref{Modes_vorticity_compar} the comparison of the growth rates of the first six modes in the pure gas and dusty case is shown. While in the dusty case, all the modes grow at $\gamma_{vort} \sim 0.22$, there is a mode selection in the pure gas case. Only $m=5$ grows at $\gamma_{vort} \sim 0.22$; larger or smaller mode numbers grow much more slowly. Finally, the saturation of the instability is also clearly different in the dusty case (see Figure \ref{Modes_vorticity}). After $t=60$ rotations, all the modes are excited and their amplitude is similar. Compared to the pure gas case, mode numbers $m>4$ have an amplitude $10$ times larger in the dusty evolution. 
	
	During this process, gas pressure and density profiles change by less than $1 \%$, an effect not large enough to generate such an obvious different evolution of the flow in the dusty case. The absence of mode selection gives also argument for concluding that it is not a shearing instability of the gas. This comparison study highlights the influence of the coupling between the two fluids, and confirms in a more quantitative way the existence of the "vortex streaming instability".

\subsubsection{ Survival of unstable vortices }

\begin{figure*}
	\begin{center}
	\begin{tabular}{ccc}
	\scriptsize{$t=200$ rotations} & \scriptsize{$t=500$ rotations} & \scriptsize{$t=1000$ rotations} \\
	\includegraphics[height=4.7cm, trim=6mm 0cm 0cm 0cm, clip=true]{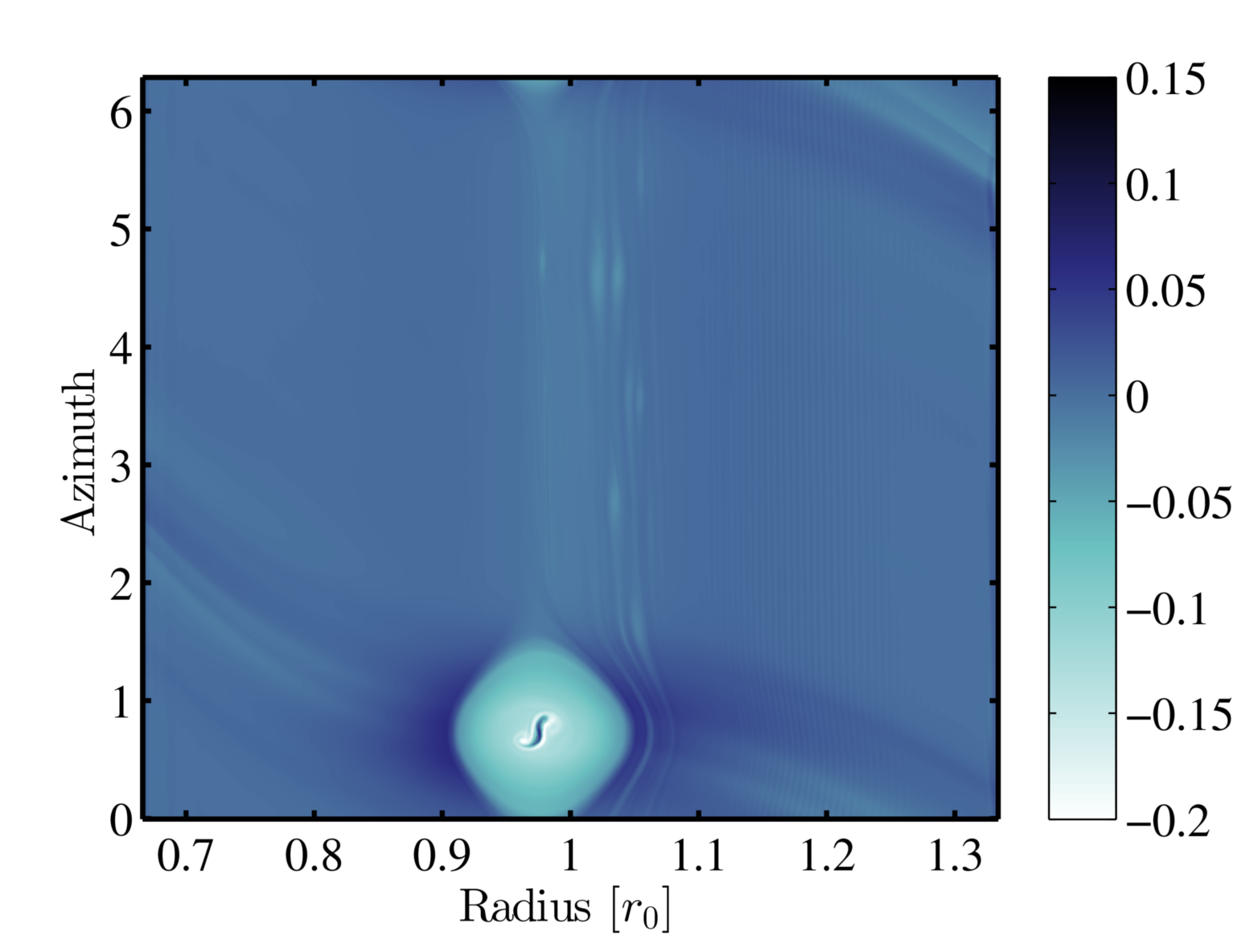} &
	\includegraphics[height=4.7cm, trim=6mm 0cm 0cm 0cm, clip=true]{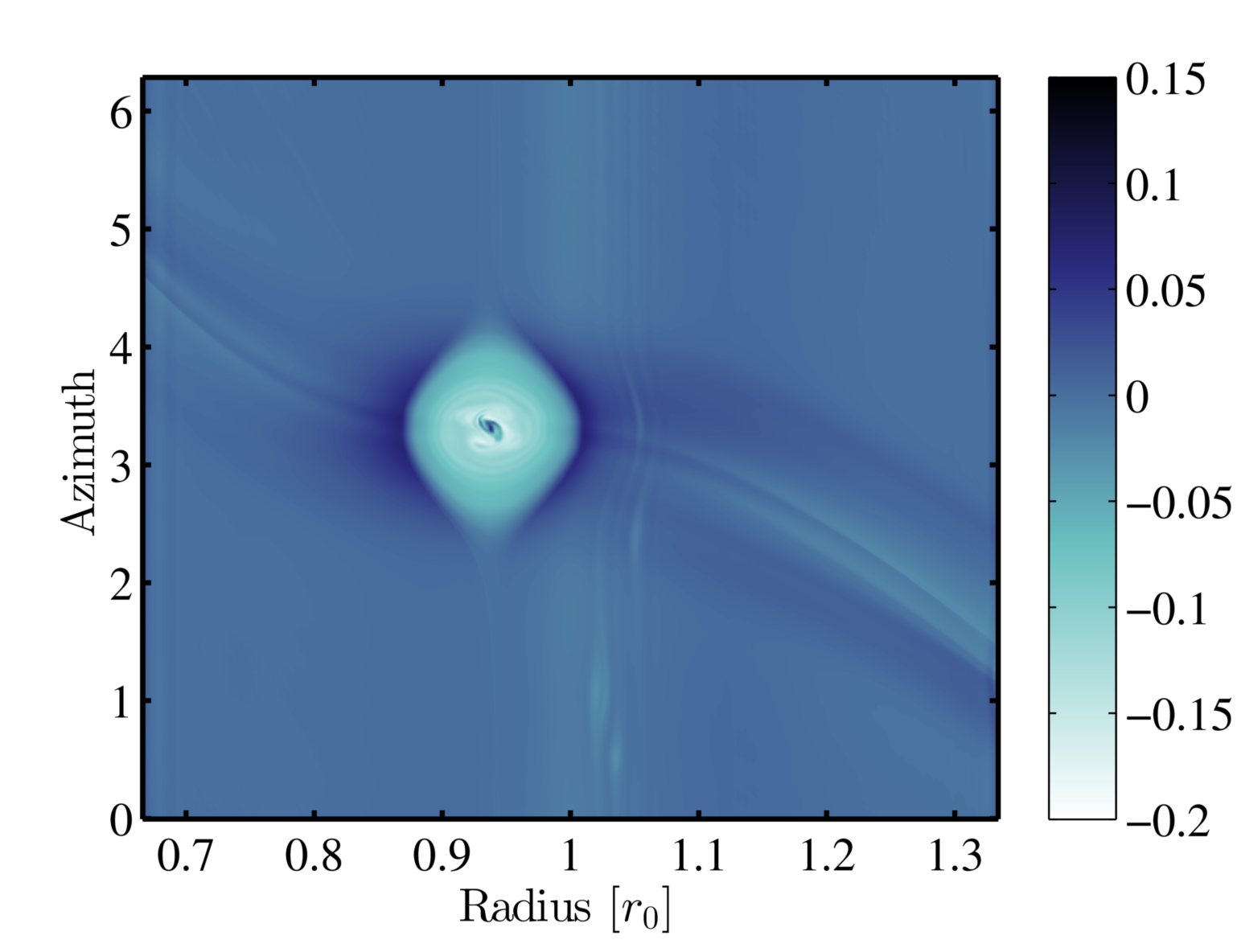} &
	\includegraphics[height=4.7cm, trim=6mm 0cm 0cm 0cm, clip=true]{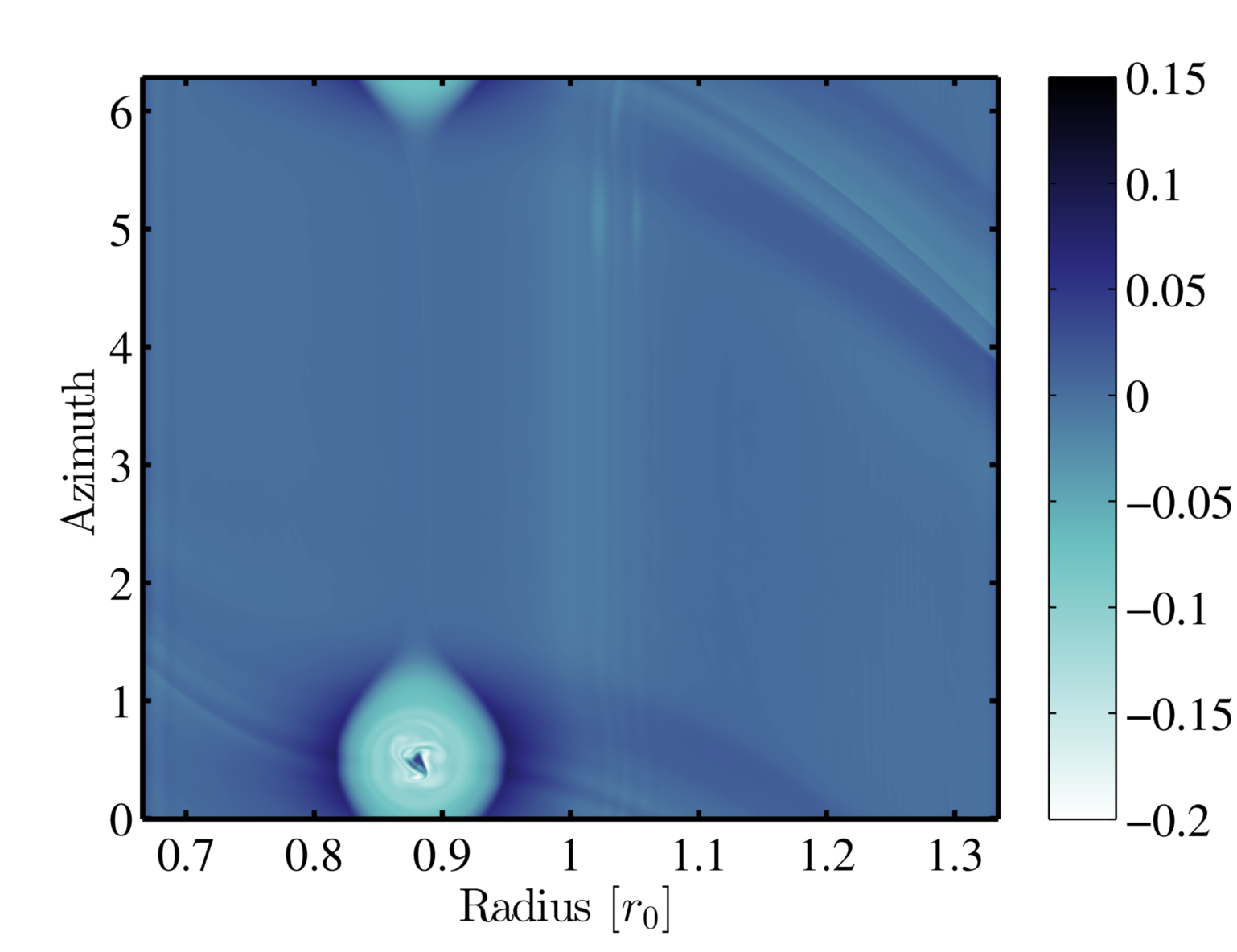} \\

	\includegraphics[height=4.7cm, trim=6mm 0cm 0cm 0cm, clip=true]{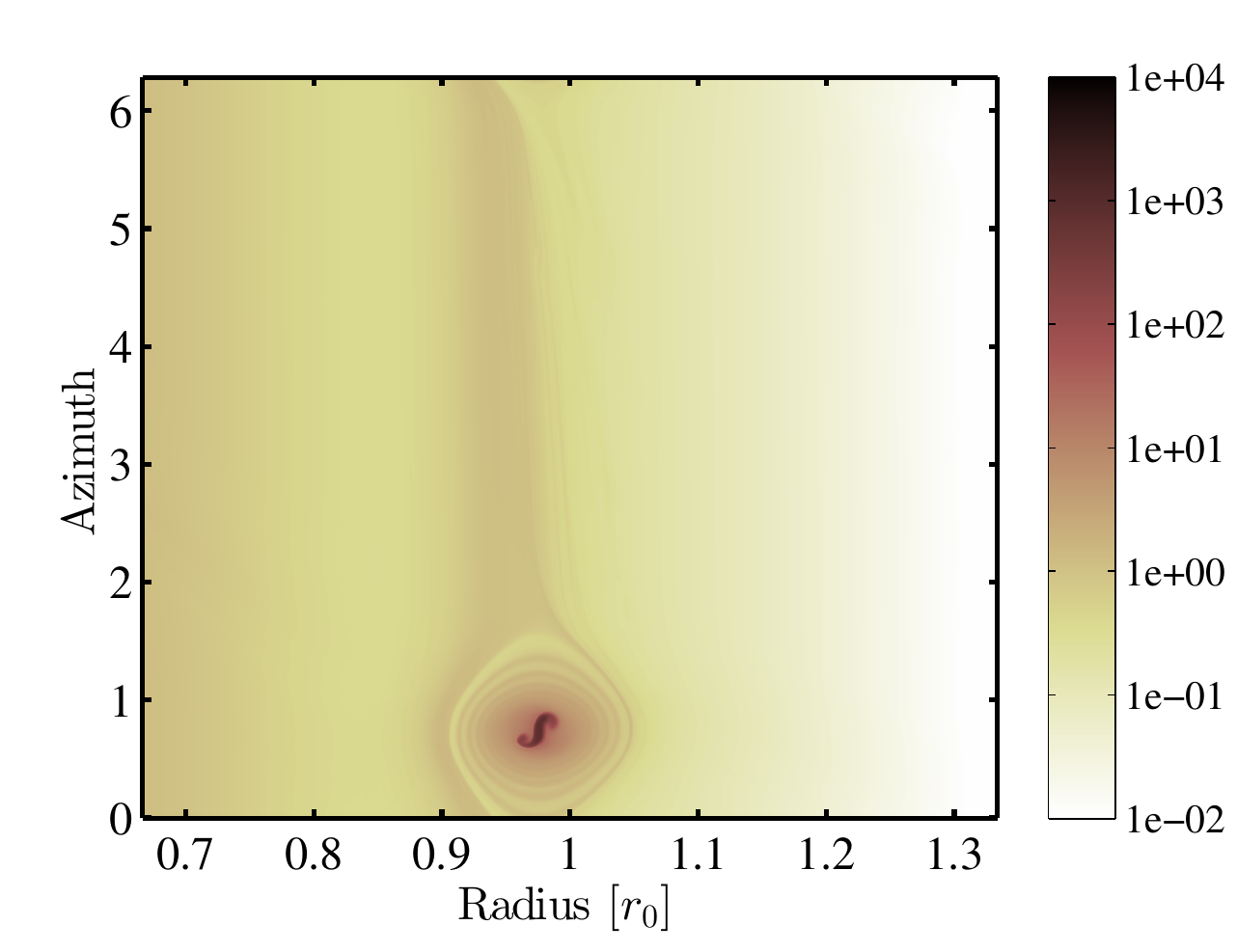} &
	\includegraphics[height=4.7cm, trim=6mm 0cm 0cm 0cm, clip=true]{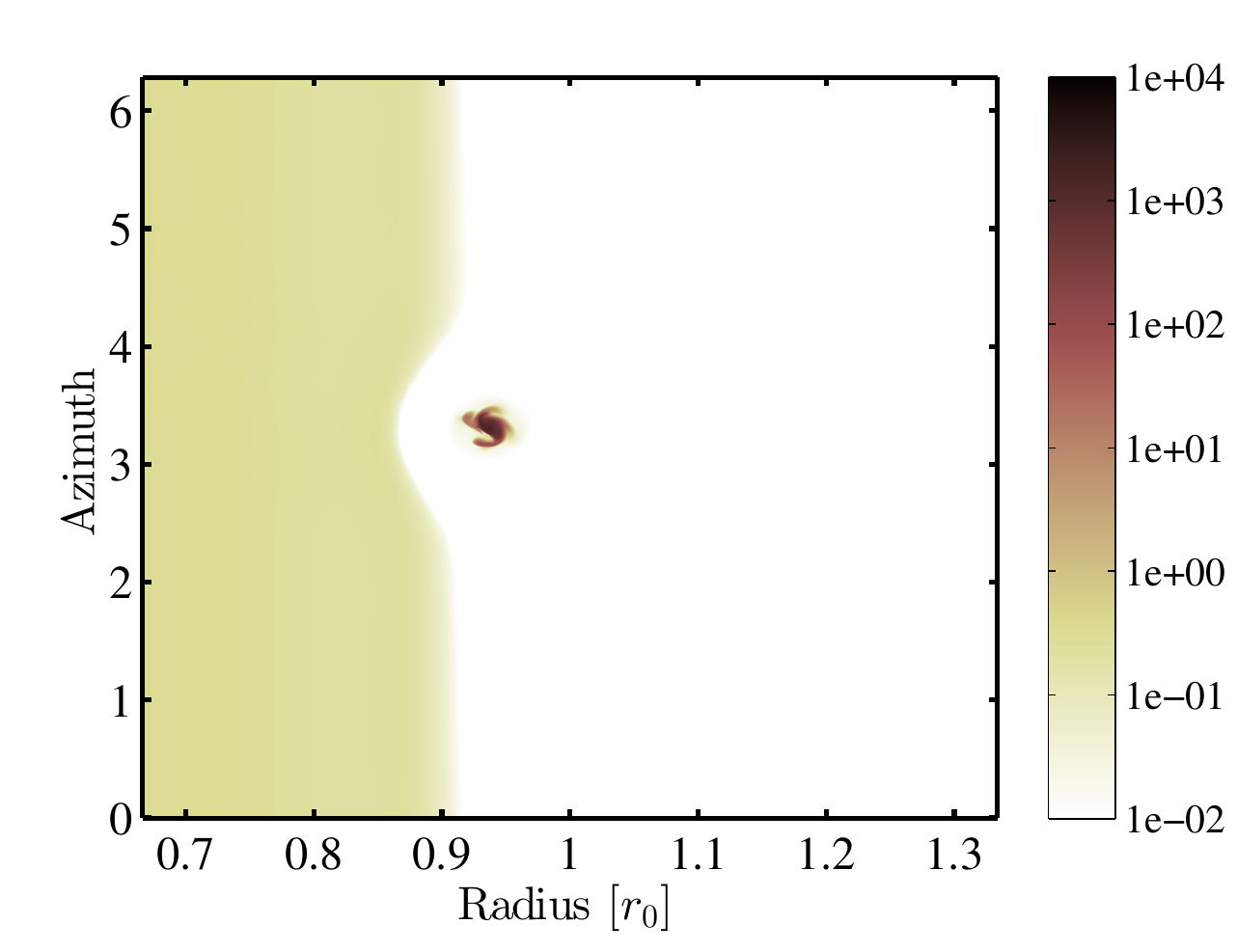} &
	\includegraphics[height=4.7cm, trim=6mm 0cm 0cm 0cm, clip=true]{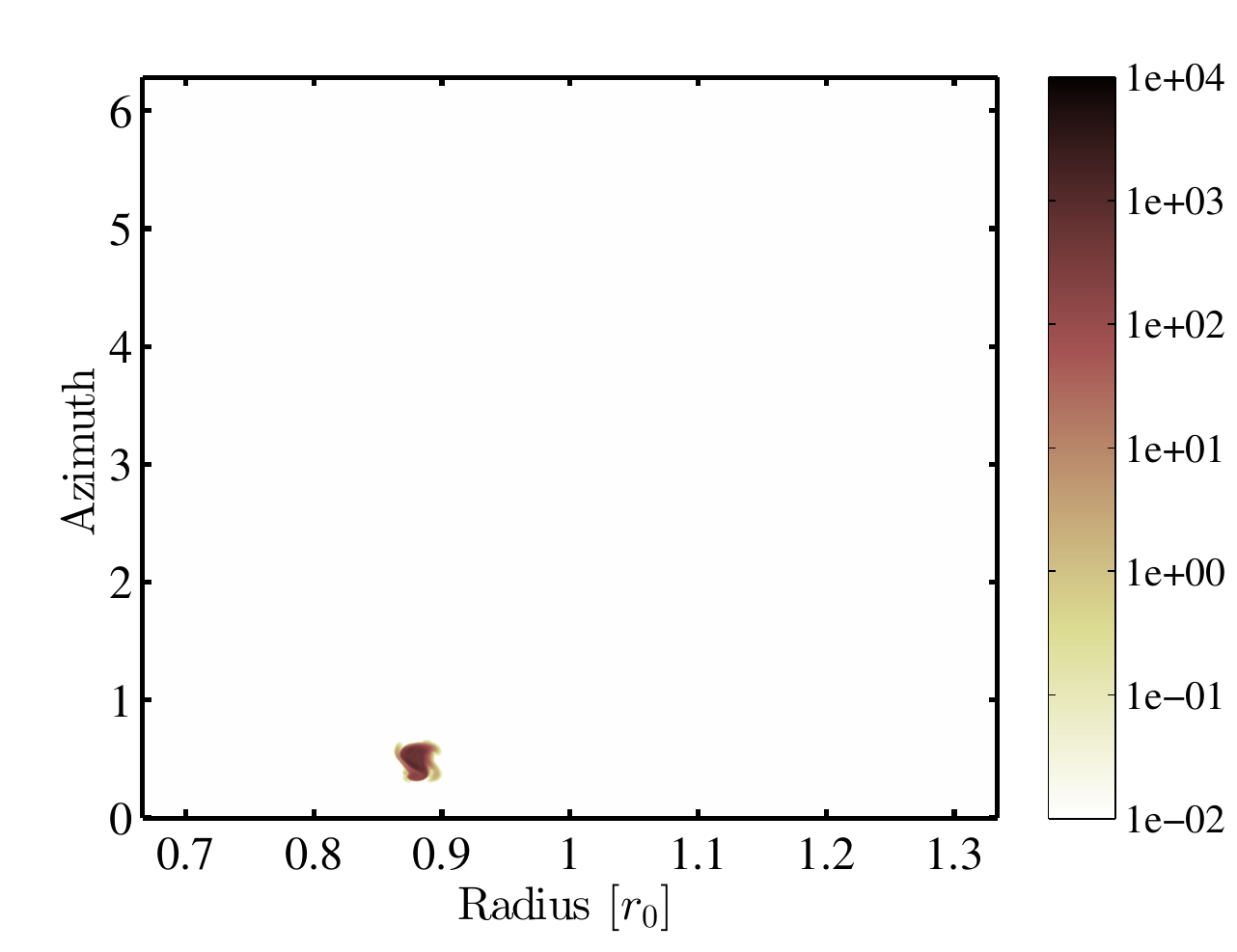} \\

	\end{tabular}
	\caption{\label{Stable_vortex_St_004} Persistence of a dusty unstable vortex for dust parameters $\epsilon=10^{-4}$ and $S_t=0.04$. We plot $(r, \; \theta)$ maps of the Rossby number ({\it{top}}) and of dust density ({\it{bottom}}) for the run using the large vortex $(R_0, \: \chi_r, \: \chi_\theta)=(-0.13, \: 0.1, \: 6.5)$. {\it{From left to right}}: snapshots at $t=200$, $t=500$, and $t=1000$ rotations, respectively. During the evolution, the local dust-to-gas ratio in the inner parts of the vortex is close to $7 \times 10^{-2}$.}
      \end{center}
\end{figure*}

\begin{figure*}
	\begin{center}
	\begin{tabular}{ccc}
	\scriptsize{$t=200$ rotations} & \scriptsize{$t=500$ rotations} & \scriptsize{$t=1000$ rotations} \\
	\includegraphics[height=4.7cm, trim=6mm 0cm 0cm 0cm, clip=true]{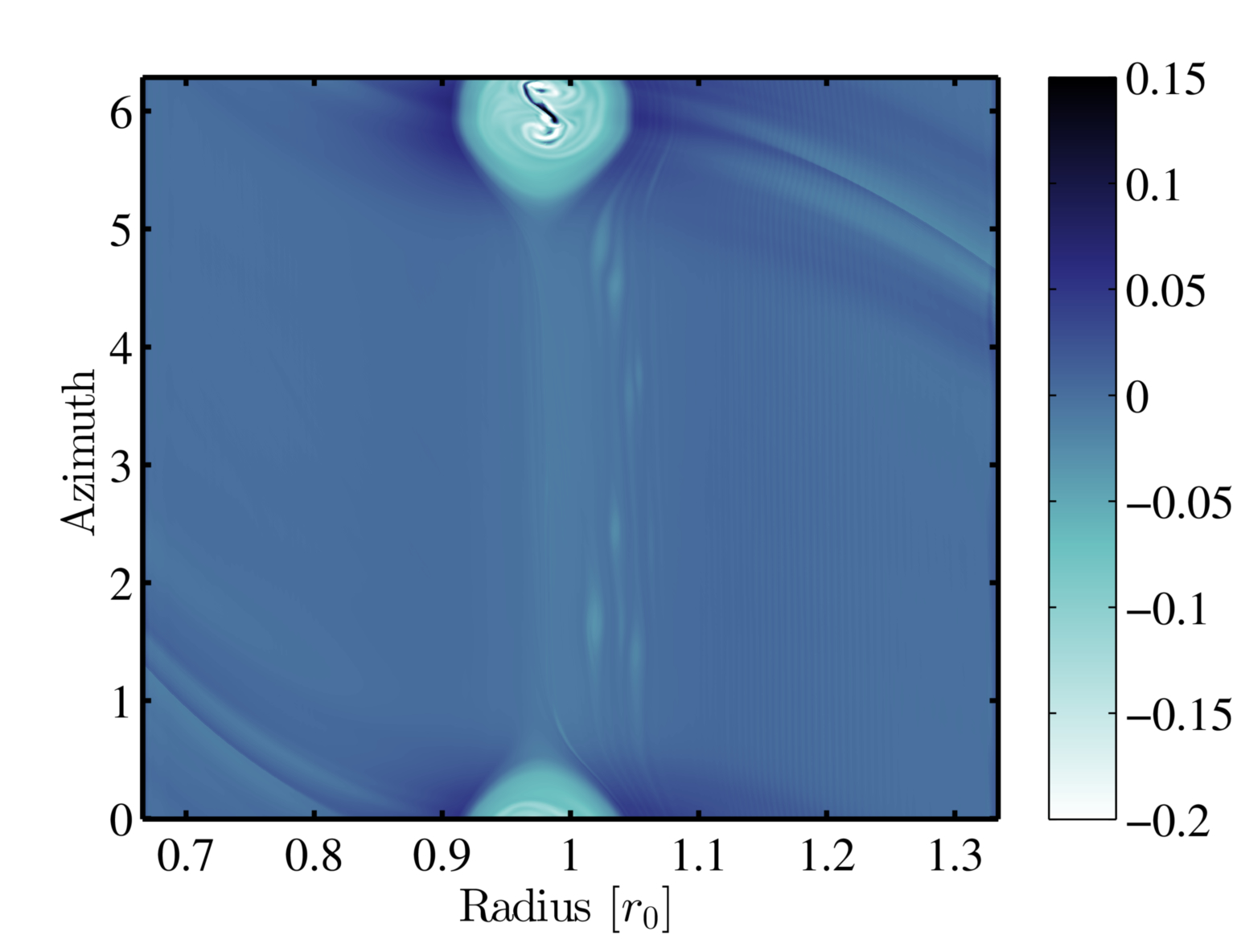} &
	\includegraphics[height=4.7cm, trim=6mm 0cm 0cm 0cm, clip=true]{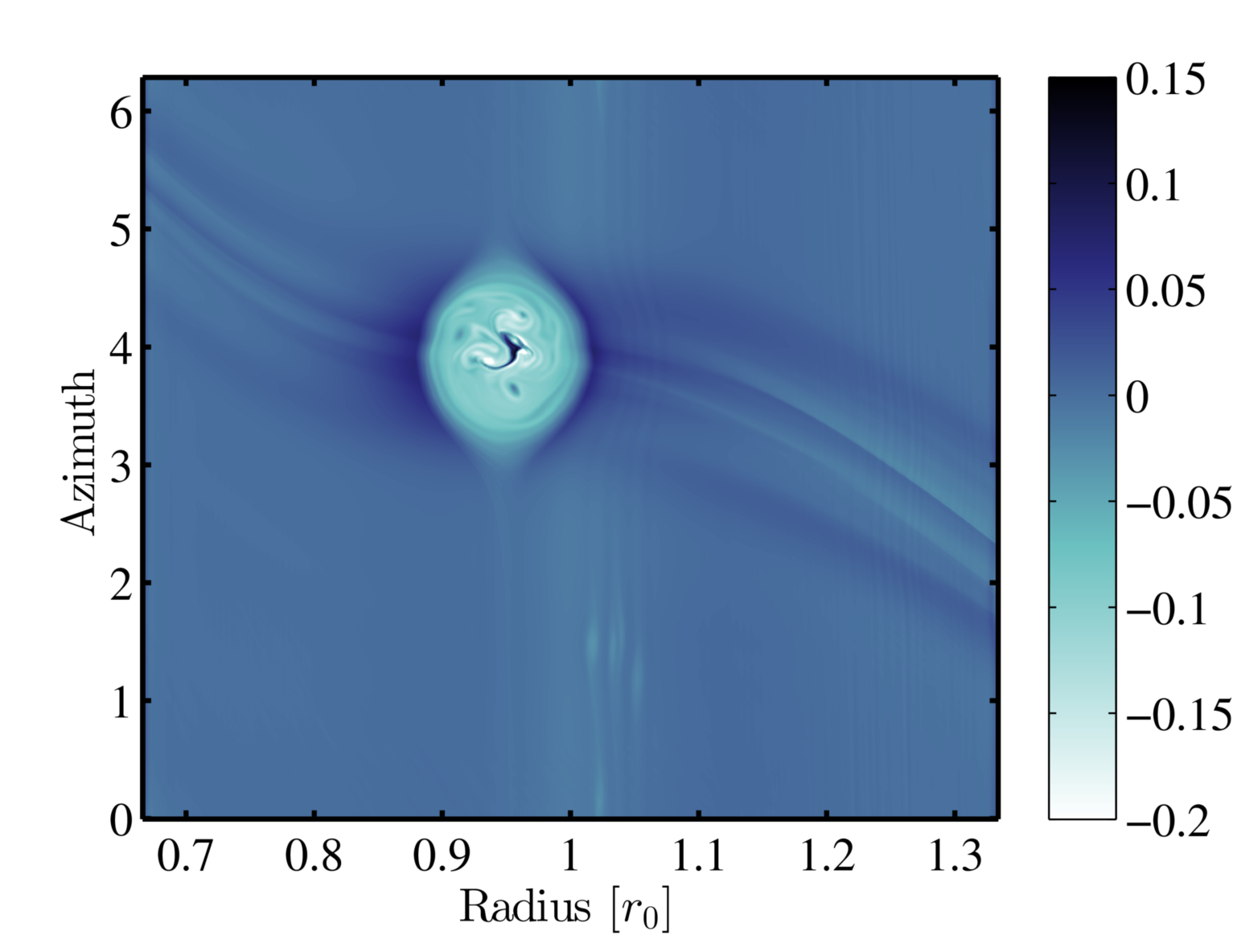} &
	\includegraphics[height=4.7cm, trim=6mm 0cm 0cm 0cm, clip=true]{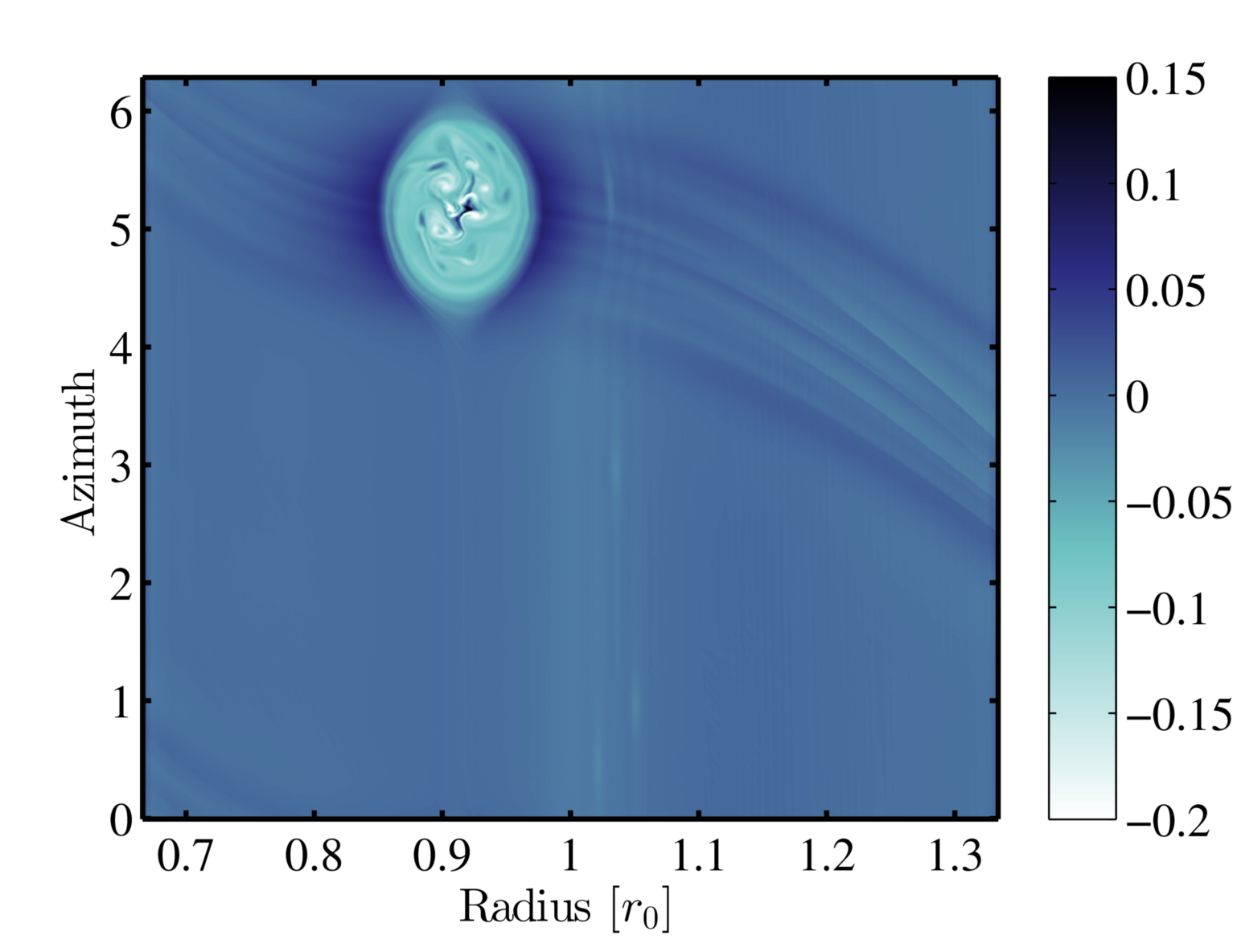} \\

	\includegraphics[height=4.7cm, trim=6mm 0cm 0cm 0cm, clip=true]{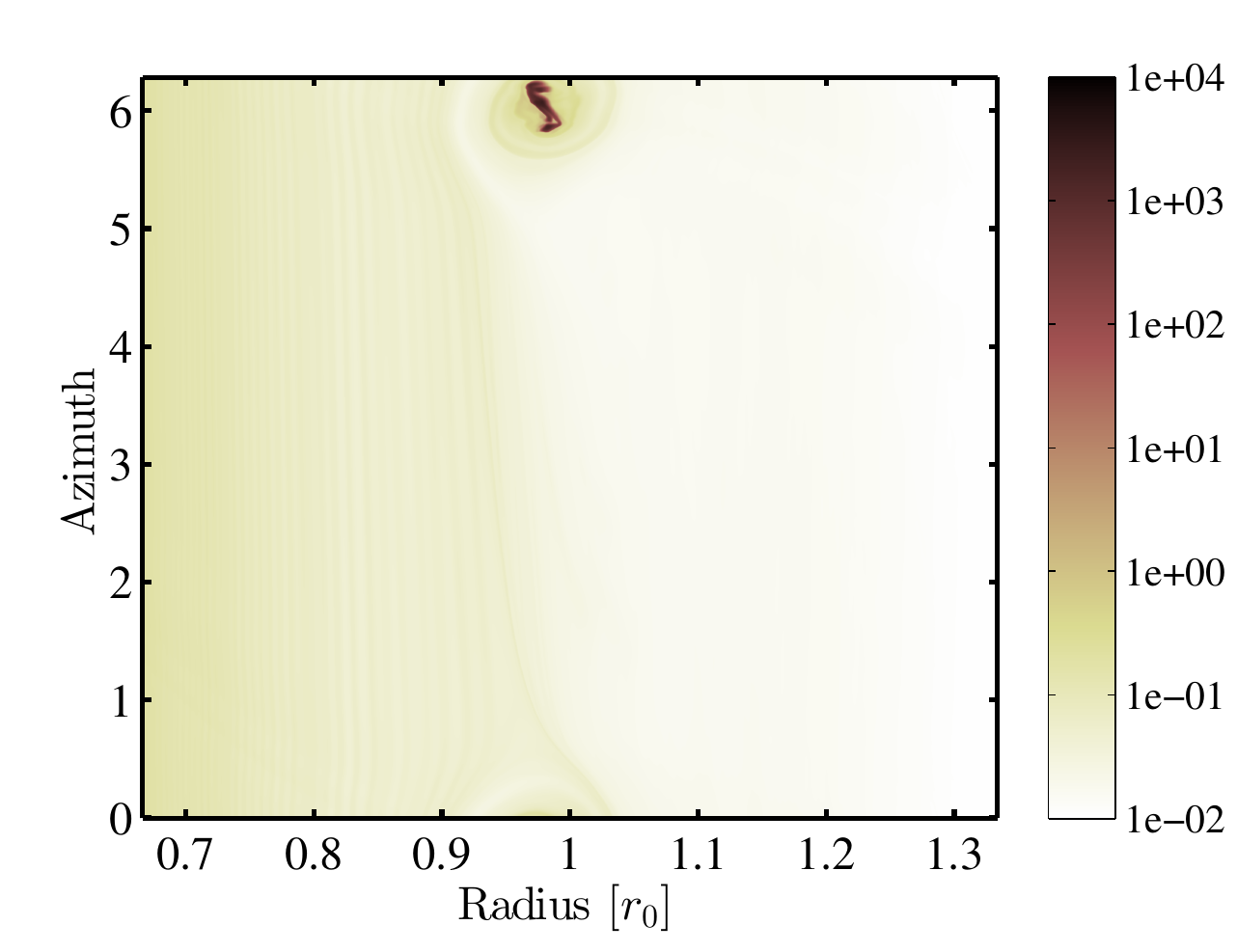} &
	\includegraphics[height=4.7cm, trim=6mm 0cm 0cm 0cm, clip=true]{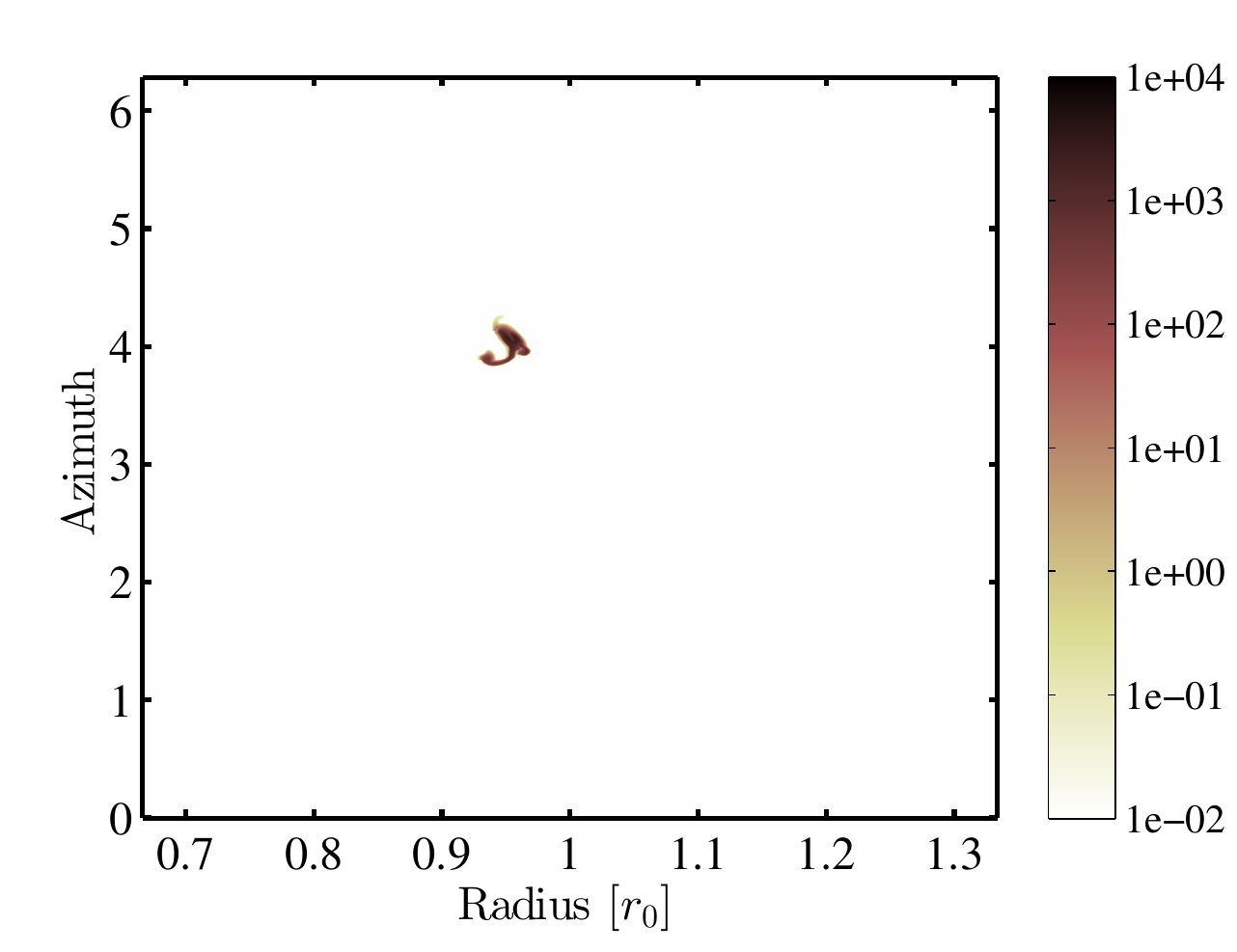} &
	\includegraphics[height=4.7cm, trim=6mm 0cm 0cm 0cm, clip=true]{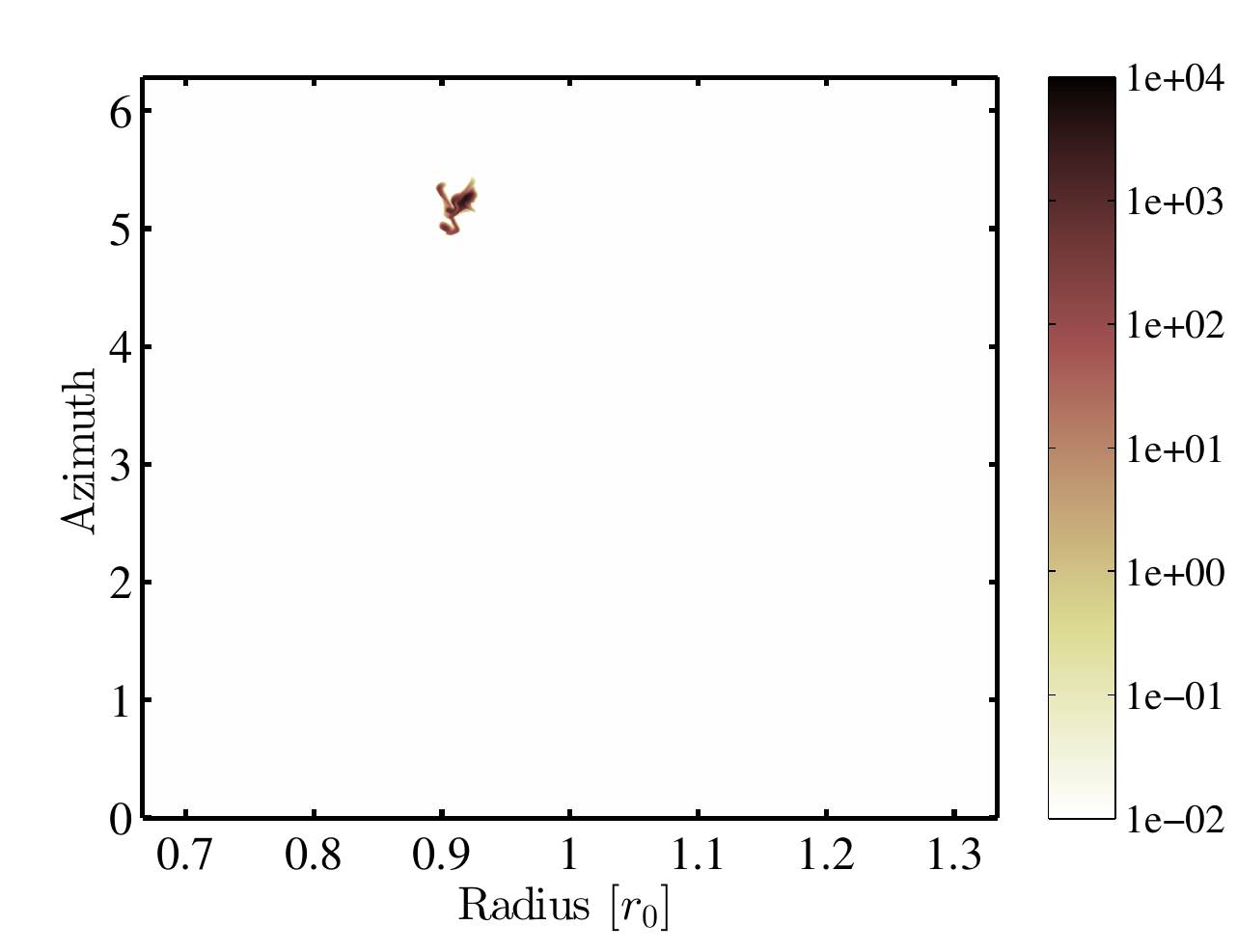} \\

	\end{tabular}
	\caption{\label{Stable_vortex_St_017} Persistence of a dusty unstable vortex for dust parameters $\epsilon=10^{-4}$ and $S_t=0.17$. We plot $(r, \; \theta)$ maps of the Rossby number ({\it{top}}) and of dust density ({\it{bottom}}) for the run using the large vortex $(R_0, \: \chi_r, \: \chi_\theta)=(-0.13, \: 0.1, \: 6.5)$. {\it{From left to right}}: snapshots at $t=200$, $t=500$, and $t=1000$ rotations, respectively. During the evolution, the local dust-to-gas ratio in the inner parts of the vortex is close to $25 \times 10^{-2}$.}
      \end{center}
\end{figure*}

	When the amount of dust captured inside the vortex is not large enough, the region where the local dust-to-gas ratio is higher than $0.1$ is reduced close to the vortex center. Thus the vortex streaming instability is confined inside the vortex, and the gas can resist and maintain the vortical structure for a long time. We observed this phenomenon in some of our setups (see Section \ref{Sect_Results}), in particular when the initial dust-to-gas ratio is $\epsilon = 10^{-4}$. The survival is reduced when the vortex radial size is smaller, as for example for setups with $\epsilon = 10^{-4}$ and $S_t=0.17$ for which the small vortex is destroyed after $\sim 600$ disk rotations but for which the big vortex survives until the end of the simulation ($\sim 1300$ rotations). In fact, the dusty and unstable portion of the vortex is relatively large for a vortex of small radial size, the surface filled by dust being similar in absolute whatever the vortex size. Increasing the Stokes number, however, fastens the streaming instability and favours the destruction of the vortex.

	The existence of these surviving vortices is interesting for planetesimal as well as planet cores formation. Keeping solid particles in a gaseous vortex increases the ability to trigger the gravitational instability firstly because of the large mass of gas and dust available, and secondly because of the compactness of the dusty region. To investigate these properties we show snapshots of the evolution of two setups where vortex survival is observed. In Figure \ref{Stable_vortex_St_004}, dust properties are $\epsilon=10^{-4}$ and $S_t=0.04$, and in Figure \ref{Stable_vortex_St_017} these parameters are $\epsilon=10^{-4}$ and $S_t=0.17$.

	In the first case, Figure \ref{Stable_vortex_St_004}, one observes the development of the vortex streaming instability on the first snapshot (left) corresponding to $t=200$ rotations. Later on, after saturation, it is clearly visible that the instability is confined in the vortex center and the gas and dust structures are quasi stable in time (regardless vortex migration). The dust-to-gas ratio in the dust packet is close to $7 \times 10^{-2}$ and does not vary significantly in time. The radial extent of the dense dust region is $18 \times 10^{-3} \; r_0$ and $20 \times 10^{-3} \; r_0$ at $t=500$ and $t=1000$ rotations, respectively. The confinement is strong and it corresponds to $\sim 0.15$ $AU$ for our disk parameters. The mass of solid particles embedded in the vortex is constant and equal $38 \times 10^{-3}$ earth masses, equivalent to $3.1$ Moon masses (with our disk parameters). By the end of the simulation, it is impossible to distinguish any evidence of a possible destruction of the vortex. Longer term simulation at higher resolution is necessary to clarify this outcome. Thus, we can only argue that this dusty unstable vortex can confine efficiently a significant amount of solids for thousands rotations.

	For larger dust grains, Figure \ref{Stable_vortex_St_017}, the vortex streaming instability is more efficient, and is sustained in a large portion of the vortex, what is visible in the vorticity maps (top panels). Surprisingly, the dusty region is not more extended than in the previous case (bottom panels), with a measured radial extension of $\sim 15 \times 10^{-3} \; r_0$, or $0.12$ $AU$ in this disk. However, the sustained maximal dust-to-gas ratio is more than 3 times larger than in the case with $S_t=0.04$. As a consequence, the total mass of solids enclosed inside the vortex is larger, about $4.1$ Moon masses. In this case, the instability influences the gas so strongly, that at $t=1000$ rotations we observe an increase of the aspect ratio of the vortex, characterizing the beginning of its destruction (as observed in other setups). Even if longer term simulations will have to carried out to confirm our expectation, we argue that this vortex will not survive in the disk thousand rotations more.
	
	At this stage, we cannot confirm without doubt that unstable vortices can survive in the disk. But they can confine easily a fraction of Earth mass in a compact region for thousands of disk rotations. It is more possible for grains of small Stokes numbers, but was not observed when the initial dust-to-gas ratio $\epsilon>10^{-4}$, indicating that it could be a marginal process. Finally, small vortices are even less subjects to survive, as the vortex streaming instability quickly affects the whole vortex and destroys it in a few hundred orbits.

\subsection{ Dust ring formation and possible planetesimal formation}
\label{Sub_sect_ring}

\begin{figure*}
	\begin{center}
	\begin{tabular}{ccc}
	\scriptsize{$t=700$ rotations} & \scriptsize{$t=1000$ rotations} & \scriptsize{$t=1300$ rotations} \\
	\includegraphics[height=4.7cm, trim=6mm 0cm 0cm 0cm, clip=true]{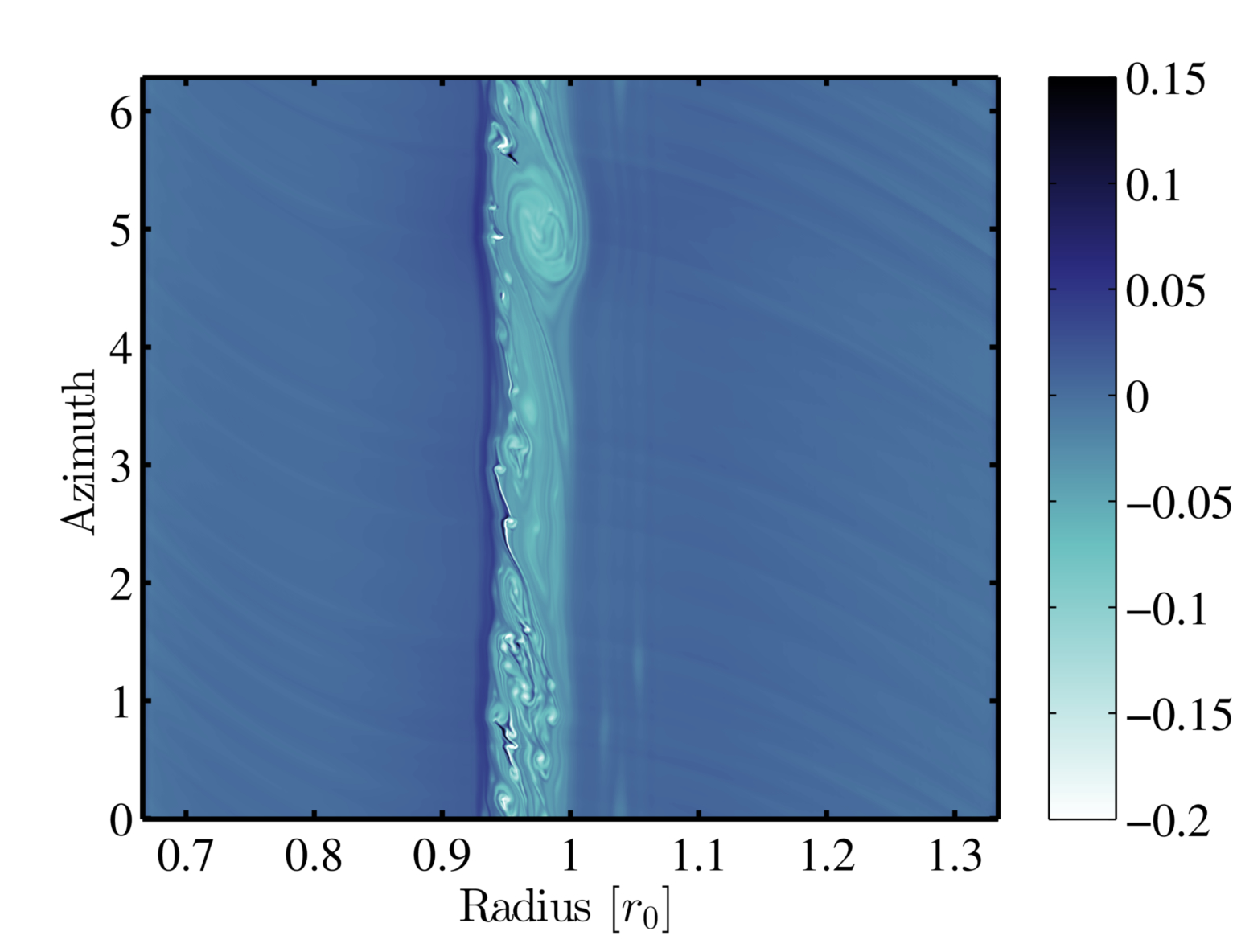} &
	\includegraphics[height=4.7cm, trim=6mm 0cm 0cm 0cm, clip=true]{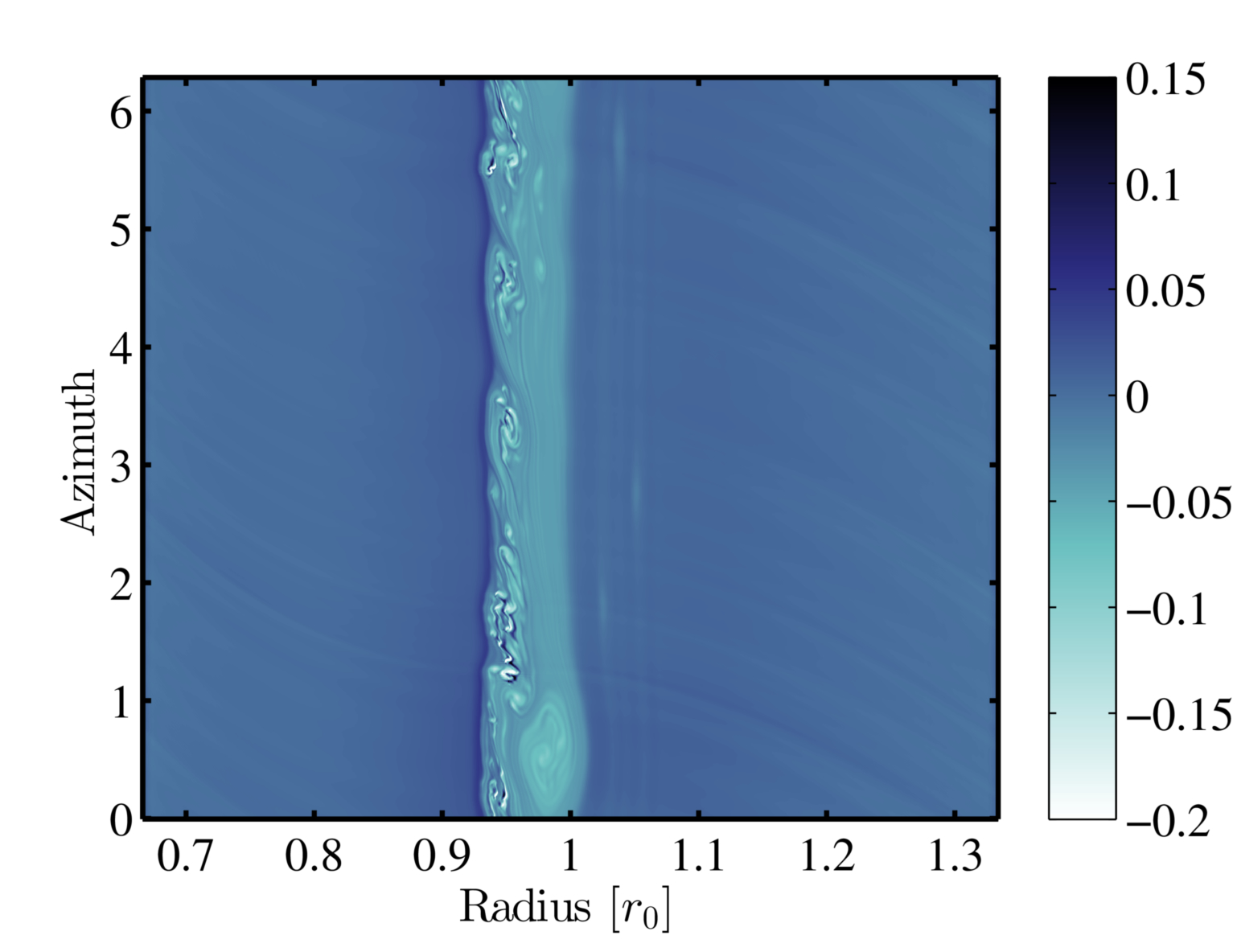} &
	\includegraphics[height=4.7cm, trim=6mm 0cm 0cm 0cm, clip=true]{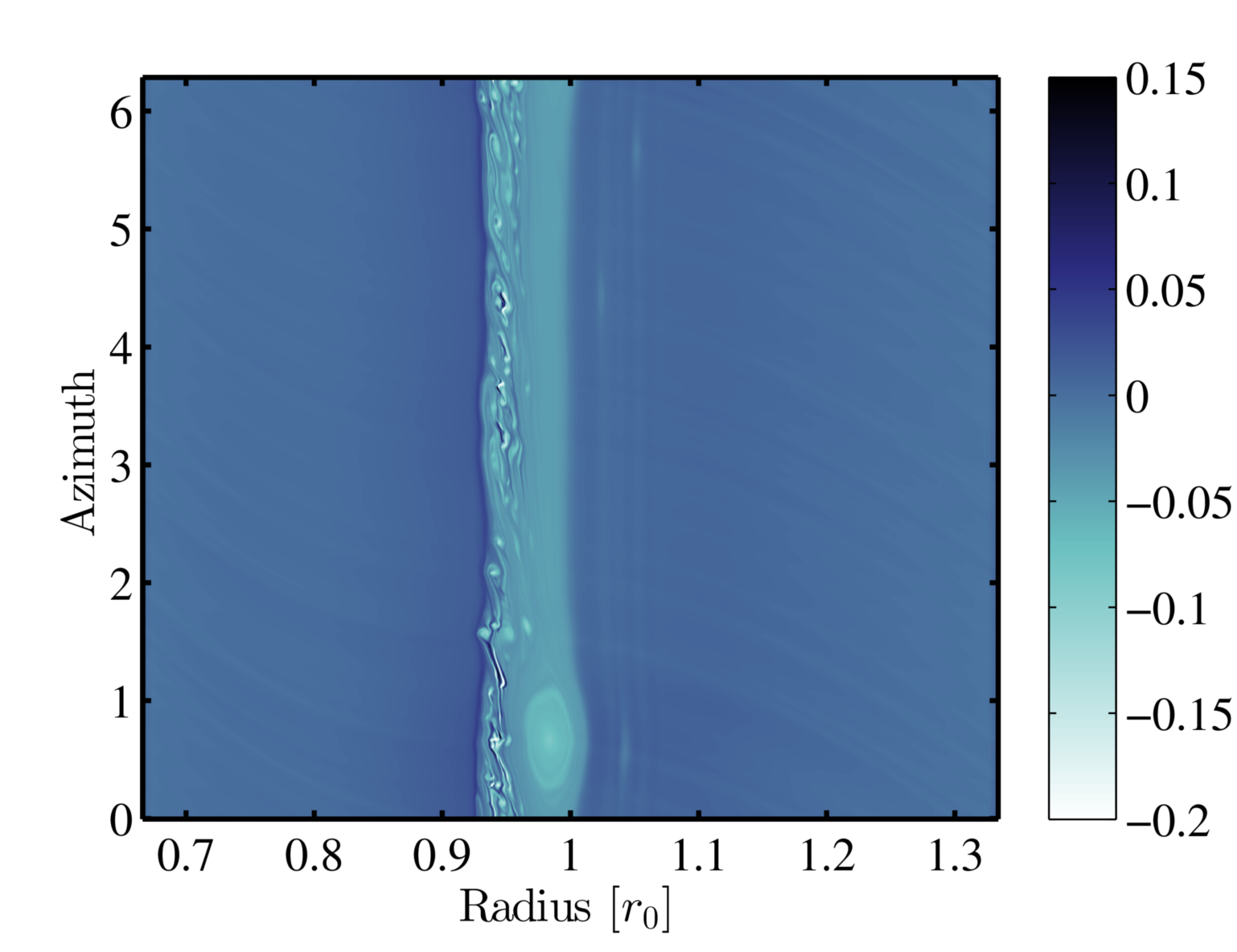} \\

	\includegraphics[height=4.7cm, trim=6mm 0cm 0cm 0cm, clip=true]{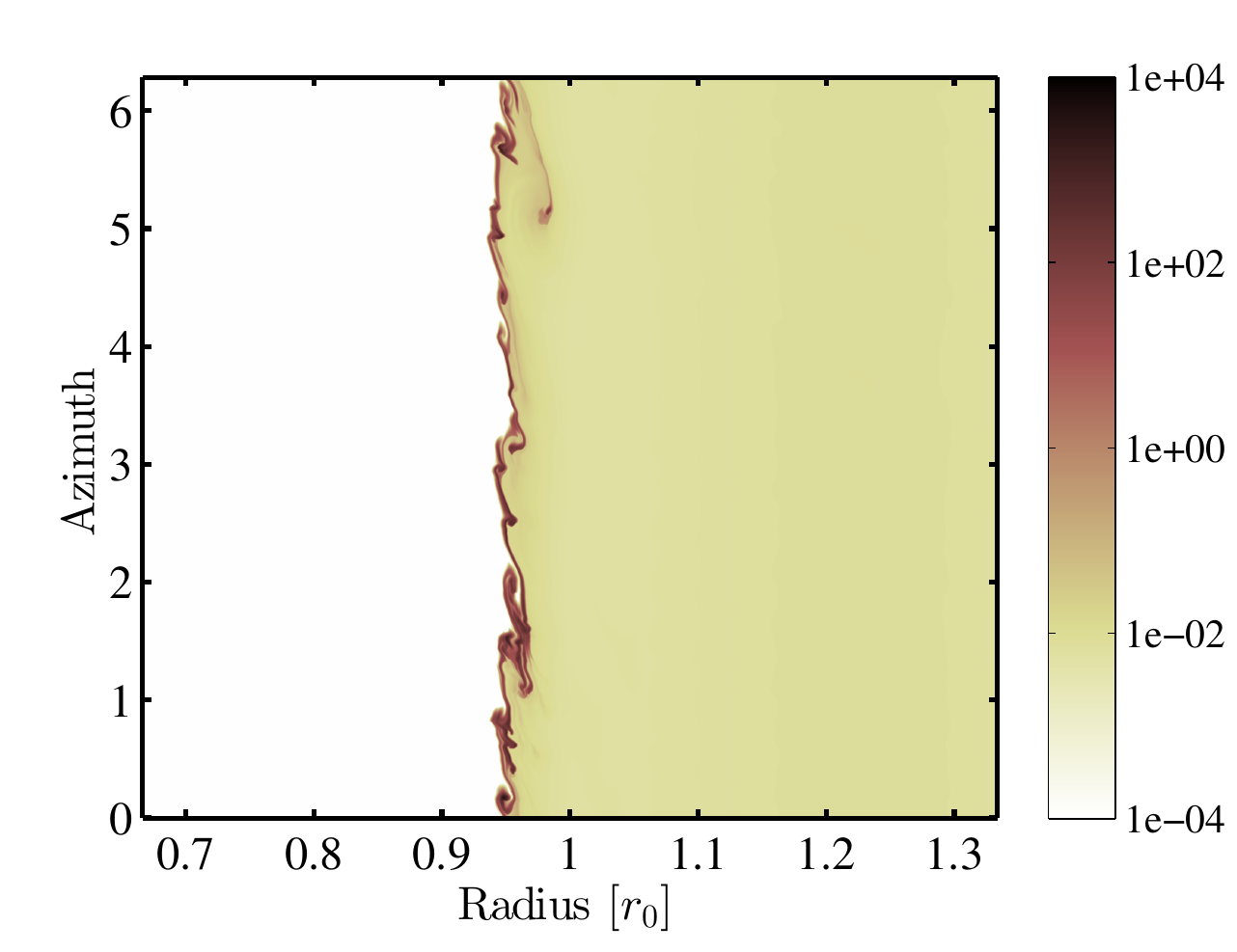} &
	\includegraphics[height=4.7cm, trim=6mm 0cm 0cm 0cm, clip=true]{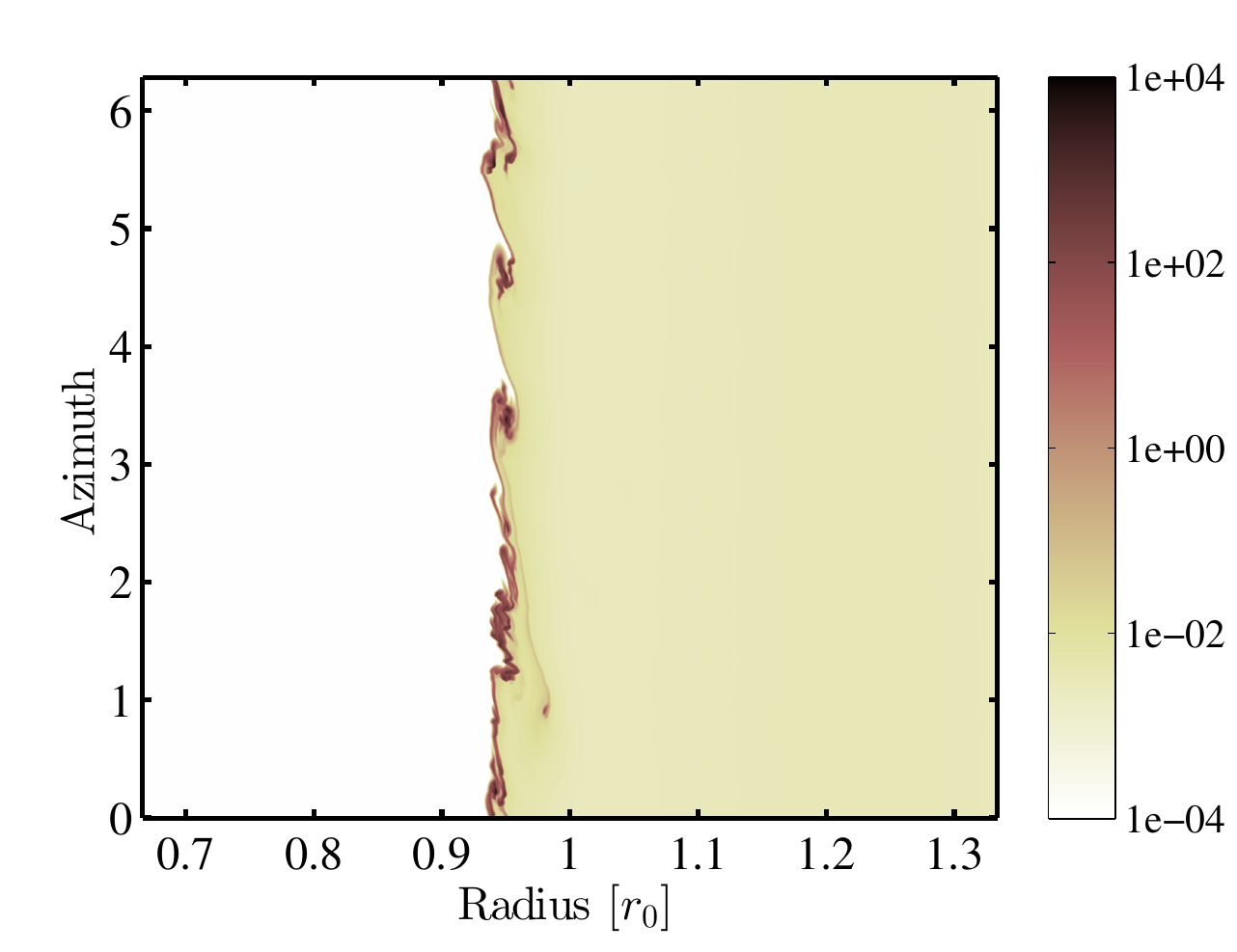} &
	\includegraphics[height=4.7cm, trim=6mm 0cm 0cm 0cm, clip=true]{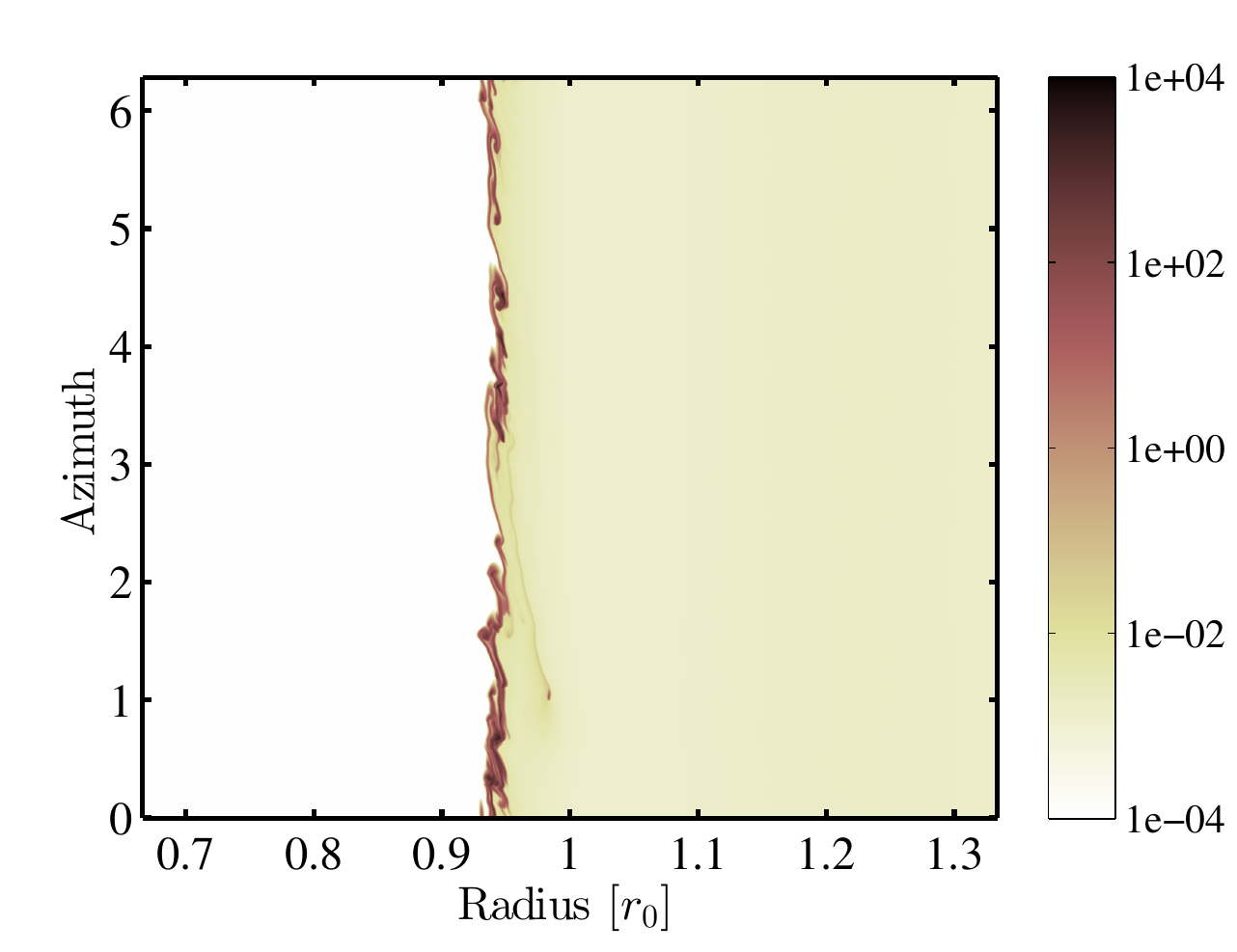} \\

	\includegraphics[height=4.7cm, trim=6mm 0cm 0cm 0cm, clip=true]{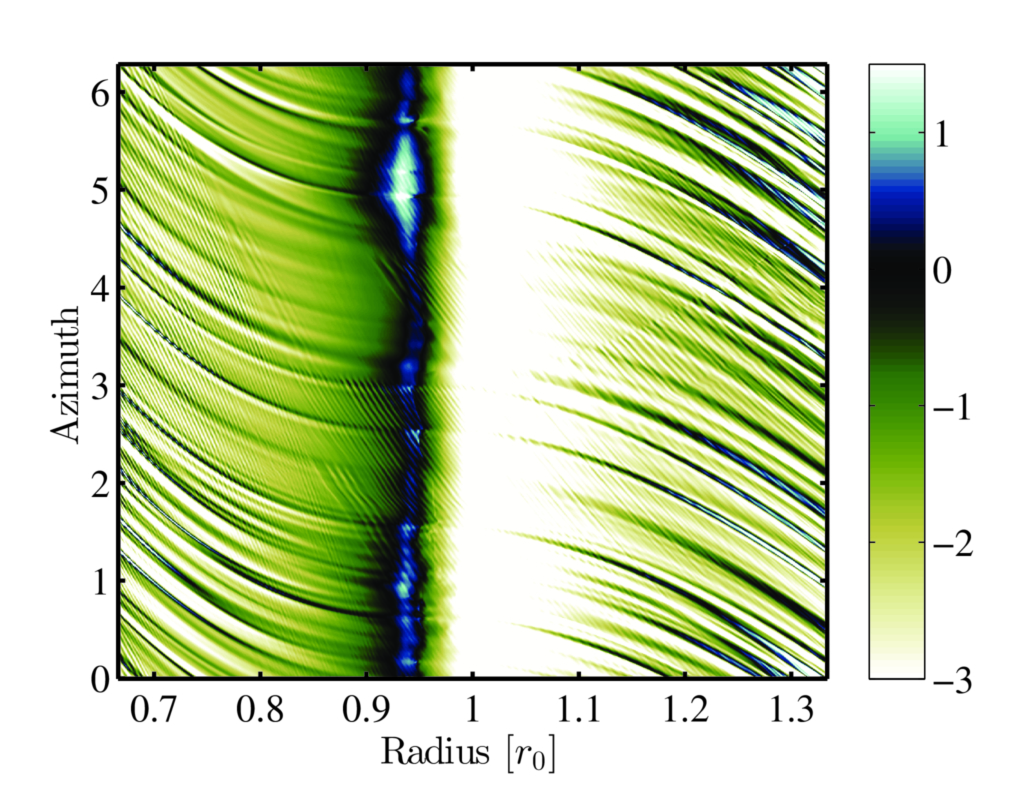} &
	\includegraphics[height=4.7cm, trim=6mm 0cm 0cm 0cm, clip=true]{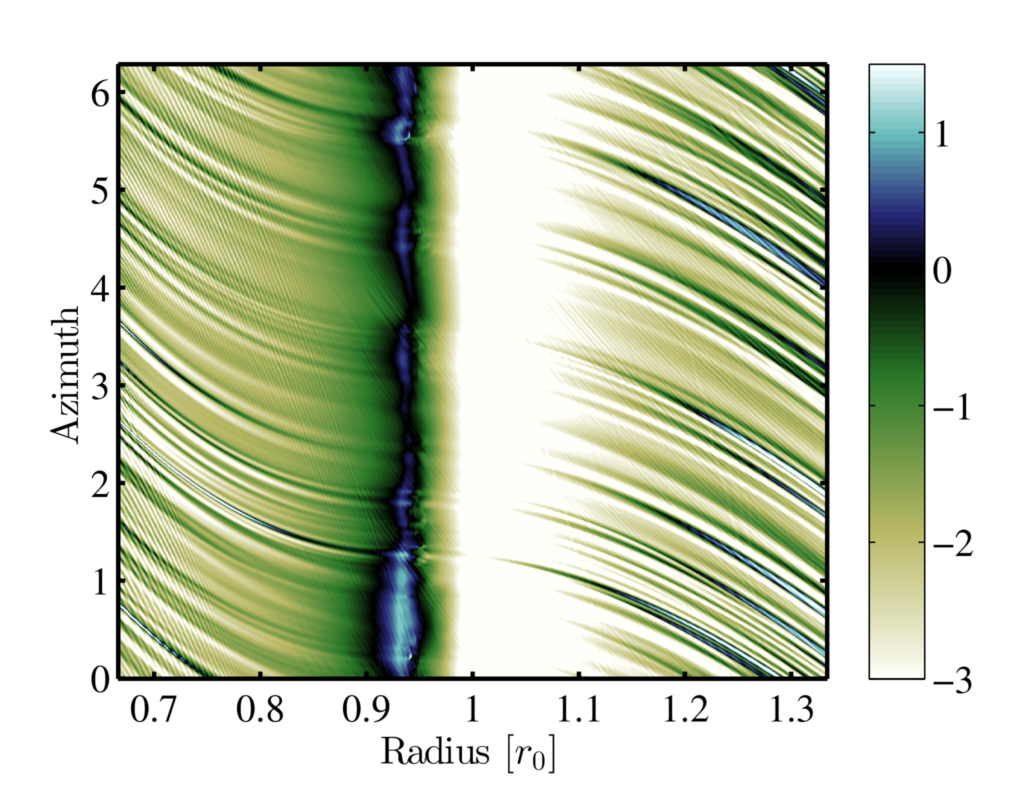} &
	\includegraphics[height=4.7cm, trim=6mm 0cm 0cm 0cm, clip=true]{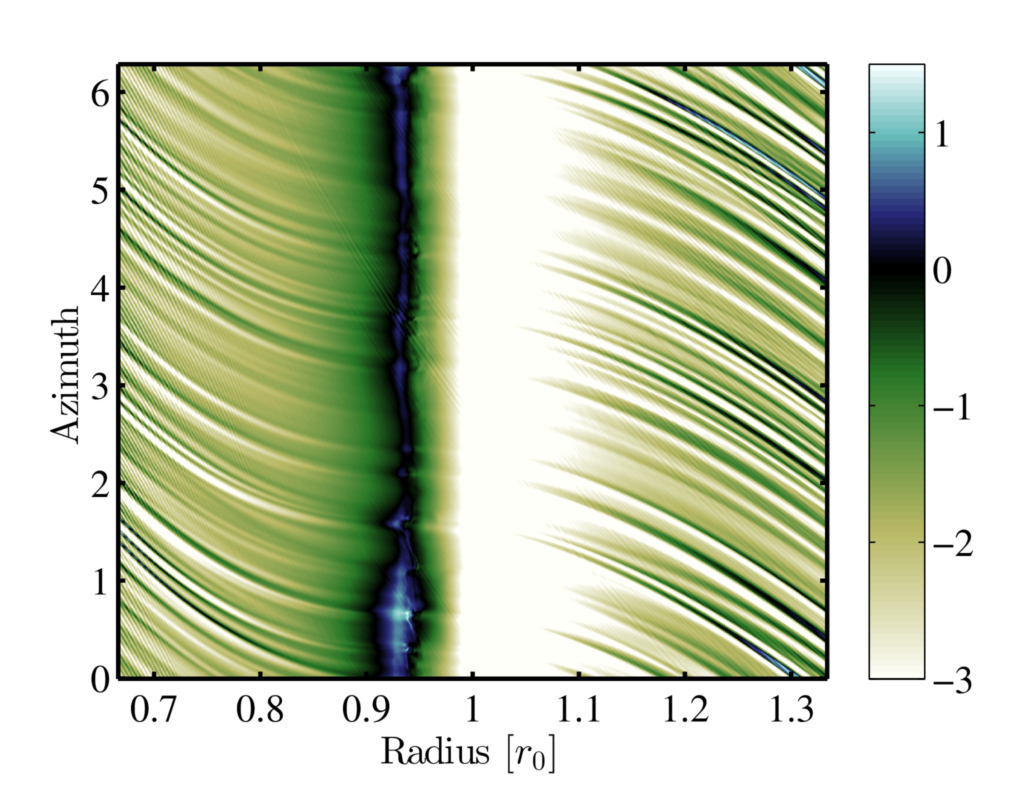} 

	\end{tabular}
	\caption{\label{Ring_Evo} Dust ring evolution for particles with $S_t=0.17$ in a disk of global dust-to-gas ratio $\epsilon=10^{-3}$. {\it{From top to bottom}}: We show the gas Rossby number, the dust density $\sigma/[\epsilon \sigma_0(r)]$, and the radial gradient of pressure as $\beta_P(r, \theta) = r P^{-1} \partial_r P$, respectively. {\it{From left to right}}: We show the evolution after $t=700$, $t=1000$, and $t=1300$ rotations, respectively. }
      \end{center}
\end{figure*}

	Perhaps the most interesting and novel result of our simulations is the formation of a long-lived dust ring after the dissipation of vortices. This is a generic feature of our simulations, although it does not form in some cases because the vortex destruction happens after the end of the simulation, as discussed previously. It also does not form in the simulations using the small vortex. To clarify the question of the ring formation, we describe the structure of these rings Figure \ref{Ring_Evo}, in the case of evolution of dust of $S_t=0.17$ and $\epsilon=10^{-3}$ captured in the large vortex of parameters $(R_0, \: \chi_r, \: \chi_\theta)=(-0.13, \: 0.1, \: 6.5)$.

	We display, every $300$ rotations from left to right, the evolution of the Rossby number, the dust density and the radial pressure gradient, from top to bottom, respectively. We observe a bump of vorticity located in the region where the vortex was orbiting. A remnant is still present in the outer parts. The most interesting feature is this narrow band of vorticity at $r \sim 0.95 \; r_0$, correlated to the position of the dust concentrations. The vortical patterns do not merge as in classical 2D turbulence, but are constantly generated by a streaming instability. Indeed, this merging is visible in the external parts of the ring, where the dust density is low ($\sim 10^{2}$). The saturated streaming instability that persists over more than $600$ rotations, and about thousand orbits in some setups (see Section \ref{Sect_Results}), is possible due to the local dust-to-gas ratio in the ring that is larger than unity in this case. A signature of the streaming instability is clear when we compare with the fiducial run of Section \ref{Sect_Test_case}, at lower resolution $(N_r, \: N_\theta) = (512, \: 1024)$. In Figure \ref{Evo_Typical} bottom plots, the ring forms but small scale dusty structures do not survive, because the streaming instability is not resolved.

	The orbital position of this ring is determined by the distribution of gas in the disk. As we can see on the maps of the radial pressure gradient (bottom plots), the dust ring is superimposed at the locations where this pressure gradient cancels. It is well known that it ensures that the radial velocity of the dust cancels (also see Eq. \ref{Eq_dust_velocity}). Then the solid component is confined in a narrow ring where the radial pressure gradient is small enough to keep the solids inside. However this condition is not fulfilled for every vortex. As our results shows, if the radial extension of the vortex is too small, the gas bump resulting from the vortex destruction is not favourable for the cancellation of the pressure gradient. Then dust can migrate into the inner parts of the disk, due to the global $\beta_P$. It is difficult to predict which vortex parameters result into the formation of a ring, but it is more probable under two conditions: large enough radial width, large enough gas density at the vortex center (strong vortex). 

\begin{figure}
	\begin{center}
	\begin{tabular}{c}
	\scriptsize{$\epsilon=10^{-2}$} \\
	\includegraphics[height=5.5cm, trim=0mm 0cm 0cm 0cm, clip=true]{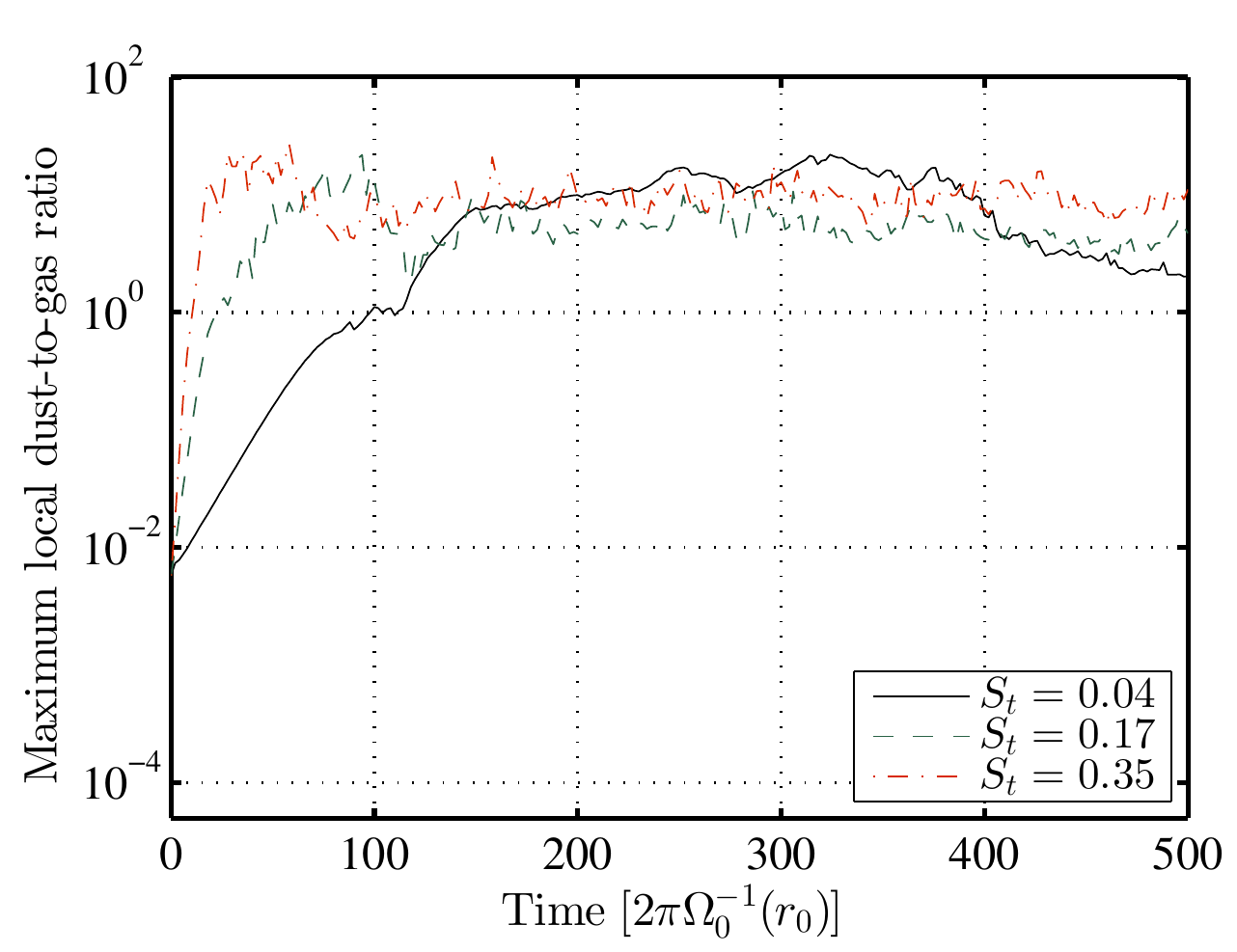} \\ \\
	\scriptsize{$\epsilon=10^{-3}$} \\
	\includegraphics[height=5.5cm, trim=0mm 0cm 0cm 0cm, clip=true]{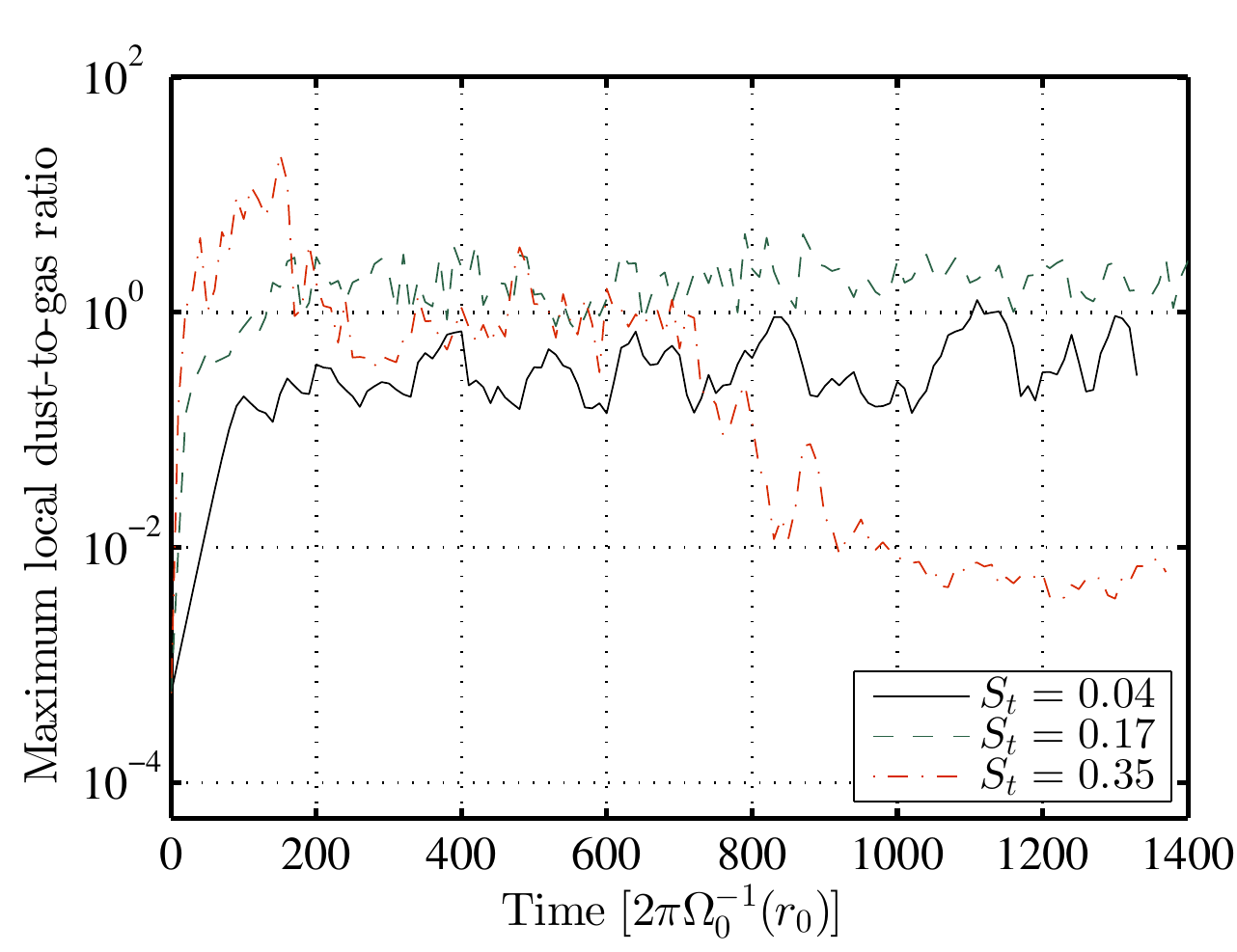} \\ \\
	\scriptsize{$\epsilon=10^{-4}$} \\
	\includegraphics[height=5.5cm, trim=0mm 0cm 0cm 0cm, clip=true]{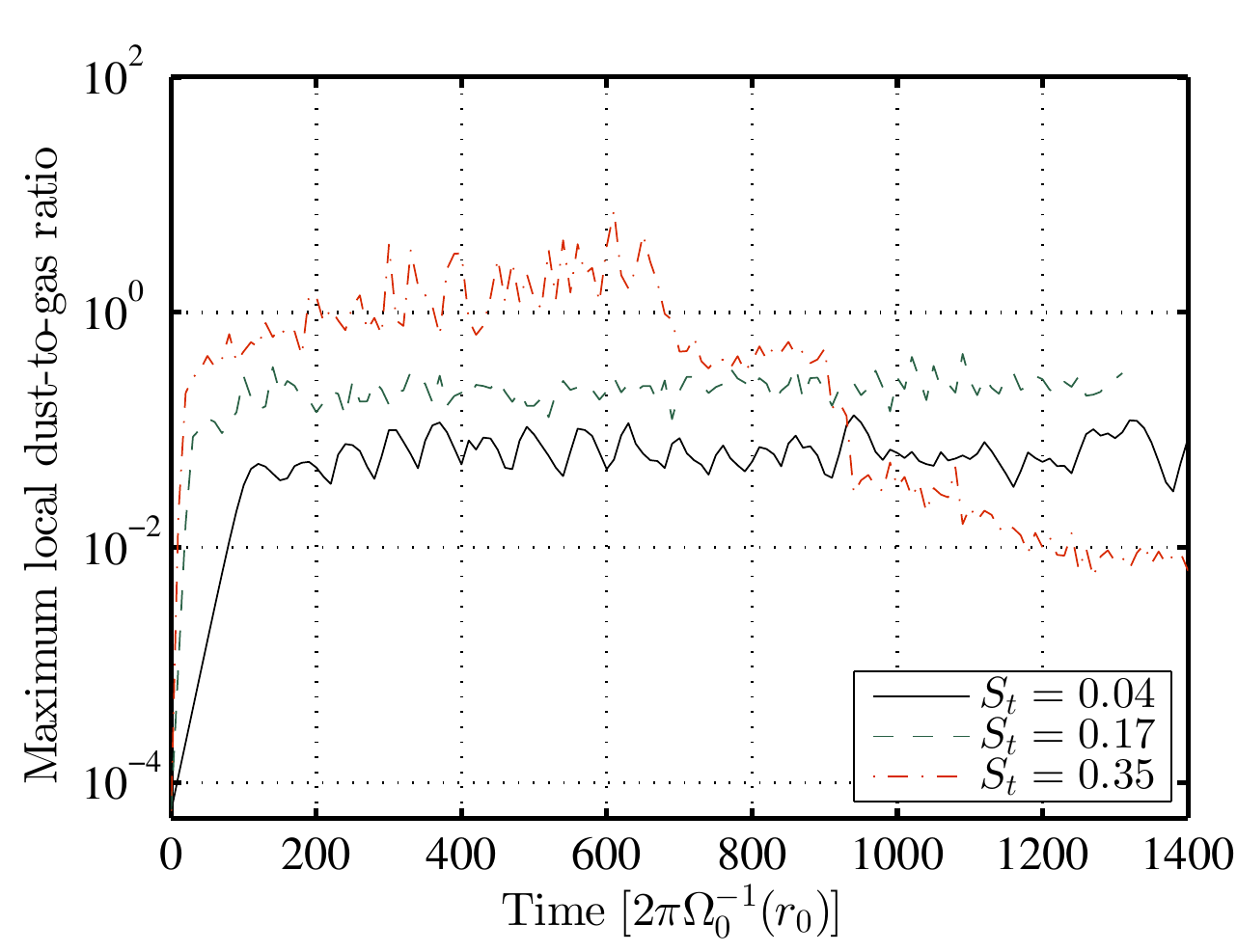} 
	\end{tabular}
	\caption{\label{Dust_gas_ratio} Evolution of the maximal value of the dust-to-gas ratio reached in the disk, for different Stokes numbers. {\it{From top to bottom}}: The global initial dust-to-gas ratio is $\epsilon=10^{-2}$,  $\epsilon=10^{-3}$, and $\epsilon=10^{-4}$ respectively. We use here the large vortex model: $(R_0, \: \chi_r, \: \chi_\theta)=(-0.13, \: 0.1, \: 6.5)$}
      \end{center}
\end{figure}

	To summarize, the dust ring exhibits significant dust density enhancement, reaching locally dust-to-gas density of order or larger than unity, hence only marginally different than the dust-to-gas density reached inside vortices in the preceding evolutionary phase. Figure \ref{Dust_gas_ratio}, we display the evolution of the maximal local dust-to-gas ratio obtained in the disk for all the setups don with the large vortex, favourable for the ring formation. Even when reducing the initial $\epsilon$ from $10^{-2}$ to $10^{-4}$ (from top to bottom), regions of dust-to-gas ratios larger than $10^{-1}$ are formed, mainly inside the ring. The dust layer appears turbulent in this ring, with obvious formation of eddies in which the dust reaches the maximum concentration, a behaviour which is reminiscent of the streaming instability. We recall that these high dust-to-gas ratios are reached from initial dust-to-gas ratios that are up to 100 times lower than those normally assumed in initial conditions of the streaming instability simulations. These dust rings can also form in other regions of the disk where radial pressure gradient cancels, like outer edge of planetary gaps or dead zones. In this case, if the vorticity is large enough, dust capture can happen, like in the RWI setup shown previously. They could indeed be frequent features, and their role in disk evolution is of high interest.

	With density enhancements of order unity, it is possible that gravitational collapse of the dust would ensue, forming planetesimals on a local dynamical timescale (which is of order of disk rotation). The turbulent nature of the dust rings, with eddies of varying size scale, implies that, if planetesimals form by gravitational collapse, they would form with a range of sizes. The mass measured inside the ring of the simulation shown in Figure \ref{Ring_Evo} is about $0.58$ Earth masses at $t=700$ rotations, $0.60$ $M_e$ at $t=1000$ rotations, and $0.61$ $M_e$ at $t=1300$ rotations, based on our disk parameters. The width of the ring is $32 \times 10^{-3} \; r_0$, $25 \times 10^{-3} \; r_0$, and $21 \times 10^{-3} \; r_0$ at the base, but $12 \times 10^{-3} \; r_0$ at half height width, which corresponds to $\sim 0.1$ $AU$ in our setup. Thus a large amount of solid material has been confined in a small region of the disk. In order to confirm the possibility of gravitational collapse, we tried to estimate the Jeans length of some eddies in the ring. Based on the surface density of the dust in these eddies, and assuming a spherical distribution in the vertical direction, we obtain a Jeans length about one order of magnitude larger than the typical size of the eddies (radius $\sim 3 \times 10^{-3} \: r_0$). As the Jeans length is proportional to the square root of the Toomre parameter of the disk background at the location of the dust ring, we expected that gravitational collapse would be unlikely ($Q_0 = 40$ at $r_0$). However, we find that the Jeans length estimate is dominated by the value of the dust density. As our results are in 2D, we may underestimate the actual density in the disk midplane, because of the vertical integration of the physical quantities. Dust sedimentation in 3D disks could increase by a large factor the dust density in the disk midplane. In combination with self gravity effects, this could easily give Jeans length smaller than the size of the dust clumps in the ring, predicting their gravitational collapse. However, 3D simulations of self-gravitating disks are necessary to confirm this claim.

	Note that the ring formation phase is the most relevant for planetesimal formation in our simulations since it is longer-lived that the vortex phase (see Section \ref{Sect_Results} and paragraph \ref{Sect_Vort_Insta}). Furthermore, vortices would be even shorter-lived when disk self-gravity is taken into account \citep{Lin2011, Lin2011a, Lin2012, Bae2015} because spiral waves are dominant rather than vortex modes. However, \cite{Lin2011} shows that some vortices can contain a reinforced density contrast, in favour of dust concentration according to the authors, but we need to explore the vorticity profile of these object to confirm this statement, as a consequence of our capture model. In this context, vortices should be seen as triggers of the dust ring formation phenomenon, and thus instrumental to planetesimal formation; they not necessary the main site for planetesimal formation. With multiple vortices in a disk multiple rings are expected to form. If planetesimals form preferentially in such rings then this could have consequences on the orbital dynamics of the planets that would later arise, possibly providing a natural explanation for closely packed systems of (low-mass) planets observed in some extrasolar planetary systems.

\subsection{ Numerical resolution issues }

\begin{figure}
	\begin{center}
	\begin{tabular}{c}
	\scriptsize{$\epsilon=10^{-3}$, $S_t=0.17$} \\
	\includegraphics[height=5.5cm, trim=0mm 0cm 0cm 0cm, clip=true]{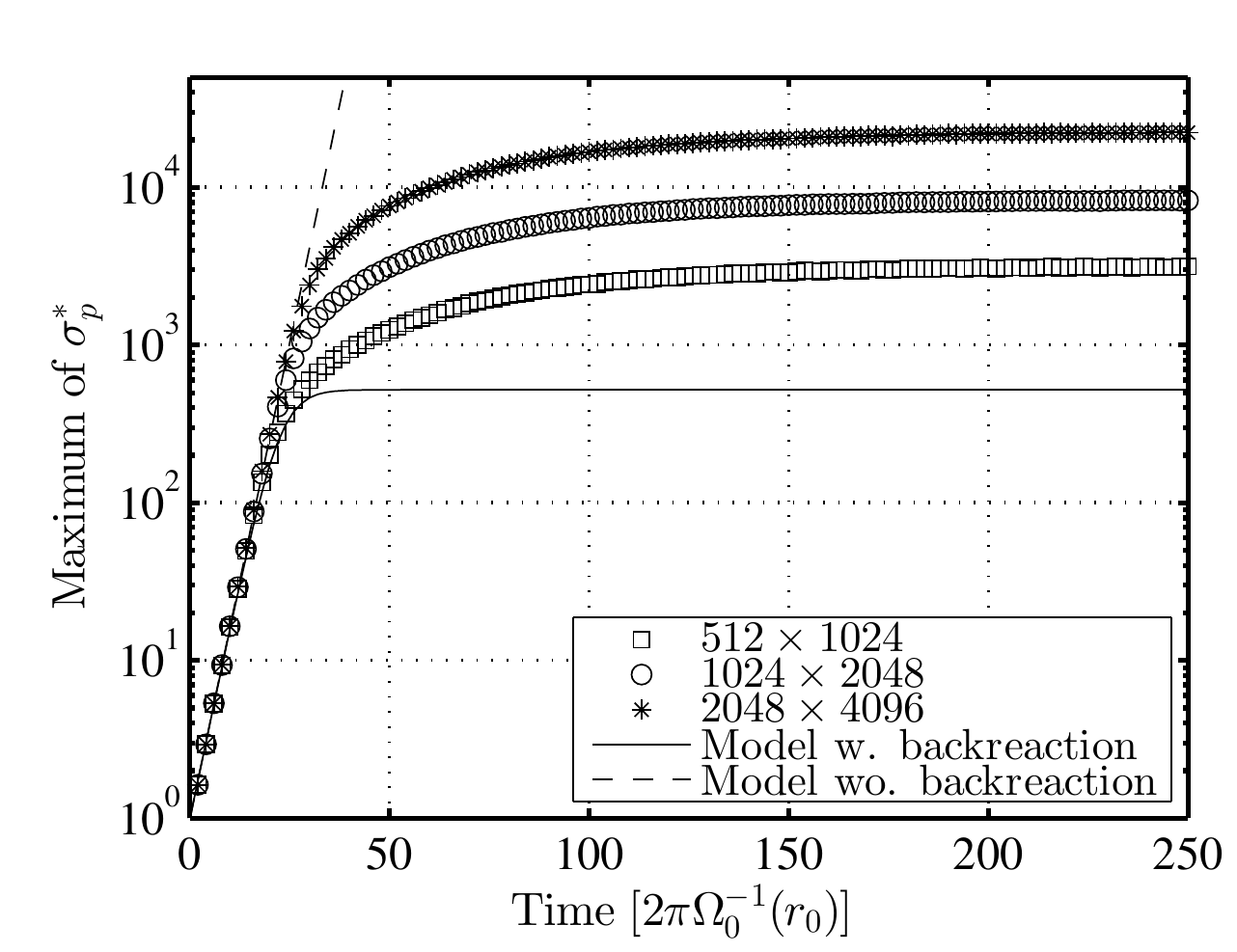} \\ \\
	\scriptsize{$\epsilon=10^{-4}$, $S_t=0.04$} \\
	\includegraphics[height=5.5cm, trim=0mm 0cm 0cm 0cm, clip=true]{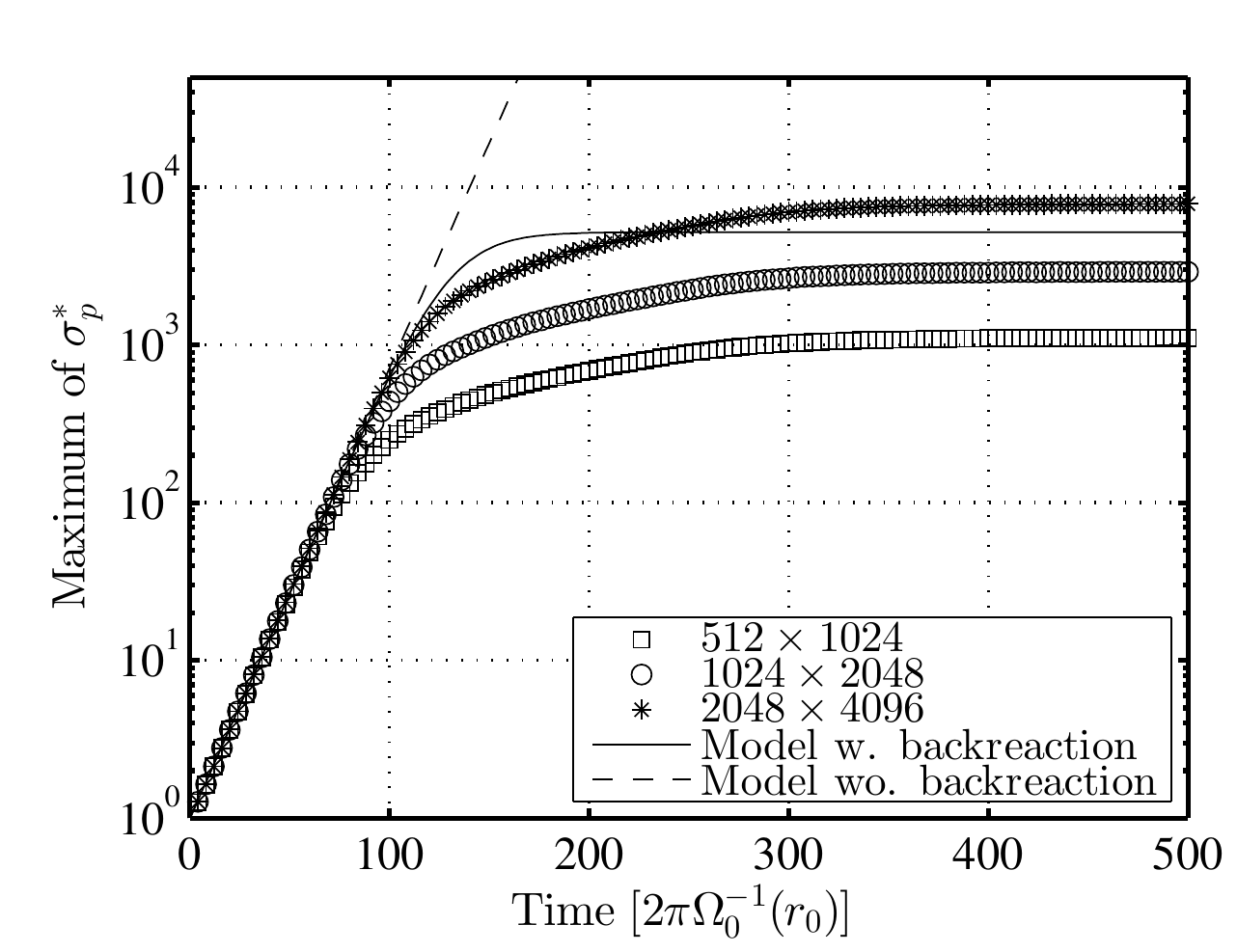} 
	\end{tabular}
	\caption{\label{No_back reaction} Dust capture without drag back reaction. We show the evolution of the maximal value of the dust density inside the large vortex model, $(R_0, \: \chi_r, \: \chi_\theta)=(-0.13, \: 0.1, \: 6.5)$, for different cases. {\it{From top to bottom}}: $\epsilon=10^{-3}$, $S_t=0.17$, and $\epsilon=10^{-4}$, $S_t=0.04$, respectively. The effect of numerical resolution (symbols) is compared with analytical models of capture: with back reaction (solid lines) and without (dashed lines). }
      \end{center}
\end{figure}

	Numerical resolution of the simulation is a critical factor for resolving the evolution of the dusty disk. As a general trend, having high resolution helps to resolve small scale structures. But in the case of dusty vortices, two other criteria are critical: {\it{(i)}} resolving large gradients of dust density during the phase of dust capture inside the vortex (and also reducing numerical diffusion), {\it{(ii)}} resolving the scale lengths of the streaming instability.

	The first criterion is necessary to obtain the right amount of dust captured inside the vortex. It is easy to check as the analytical model of capture is accurate and compares well with the high resolution simulations. To confirm this, we made a test case of capture without drag back reaction onto the gas. By this way, the dust should be constantly captured as the gas vortex is not affected. By changing the numerical resolution of the simulation, we find which minimum resolution is required to resolve the capture phase with enough accuracy. Figure \ref{No_back reaction} presents two cases: $\epsilon=10^{-3}$, $S_t=0.17$ (top), and $\epsilon=10^{-4}$, $S_t=0.04$ (bottom). In each case, we change the resolution from $(N_r, \: N_\theta) = (512, \: 1024)$ to $(N_r, \: N_\theta) = (2048, \: 4096)$ by factors of two, and compare the dust density maximum as a function of time with the capture model with back reaction (solid lines) and without (dashed lines). 

	In the first case (top), the value of the dust density maximum deviates from the expected capture model when $\sigma^*_p > 3 \times 10^2$, $\sigma^*_p > 10^3$, and $\sigma^*_p > 3 \times 10^3$, for increasing resolutions. After about $150$ disk rotations, a saturated state is reached, due to the numerical diffusion of the scheme, which makes difficult maxima to increase. The saturated values are however much larger for high resolutions: $\sigma^*_p = 3 \times 10^3$ for the lowest and $\sigma^*_p = 2.2 \times 10^4$ for the highest resolution. This discrepancy is in accordance with the factor of 8 between these two resolutions. Despite this artificial saturated evolution of the density maximum, the three resolutions produce an evolution accurate enough to resolve the capture with back reaction (solid line). In fact, the saturation due to the coupled evolution of the gas and the particle fluids overcomes the one due to numerical diffusion.
	
	It is not the case for the second setup (bottom) for which the global dust-to-gas ratio is smaller as well as the Stokes number of the particles. The results are similar to the ones obtained in the first setup in terms of numerical diffusion and the ranking of the resolutions. However, the saturation due to numerical effects is stronger than the expected saturation due to physical effects of the drag force back reaction (solid line). Only the highest resolution could be used in this case as it shows an evolution similar to the model, although it is not due to physical diffusion terms.

	These examples show that in order to explore the capture of diluted particle fluids (small $\epsilon$, typically $\epsilon<10^{-4}$) and of small size (small $S_t$, typically $S_t<0.04$), high numerical resolution is necessary. The comparison with our analytical model is an efficient benchmark to confirm if the resolution used is high enough for given properties of a particle fluid. Finally, we show that the resolution in the disk midplane $(r, \: \theta)$ is critical, and producing accurate 3D simulations of dusty disks will be challenging as $(N_r, \: N_\theta) = (512, \: 1024)$ already limits to $\epsilon > 10^{-3}$ and $S_t > 0.1$. 

	The evolution of the dust ring, undergoing a streaming instability, is also very sensitive to the numerical resolution. As we show in Figure \ref{Evo_Typical} (d), using $(N_r, \: N_\theta) = (512, \: 1024)$ is not enough to resolve the ring evolution. Compared to the case at $(N_r, \: N_\theta) = (2048, \: 4096)$, Figure \ref{Ring_Evo}, the dust ring has a completely different evolution. This kind of streaming instability, in the $(r, \: \theta)$ plane, has to be studied in details to define the scale length of the modes and the necessary resolution to resolve it, as it is already done in the classical study in the $(r, \: z)$ plane \citep{Youdin2005, Youdin2007, Kowalik2013}

\subsection{Caveats: role of viscosity and self-gravity}

	Our simulations are by construction inviscid. As discussed by various authors (see Section \ref{Sect_Intro}), the lifetime of vortices would be affected by viscosity, hence we should regard our results as upper limits on the vortex duration. However, one may ask how much the numerical viscosity of our scheme could matter in this context.
This is relevant also to the interesting phase of dust ring formation. We have estimated the magnitude of our numerical viscosity using the approach of \cite{Flock2011}, in which the alpha-viscosity is defined by:

\begin{equation}
	\alpha = \frac{\int \sigma U_g (V_g - \overline{V_g} ) \, c_s^{-2} \, dS}{\int \sigma dS} \: ,
\end{equation}
where $U_g$ is the radial gas velocity and $\overline{V_g}$ the azimuthal average of the azimuthal gas velocity.

	This effectively computes the effective stress tensor of a viscous fluid. We find that the equivalent $\alpha$ viscosity in our simulations is $5 \times 10^{-5}$ or smaller. The low viscosity is a direct consequence of using a well-balanced scheme for the hydro equations. Note that recent MHD simulations of dusty disks with non-ideal MHD effect such as ambipolar diffusion find that the equivalent $\alpha$ viscosity is at most $10^{-3}$ rather than $> 10^{-2}$ as often found in ideal MHD simulations of the MRI in accretion discs \citep{Zhu2014a}. While our numerical viscosity is even smaller, the difference is only an order of magnitude and implies a viscous timescale of $\tau_{visc} > 10^6$ yr at $r=1-10$ AU, where $\tau_{visc} = r^2/\nu$ and $\nu = \alpha c_s H$. This is still very long relative to the orbital time, suggesting that a realistic viscosity should not affect our results significantly. In particular, it should not dissipate structures such as our dusty rings on timescales so short as to impact the general conclusions of Section \ref{Sub_sect_ring}.

\section{ Conclusions }

	In this paper we have presented a study of the long-term dust evolution in presence of vortices in protoplanetary disks. Our simulations, which treat dust and gas as two coupled fluids, are two-dimensional but are global and in some cases reach a resolution that is at the state-of-the-art of what is currently possible in this type of calculations owing to the ability of our code, RoSSBi, to run efficiently on some of the largest supercomputers. A major thrust of our work, that sets it apart from previous literature, is the ability to follow the coupled dust and gas evolution for very long timescales, exceeding even a thousand disk orbits, and starting even from very small dust-to-gas ratios ($\epsilon=10^{-4}$). This has allowed us to uncover new phenomena such as the formation of large-scale dust rings after vortex dissipation, which are then subject to a streaming instability and can become a site of rapid planetesimal formation, at least as relevant as the region inside the vortex itself. While we cannot confirm that planetesimals do actually form in any of the instances in which very dense gas clumps exceeding a dust-to-gas ratio of order unity form, due to the lack of self-gravity, the phenomenology of the flow and dust-to-gas ratios achieved both in the vortex phase and in the ring phase are reminiscent of those found in standard studies of the streaming instability, in which rapid planetesimal formation has been demonstrated. 

We also developed a powerful analytical model that allows to understand and reproduce the results of numerical simulations in the linear capture phase. In summary, the main results obtained in our work are the following:

\begin{itemize}

\item{ Vortices are confirmed to be extremely efficient at concentrating dust on relatively short timescales in the regions with highest vorticity. Dust capture drives the dust-to-gas ratios to values of order unity in the vortex region in a manner quite irrespective of the initial dust-to-gas ratio, which was varied by orders of magnitude in our initial conditions. }

\item{ Vortices are dissipated by a two-fluid 'vortex streaming instability', and their legacy is a disk-wide dust ring which is also subject to the streaming instability, as shown by mode analysis, when the resolution of the simulation is sufficiently high to capture it. }

\item{ Vortex dissipation timescales depend strongly on vortex size, as small vortices are dissipated faster than large vortices. Also, small vortices do not always trigger dust ring formation, because the radial pressure gradient does not cancel in the leftover gas structure. }

\item{ The largest increases in the dust-to-gas ratios, exceeeding unity, occur during the vortex instability phase and in the dust ring phase rather than in the initial linear capture phase. The dust ring phase lasts as long as the simulations are run for, hence longer than the vortex instability phase which lasts only a few hundred orbits. }

\item{ The largest mass in high density dust clumps is collected in the dust ring formation phase, estimated to an Earth mass as opposed to a few Moon masses in the linear capture phase. The dust ring indeed encompasses the largest surface area of all the phases in which dust concentration occurs, being a global rather than a local feature in the disk. }

\item{ The evolution of dusty structures is not very sensitive on initial dust-to-gas ratio and Stokes number, but it depends sensitively on vortex size. }

\item{ Our analytical model predicts correctly the saturation of the dust-to-gas ratio during the linear capture phase in simulations with sufficiently high resolution, serving as a benchmark to uncover numerical effect due to lack of resolution. The model is robust for the whole range of parameters that we changed, which suggests it can be used to predict behaviour of a variety of initial conditions without running actual simulations, up to the point when the approximation brakes down as the vorticity cancels at the center and the vortex enters the unstable regime. }

\end{itemize}

	Future work will have to explore further the two-fluid streaming instability phase occurring both in the final stages of the vortex lifespan and, even more, inside the dust ring. We note that standard streaming instability consider a vertical column of gas across the disk, while here the two-fluid instabilities are operating in the $(r, \: \theta)$ plane, hence the role of shear will be different, for example. This entails a focused study of the streaming instability in this configuration, which we intend to carry out in future work. 

	Furthermore, in order to address planetesimal formation properly we will have to include self-gravity and eventually move to three-dimensional setups. In the latter, dust sedimentation towards the disk midplane will be an additional important process that may boost dust densities significantly, making dust clumps finally Jeans unstable. Starting from a disk only a few times massive, hence still within the upper limit of the minimum mass solar nebula model, will also enhance dust densities and promote local collapse. It is now more accepted that the MMSN is a too restrictive approximation, and that disks may easily be more massive, up to ten times the MMSN \citep{Schlichting2014}. Such a massive disk, which is likely based on the observed distribution of masses in T Tauri disks, could lock enough mass inside the dust ring to allow assemble in situ the core of a gas giant planet as the mass of the ring scaled-up by the disk mass would amount to a few Earth masses. In such a disk, gas accretion would also be very efficient and rapidly lead to runaway gas accretion, ending to the formation of a gas giant on timescales not exceeding 1 Myr \citep{Helled2014}.

	Yet we stress that our simulations achieve the important goal of obtaining naturally the very high dust-to-gas ratios required for the streaming instability even when starting from ratios as low as $10^{-4}$. Hence we can consider our results as providing naturally the conditions that elsewhere in the literature \citep{Youdin2005, Youdin2007, Johansen2007, Kowalik2013} have been shown to lead to rapid planetesimal formation. In this context, vortices, from main sites of dust accumulation as argued in past work, become predominantly a trigger. Finally, the fact that these high dust-to-gas ratios are obtains almost regardless the initial dust proportion shows that concentration of dust (and therefore also the formation of planetesimals and terrestrial planets) is not very sensitive to the disk (and therefore stellar) metallicity. This is consistent with observations of small planets for which the occurrence rate is almost flat with stellar metallicity \citep{Dressing2013, Wang2015}.

	There are still several shortcoming of our simulations, such as the fact that we considered only one vortex at a time and only one dust grain size at a time. Furthermore, gas is treated as inviscid, which is clearly an oversimplification given that transport processes such as MRI required to generate an accretion flow towards the star will also generate a turbulent viscosity. However, for the latter we have shown in the previous section that the effect should be small enough to not alter strongly our findings. Multiple vortices and multiple dust grain sizes will be introduced soon in our next set of simulations. Considering multiple grain sizes simultaneously will help to compare with observations of protoplanetary disks at various stages of evolution, which potentially offer constraints on the efficiency of the various dust capture and accumulation phases described in this work. For example, most protoplanetary disks remain opaque, out to several $AU$, at wavelengths ranging from microns to submm for several $Myr$. It is thus important that adequate small grains must be retained to maintain such a state. Especially for the dust ring phase, which lasts longer, these concerns will have to be taken into account. Alternatively, planetesimals may form efficiently in the ring but also collide and shatter frequently after formation, returning a fraction of the dust to the diffuse phase. The turbulent velocities in the flow are of only slightly smaller than the thermal sound speed within the ring ($\sim 20\% \: c_s$), which suggests collisions soon after formation might be important.

	If planetesimal formation occurs efficiently within the dust ring, this could be a natural way to explain the in-situ formation of closely packed systems of low mass planets, one of the unexpected findings of the Kepler mission \citep{Lissauer2011}, which is hard to explain with conventional models of rocky planet formation starting from an homogeneous distribution of embryos in the disk.

\begin{acknowledgements}

	This work has been carried out within the frame of the National center for Competence in Research {\it{PlanetS}} supported by the Swiss National Science Foundation (SNSF). The authors acknowledge the financial support of the SNSF. Numerical simulations were performed on the {\it{Piz Daint}} Cray XC30 system of the Swiss National Supercomputing Center (CSCS). Data analysis and plots were produced using Matlab. All color maps are vector images that can be zoomed in to reveal details down to the grid cell size of the simulations. Clément Surville would like to thank R. Helled, C. Dullemond, M.M. Mac Law, and H. Klahr for their comments and feedback on this project.

\end{acknowledgements}

\bibliographystyle{aa}
\bibliography{Biblio}

\end{document}